\documentclass[sort&compress,english,preview,5p]{elsarticle}
\usepackage{graphicx}
\usepackage{dcolumn}
\usepackage{bm}
\usepackage{amsmath}
\usepackage[version=4]{mhchem}
\usepackage{comment}
\usepackage{epstopdf} 
\usepackage{xr-hyper}
\usepackage{lineno,hyperref}
\usepackage{color} 
\usepackage[colorinlistoftodos]{todonotes}
\usepackage{soul} 
\usepackage{subcaption}

\makeatletter

\usepackage{amssymb}

\usepackage{lineno,hyperref}
\modulolinenumbers[5]

\journal{Ultramicroscopy}

\begin{document}

\begin{frontmatter}



\title{Pushing the Dose Limit of Atomic-Resolution Imaging: A 4D-STEM case study of NaCl}


\author[emat,nanolab]{Tamazouzt Chennit\corref{corresp}}
\author[emat,nanolab]{Arno Annys}
\author[emat, nanolab]{Songge Li}
\author[emat,nanolab]{Nicolas Gauquelin}
\author[emat,nanolab]{Hoelen L. Lalandec Robert}
\author[emat,nanolab]{Jo Verbeeck}

\cortext[corresp]{Corresponding author: tamazouzt.chennit@uantwerpen.be}
\address[emat]{EMAT, University of Antwerp, Groenenborgerlaan 171, 2020 Antwerp, Belgium}
\address[nanolab]{NANOlight Center of Excellence, University of Antwerp, Groenenborgerlaan 171, 2020 Antwerp, Belgium}

\begin{abstract}
\label{sec: abstract}

    Beam-induced damage fundamentally limits the characterization of beam-sensitive materials by scanning transmission electron microscopy (STEM), since structural information must be recorded before the electron beam irreversibly modifies the specimen. Here, sodium chloride (NaCl) is employed as a model beam-sensitive ionic crystal to investigate beam-induced structural evolution and, more importantly, the low-dose regime in which useful structural information can be recovered prior to significant damage. Four-dimensional STEM (4D-STEM) datasets were acquired at 200~kV using a Timepix3 direct electron detector and reconstructed using real-time integrated centre of mass (riCoM) imaging. We first establish the characteristic damage behavior of NaCl by intentionally using high electron doses and by following its evolution under continued irradiation. Across different dwell times and raster scan orientations, damage develops reproducibly into square-faceted voids whose boundaries remain aligned with the $\langle100\rangle$ crystallographic directions of the rock-salt lattice, demonstrating that the resulting morphology is governed predominantly by the intrinsic crystallography rather than by the scan geometry. Having established this characteristic damage response, we subsequently investigate the low-dose imaging limit of NaCl. A multiframe dose-fractionated acquisition series is used to follow the beam-induced structural degradation using a dose of 130~e$^{-}$\AA$^{-2}$. A normalized cross-correlation with respect to the first frame provides a quantitative measure of the progressive damage evolution. This furthermore enables evaluating the onset of measurable damage as a function of accumulated dose. Once this onset is found, we acquire another frame series with an electron dose of 32~e$^{-}$\AA$^{-2}$. These results demonstrate how dose-efficient 4D-STEM acquisition, combined with the riCoM technique, can extend the accessible imaging regime of highly beam-sensitive ionic materials while maintaining atomic resolution.
    
\end{abstract}


\end{frontmatter}

\section{Introduction}
\label{sec: intro}

    Beam-induced damage represents a fundamental limitation for high-resolution electron microscopy. The interaction of energetic electrons with a beam-sensitive specimen can induce irreversible structural and chemical modifications through mechanisms including knock-on displacement, radiolysis, charging, and local heating \cite{egerton_mechanisms_2012, egerton_radiation_2019}. As a result, the spatial resolution attainable in an electron microscope is determined not only by the electron-optical performance of the instrument but also by the electron exposure that the specimen can withstand before its structure is significantly altered \cite{egerton_limits_2007, egerton_damage-limited_2026, lalandec_robert_benchmarking_2025}. For highly beam-sensitive materials, the central experimental challenge is therefore not simply to maximize the theoretical resolution through correcting aberrations, but also to determine how much information can be recovered before the imaging process itself modifies the structure under investigation.

    Sodium chloride (NaCl) provides a particularly well-defined model system for investigating this dose–damage relationship. Its simple and well-characterized crystal structure facilitates the identification of irradiation-induced structural changes, while its pronounced sensitivity makes it an experimentally relevant test case for low-dose imaging. Prior work has established that electron irradiation of alkali halides such as NaCl predominantly induces radiolytic defects, including F- and H-centres, whose subsequent evolution can result in the formation of metallic sodium colloids, vacancy clusters, and ultimately extensive structural degradation \cite{soppe_jain-lidiard_1992, hobbs_radiation_1994}. Despite this body of work, direct high-resolution imaging of unencapsulated NaCl remains challenging, as the crystal can evolve rapidly under the electron beam. Here, an acceleration voltage of 200~kV was selected as a compromise between reducing the probability of inelastic interactions associated with radiolysis and maintaining sufficiently stable imaging conditions. 

    Several strategies have been explored to overcome the pronounced beam sensitivity of NaCl. Palacio \textit{et al.} \cite{palacio_ultra-thin_nodate} demonstrated that NaCl can serve as a sacrificial protective layer to investigate a 2D material specimen, such as graphene. Inversely, Lehnert \textit{et al.} \cite{lehnert_quasi-two-dimensional_2021} showed that an encapsulation process, using graphene as well, enables atomic-scale imaging of NaCl by high-resolution transmission electron microscopy (HRTEM). In the latter case, the encapsulating layers substantially increase the stability of the NaCl, extending the time available for imaging and enabling the use of relatively high cumulative electron doses ($10^{3}$--$10^{7}$~e$^{-}$/\AA$^{2}$), despite the additional scattering introduced by the encapsulation. Rather than increasing the tolerance of the specimen to electron irradiation, an alternative strategy is to reduce the exposure required to recover the desired structural information, through ensuring an optimal usage of the total detected counts. This approach avoids the need for additional stabilizing layers and instead exploits the limited dose window over which the pristine structure remains sufficiently intact for imaging.

    Conventional high-angle annular dark-field scanning transmission electron microscopy (HAADF-STEM) \cite{pennycook_high-resolution_1991, carter_imaging_2016} provides a robust and intuitive incoherent contrast mechanism and is particularly effective for visualizing mass loss, defect formation and other manifestations of radiation damage. However, HAADF-STEM forms the image from electrons scattered into a restricted, and poorly populated, angular range, thus leaving most of the scattering signal unused. This becomes increasingly limiting as the available electron budget is reduced. On the other hand, the development of fast direct electron detectors \cite{mcmullan_electron_2007, llopart_timepix_2007, ballabriga_medipix3_2011, poikela_timepix3_2014, mir_characterisation_2017, ryll_pnccd-based_2016, philipp_very-high_2022, zambon_kite_2023, ercius_4d_2024, llopart_timepix4_2022} has enabled four-dimensional scanning transmission electron microscopy (4D-STEM) \cite{yang_4d_2015}, in which a complete convergent beam electron diffraction (CBED) pattern is recorded at each probe position. Obtaining the full diffraction distribution preserves substantially more of the available information and enables multiple imaging modalities from the same acquisition. This includes differential phase contrast and centre of mass methods \cite{muller_atomic_2014,lazic_phase_2016,yucelen_phase_2018,muller-caspary_comparison_2019}, as well as phase retrieval approaches such as electron ptychography \cite{wang_ptychography_2025}. This ability to redistribute and computationally process the recorded signal makes 4D-STEM particularly attractive for dose-limited specimens.
    
    Among recently introduced approaches, real-time integrated centre of mass (riCoM) imaging measures the potential of the specimen based on the first moment of the CBED pattern and has demonstrated strong dose efficiency in 4D-STEM experiments \cite{yu_real-time_2022}. When combined with event-based direct electron detection, riCoM provides a route towards extracting structural information from a very small number of detected electrons \cite{poikela_timepix3_2014, frojdh_timepix3_2015, annys_removing_2025, robert_guided_2026}. This makes it possible to investigate whether atomic-scale information can be recovered from unencapsulated NaCl at doses several orders of magnitude lower than what has been reported for conventional high-resolution imaging.
    
\section{Methods}
\label{sec: methods}

\subsection{Sample preparation}
\label{subsec: sample prep}

    Two NaCl specimen geometries were investigated. Comparatively thick NaCl crystals, with an estimated thickness of approximately 100-150~nm, were prepared by dispersing 10~mg of ultrapure NaCl powder (Sigma-Aldrich) in 40~mL of isopropanol. A 3~$\mu$L aliquot of the resulting suspension was subsequently drop-casted onto home-made ultrathin carbon film TEM grids, with a carbon film thickness of approximately 3~nm. Owing to their comparatively high mass-thickness contrast and the clear visibility of beam-induced void formation, these specimens were primarily used to establish the characteristic real-space damage evolution and to investigate the influence of acquisition conditions and scan orientation.

    A thinner NaCl specimen, with an estimated thickness of approximately 30~nm, was prepared by focused ion beam (FIB) milling, using a Thermo Fisher Helios NanoLab 650 dual-beam FIB-SEM. Prior to milling, the region of interest was protected by carbon thread evaporation followed by sequential electron beam and ion beam-assisted Pt deposition. The final specimen thickness was estimated from SEM imaging during the thinning procedure. The thinner specimen was used for the low-dose 4D-STEM measurements. Reducing the specimen thickness limits multiple scattering and thickness-dependent contrast variations \cite{close_towards_2015, addiego_thickness_2020, robert_dynamical_2022-1, clark_effect_2023, gao_central_2024}, which facilitates the recovery and interpretation of high-resolution phase images.

\subsection{4D-STEM acquisition and event-based detection}
\label{subsec: 4dstem acquisition}

    Experiments were performed on an aberration-corrected Titan Themis 60-300 instrument (Thermo Fisher Scientific), operated at 200~kV with a probe convergence semi-angle of 18~mrad. At this convergence angle, the corresponding spatial frequency extends to approximately 7.2~nm$^{-1}$ at $\alpha$, equivalent to a theoretical resolution of $\sim$1.4~\AA. Assuming ptychographic information transfer up to $2\alpha$, the theoretical resolution limit would extend to $\sim$70~pm. NaCl crystals were aligned along the [001] zone axis by tilting the sample stage. To minimize pre-irradiation of the regions of interest, all focusing, alignment, and stage tilting were performed on a separate region of the specimen.
    
    4D-STEM datasets were acquired using a Timepix3 chip, based on an Advacam Advapix direct electron detector prototype \cite{poikela_timepix3_2014, frojdh_timepix3_2015, jannis_event_2022, annys_removing_2025, lalandec_robert_guided_2026}, recording an equivalent 256~$\times$~256 pixel CBED pattern at each scan position. The Timepix3 chip was operated in an event-based mode, recording individual electron events together with their detector coordinates and time of arrival. This provides single-electron sensitivity while retaining the temporal information associated with each count, resulting in a sparse representation of the diffraction signal.

\subsection{riCoM reconstruction and image registration}
\label{subsec: reconstruction and registration}

    Micrographs were reconstructed from the 4D-STEM datasets using riCoM \cite{yu_real-time_2022}. Rather than converting the event-driven Timepix3 acquisitions into dense four-\\dimensional arrays, the data were processed directly from the recorded event stream using the \texttt{EvenTem} software \cite{annys_removing_2025, lalandec_robert_guided_2026}. This approach enables live imaging feedback while substantially reducing the data volume associated with storing the equivalent dense datasets.
    
    The successively obtained images were rigidly registered prior to quantitative comparison to compensate for global specimen drift during the acquisition series, as was used to characterize the response of NaCl to electron irradiation, presented in Figures~\ref{fig:fig02}--\ref{fig:fig04}. For datasets exhibiting progressive beam-induced damage, the alignment was instead determined from a reference region outside the evolving irradiated area. This prevented the structural changes associated with damage from influencing the registration and introducing an artificial contribution to the measured image evolution. For the remaining datasets recovered in this work, and acquired at lower electron doses, registration was not applied as the associated specimen drift was sufficiently small.

\section{Direct visualization of beam-induced damage in NaCl}
\label{sec: direct visualization}

\begin{figure}[h!]
        \centering
        \includegraphics[width=0.5\textwidth]{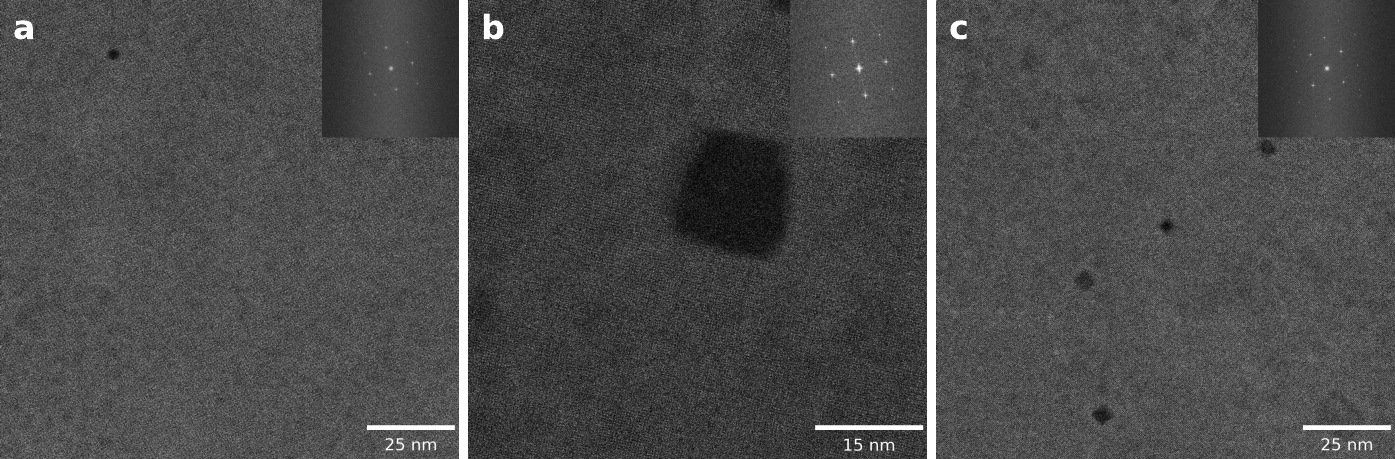}
        \caption{HAADF-STEM visualization of beam-induced damage in NaCl following 4D-STEM irradiation. (a) Low-magnification overview of an initially undisturbed crystalline region; the dark feature in the upper-left corresponds to a beam-parking position. (b) Enlarged view of a localized, square-shaped damaged region produced during a high-magnification 4D-STEM acquisition. (c) Low-magnification overview after multiple 4D-STEM acquisitions, showing several square-shaped damaged regions at positions corresponding to previously irradiated scan areas.}
        \label{fig:fig01}
    \end{figure}

    Before examining the ultra-low-dose regime, we first characterized the response of NaCl to electron irradiation under conventional high-resolution STEM conditions. The irradiated areas were subsequently examined in real space using HAADF-STEM. Figure~\ref{fig:fig01} shows HAADF-STEM images acquired following the 4D-STEM dose-series experiments. These images were recorded at lower magnification and with a higher probe current of approximately 5~pA than used for the preceding 4D-STEM acquisitions, providing sufficient signal for post-acquisition inspection while minimizing additional electron exposure.

    Prior to irradiation, the selected region appears homogeneous, with no discernible evidence of beam-induced modification (Figure~\ref{fig:fig01}a). Following a single high-magnification 4D-STEM acquisition, a localized square-shaped feature is already visible within the irradiated field of view (Figure~\ref{fig:fig01}b), indicating mass-thickness loss. After multiple successive acquisitions, several such damaged regions are visible in the low-magnification overview, each being spatially correlated with a previously scanned area (Figure~\ref{fig:fig01}c).
    
    Interestingly, the damage also exhibits a reproducible morphology. The modified regions consistently develop some approximately square boundaries, with their edges closely following the $\langle100\rangle$ crystallographic directions of the rock-salt lattice. This observation suggests that the final morphology is governed primarily by the underlying crystal structure rather than by the geometry of the raster scan itself. Next, to investigate how this morphology develops with increasing exposure, we examine the individual frames recorded during the 4D-STEM acquisition series.
    
\begin{figure*}[h!]
        \centering
        \includegraphics[width=\textwidth]{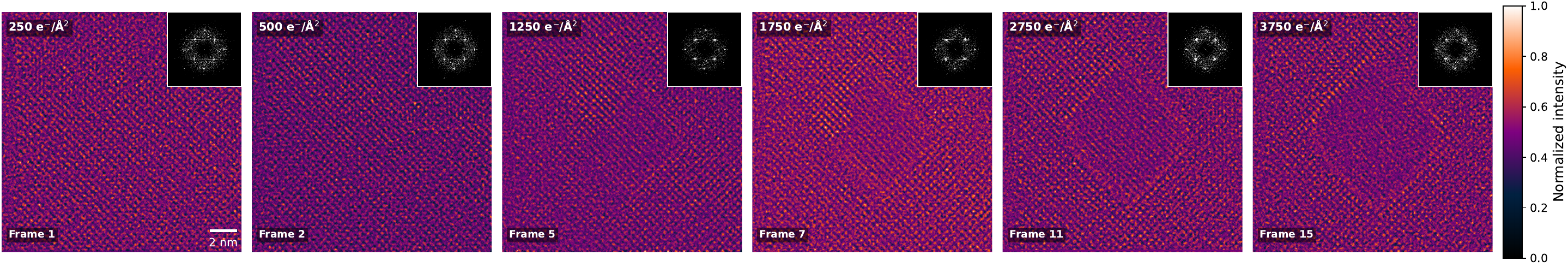}
        \caption{Sequential riCOM reconstructions from a 4D-STEM acquisition series showing the evolution of beam-induced damage in NaCl with increasing accumulated electron dose. Localized structural modifications progressively develop into a square-faceted damaged region. The data were acquired at 200~kV with a convergence semi-angle of 18~mrad, a pixel dwell time of 1~$\mu$s, a probe current of 1.2~pA, and a scan step of 17.7~pm over an 18~nm field of view. Each frame corresponds to a dose of 250~e$^{-}$\AA$^{-2}$, with the total accumulated dose indicated on each reconstruction. The square root of the Fourier-transform amplitude is shown as an inset.
}
        \label{fig:fig02}
    \end{figure*}

    Figure~\ref{fig:fig02} shows a selection of sequentially reconstructed riCOM images from the 4D-STEM acquisition series. While the complete series contains additional intermediate frames, only a subset is shown here to highlight the progressive development of beam-induced damage in a simple manner. Each reconstruction corresponds to a defined stage of the irradiation history and therefore to a specific accumulated electron dose, enabling the evolution of the damage to be followed as a function of dose.
    
    The sequence reveals a progressive transition from an initially intact structure towards extensive local material loss resulting from continuous electron beam irradiation. At low accumulated dose, the lattice remains largely preserved, with only minor local changes. As the experiment progresses, these modifications become increasingly pronounced and expand spatially. At later stages, the affected area develops into a well-defined, square-faceted, region whose boundaries preferentially follow the $\langle100\rangle$ directions. This is consistent with the morphology observed independently in the post-acquisition HAADF-STEM images (Figure~\ref{fig:fig01}), directly connecting the damage evolution observed during 4D-STEM acquisition with the final real-space morphology. As was hinted by the HAADF-STEM results, the development of crystallographically defined facets indicates that the damage does not simply propagate according to the spatial distribution of the incident probe and scan pattern. Instead, the underlying symmetry of the NaCl lattice appears to play a dominant role in determining how the irradiated region evolves once structural modification is initiated \cite{izumi_effects_1969,seinen_radiation_1994}.
    
\begin{figure*}[h!]
        \centering
        \includegraphics[width=\textwidth]{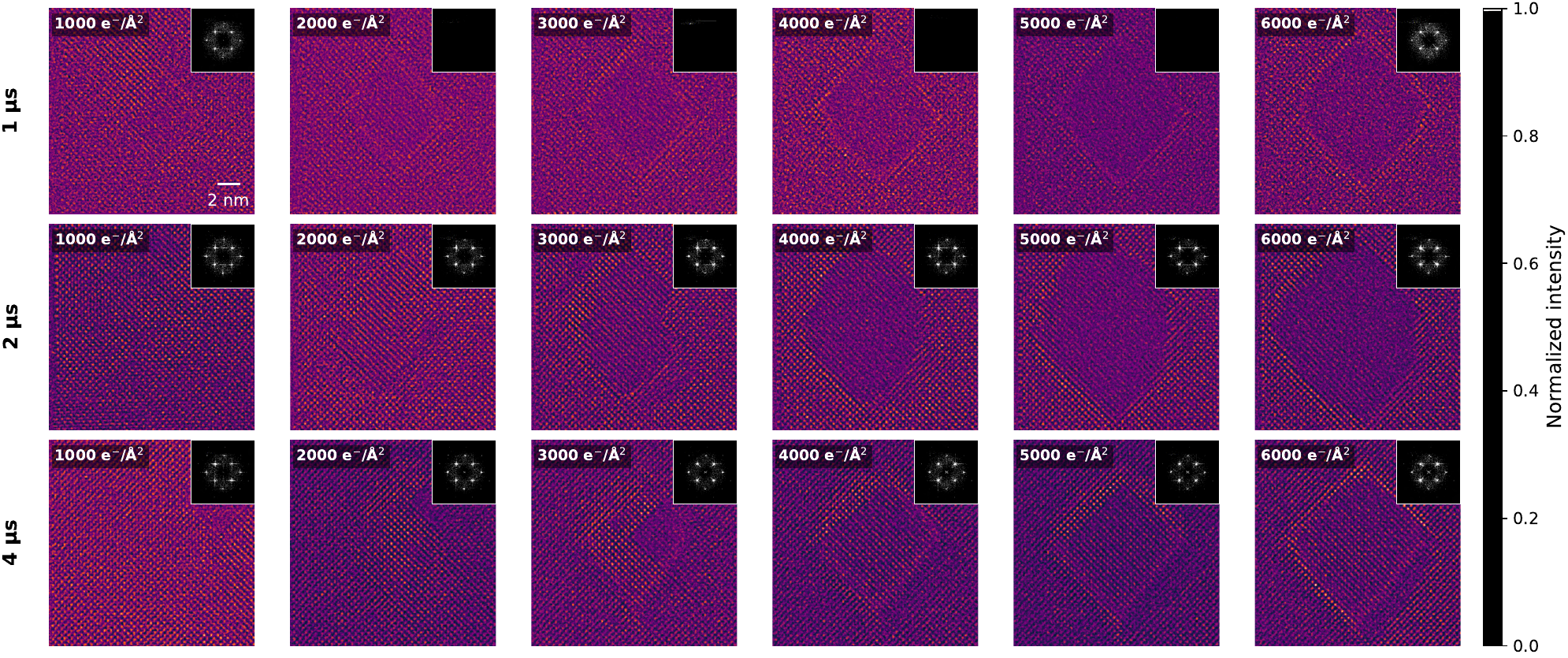}
        \caption{Dose-matched comparison of riCoM reconstructions acquired with pixel dwell times of 1, 2, and 4~$\mu$s. Representative frames were selected such that each row corresponds approximately to the same accumulated electron dose, allowing qualitative comparison of structural evolution across different acquisition timescales. Each frame in the 1~$\mu$s series corresponds to a dose of 250~e$^{-}$\AA$^{-2}$, while the 2 and 4~$\mu$s series correspond to 500 and 1000~e$^{-}$\AA$^{-2}$ per frame, respectively. The square root of the Fourier-transform amplitude is shown as an inset.
        }
        \label{fig:fig03}
    \end{figure*}

\begin{figure*}[h!]
        \centering
        \includegraphics[width=\textwidth]{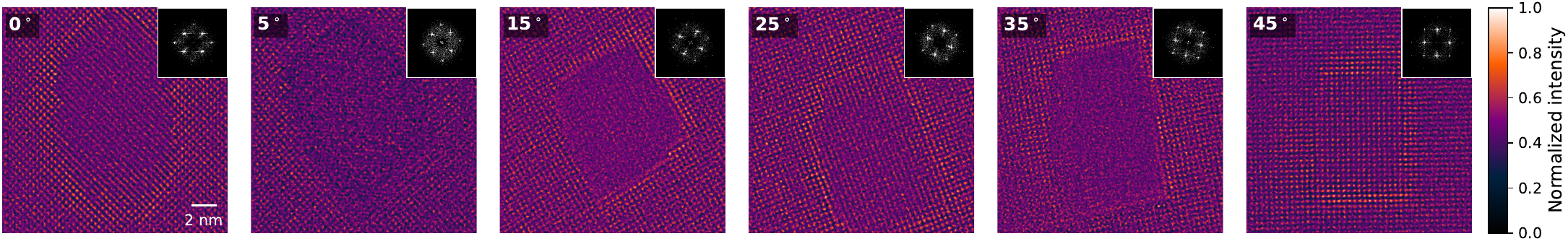}
        \caption{Comparison of NaCl damage evolution under rotated 4D-STEM scan geometries. The final frame from each acquisition series is shown. Each frame corresponds to a total electron dose of 725~e$^{-}$\AA$^{-2}$, resulting in an accumulated dose of 5800~e$^{-}$\AA$^{-2}$ after eight frames, over a field of view of 18~nm. The square root of the Fourier-transform amplitude is shown as an inset.}
        \label{fig:fig04}
    \end{figure*}

    We next examined whether the observed evolution is governed solely by the accumulated electron dose or whether the timescale over which the dose is delivered also influences the response. Additional 4D-STEM datasets were thus acquired from separate regions using different pixel dwell times while maintaining approximately the same accumulated dose. Apart from the dwell time, all other acquisition conditions were kept consistent with those described for Figure~\ref{fig:fig02}. The corresponding riCOM reconstructions are shown in Figure~\ref{fig:fig03}, with each column representing a different dwell time and each row corresponding approximately to the same accumulated electron dose. This arrangement allows the structural evolution to be compared at similar doses but over different irradiation timescales.
    
    Despite this variation, the progression of beam-induced damage remains remarkably consistent across the three series. At comparable accumulated doses, the same characteristic stages of structural degradation are observed, with no systematic evidence that changing the dwell time accelerates, delays, or otherwise modifies the overall evolution of the damaged region. Nevertheless, individual frames reveal that this process is not strictly monotonic at the local scale. For example, in the 4~$\mu$s series, a defect develops along a crystal edge at an accumulated dose of approximately 2000~e$^{-}$/\AA$^{2}$, while in the subsequent frame the damaged region partially recovers on the side where the defect initially formed and simultaneously extends along the adjacent lower edge. Such local rearrangements are consistent with the dynamic evolution of irradiation-induced defects previously reported in electron-irradiated NaCl, where the generated point defects can migrate, recombine, and aggregate before progressing towards more extensive structural degradation \cite{izumi_effects_1969,seinen_radiation_1994}.
    
    With further irradiation, the defect continues to expand and the associated void becomes progressively larger, reaching a more pronounced state at approximately \\
    6000~e$^{-}$/\AA$^{2}$. This progression from local rearrangement to sustained defect growth is therefore consistent with the defect-mediated damage evolution previously described for alkali halide crystals \cite{izumi_effects_1969,seinen_radiation_1994}. Thus, while the overall progression of structural degradation is primarily correlated with accumulated electron dose under the conditions investigated here, the frame-to-frame evolution indicates that its finer development entails complex dynamics, which are not strictly cumulative.

    The characteristic damage morphology was further investigated by varying the raster scan orientation. The 4D-STEM acquisition was repeated for different raster tilt angles while maintaining identical probe current, field of view, and, consequently, pixel step size to those used for Figure~\ref{fig:fig02}. For each orientation, eight sequential frames were acquired with a pixel dwell time of 3~$\mu$s. The final frame from each series, corresponding to the same total accumulated electron dose of 5800~e$^{-}$\AA$^{-2}$, was selected for direct comparison of the resulting modified specimen structure.
   
    Despite the change in raster scan orientation, directly visible in the FFT insets, the damage morphology remains essentially unchanged. In all cases, the altered regions develop preferentially along $\langle100\rangle$ directions rather than following the direction of the raster scan. This insensitivity confirms the dominant role of the crystalline lattice over the geometry of dose delivery.
    
    One minor difference can nevertheless still be noted, in the edge of the damaged areas. While some regions develop well-defined square-shaped facets, others show stepped boundaries, though both remain aligned with the lattice. This suggests that, once the irradiation-induced structural modification is initiated, defect propagation remains wholly constrained by the crystal, with the extended facets and discrete steps reflecting preferential evolution along specific axes. Arguably, the specific process may further reflect local differences in irradiation history, priorly existing defects or strain, which are likely to influence the local progression of damage while preserving the overall crystallographic preference.

    Taken together, these observations establish the characteristic high-exposure response of unencapsulated NaCl under the present experimental conditions. Most importantly, an accumulated dose can progressively transform a collection of initially independent and localized, structural modifications into a well-defined, crystallographically faceted region of material loss. Moreover, the persistence of this morphology across different dose rates and raster orientations demonstrates that the damage cannot be explained simply as a geometrical consequence of scanning trajectory or dose rate. This high-exposure response provides an important reference for the subsequent investigation of the ultra-low dose regime, whose objective is to recover structural information before a pronounced specimen modification occurs.

\section{Dose-dependent recovery of atomic-scale structure}
\label{sec: results: dose-dependent-recovery}

\begin{figure*}[h!]
        \centering
        \includegraphics[width=\textwidth]{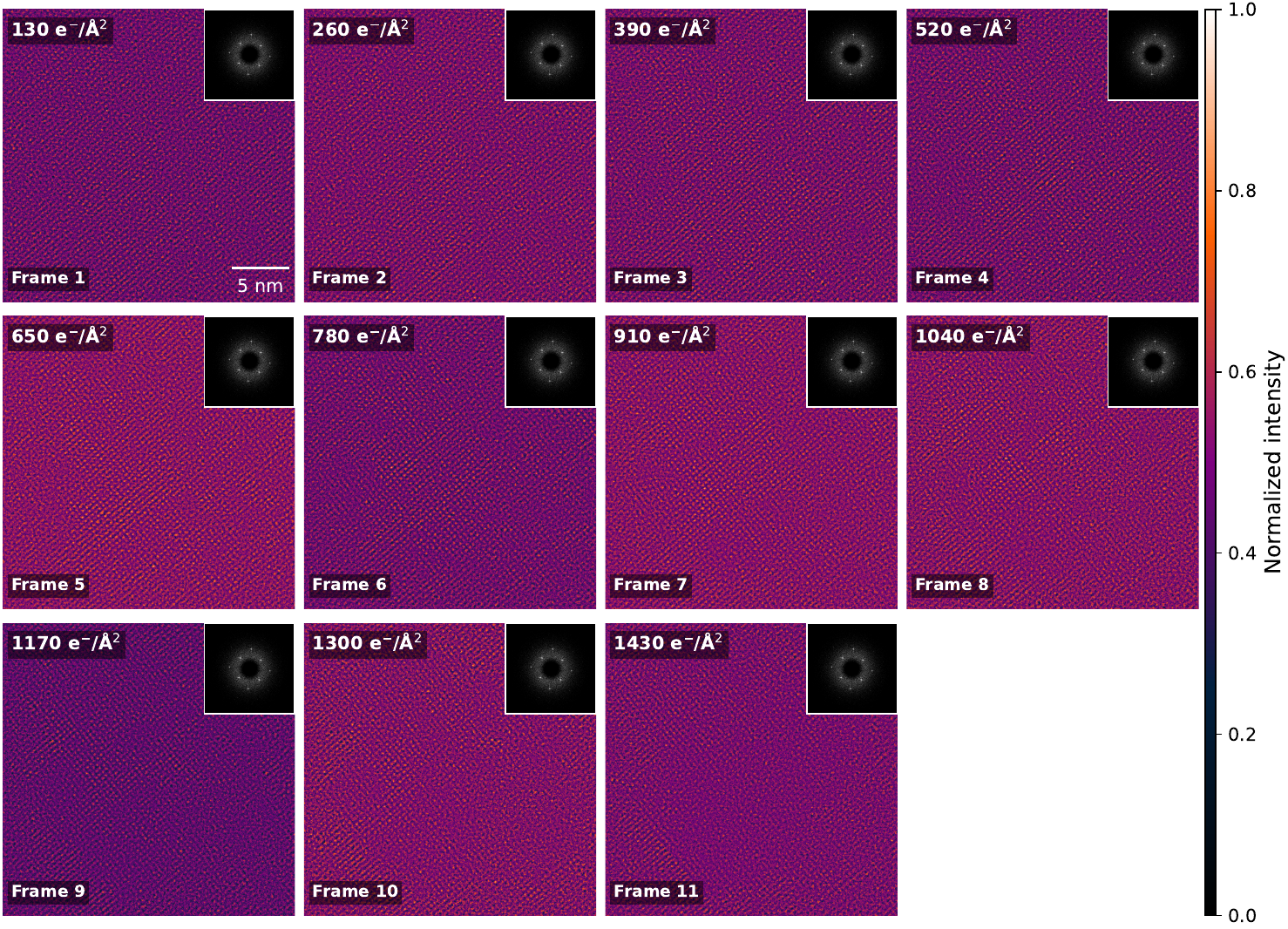}
        \caption{Representative riCOM reconstructions from a repeated acquisition series, with approximately 130~e$^{-}$\AA$^{-2}$ delivered per frame at a probe current of 1.2~pA, a pixel dwell time of $4~\mu$s and a scan step size of 24.94~pm over a 25.55~nm field of view. The accumulated electron dose and scale bar are indicated for each reconstruction, while the square root of the Fourier-transform amplitude is shown as an inset.}
        \label{fig:highdose_series}
    \end{figure*}

    Having established the evolution of beam-induced damage at higher exposure, we next investigated the extent to which the electron dose could be reduced while retaining useful structural information in the measurements. A first series of 4D-STEM acquisitions was performed on the 30~nm-thick sample mentioned in subsection \ref{subsec: sample prep}, while delivering approximately 130~e$^{-}$\AA$^{-2}$ per frame. Its results are shown in figure~\ref{fig:highdose_series}. After focusing and zone-axis alignment on a separate region, eleven consecutive frames were recorded from an initially undamaged area. 
    
    This series provides a complementary view of the transition from an initially intact crystal towards pronounced structural degradation. The first acquisitions retain clear atomic-scale information, whereas substantial modifications become discernible from frame~4 onwards, corresponding to an accumulated dose of approximately 520~e$^{-}$\AA$^{-2}$. At higher accumulated exposure, the damaged regions progressively expand, eventually resulting in substantial material loss. This observation provides an experimental reference for the dose range over which beam-induced modifications become readily detectable.
    
\begin{figure}[h!]
        \centering
        \includegraphics[width=0.5\textwidth]{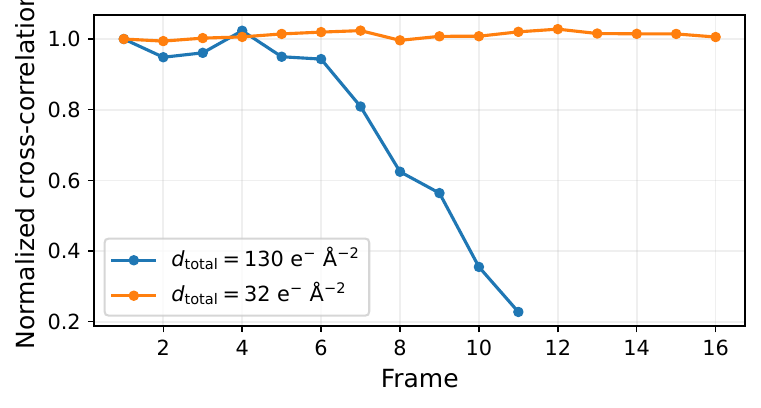}
        \caption{Normalized cross-correlation of the reconstructed images with the first frame of each acquisition series as a function of frame index. The series were acquired at a pixel dwell time of $4~\mu$s, with total doses of $32$ and $130~e^{-}$\AA$^{-2}$ per frame.}
        \label{fig:cross-correlation}
    \end{figure}

    For this series, we used the normalized cross-correlation to track changes in the riCOM images relative to the first frame of each acquisition shown in Figure~\ref{fig:cross-correlation}. A decrease in the correlation coefficient indicates that the image is becoming less similar to the initial state. The correlation is therefore used as an indicator of image evolution rather than a direct measure of crystallinity. The marked decrease from frame~4 onwards agrees with the structural degradation observed in the corresponding riCOM reconstructions.

    Notably, in this criterion, additional small-scale, frame-to-frame, variations may arise from electron shot noise, residual scan distortions, or local structural fluctuations. Considering the wider field of view therefore allows these fine local disturbances to be distinguished from systematic changes indicative of the large-scale evolution in damage.\\

    \begin{figure}[h!]
        \centering
        \includegraphics[width=0.5\textwidth]{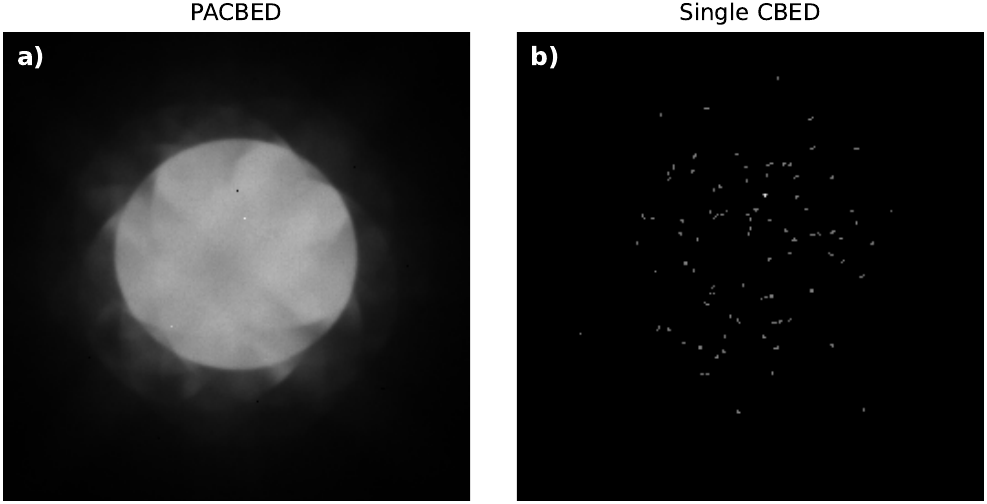}
        \caption{Reformation of 2D diffraction frames from the experimental event-based data acquired from the NaCl crystal specimen. (a) Single CBED pattern and (b) PACBED pattern with a total dose of 130~e$^{-}$\AA$^{-2}$}
        \label{fig:highdose-cbed}
    \end{figure}
    
    \begin{figure*}[h!]
        \centering
        \includegraphics[width=\textwidth]{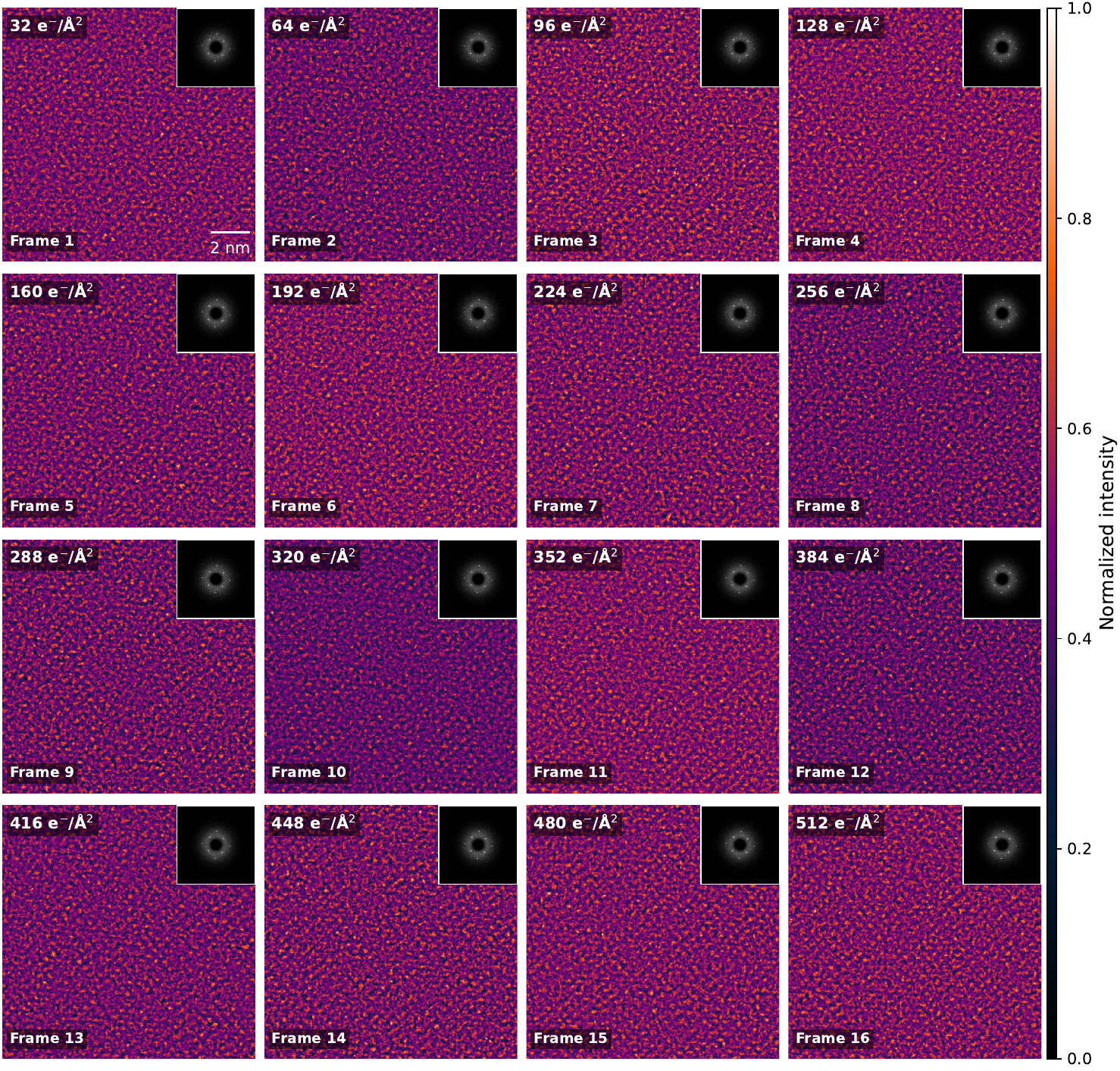}
        \caption{Representative riCOM reconstructions from a repeated acquisition series, with approximately 32~e$^{-}$\AA$^{-2}$ delivered per frame at a probe current of 1.2~pA and a scan step size of 49.89~pm over a 50.59~nm field of view. The accumulated electron dose and scale bar are indicated for each reconstruction, while the square root of the Fourier-transform amplitude is shown as an inset.}
        \label{fig:lowdose_series}
    \end{figure*}
    
    Following the experiment described above, a new series of sixteen consecutive 4D-STEM frames was acquired on another region of the 30~nm-thick specimen, using a larger field of view and reducing the dose to approximately 32~e$^{-}$\AA$^{-2}$ per frame. This acquisition was designed to assess the preservation of structural information under substantially reduced electron exposure. To provide an objective criterion for such atomic-scale information recovery, we define the resolution in this work based on the highest spatial frequency for which crystalline reflections remain experimentally detectable in the Fourier transform of the reconstruction. In the present case, the third-order Bragg reflections are clearly recovered, corresponding to a spatial period of approximately 2.11~\AA. Under those considerations, the present result demonstrates the recovery of atomically resolved structural information from unencapsulated NaCl at a dose of approximately 32~e$^{-}$\AA$^{-2}$ per frame. To our knowledge, this represents the lowest electron dose at which such an atomic-scale measurement has been achieved, with an unencapsulated NaCl specimen.\\
    
    \begin{figure}[h!]
        \centering
        \includegraphics[width=0.5\textwidth]{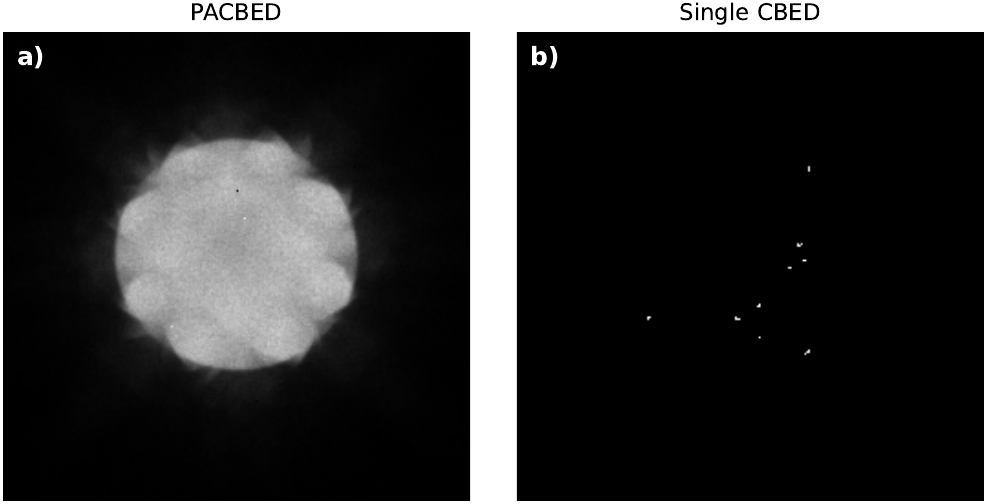}
        \caption{Reformation of 2D diffraction frames from the experimental event-based data acquired from the NaCl crystal specimen. (a) Single CBED pattern and (b) PACBED pattern with a total dose of 32~e$^{-}$\AA$^{-2}$}
        \label{fig:lowdose-cbed}
    \end{figure}
    
    Representative riCoM reconstructions from this series are shown in Figure~\ref{fig:lowdose_series}. Given the large field of view over which this series was acquired, only 1/4 of the whole recorded and reconstructed FoV is shown for a better visualization of the atomic features, the frame series of the whole FoV are available in the supplementary materials section \ref{supplementary}.
    
    In contrast to the pronounced degradation observed in the higher-dose series shown in Figure~\ref{fig:highdose_series}, this sequence does not exhibit a sharply defined transition from an intact structure to a clear damage-dominated state. Atomic-level contrast remains visible throughout much of the sequence, while the changes between successive reconstructions are comparatively gradual. The normalized cross-correlation with the first reconstruction, shown in Figure~\ref{fig:cross-correlation}, supports this observation. Although this coefficient fluctuates from frame to frame, it does not show an unambiguous abrupt decrease that could be associated with a distinct damage onset. This behavior also contrasts with the series acquired at approximately 130~e$^{-}$\AA$^{-2}$, for which the cross-correlation exhibited a pronounced decrease beginning at frame~4.

\section{Dose-efficient recovery of atomic-scale information}
\label{sec: dose efficient}
    
    The results presented above illustrate the central compromise in imaging beam-sensitive materials such as NaCl. While increasing the dose improves counting statistics, it also amplifies the likelihood of irreversible structural modification. The acquisition at approximately 32~e$^{-}$\AA$^{-2}$ per frame therefore provides a favorable regime in which atomic-scale structural information can be recovered while limiting specimen modification. It is then interesting to explore whether the information contained in this low-dose reconstruction could be further enhanced computationally, without additional electron exposure.
    
\begin{figure}[h!]
        \centering
        \includegraphics[width=0.5\textwidth]{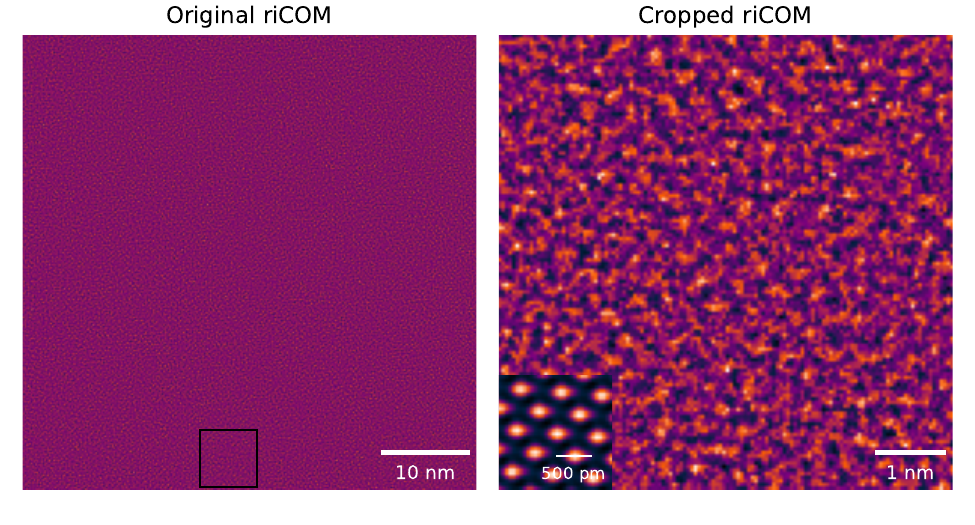}
        \caption{Template-based recovery of atomic-scale information from the approximately 32~e$^{-}$\AA$^{-2}$ riCoM reconstruction.}
        \label{fig:template-matching-32}
    \end{figure}

    To that end, we exploited the inherent spatial redundancy of the crystalline lattice. A representative atomic-scale motif was selected from the riCoM reconstruction and used to identify equivalent regions throughout the field of view. These regions were aligned and averaged, thus combining multiple spatially separated observations of the same structural motif. As illustrated in Figure~\ref{fig:template-matching-32}, this averaging reinforces the reproducible structural contrast while reducing uncorrelated noise, thereby improving the effective signal-to-noise ratio without a need for further electron exposure. The approach is conceptually related to single-particle analysis (SPA), where multiple noisy images of identical or structurally related particles are aligned and combined to enhance their effective signal-to-noise ratio, all the while recovering a three-dimensional model \cite{ cheng_primer_2015, pei_cryogenic_2023}.

    In the present case, however, the redundancy arises from the periodicity of the crystal rather than from repeated observations of individual and isolated, particles. Increasing the field of view consequently provides proportionally more motifs for averaging, all structurally equivalent prior to the damage onset. Hence, this approach readily offers a mean of improving the statistical precision of the recovered structure while maintaining the same dose per included motif. This demonstrates that dose efficiency can be enhanced not only through the acquisition conditions but also thanks to a computational exploitation of the structural redundancy already contained in the measurement.

\begin{figure*}[h!]
        \centering
        \includegraphics[width=\textwidth]{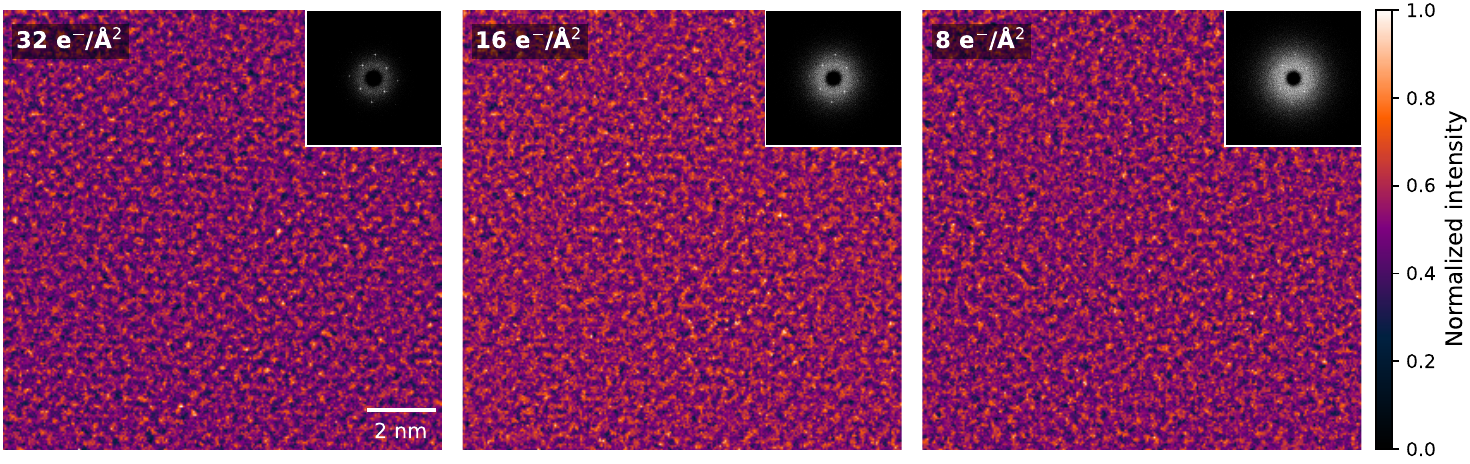}
        \caption{Cropped riCoM reconstructions obtained at progressively reduced electron doses of 32, 16, and 8~e$^{-}$\AA$^{-2}$ per frame.}
        \label{fig:dose-reduction}
    \end{figure*}

    Next, having demonstrated this principle of redundancy exploitation, we examined how much of the atomic-scale information remains accessible when the number of detected electron events is further reduced. To achieve this, a fraction of the recorded electron events in the frame series shown in figure~\ref{fig:lowdose_series}, was randomly discarded from the original dataset, effectively reducing the number of electrons contributing to the reconstruction and generating datasets corresponding to approximately 16 and 8~e$^{-}$\AA$^{-2}$ per frame. The resulting riCoM reconstructions and their corresponding Fourier transforms are compared in Figure~\ref{fig:dose-reduction}. As a reminder, at 32~e$^{-}$\AA$^{-2}$, the reconstruction exhibits clear atomic-scale periodic contrast, accompanied by well-defined Bragg reflections extending to the third order in the Fourier transform, corresponding to a resolution of 2.11~\AA. Remarkably, reducing the dose by a factor of two, to approximately 16~e$^{-}$\AA$^{-2}$, still preserves the characteristic reflections up to the second-order spots, corresponding to 2.82~\AA. At 8~e$^{-}$\AA$^{-2}$, the reduced electron statistics result in a visibly noisier riCoM reconstruction and a corresponding loss of high-frequency information in the Fourier transform. The second-order Bragg reflections are still faintly visible. Considering only the first-order ones, the effective resolution in the micrograph therefore decreases from 2.11~\AA{}, at 32~e$^{-}$\AA$^{-2}$, to approximately 5.64~\AA{}, at 8~e$^{-}$\AA$^{-2}$.
  
\section{Discussion}
\label{sec: discussion}

\subsection{Crystallographically constrained evolution of \\beam-induced damage}
\label{subsec: discuss:damage-pathways}

    The damage experiments shown in Figures~\ref{fig:fig02}--\ref{fig:fig04} provide insight into how structural degradation develops once irradiation-induced defects are introduced into unencapsulated NaCl. Although the electron beam is responsible for initiating the damage, the resulting morphology was not dictated by the direction of the raster scan. Instead, damaged regions consistently developed facets and stepped edges aligned with the $\langle100\rangle$ axes of the lattice, including when the scan orientation was varied. This demonstrates that the subsequent evolution of the damaged region is governed by the properties of the specimen rather than by the trajectory of the probe. This observed crystallographically determined process, in the migration and reconfiguration of beam-induced defects, is consistent with prior work on alkali halide crystals, where irradiation-generated point defects evolve into extended structures with preferred crystallographic orientations \cite{izumi_effects_1969,seinen_radiation_1994}.
    
    The dwell time series provides a further constraint on this interpretation. Within the range investigated here, varying it did not produce a systematic change in the overall progression of the damage at comparable accumulated doses. This suggests that, under the present conditions, the number of electrons delivered to the specimen provides a direct descriptor of the observed macroscopic evolution. However, this observation does not disprove the existence of finer dynamic effects in this process. Indeed, the frame-to-frame evolution of individual damaged regions shows that the structural response is not strictly monotonic. Local defects can, at least partially, deplete around one location while extending along a neighbouring crystallographic direction, indicating that irradiation does not simply accumulate damage at fixed positions. Instead, the defect population can redistribute as irradiation proceeds, consistent with the migration, aggregation, and recombination processes reported for alkali halides \cite{seinen_radiation_1994}. The present measurements cannot unambiguously distinguish these individual processes, but they demonstrate that the observed void formation is the result of an evolving defect population rather than a direct imprint of the electron beam, and scan pattern, geometry.\\
    
    This interpretation also permits to reconcile the apparently different damage morphologies observed across the specimen. Square-shaped facets and stepped edges represent different local manifestations of damage while retaining the same crystallographic preference. Their specificities may therefore arise from pre-existing defects, thickness, strain or irradiation history, rather than from a fundamentally different damage pathway. Taken together, these observations indicate that the beam-induced response of NaCl is reproducible with respect to the irradiation conditions investigated here. For clarity, however, this behavior should not be interpreted as evidence that the damage kinetics are always independent of all experimental parameters, as this was, of course, only shown in the specific conditions set by this work. Further experiments will be necessary to establish a more general understanding.

\subsection{Defining a useful dose window for structural imaging}
\label{subsec: discuss:dose-window}

    The dose series experiments highlight a distinction that is particularly important for beam-sensitive materials: the dose required to obtain a measurable signal is not necessarily the dose at which the resulting image remains representative of the intact structure. In the present case, increasing exposure initially provides additional counting statistics, but the same electrons that improve the measurement also drive structural evolution. Once irradiation-induced structural changes become significant, additional electrons may still increase the nominal signal while simultaneously reducing the fidelity of the information being measured. For beam-sensitive materials, therefore, the relevant imaging limit is not determined by the ability to detect additional electrons but by the point at which the information they provide becomes obscured by a non-negligible degree of specimen deterioration.
    
    This notion provides a more useful interpretation, to the results of the dose series, than simply trying to extract a single critical dose to NaCl. The onset of readily detectable structural modification observed in the higher-dose experiment, presented in figure~\ref{fig:highdose_series}, defines an experimental upper boundary for measurements aimed at characterizing the initial lattice under the conditions used here. It does not, however, represent a universal threshold. In particular, the higher- and lower-dose measurements were performed from different regions of the specimen. Their interest, instead, is then to establish a practical dose range. The local thickness, the zone axis, strain, pre-existing defects and irradiation history may all have an influence on the practical damage onset.

    Importantly, the preservation of high-order Bragg reflections at a dose of 32~e$^{-}$\AA$^{-2}$ per frame shows that atomic-scale periodic information can be recovered before a pronounced structural degradation becomes apparent. What makes this relevant is that spatial frequencies associated with the crystal lattice remain experimentally detectable, despite the limited number of electrons, and thus even when the corresponding image is, arguably, to noisy too permit a direct and simple interpretation of atomic contrast. Hence, defining this highest recoverable spatial frequency, associated to the structure of interest, provides a more meaningful criterion than the visual sharpness of the micrograph alone.
    
    The comparison with the case of graphene-encapsulated NaCl further illustrates that there are different, perhaps complementary, ways of approaching the same dose-limited imaging problem. Encapsulation can increase the resistance of NaCl to irradiation, by modifying the environment of displaced atoms and their subsequent evolution \cite{lehnert_quasi-two-dimensional_2021}. The approach adopted in this work, instead, reduces the amount of electron counts that is fundamentally required to recover the structural information from an unencapsulated specimen, by exploiting an imaging modality that shows better performance. This solution is particularly interesting, since reducing the required dose preserves not only the specimen itself but also the physical meaning of the measurement.

    More generally, the appropriate dose should be considered in relation to the structural question being asked. Recovering the periodicity of an intact lattice may remain possible after a dose at which subtle defects or local distortions are already introduced. The dose window, for useful and targeted imaging, is therefore bounded not simply by the conditions-specific onset of damage and void formation, but rather by the resilience of the structure of interest.

\subsection{Exploiting crystalline redundancy for dose-efficient information recovery}
\label{subsec: discuss:dose-efficiency}

    The low-dose measurements also reveal a route to improving the reliability of structural information without increasing the exposure of an individual region. At a reduced dose, the limiting factor is the statistical uncertainty associated with the finite number of detected electrons. For a crystalline specimen, however, the same structural motif occurs repeatedly throughout the field of view. These repetitions provide independent measurements of equivalent portions of the lattice, which can be combined computationally, provided that a corresponding degradation has not occurred yet.
    
    The template matching procedure presented in figure\ref{fig:template-matching-32}, aims to exploit this redundancy. Hence, rather than acquiring additional electrons from the same structural motif, equivalent ones that are already present in the accessible field of view are aligned and averaged. As noted prior, this is conceptually related to the principle underlying single-particle analysis, where multiple noisy observations are aligned to recover structural information that is poorly represented in an individual observation \cite{moecking_cryo-em_2025}. In the present case, however, the redundancy is generated by translational periodicity within a crystal rather than by a collection of nominally identical, or structurally related, particles.

    This distinction is important because it determines both the strength and the limitations of the approach. Crystalline periodicity provides a large number of equivalent observations, but only insofar as these observations genuinely represent the same underlying structure. Variations in thickness, orientation, local damage state, strain, or reconstruction quality can therefore limit the extent to which motifs can be combined to produce a significant result. The ability to exploit this spatial redundancy also relies on the statistical behaviour of the underlying reconstruction. In the present case, the riCoM reconstruction provides a direct and comparatively well-behaved representation of the measured signal, with sufficiently consistent contrast and noise characteristics across the field of view for equivalent motifs to be meaningfully aligned and averaged. This makes it possible for the structural signal to reinforce coherently across multiple regions while uncorrelated noise is progressively suppressed. In contrast, applying the same averaging strategy directly to independently reconstructed regions from an iterative ptychographic reconstruction would not necessarily provide the same benefit, as reconstruction-dependent variations, bias, and correlated artefacts could prevent a common low-noise structural motif from emerging through simple averaging\cite{maiden_further_2017, chennit_investigating_2025}. The suitability of riCoM for this type of spatially redundant information recovery is therefore an important aspect of the approach. On the other hand, provided sufficient homogeneity across the illuminated crystalline area, increasing the field of view allows a larger number of equivalent observations to be obtained under the same exposure conditions.
    
    The approach also alters the usual trade-off between counting statistics and specimen damage. While increasing the electron dose improves the statistics of an individual measurement, this remains inseparable from the possibility of structural modification in a beam-sensitive material. Spatial averaging provides a different route: it increases the statistical support for a structural feature by using information already distributed across the measured field of view. The resulting gain is therefore not a new set of information, but a more complete use of the information already encoded in the micrograph.

\subsection{Dose-dependent limits of spatial-frequency recovery}
\label{dose and frequency recovery}

    The dose-reduction experiment shown in figure~\ref{fig:dose-reduction} provides a direct assessment of how the available electron statistics affect the recovery of spatial information in riCoM. Because the lower-dose datasets were generated by randomly discarding events from the same original acquisition, the specimen state, scan conditions, and detector response remain unchanged. The observed differences can therefore be attributed primarily to the reduction in the number of detected electrons contributing to the reconstruction. This provides a controlled way of examining the statistical limit of spatial-information recovery without introducing additional variations in the experimental conditions.

    The results show that reducing the dose does not lead to an immediate loss of all structural information. Instead, the loss occurs progressively, with the highest spatial frequencies becoming increasingly difficult to distinguish as the number of detected events decreases. This behaviour is expected for a counting-statistics-limited measurement, for which the shot-noise contribution increases relative to the structural signal as the number of detected electrons is reduced. The dose dependence is therefore spatial-frequency dependent: weaker high-frequency components reach the noise level before the stronger, lower-frequency components.

    Importantly, the 16~e$^{-}$\AA$^{-2}$ dataset retains substantial atomic-scale structural information despite containing only half the electron events of the original dataset. This indicates that the reconstruction does not require the full electron budget of the 32~e$^{-}$\AA$^{-2}$ acquisition to recover meaningful periodic structural information. At the lower dose of 8~e$^{-}$\AA$^{-2}$, the further increase in noise is accompanied by a more pronounced reduction in the highest recoverable spatial frequencies. The experiment therefore demonstrates a gradual reduction in the accessible spatial-frequency bandwidth rather than a sharp transition between image formation and complete loss of structural information.

    The random event subsampling also highlights that the relevant dose limit depends on the spatial frequencies that are required for a particular imaging task. A dose sufficient to recover lower spatial frequencies may be insufficient to recover weaker higher-frequency components, even though the same specimen and imaging conditions are used. Consequently, the dose required for a given structural resolution cannot be considered independently of the signal strength at the corresponding spatial frequency. In the present case, the results establish the dose range over which atomic-scale periodic information remains detectable under the specific riCoM acquisition and reconstruction conditions used for unencapsulated NaCl.

    These observations are also relevant to the interpretation of the dose limit in the context of the preceding redundancy-based analysis. The template-matching experiment demonstrates that repeated structural motifs within the crystal can be combined to improve the representation of the underlying signal without subjecting the specimen to additional electron exposure, whereas the dose-reduction experiment demonstrates the extent to which that information remains reliably present as the available electron statistics are reduced. Together, these results show that both the electron dose and the spatial redundancy of the specimen influence the practical detectability of structural information. For beam-sensitive crystalline materials, exploiting such redundancy can therefore complement dose reduction by improving the reliability of the recovered structural signal without requiring additional exposure.

\section{Conclusion and outlook}
\label{sec: conclusion}

    In this work, we established a dose-efficient 4D-STEM approach for recovering atomic-scale structural information from unencapsulated NaCl, while explicitly accounting for the particular effects of electron exposure and specimen modification. By combining damage characterization with low-dose phase contrast imaging, we identified an experimentally accessible regime in which crystallographic information can be recovered at substantially reduced exposure, while the accompanying dose series provides a reference for assessing when irradiation-induced structural changes become significant. Importantly, employing an event-driven detector had two major advantages here. First, it offered the acquisition speed, and few-$\mu$s dwell times, necessary to conduct the series. Second, it enabled an efficient data representation for sparse CBED patterns, thus permitting the fast and flexible usage of recorded electrons.
    
    The systematic reduction of the dose, from approximately 32 to 16 and 8~e$^{-}$\AA$^{-2}$ per frame, furthermore showed how the recoverable structural information evolves as the available electron statistics decrease. A two-fold reduction to 16~e$^{-}$\AA$^{-2}$ still preserved substantial high-spatial frequency components, whereas a reduction to 8~e$^{-}$\AA$^{-2}$ resulted in their clear loss. Additionally, a template-based recovery of equivalent structural motifs was explored, which showed a promising capacity for their extraction. These results illustrate that dose-efficiency depends not only on the number of electrons invested in an individual measurement but also on how effectively they can be used as well as on the specific process of deterioration for the specimen. Combining event-driven acquisition, phase-sensitive riCoM reconstruction, and the exploitation of crystal periodicity therefore provides a general platform to maximize extractable structural information, all within a limited electron budget. This framework highlights a pathway towards extending quantitative 4D-STEM-based techniques to increasingly dose-sensitive materials and towards atomic-scale imaging under conditions where increasing electron exposure is no longer a viable route for improved structural information.
    
    Several directions could build upon these results. A broader investigation of the interplay between irradiation conditions, specimen properties, and scan parameters across different beam-sensitive materials would help determine which aspects of the observed response are material-specific and which can be generalized to other systems. Complementary chemical and structural characterization could further clarify the microscopic processes responsible for the observed damage evolution and help distinguish the contributions of different irradiation-induced mechanisms. Future work could also explore how spatial redundancy-based recovery performs for more complex crystals, lower-symmetry structures, and specimens exhibiting greater structural heterogeneity.
    
\section*{CRediT authorship contribution}
    
    \textbf{T.C.}: Conceptualization, Sample preparation, Data curation, Investigation, Formal analysis, Writing – original draft, review, editing. \textbf{A.A.}: Software, Investigation, Formal analysis, Writing – review editing. \textbf{S.L.}: Software, Investigation, Formal analysis, review. \textbf{N.G.}: Data curation, Investigation, Formal analysis, Writing – review editing. \textbf{H.L.L.R}: Investigation, Formal analysis, Supervision, Writing – review editing. \textbf{J.V.}: Funding acquisition, Supervision, Investigation, Formal analysis, Writing – review editing.

\section*{Declaration of competing interest}
    
    The authors declare that they have no known competing financial interests or personal relationships that could have appeared to influence the work reported in this paper.

\section*{Data availability}
    
    The original data supporting the findings of this study are openly available in Zenodo at:\\
    https://doi.org/10.5281/zenodo.22729645

\section*{Funding}

    \textbf{T.C.} and \textbf{J.V.} acknowledge funding from the Flemish government (iBOF project PERsist). \textbf{T.C.}, \textbf{S.L.} and \textbf{J.V.} acknowledge funding from the Research Foundation - Flanders (FWO, Belgium), under granted project No. G013122N. \text{H.L.L.R.} acknowledges funding from the Research Foundation - Flanders (FWO, Belgium), under granted postdoctoral fellowship No. 12A9V27N.
    
    \textbf{H.L.L.R.} and \textbf{J.V.} acknowledge funding from the Horizon 2020 research and innovation programme (European Union), under grant agreement No. 101017720 (FET-Proactive EBEAM). \textbf{H.L.L.R.}, \textbf{A.A.} and \textbf{J.V.} acknowledge funding from the Horizon Europe programme (European Union) under grant agreement No. 101094299 (Impress). \textbf{N.G.} and \textbf{J.V.} acknowledge funding from the Horizon 2020 research and innovation programme (European Union), under grant agreement No. 101130652 (RIANA). \\
    Views and opinions expressed are however those of the authors only and do not necessarily reflect those of the European Union or the European Research Executive Agency (REA). Neither the European Union nor the granting authority can be held responsible for them.
    
\section*{Appendix A. Supplementary material}
\label{supplementary}
    Supplementary material related to this article can be found online.

@article{hofer_phase_2024,
	title = {Phase offset method of ptychographic contrast reversal correction},
	volume = {258},
	issn = {0304-3991},
	url = {https://www.sciencedirect.com/science/article/pii/S0304399124000019},
	doi = {10.1016/j.ultramic.2024.113922},
	abstract = {The contrast transfer function of direct ptychography methods such as the single side band (SSB) method are single signed, yet these methods still sometimes exhibit contrast reversals, most often where the projected potentials are strong. In thicker samples central focusing often provides the best ptychographic contrast as this leads to defocus variations within the sample canceling out. However focusing away from the entrance surface is often undesirable as this degrades the annular dark field (ADF) signal. Here we discuss how phase wrap asymptotes in the frequency response of SSB ptychography give rise to contrast reversals, without the need for dynamical scattering, and how these can be counteracted by manipulating the phases such that the asymptotes are either shifted to higher frequencies or damped via amplitude modulation. This is what enables post collection defocus correction of contrast reversals. However, the phase offset method of counteracting contrast reversals we introduce here is generally found to be superior to post collection application of defocus, with greater reliability and generally stronger contrast. Importantly, the phase offset method also works for thin and thick samples where central focusing does not. Finally, the independence of the method from focus is useful for optical sectioning involving ptychography, improving interpretability by better disentangling the effects of strong potentials and focus.},
	urldate = {2024-03-02},
	journal = {Ultramicroscopy},
	author = {Hofer, Christoph and Gao, Chuang and Chennit, Tamazouzt and Yuan, Biao and Pennycook, Timothy J.},
	month = apr,
	year = {2024},
	keywords = {4D STEM, Electron ptychography, Phase wrap},
	pages = {113922},
}

@article{pennycook_high_2019,
	title = {High dose efficiency atomic resolution imaging via electron ptychography},
	volume = {196},
	issn = {0304-3991},
	url = {https://www.sciencedirect.com/science/article/pii/S0304399118302316},
	doi = {10.1016/j.ultramic.2018.10.005},
	abstract = {Radiation damage places a fundamental limitation on the ability of microscopy to resolve many types of materials at high resolution. Here we evaluate the dose efficiency of phase contrast imaging with electron ptychography. The method is found to be far more resilient to temporal incoherence than conventional and spherical aberration optimized phase contrast imaging, resulting in significantly greater clarity at a given dose. This robustness is explained by the presence of achromatic lines in the four dimensional ptychographic dataset.},
	urldate = {2024-03-02},
	journal = {Ultramicroscopy},
	author = {Pennycook, Timothy J. and Martinez, Gerardo T. and Nellist, Peter D. and Meyer, Jannik C.},
	month = jan,
	year = {2019},
	keywords = {Dose efficiency, Phase contrast, Ptychography, STEM, TEM},
	pages = {131--135},
	file = {Full Text:files/18/Pennycook et al. - 2019 - High dose efficiency atomic resolution imaging via.pdf:application/pdf},
}

@article{yang_enhanced_2016,
	title = {Enhanced phase contrast transfer using ptychography combined with a pre-specimen phase plate in a scanning transmission electron microscope},
	volume = {171},
	issn = {0304-3991},
	url = {https://www.sciencedirect.com/science/article/pii/S0304399116301966},
	doi = {10.1016/j.ultramic.2016.09.002},
	abstract = {The ability to image light elements in both crystalline and noncrystalline materials at near atomic resolution with an enhanced contrast is highly advantageous to understand the structure and properties of a wide range of beam sensitive materials including biological specimens and molecular hetero-structures. This requires the imaging system to have an efficient phase contrast transfer at both low and high spatial frequencies. In this work we introduce a new phase contrast imaging method in a scanning transmission electron microscope (STEM) using a pre-specimen phase plate in the probe forming aperture, combined with a fast pixelated detector to record diffraction patterns at every probe position, and phase reconstruction using ptychography. The phase plate significantly enhances the contrast transfer of low spatial frequency information, and ptychography maximizes the extraction of the phase information at all spatial frequencies. In addition, the STEM probe with the presence of the phase plate retains its atomic resolution, allowing simultaneous incoherent Z-contrast imaging to be obtained along with the ptychographic phase image. An experimental image of Au nanoparticles on a carbon support shows high contrast for both materials. Multislice image simulations of a DNA molecule shows the capability of imaging soft matter at low dose conditions, which implies potential applications of low dose imaging of a wide range of beam sensitive materials.},
	urldate = {2024-03-02},
	journal = {Ultramicroscopy},
	author = {Yang, Hao and Ercius, Peter and Nellist, Peter D. and Ophus, Colin},
	month = dec,
	year = {2016},
	keywords = {Phase contrast, Ptychography, STEM, PCTF, Phase plate, Pixelated detectors},
	pages = {117--125},
	file = {Full Text:files/22/Yang et al. - 2016 - Enhanced phase contrast transfer using ptychograph.pdf:application/pdf},
}

@article{jannis_event_2022,
	title = {Event driven {4D} {STEM} acquisition with a {Timepix3} detector: {Microsecond} dwell time and faster scans for high precision and low dose applications},
	volume = {233},
	issn = {0304-3991},
	shorttitle = {Event driven {4D} {STEM} acquisition with a {Timepix3} detector},
	url = {https://www.sciencedirect.com/science/article/pii/S0304399121001996},
	doi = {10.1016/j.ultramic.2021.113423},
	abstract = {Four dimensional scanning transmission electron microscopy (4D STEM) records the scattering of electrons in a material in great detail. The benefits offered by 4D STEM are substantial, with the wealth of data it provides facilitating for instance high precision, high electron dose efficiency phase imaging via centre of mass or ptychography based analysis. However the requirement for a 2D image of the scattering to be recorded at each probe position has long placed a severe bottleneck on the speed at which 4D STEM can be performed. Recent advances in camera technology have greatly reduced this bottleneck, with the detection efficiency of direct electron detectors being especially well suited to the technique. However even the fastest frame driven pixelated detectors still significantly limit the scan speed which can be used in 4D STEM, making the resulting data susceptible to drift and hampering its use for low dose beam sensitive applications. Here we report the development of the use of an event driven Timepix3 direct electron camera that allows us to overcome this bottleneck and achieve 4D STEM dwell times down to 100 ns; orders of magnitude faster than what has been possible with frame based readout. We characterize the detector for different acceleration voltages and show that the method is especially well suited for low dose imaging and promises rich datasets without compromising dwell time when compared to conventional STEM imaging.},
	urldate = {2024-03-02},
	journal = {Ultramicroscopy},
	author = {Jannis, D. and Hofer, C. and Gao, C. and Xie, X. and Béché, A. and Pennycook, T. J. and Verbeeck, J.},
	month = mar,
	year = {2022},
	keywords = {4D STEM, Event based detection, Low dose, Scanning transmission electron microscopy},
	pages = {113423},
	file = {Submitted Version:files/30/Jannis et al. - 2022 - Event driven 4D STEM acquisition with a Timepix3 d.pdf:application/pdf},
}

@article{gao_central_2024,
	title = {On central focusing for contrast optimization in direct electron ptychography of thick samples},
	volume = {256},
	issn = {0304-3991},
	url = {https://www.sciencedirect.com/science/article/pii/S0304399123001961},
	doi = {10.1016/j.ultramic.2023.113879},
	abstract = {Ptychography provides high dose efficiency images that can reveal light elements next to heavy atoms. However, despite ptychography having an otherwise single signed contrast transfer function, contrast reversals can occur when the projected potential becomes strong for both direct and iterative inversion ptychography methods. It has recently been shown that these reversals can often be counteracted in direct ptychography methods by adapting the focus. Here we provide an explanation of why the best contrast is often found with the probe focused to the middle of the sample. The phase contribution due to defocus at each sample slice above and below the central plane in this configuration effectively cancels out, which can prevent contrast reversals when dynamical scattering effects are not overly strong. In addition we show that the convergence angle can be an important consideration for removal of contrast reversals in relatively thin samples.},
	urldate = {2024-03-02},
	journal = {Ultramicroscopy},
	author = {Gao, C. and Hofer, C. and Pennycook, T. J.},
	month = feb,
	year = {2024},
	keywords = {4D STEM, Electron ptychography, Low dose, Scanning transmission electron microscopy},
	pages = {113879},
	file = {Submitted Version:files/32/Gao et al. - 2024 - On central focusing for contrast optimization in d.pdf:application/pdf},
}

@article{oster_optimized_2023,
	title = {Optimized detector configurations for the reconstruction of phase-contrast images in scanning transmission electron microscopy},
	volume = {246},
	issn = {0304-3991},
	url = {https://www.sciencedirect.com/science/article/pii/S0304399122001899},
	doi = {10.1016/j.ultramic.2022.113670},
	abstract = {Using the differential phase contrast mechanism and anti-symmetric detector geometries it is possible to image distributions of electric and magnetic fields in a scanning transmission electron microscope. Different detector geometries can be used for imaging and, due to their efficiency, mainly ring quadrant detectors and pixelated detectors have been used in recent high resolution differential phase contrast experiments. In 4D-Scanning Transmission Electron Microscopy one uses a pixelated (2D) detector to obtain the complete scattering distribution for every (2D) image point. The accuracy of pixelated detectors increases with an increasing number of pixels, which in turn also leads to a larger amount of data that needs to be evaluated. To reduce the required numerical effort, we are looking for alternative detector geometries by further segmenting ring quadrant detectors. To compare the different geometries, their signal-to-noise ratios are calculated for an ideal STEM and several weak phase objects. Images can be obtained by combining the data of different detector pixels using a scheme similar to a reconstruction from a focal series. The procedure can be interpreted as the simplest example of ptychography including only the first-order diffraction disks. Our results show that a 50-segment annular bright-field detector can reach a signal-to-noise ratio close to that of a 128 × 128 pixelated detector, while having a significantly lower number of segments that need to be evaluated.},
	urldate = {2024-03-02},
	journal = {Ultramicroscopy},
	author = {Oster, Alex and Kohl, Helmut},
	month = apr,
	year = {2023},
	keywords = {4D-STEM, Image formation, Optimized detector geometries, Phase contrast methods, Phase contrast transfer, Pixelated detector, Scanning Transmission Electron Microscopy, Signal-to-noise ratio},
	pages = {113670},
}

@article{mir_characterisation_2017,
	title = {Characterisation of the {Medipix3} detector for 60 and 80   {keV} electrons},
	volume = {182},
	issn = {0304-3991},
	url = {https://www.sciencedirect.com/science/article/pii/S0304399116303989},
	doi = {10.1016/j.ultramic.2017.06.010},
	abstract = {In this paper we report quantitative measurements of the imaging performance for the current generation of hybrid pixel detector, Medipix3, used as a direct electron detector. We have measured the modulation transfer function and detective quantum efficiency at beam energies of 60 and 80keV. In single pixel mode, energy threshold values can be chosen to maximize either the modulation transfer function or the detective quantum efficiency, obtaining values near to, or exceeding those for a theoretical detector with square pixels. The Medipix3 charge summing mode delivers simultaneous, high values of both modulation transfer function and detective quantum efficiency. We have also characterized the detector response to single electron events and describe an empirical model that predicts the detector modulation transfer function and detective quantum efficiency based on energy threshold. Exemplifying our findings we demonstrate the Medipix3 imaging performance recording a fully exposed electron diffraction pattern at 24-bit depth together with images in single pixel and charge summing modes. Our findings highlight that for transmission electron microscopy performed at low energies (energies {\textless}100keV) thick hybrid pixel detectors provide an advantageous architecture for direct electron imaging.},
	urldate = {2024-03-02},
	journal = {Ultramicroscopy},
	author = {Mir, J. A. and Clough, R. and MacInnes, R. and Gough, C. and Plackett, R. and Shipsey, I. and Sawada, H. and MacLaren, I. and Ballabriga, R. and Maneuski, D. and O'Shea, V. and McGrouther, D. and Kirkland, A. I.},
	month = nov,
	year = {2017},
	keywords = {Charge summing mode, DQE, Medipix3, Nyquist frequency, Single pixel mode, TEM, MTF},
	pages = {44--53},
	file = {Full Text:files/39/Mir et al. - 2017 - Characterisation of the Medipix3 detector for 60 a.pdf:application/pdf},
}

@article{hofer_reliable_2023,
	title = {Reliable phase quantification in focused probe electron ptychography of thin materials},
	volume = {254},
	issn = {0304-3991},
	url = {https://www.sciencedirect.com/science/article/pii/S0304399123001468},
	doi = {10.1016/j.ultramic.2023.113829},
	abstract = {Electron ptychography provides highly sensitive, dose efficient phase images which can be corrected for aberrations after the data has been acquired. This is crucial when very precise quantification is required, such as with sensitivity to charge transfer due to bonding. Drift can now be essentially eliminated as a major impediment to focused probe ptychography, which benefits from the availability of easily interpretable simultaneous Z-contrast imaging. However challenges have remained when quantifying the ptychographic phases of atomic sites. The phase response of a single atom has a negative halo which can cause atoms to reduce in phase when brought closer together. When unaccounted for, as in integrating methods of quantification, this effect can completely obscure the effects of charge transfer. Here we provide a new method of quantification that overcomes this challenge, at least for 2D materials, and is robust to experimental parameters such as noise, sample tilt.},
	urldate = {2024-03-02},
	journal = {Ultramicroscopy},
	author = {Hofer, Christoph and Pennycook, Timothy J.},
	month = dec,
	year = {2023},
	keywords = {Phase retrieval, Quantification, Single side band ptychography},
	pages = {113829},
	file = {Submitted Version:files/41/Hofer and Pennycook - 2023 - Reliable phase quantification in focused probe ele.pdf:application/pdf},
}

@article{hue_wave-front_2010,
	title = {Wave-front phase retrieval in transmission electron microscopy via ptychography},
	volume = {82},
	url = {https://link.aps.org/doi/10.1103/PhysRevB.82.121415},
	doi = {10.1103/PhysRevB.82.121415},
	abstract = {There are many different strategies that allow the solving of the well-known phase problem corresponding to the loss of phase information during a physical measurement. In microscopy, and, in particular, in transmission electron microscopy, most of these strategies focus on the retrieval of high-resolution information with the importance of lower resolution data often overlooked. Ptychography offers a means to investigate such data. Ptychography is a robust diffractive imaging technique with fast convergence for phase retrieval but, until now, has not been applied at the nanoscale. In this paper, we use the ptychographical iterative engine to retrieve the phase change at the exit plane of metallic nanoparticles using a conventional transmission electron microscope. Ptychographical reconstructions yielded images with a phase resolution of π/10 and a spatial resolution of 1 nm. These results stand as a first step toward aberration-free lensless imaging. The technique lends itself to be an alternative to off-axis electron holography or focal series reconstruction.},
	number = {12},
	urldate = {2024-03-02},
	journal = {Physical Review B},
	publisher = {American Physical Society},
	author = {Hüe, F. and Rodenburg, J. M. and Maiden, A. M. and Sweeney, F. and Midgley, P. A.},
	month = sep,
	year = {2010},
	pages = {121415},
	file = {Accepted Version:files/43/Hüe et al. - 2010 - Wave-front phase retrieval in transmission electro.pdf:application/pdf;APS Snapshot:files/44/PhysRevB.82.html:text/html},
}

@misc{varnavides_iterative_2023,
	title = {Iterative {Phase} {Retrieval} {Algorithms} for {Scanning} {Transmission} {Electron} {Microscopy}},
	url = {http://arxiv.org/abs/2309.05250},
	doi = {10.48550/arXiv.2309.05250},
	abstract = {Scanning transmission electron microscopy (STEM) has been extensively used for imaging complex materials down to atomic resolution. The most commonly employed STEM imaging modality of annular dark field produces easily-interpretable contrast, but is dose-inefficient and produces little to no contrast for light elements and weakly-scattering samples. An alternative is to use phase contrast STEM imaging, enabled by high speed detectors able to record full images of a diffracted STEM probe over a grid of scan positions. Phase contrast imaging in STEM is highly dose-efficient, able to measure the structure of beam-sensitive materials and even biological samples. Here, we comprehensively describe the theoretical background, algorithmic implementation details, and perform both simulated and experimental tests for three iterative phase retrieval STEM methods: focused-probe differential phase contrast, defocused-probe parallax imaging, and a generalized ptychographic gradient descent method implemented in two and three dimensions. We discuss the strengths and weaknesses of each of these approaches using a consistent framework to allow for easier comparison. This presentation of STEM phase retrieval methods will make these methods more approachable, reproducible and more readily adoptable for many classes of samples.},
	urldate = {2024-03-02},
	publisher = {arXiv},
	author = {Varnavides, Georgios and Ribet, Stephanie M. and Zeltmann, Steven E. and Yu, Yue and Savitzky, Benjamin H. and Dravid, Vinayak P. and Scott, Mary C. and Ophus, Colin},
	month = sep,
	year = {2023},
	note = {arXiv:2309.05250 [cond-mat]},
	keywords = {Condensed Matter - Materials Science},
	file = {arXiv Fulltext PDF:files/49/Varnavides et al. - 2023 - Iterative Phase Retrieval Algorithms for Scanning .pdf:application/pdf;arXiv.org Snapshot:files/50/2309.html:text/html},
}

@article{jiang_electron_2018,
	title = {Electron ptychography of {2D} materials to deep sub-ångström resolution},
	volume = {559},
	copyright = {2018 Macmillan Publishers Ltd., part of Springer Nature},
	issn = {1476-4687},
	url = {https://www.nature.com/articles/s41586-018-0298-5},
	doi = {10.1038/s41586-018-0298-5},
	abstract = {Aberration-corrected optics have made electron microscopy at atomic resolution a widespread and often essential tool for characterizing nanoscale structures. Image resolution has traditionally been improved by increasing the numerical aperture of the lens (α) and the beam energy, with the state-of-the-art at 300 kiloelectronvolts just entering the deep sub-ångström (that is, less than 0.5 ångström) regime. Two-dimensional (2D) materials are imaged at lower beam energies to avoid displacement damage from large momenta transfers, limiting spatial resolution to about 1 ångström. Here, by combining an electron microscope pixel-array detector with the dynamic range necessary to record the complete distribution of transmitted electrons and full-field ptychography to recover phase information from the full phase space, we increase the spatial resolution well beyond the traditional numerical-aperture-limited resolution. At a beam energy of 80 kiloelectronvolts, our ptychographic reconstruction improves the image contrast of single-atom defects in MoS2 substantially, reaching an information limit close to 5α, which corresponds to an Abbe diffraction-limited resolution of 0.39 ångström, at the electron dose and imaging conditions for which conventional imaging methods reach only 0.98 ångström.},
	language = {en},
	number = {7714},
	urldate = {2024-03-03},
	journal = {Nature},
	publisher = {Nature Publishing Group},
	author = {Jiang, Yi and Chen, Zhen and Han, Yimo and Deb, Pratiti and Gao, Hui and Xie, Saien and Purohit, Prafull and Tate, Mark W. and Park, Jiwoong and Gruner, Sol M. and Elser, Veit and Muller, David A.},
	month = jul,
	year = {2018},
	keywords = {Transmission electron microscopy, Imaging techniques, Two-dimensional materials},
	pages = {343--349},
	file = {Submitted Version:files/65/Jiang et al. - 2018 - Electron ptychography of 2D materials to deep sub-.pdf:application/pdf},
}

@article{chen_mixed-state_2020,
	title = {Mixed-state electron ptychography enables sub-angstrom resolution imaging with picometer precision at low dose},
	volume = {11},
	copyright = {2020 The Author(s)},
	issn = {2041-1723},
	url = {https://www.nature.com/articles/s41467-020-16688-6},
	doi = {10.1038/s41467-020-16688-6},
	abstract = {Both high resolution and high precision are required to quantitatively determine the atomic structure of complex nanostructured materials. However, for conventional imaging methods in scanning transmission electron microscopy (STEM), atomic resolution with picometer precision cannot usually be achieved for weakly-scattering samples or radiation-sensitive materials, such as 2D materials. Here, we demonstrate low-dose, sub-angstrom resolution imaging with picometer precision using mixed-state electron ptychography. We show that correctly accounting for the partial coherence of the electron beam is a prerequisite for high-quality structural reconstructions due to the intrinsic partial coherence of the electron beam. The mixed-state reconstruction gains importance especially when simultaneously pursuing high resolution, high precision and large field-of-view imaging. Compared with conventional atomic-resolution STEM imaging techniques, the mixed-state ptychographic approach simultaneously provides a four-times-faster acquisition, with double the information limit at the same dose, or up to a fifty-fold reduction in dose at the same resolution.},
	language = {en},
	number = {1},
	urldate = {2024-03-04},
	journal = {Nature Communications},
	publisher = {Nature Publishing Group},
	author = {Chen, Zhen and Odstrcil, Michal and Jiang, Yi and Han, Yimo and Chiu, Ming-Hui and Li, Lain-Jong and Muller, David A.},
	month = jun,
	year = {2020},
	keywords = {Imaging techniques, Microscopy, Nanoscale materials},
	pages = {2994},
	file = {Full Text PDF:files/67/Chen et al. - 2020 - Mixed-state electron ptychography enables sub-angs.pdf:application/pdf},
}

@misc{maiden_wasp_2024,
	title = {{WASP}: {Weighted} {Average} of {Sequential} {Projections} for ptychographic phase retrieval},
	shorttitle = {{WASP}},
	url = {https://preprints.opticaopen.org/articles/preprint/WASP_Weighted_Average_of_Sequential_Projections_for_ptychographic_phase_retrieval/24894489/1},
	doi = {10.1364/opticaopen.24894489.v1},
	abstract = {We introduce the Weighted Average of Sequential Projections, or WASP, an algorithm for ptychography. Using both simulations and real-world experiments, we test this new approach and compare performance against several alternative algorithms. These tests indicate that WASP effectively combines the benefits of its competitors, with a rapid initial convergence rate, robustness to noise and poor initial conditions, a small memory footprint, easy tuning, and the ability to reach a global minimum when provided with noiseless data. We also show how WASP can be parallelised to split operation across several different computation nodes.},
	language = {en},
	urldate = {2024-03-15},
	publisher = {Optica Open},
	author = {Maiden, Andrew and Mei, Wenjie and Li, Peng},
	month = jan,
	year = {2024},
	file = {Full Text PDF:files/121/Maiden et al. - 2024 - WASP Weighted Average of Sequential Projections f.pdf:application/pdf},
}

@article{ooe_direct_2023,
	title = {Direct imaging of local atomic structures in zeolite using optimum bright-field scanning transmission electron microscopy},
	volume = {9},
	url = {https://www.science.org/doi/full/10.1126/sciadv.adf6865},
	doi = {10.1126/sciadv.adf6865},
	abstract = {Zeolites are used in industries as catalysts, ion exchangers, and molecular sieves because of their unique porous atomic structures. However, direct observation of zeolitic local atomic structures via electron microscopy is difficult owing to low electron irradiation resistance. Subsequently, their fundamental structure-property relationships remain unclear. A low-electron-dose imaging technique, optimum bright-field scanning transmission electron microscopy (OBF STEM), has recently been developed. It reconstructs images with a high signal-to-noise ratio and a dose efficiency approximately two orders of magnitude higher than that of conventional methods. Here, we performed low-dose atomic-resolution OBF STEM observations of two types of zeolite, effectively visualizing all atomic sites in their frameworks. In addition, we visualized the complex local atomic structure of the twin boundaries in a faujasite (FAU)–type zeolite and Na+ ions with low occupancy in eight-membered rings in a Na-Linde Type A (LTA) zeolite. The results of this study facilitate the characterization of local atomic structures in many electron beam–sensitive materials.},
	number = {31},
	urldate = {2024-04-21},
	journal = {Science Advances},
	publisher = {American Association for the Advancement of Science},
	author = {Ooe, Kousuke and Seki, Takehito and Yoshida, Kaname and Kohno, Yuji and Ikuhara, Yuichi and Shibata, Naoya},
	month = aug,
	year = {2023},
	pages = {eadf6865},
	file = {Full Text PDF:files/126/Ooe et al. - 2023 - Direct imaging of local atomic structures in zeoli.pdf:application/pdf},
}

@article{gao_overcoming_2022,
	title = {Overcoming contrast reversals in focused probe ptychography of thick materials: {An} optimal pipeline for efficiently determining local atomic structure in materials science},
	volume = {121},
	issn = {0003-6951},
	shorttitle = {Overcoming contrast reversals in focused probe ptychography of thick materials},
	url = {https://doi.org/10.1063/5.0101895},
	doi = {10.1063/5.0101895},
	abstract = {Ptychography provides highly efficient imaging in scanning transmission electron microscopy (STEM), but questions have remained over its applicability to strongly scattering samples such as those most commonly seen in materials science. Although contrast reversals can appear in ptychographic phase images as the projected potentials of the sample increase, we show here how these can be easily overcome by a small amount of defocus. The amount of defocus is small enough that it not only can exist naturally when focusing using the annular dark field (ADF) signal but can also be adjusted post acquisition. The ptychographic images of strongly scattering materials are clearer at finite doses than other STEM techniques and can better reveal light atomic columns within heavy lattices. In addition, data for ptychography can now be collected simultaneously with the fastest of ADF scans. This combination of sensitivity and interpretability presents an ideal workflow for materials science.},
	number = {8},
	urldate = {2024-04-23},
	journal = {Applied Physics Letters},
	author = {Gao, C. and Hofer, C. and Jannis, D. and Béché, A. and Verbeeck, J. and Pennycook, T. J.},
	month = aug,
	year = {2022},
	pages = {081906},
	file = {Full Text PDF:files/128/Gao et al. - 2022 - Overcoming contrast reversals in focused probe pty.pdf:application/pdf;Snapshot:files/129/Overcoming-contrast-reversals-in-focused-probe.html:text/html},
}

@article{li_role_2023,
	title = {The role of ion migration, octahedral tilt, and the {A}-site cation on the instability of {Cs1}-{xFAxPbI3}},
	volume = {14},
	copyright = {2023 The Author(s)},
	issn = {2041-1723},
	url = {https://www.nature.com/articles/s41467-023-44235-6},
	doi = {10.1038/s41467-023-44235-6},
	abstract = {Organic-inorganic hybrid perovskites are promising materials for the next generation photovoltaics and optoelectronics; however, their practical application has been hindered by poor structural stability mainly caused by ion migration and external stimuli. Understanding the mechanism(s) of ion migration and structure decomposition is thus critical. Here we observe the sequence of structural changes at the atomic level that precede structural decomposition in the technologically important Cs1-xFAxPbI3 using ultralow dose transmission electron microscopy. We find that these changes differ, depending upon the A-site composition. Initially, there is a random loss of FA+, complemented by the loss of I-. The remaining FA+ and I- ions then migrate, unit cell by unit cell, into an ordered and more stable phase with a √2 x √2 superstructure. Further ion loss is accompanied by A-site dependent octahedral tilt modes and associated tetragonal phases with different stabilities. These observations of the loss of FA+/I- ion pairs, ion migration, octahedral tilt modes, and the role of the A-cation, provide insights into the atomic-scale structural mechanisms that drive and block ion loss and ion migration, opening pathways to inhibit ion loss, migration and improve structural stability.},
	language = {en},
	number = {1},
	urldate = {2024-04-28},
	journal = {Nature Communications},
	publisher = {Nature Publishing Group},
	author = {Li, Weilun and Hao, Mengmeng and Baktash, Ardeshir and Wang, Lianzhou and Etheridge, Joanne},
	month = dec,
	year = {2023},
	keywords = {Quantum dots, Solar cells},
	pages = {8523},
	file = {Full Text PDF:files/167/Li et al. - 2023 - The role of ion migration, octahedral tilt, and th.pdf:application/pdf},
}

@article{thibault_high-resolution_2008,
	title = {High-{Resolution} {Scanning} {X}-ray {Diffraction} {Microscopy}},
	volume = {321},
	url = {https://www.science.org/doi/10.1126/science.1158573},
	doi = {10.1126/science.1158573},
	abstract = {Coherent diffractive imaging (CDI) and scanning transmission x-ray microscopy (STXM) are two popular microscopy techniques that have evolved quite independently. CDI promises to reach resolutions below 10 nanometers, but the reconstruction procedures put stringent requirements on data quality and sample preparation. In contrast, STXM features straightforward data analysis, but its resolution is limited by the spot size on the specimen. We demonstrate a ptychographic imaging method that bridges the gap between CDI and STXM by measuring complete diffraction patterns at each point of a STXM scan. The high penetration power of x-rays in combination with the high spatial resolution will allow investigation of a wide range of complex mesoscopic life and material science specimens, such as embedded semiconductor devices or cellular networks.},
	number = {5887},
	urldate = {2024-05-02},
	journal = {Science},
	publisher = {American Association for the Advancement of Science},
	author = {Thibault, Pierre and Dierolf, Martin and Menzel, Andreas and Bunk, Oliver and David, Christian and Pfeiffer, Franz},
	month = jul,
	year = {2008},
	pages = {379--382},
	file = {Full Text PDF:files/230/Thibault et al. - 2008 - High-Resolution Scanning X-ray Diffraction Microsc.pdf:application/pdf;High-Resolution_Scanning_X-Ray_Diffraction_Microsc.pdf:files/1225/High-Resolution_Scanning_X-Ray_Diffraction_Microsc.pdf:application/pdf},
}

@article{luke_relaxed_2004,
	title = {Relaxed averaged alternating reflections for diffraction imaging},
	volume = {21},
	issn = {0266-5611},
	url = {https://dx.doi.org/10.1088/0266-5611/21/1/004},
	doi = {10.1088/0266-5611/21/1/004},
	abstract = {We report on progress in algorithms for iterative phase retrieval. The theory of convex optimization is used to develop and to gain insight into counterparts for the nonconvex problem of phase retrieval. We propose a relaxation of averaged alternating reflectors and determine the fixed-point set of the related operator in the convex case. A numerical study supports our theoretical observations and demonstrates the effectiveness of the algorithm compared to the current state of the art.},
	language = {en},
	number = {1},
	urldate = {2024-05-02},
	journal = {Inverse Problems},
	author = {Luke, D. Russell},
	month = nov,
	year = {2004},
	pages = {37},
	file = {IOP Full Text PDF:files/234/Luke - 2004 - Relaxed averaged alternating reflections for diffr.pdf:application/pdf},
}

@misc{gilgenbach_sampling_2023,
	title = {Sampling metrics for robust reconstructions in multislice ptychography: {Theory} and experiment},
	shorttitle = {Sampling metrics for robust reconstructions in multislice ptychography},
	url = {http://arxiv.org/abs/2311.15181},
	abstract = {While multislice electron ptychography can provide thermal-vibration limited resolution and 3D information, it relies on the proper selection of many intertwined experimental and computational parameters. Here, we outline a theoretical basis for selecting experimental parameters to enable robust ptychographic reconstructions. We develop a series of physically informed metrics to describe the selection of experimental parameters in multislice ptychography. Image simulations are used to comprehensively evaluate the validity of these metrics over a broad range of experimental conditions. We develop two metrics, areal oversampling and Ronchigram magnification, which predict reconstruction success with high accuracy. Lastly, we validate these conclusions with experimental ptychographic data, and demonstrate close agreement between trends in simulated and experimental data. Using these metrics, we achieve experimental multislice reconstructions at a scan step of \$1.0 {\textbackslash}unit\{{\textbackslash}AA/px\}\$, enabling large field-of-view (\${\textgreater}{\textbackslash}!18{\textbackslash}unit\{nm\}\$), data-efficient reconstructions. These experimental design principles enable the routine and reliable use of multislice ptychography for materials characterization.},
	language = {en},
	urldate = {2024-06-06},
	publisher = {arXiv},
	author = {Gilgenbach, Colin and Chen, Xi and LeBeau, James M.},
	month = nov,
	year = {2023},
	note = {arXiv:2311.15181 [cond-mat]},
	keywords = {Condensed Matter - Materials Science},
	file = {Gilgenbach et al. - 2023 - Sampling metrics for robust reconstructions in mul.pdf:files/257/Gilgenbach et al. - 2023 - Sampling metrics for robust reconstructions in mul.pdf:application/pdf},
}

@article{madsen_accuracy_2017,
	title = {Accuracy of surface strain measurements from transmission electron microscopy images of nanoparticles},
	volume = {3},
	issn = {2198-0926},
	url = {https://doi.org/10.1186/s40679-017-0047-0},
	doi = {10.1186/s40679-017-0047-0},
	abstract = {Strain analysis from high-resolution transmission electron microscopy (HRTEM) images offers a convenient tool for measuring strain in materials at the atomic scale. In this paper we present a theoretical study of the precision and accuracy of surface strain measurements directly from aberration-corrected HRTEM images. We examine the influence of defocus, crystal tilt and noise, and find that absolute errors of at least 1–2\% strain should be expected. The model structures include surface relaxations determined using molecular dynamics, and we show that this is important for correctly evaluating the errors introduced by image aberrations.},
	number = {1},
	urldate = {2024-06-17},
	journal = {Advanced Structural and Chemical Imaging},
	author = {Madsen, Jacob and Liu, Pei and Wagner, Jakob B. and Hansen, Thomas W. and Schiøtz, Jakob},
	month = oct,
	year = {2017},
	keywords = {High-resolution transmission electron microscopy, Nanoparticles, Strain mapping, Surface strain},
	pages = {14},
	file = {Full Text PDF:files/264/Madsen et al. - 2017 - Accuracy of surface strain measurements from trans.pdf:application/pdf;Snapshot:files/265/s40679-017-0047-0.html:text/html},
}

@article{velazco_reducing_2022,
	title = {Reducing electron beam damage through alternative {STEM} scanning strategies, {Part} {I}: {Experimental} findings},
	volume = {232},
	issn = {0304-3991},
	shorttitle = {Reducing electron beam damage through alternative {STEM} scanning strategies, {Part} {I}},
	url = {https://www.sciencedirect.com/science/article/pii/S0304399121001777},
	doi = {10.1016/j.ultramic.2021.113398},
	abstract = {The highly energetic electrons in a transmission electron microscope (TEM) can alter or even completely destroy the structure of samples before sufficient information can be obtained. This is especially problematic in the case of zeolites, organic and biological materials. As this effect depends on both the electron beam and the sample and can involve multiple damage pathways, its study remained difficult and is plagued with irreproducibility issues, circumstantial evidence, rumors, and a general lack of solid data. Here we take on the experimental challenge to investigate the role of the STEM scan pattern on the damage behavior of a commercially available zeolite sample with the clear aim to make our observations as reproducible as possible. We make use of a freely programmable scan engine that gives full control over the tempospatial distribution of the electron probe on the sample and we use its flexibility to obtain multiple repeated experiments under identical conditions comparing the difference in beam damage between a conventional raster scan pattern and a newly proposed interleaved scan pattern that provides exactly the same dose and dose rate and visits exactly the same scan points. We observe a significant difference in beam damage for both patterns with up to 11 \% reduction in damage (measured from mass loss). These observations demonstrate without doubt that electron dose, dose rate and acceleration voltage are not the only parameters affecting beam damage in (S)TEM experiments and invite the community to rethink beam damage as an unavoidable consequence of applied electron dose.},
	urldate = {2024-06-19},
	journal = {Ultramicroscopy},
	author = {Velazco, A. and Béché, A. and Jannis, D. and Verbeeck, J.},
	month = jan,
	year = {2022},
	keywords = {Scanning transmission electron microscopy, Electron beam damage, Scanning strategies},
	pages = {113398},
	file = {ScienceDirect Snapshot:files/270/S0304399121001777.html:text/html;Submitted Version:files/271/Velazco et al. - 2022 - Reducing electron beam damage through alternative .pdf:application/pdf},
}

@misc{craig_considerations_2024,
	title = {Considerations for extracting moir{\textbackslash}'e-level strain from dark field intensities in transmission electron microscopy},
	url = {http://arxiv.org/abs/2406.04515},
	abstract = {Bragg interferometry (BI) is an imaging technique based on four-dimensional scanning electron microscopy (4D-STEM) wherein the intensities of select overlapping Bragg disks are fit or more qualitatively analyzed in the context of simple trigonometric equations to determine local stacking order. In 4D-STEM based approaches, the collection of full diffraction patterns at each real-space position of the scanning probe allows the use of precise virtual apertures much smaller and more variable in shape than those used in conventional dark field imaging, such that even buried interfaces marginally twisted from other layers can be targeted. A coarse-grained form of dark field ptychography, BI uses simple physically derived fitting functions to extract the average structure within the illumination region and is therefore viable over large fields of view. BI has shown a particular advantage for selectively investigating the interlayer stacking and associated moir{\textbackslash}'e reconstruction of bilayer interfaces within complex multi-layered structures. This has enabled investigation of reconstruction and substrate effects in bilayers through encapsulating hexagonal boron nitride and of select bilayer interfaces within trilayer stacks. However, the technique can be improved to provide a greater spatial resolution and probe a wider range of twisted structures, for which current limitations on acquisition parameters can lead to large illumination regions and the computationally involved post-processing can fail. Here we analyze these limitations and the computational processing in greater depth, presenting a few methods for improvement over previous works, discussing potential areas for further expansion, and illustrating the current capabilities of this approach for extracting moir{\textbackslash}'e-scale strain.},
	language = {en},
	urldate = {2024-06-19},
	publisher = {arXiv},
	author = {Craig, Isaac M. and Van Winkle, Madeline and Ophus, Colin and Bediako, D. Kwabena},
	month = jun,
	year = {2024},
	note = {arXiv:2406.04515 [cond-mat, physics:physics]},
	keywords = {Condensed Matter - Materials Science, Physics - Optics},
	file = {Craig et al. - 2024 - Considerations for extracting moir'e-level strain.pdf:files/272/Craig et al. - 2024 - Considerations for extracting moir'e-level strain.pdf:application/pdf},
}

@misc{han_atomically_2024,
	title = {Atomically resolved imaging of radiation-sensitive metal-organic frameworks via electron ptychography},
	copyright = {https://creativecommons.org/licenses/by/4.0/},
	url = {https://www.researchsquare.com/article/rs-4505545/v1},
	doi = {10.21203/rs.3.rs-4505545/v1},
	abstract = {Abstract
          
            Electron ptychography, recognized as an ideal technique for low-dose imaging, consistently achieves deep sub-angstrom resolution in low-dimensional materials at electron doses of several thousand electrons per square angstrom (e
            -
            /Å
            2
            ). Despite its proven efficacy, the application of electron ptychography at even lower doses—necessary for materials highly sensitive to electron beams—raises questions regarding its feasibility and the attainable resolution under such stringent conditions. Herein, we demonstrate the successful implementation of electron ptychography reconstruction at an unprecedentedly low electron dose of {\textasciitilde}100 e
            -
            /Å
            2
            , for metal-organic frameworks (MOFs), which are known for their extreme sensitivity. The reconstructed images, achieving a resolution of {\textasciitilde}2 Å, clearly resolve organic linkers, metal clusters, and even atomic columns within these clusters, while unravelling various local structural features in MOFs, including missing linkers, extra clusters, and surface termination modes. By combining the findings from simulations and experiments, we have identified that employing a small convergence semi-angle during data acquisition is crucial for effective iterative ptychographic reconstruction under such low-dose conditions. This important insight advances our understanding of the rapidly evolving electron ptychography technique and provides a novel approach to high-resolution imaging of various sensitive materials.},
	language = {en},
	urldate = {2024-06-20},
	author = {Han, Yu and Li, Guanxing and Xu, Ming and Tang, Wen-Qi and Liu, Ying and Chen, Cailing and Zhang, Daliang and Liu, Lingmei and Ning, Shoucong and Zhang, Hui and Gu, Zhi-Yuan and Lai, Zhiping and Muller, David},
	month = jun,
	year = {2024},
	file = {Han et al. - 2024 - Atomically resolved imaging of radiation-sensitive.pdf:files/292/Han et al. - 2024 - Atomically resolved imaging of radiation-sensitive.pdf:application/pdf},
}

@article{pei_cryogenic_2023,
	title = {Cryogenic electron ptychographic single particle analysis with wide bandwidth information transfer},
	volume = {14},
	copyright = {2023 The Author(s)},
	issn = {2041-1723},
	url = {https://www.nature.com/articles/s41467-023-38268-0},
	doi = {10.1038/s41467-023-38268-0},
	abstract = {Advances in cryogenic transmission electron microscopy have revolutionised the determination of many macromolecular structures at atomic or near-atomic resolution. This method is based on conventional defocused phase contrast imaging. However, it has limitations of weaker contrast for small biological molecules embedded in vitreous ice, in comparison with cryo-ptychography, which shows increased contrast. Here we report a single-particle analysis based on the use of ptychographic reconstruction data, demonstrating that three dimensional reconstructions with a wide information transfer bandwidth can be recovered by Fourier domain synthesis. Our work suggests future applications in otherwise challenging single particle analyses, including small macromolecules and heterogeneous or flexible particles. In addition structure determination in situ within cells without the requirement for protein purification and expression may be possible.},
	language = {en},
	number = {1},
	urldate = {2024-06-28},
	journal = {Nature Communications},
	publisher = {Nature Publishing Group},
	author = {Pei, Xudong and Zhou, Liqi and Huang, Chen and Boyce, Mark and Kim, Judy S. and Liberti, Emanuela and Hu, Yiming and Sasaki, Takeo and Nellist, Peter D. and Zhang, Peijun and Stuart, David I. and Kirkland, Angus I. and Wang, Peng},
	month = may,
	year = {2023},
	keywords = {Cryoelectron microscopy, Transmission electron microscopy},
	pages = {3027},
	file = {Full Text PDF:files/310/Pei et al. - 2023 - Cryogenic electron ptychographic single particle a.pdf:application/pdf},
}

@misc{ning_robust_2024,
	title = {Robust {Ptychographic} {Reconstruction} with an {Out}-of-{Focus} {Electron} {Probe}},
	url = {http://arxiv.org/abs/2406.15879},
	doi = {10.48550/arXiv.2406.15879},
	abstract = {As a burgeoning technique, out-of-focus electron ptychography offers the potential for rapidly imaging atomic-scale large fields of view (FoV) using a single diffraction dataset. However, achieving robust out-of-focus ptychographic reconstruction poses a significant challenge due to the inherent scan instabilities of electron microscopes, compounded by the presence of unknown aberrations in the probe-forming lens. In this study, we substantially enhance the robustness of out-of-focus ptychographic reconstruction by extending our previous calibration method (the Fourier method), which was originally developed for the in-focus scenario. This extended Fourier method surpasses existing calibration techniques by providing more reliable and accurate initialization of scan positions and electron probes. Additionally, we comprehensively explore and recommend optimized experimental parameters for robust out-of-focus ptychography, includingaperture size and defocus, through extensive simulations. Lastly, we conduct a comprehensive comparison between ptychographic reconstructions obtained with focused and defocused electron probes, particularly in the context of low-dose and precise phase imaging, utilizing our calibration method as the basis for evaluation.},
	urldate = {2024-07-02},
	publisher = {arXiv},
	author = {Ning, Shoucong and Xu, Wenhui and Sheng, Pengju and Loh, Leyi and Pennycook, Stephen and Zhang, Fucai and Bosman, Michel and He, Qian},
	month = jun,
	year = {2024},
	note = {arXiv:2406.15879 [cond-mat, physics:physics]},
	keywords = {Condensed Matter - Materials Science, Physics - Optics},
	file = {arXiv Fulltext PDF:files/313/Ning et al. - 2024 - Robust Ptychographic Reconstruction with an Out-of.pdf:application/pdf;arXiv.org Snapshot:files/314/2406.html:text/html},
}

@article{xu_high-performance_nodate,
	title = {A high-performance reconstruction method for partially coherent ptychography},
	abstract = {Ptychography is now integrated as a tool in mainstream microscopy allowing quantitative and high-resolution imaging capabilities over a wide field of view. However, its ultimate performance is inevitably limited by the available coherent flux when implemented using electrons or laboratory X-ray sources. We present a universal reconstruction algorithm with high tolerance to low coherence for both far-field and near-field ptychography. The approach is practical for partial temporal and spatial coherence and requires no prior knowledge of the source properties. Our initial visible-light and electron data show that the method can dramatically improve the reconstruction quality and accelerate the convergence rate of the reconstruction. The approach also integrates well into existing ptychographic engines. It can also improve mixed-state and numerical monochromatisation methods, requiring a smaller number of coherent modes or lower dimensionality of Krylov subspace while providing more stable and faster convergence. We propose that this approach could have significant impact on ptychography of weakly scattering samples.},
	language = {en},
	author = {Xu, Wenhui and Ning, Shoucong and Sheng, Pengju and Lin, Huixiang and Kirkland, Angus I and Peng, Yong and Zhang, Fucai},
	file = {Xu et al. - A high-performance reconstruction method for parti.pdf:files/326/Xu et al. - A high-performance reconstruction method for parti.pdf:application/pdf},
}

@article{van_dyck_persistent_2011,
	title = {Persistent misconceptions about incoherence in electron microscopy},
	volume = {111},
	copyright = {https://www.elsevier.com/tdm/userlicense/1.0/},
	issn = {03043991},
	url = {https://linkinghub.elsevier.com/retrieve/pii/S0304399111000222},
	doi = {10.1016/j.ultramic.2011.01.007},
	abstract = {Incoherence in electron microscopic imaging occurs when during the observation the microscope and the object are subject to ﬂuctuations. In order to speed up the computer simulation of the images, approximations are used that are considered as valid. In this paper we will question the validity of these approximations and show that in speciﬁc cases they can lead to erroneous results.},
	language = {en},
	number = {7},
	urldate = {2024-09-03},
	journal = {Ultramicroscopy},
	author = {Van Dyck, D.},
	month = jun,
	year = {2011},
	pages = {894--900},
	file = {Van Dyck - 2011 - Persistent misconceptions about incoherence in ele.pdf:files/368/Van Dyck - 2011 - Persistent misconceptions about incoherence in ele.pdf:application/pdf},
}

@article{katkovnik_sparse_2013,
	title = {Sparse ptychographical coherent diffractive imaging from noisy measurements},
	volume = {30},
	copyright = {https://doi.org/10.1364/OA\_License\_v1\#VOR},
	issn = {1084-7529, 1520-8532},
	url = {https://opg.optica.org/abstract.cfm?URI=josaa-30-3-367},
	doi = {10.1364/JOSAA.30.000367},
	language = {en},
	number = {3},
	urldate = {2024-09-11},
	journal = {Journal of the Optical Society of America A},
	author = {Katkovnik, Vladimir and Astola, Jaakko},
	month = mar,
	year = {2013},
	pages = {367},
	file = {Katkovnik and Astola - 2013 - Sparse ptychographical coherent diffractive imagin.pdf:files/378/Katkovnik and Astola - 2013 - Sparse ptychographical coherent diffractive imagin.pdf:application/pdf},
}

@article{scheid_electron_2023,
	title = {Electron {Ptychographic} {Phase} {Imaging} of {Beam}-sensitive {All}-inorganic {Halide} {Perovskites} {Using} {Four}-dimensional {Scanning} {Transmission} {Electron} {Microscopy}},
	volume = {29},
	copyright = {https://creativecommons.org/licenses/by/4.0/},
	issn = {1431-9276, 1435-8115},
	url = {https://academic.oup.com/mam/article/29/3/869/7131445},
	doi = {10.1093/micmic/ozad017},
	abstract = {Halide perovskites (HPs) are promising candidates for optoelectronic devices, such as solar cells or light-emitting diodes. Despite recent progress in performance optimization and low-cost manufacturing, their commercialization remains hindered due to structural instabilities. While essential to the development of the technology, the relation between the microscopic properties of HPs and the relevant degradation mechanisms is still not well understood. The sensitivity of HPs toward electron-beam irradiation poses significant challenges for transmission electron microscopy (TEM) investigations of structure and degradation mechanisms at the atomic scale. However, technological advances and the development of direct electron cameras (DECs) have opened up a completely new field of electron microscopy: four-dimensional scanning TEM (4D-STEM). From a 4D-STEM dataset, it is possible to extract not only the intensity signal for any STEM detector geometry but also the phase information of the specimen. This work aims to show the potential of 4D-STEM, in particular, electron exit-wave phase reconstructions via focused probe ptychography as a low-dose and dose-efficient technique to image the atomic structure of beam-sensitive HPs. The damage mechanism under conventional irradiation is described and atomically resolved almost aberration-free phase images of three all-inorganic HPs, CsPbBr3, CsPbIBr2, and CsPbI3, are presented with a resolution down to the aperture-constrained diffraction limit.},
	language = {en},
	number = {3},
	urldate = {2024-09-18},
	journal = {Microscopy and Microanalysis},
	author = {Scheid, Anna and Wang, Yi and Jung, Mina and Heil, Tobias and Moia, Davide and Maier, Joachim and Van Aken, Peter A},
	month = jun,
	year = {2023},
	pages = {869--878},
	file = {Scheid et al. - 2023 - Electron Ptychographic Phase Imaging of Beam-sensi.pdf:files/465/Scheid et al. - 2023 - Electron Ptychographic Phase Imaging of Beam-sensi.pdf:application/pdf},
}

@book{kirkland_advanced_2010,
	address = {Boston, MA},
	title = {Advanced {Computing} in {Electron} {Microscopy}},
	copyright = {https://www.springernature.com/gp/researchers/text-and-data-mining},
	isbn = {978-1-4419-6532-5 978-1-4419-6533-2},
	url = {https://link.springer.com/10.1007/978-1-4419-6533-2},
	doi = {10.1007/978-1-4419-6533-2},
	language = {en},
	urldate = {2024-09-19},
	publisher = {Springer US},
	author = {Kirkland, Earl J.},
	year = {2010},
	file = {Kirkland - 2010 - Advanced Computing in Electron Microscopy.pdf:files/468/Kirkland - 2010 - Advanced Computing in Electron Microscopy.pdf:application/pdf},
}

@book{hawkes_springer_2019,
	address = {Cham},
	series = {Springer {Handbooks}},
	title = {Springer {Handbook} of {Microscopy}},
	copyright = {http://www.springer.com/tdm},
	isbn = {978-3-030-00068-4 978-3-030-00069-1},
	url = {http://link.springer.com/10.1007/978-3-030-00069-1},
	doi = {10.1007/978-3-030-00069-1},
	language = {en},
	urldate = {2024-09-19},
	publisher = {Springer International Publishing},
	editor = {Hawkes, Peter W. and Spence, John C. H.},
	year = {2019},
	file = {Hawkes and Spence - 2019 - Springer Handbook of Microscopy.pdf:files/470/Hawkes and Spence - 2019 - Springer Handbook of Microscopy.pdf:application/pdf},
}

@misc{ibanez_retrieval_2024,
	title = {Retrieval of phase information from low-dose electron microscopy experiments: are we at the limit yet?},
	shorttitle = {Retrieval of phase information from low-dose electron microscopy experiments},
	url = {http://arxiv.org/abs/2408.10590},
	abstract = {The challenge of imaging low-density objects in an electron microscope without causing beam damage is significant in modern TEM. This is especially challenging for life science imaging, where the sample, rather than the instrument, still determines the resolution limit. Here, we explore whether we have to accept this or can progress further in this area. To do this, we use numerical simulations to see how much information we can obtain from a weak phase object at different electron doses. Starting from a model with four phase values, we compare Zernike phase contrast with measuring diffracted intensity under multiple random phase illuminations to solve the inverse problem. Our simulations have shown that diffraction-based methods perform better than the Zernike method, as we have found and addressed a normalization issue that, in some other studies, led to an overly optimistic representation of the Zernike setup. We further validated this using more realistic 2D objects and found that random phase illuminated diffraction can be up to five times more efficient than an ideal Zernike implementation. These findings suggest that diffraction-based methods could be a promising approach for imaging beam-sensitive materials and that current low-dose imaging methods are not yet at the quantum limit.},
	language = {en},
	urldate = {2024-09-23},
	publisher = {arXiv},
	author = {Ibáñez, Francisco Vega and Verbeeck, Jo},
	month = aug,
	year = {2024},
	note = {arXiv:2408.10590 [physics]},
	keywords = {Physics - Optics, Physics - Applied Physics},
	file = {Ibáñez and Verbeeck - 2024 - Retrieval of phase information from low-dose elect.pdf:files/538/Ibáñez and Verbeeck - 2024 - Retrieval of phase information from low-dose elect.pdf:application/pdf},
}

@article{pennycook_efficient_2015,
	series = {Special {Issue}: 80th {Birthday} of {Harald} {Rose}; {PICO} 2015 – {Third} {Conference} on {Frontiers} of {Aberration} {Corrected} {Electron} {Microscopy}},
	title = {Efficient phase contrast imaging in {STEM} using a pixelated detector. {Part} 1: {Experimental} demonstration at atomic resolution},
	volume = {151},
	issn = {0304-3991},
	shorttitle = {Efficient phase contrast imaging in {STEM} using a pixelated detector. {Part} 1},
	url = {https://www.sciencedirect.com/science/article/pii/S0304399114001934},
	doi = {10.1016/j.ultramic.2014.09.013},
	abstract = {We demonstrate a method to achieve high efficiency phase contrast imaging in aberration corrected scanning transmission electron microscopy (STEM) with a pixelated detector. The pixelated detector is used to record the Ronchigram as a function of probe position which is then analyzed with ptychography. Ptychography has previously been used to provide super-resolution beyond the diffraction limit of the optics, alongside numerically correcting for spherical aberration. Here we rely on a hardware aberration corrector to eliminate aberrations, but use the pixelated detector data set to utilize the largest possible volume of Fourier space to create high efficiency phase contrast images. The use of ptychography to diagnose the effects of chromatic aberration is also demonstrated. Finally, the four dimensional dataset is used to compare different bright field detector configurations from the same scan for a sample of bilayer graphene. Our method of high efficiency ptychography produces the clearest images, while annular bright field produces almost no contrast for an in-focus aberration-corrected probe.},
	urldate = {2024-09-23},
	journal = {Ultramicroscopy},
	author = {Pennycook, Timothy J. and Lupini, Andrew R. and Yang, Hao and Murfitt, Matthew F. and Jones, Lewys and Nellist, Peter D.},
	month = apr,
	year = {2015},
	keywords = {Phase contrast, STEM, Pixelated detectors, ABF, Chromatic aberrations, DPC, Ptycography},
	pages = {160--167},
	file = {ScienceDirect Snapshot:files/560/S0304399114001934.html:text/html},
}

@article{cao_image_2018,
	title = {Image feature delocalization in defocused probe electron ptychography.},
	volume = {187},
	url = {https://www.semanticscholar.org/paper/a503874521042293a02ad3905fb7cd9251482094},
	doi = {10.1016/j.ultramic.2018.01.006},
	abstract = {S2 TL;DR: Factors that result in the delocalization effect when a defocused probe is used for the ptychographic data collection are explored: source incoherence, the effects of detector faults, data truncation and a poorly calibrated illumination step size (or camera length).},
	journal = {Ultramicroscopy},
	author = {Cao, S. and Maiden, A. and Rodenburg, J.},
	year = {2018},
	pages = {71--83},
}

@article{sha_deep_2022,
	title = {Deep sub-angstrom resolution imaging by electron ptychography with misorientation correction},
	volume = {8},
	url = {https://www.semanticscholar.org/paper/004d829f926b96de981ba98ba5f797e21883f83d},
	doi = {10.1126/sciadv.abn2275},
	abstract = {Superresolution imaging of solids is essential to explore local symmetry breaking and derived material properties. Electron ptychography is one of the most promising schemes to realize superresolution imaging beyond aberration correction. However, to reach both deep sub-angstrom resolution imaging and accurate measurement of atomic structures, it is still required for the electron beam to be nearly parallel to the zone axis of crystals. Here, we report an efficient and robust method to correct the specimen misorientation in electron ptychography, giving deep sub-angstrom resolution for specimens with large misorientations. The method largely reduces the experimental difficulties of electron ptychography and paves the way for widespread applications of ptychographic deep sub-angstrom resolution imaging.},
	journal = {Science Advances},
	author = {Sha, Haozhi and Cui, Jizhe and Yu, Rong},
	year = {2022},
	pages = {null},
}

@article{wang_electron_2017,
	title = {Electron {Ptychographic} {Diffractive} {Imaging} of {Boron} {Atoms} in {LaB6} {Crystals}},
	volume = {7},
	url = {https://www.semanticscholar.org/paper/0bfcb2f5c6b9593fc4014c63acee22f092a3528c},
	doi = {10.1038/s41598-017-02778-x},
	abstract = {S2 TL;DR: A ptychographic reconstruction of a LaB6 crystal is demonstrated in which light B atoms were clearly resolved together with the heavy La atoms in the reconstructed phase and offers an alternative future basis for imaging the atomic structure of materials, particularly those containing low atomic number elements.},
	journal = {Scientific Reports},
	author = {Wang, Peng and Zhang, Fucai and Gao, Si and Zhang, Mian and Kirkland, A.},
	year = {2017},
	pages = {null},
}

@article{yang_simultaneous_2016,
	title = {Simultaneous atomic-resolution electron ptychography and {Z}-contrast imaging of light and heavy elements in complex nanostructures},
	volume = {7},
	url = {https://www.semanticscholar.org/paper/b2f568bcdf2b7b6e97af0acdad28d3d6dc352b78},
	doi = {10.1038/ncomms12532},
	abstract = {S2 TL;DR: Recent developments in fast electron detectors and data processing capability is shown to enable electron ptychography, to extend the capability of the aberration-corrected scanning transmission electron microscope by allowing quantitative phase images to be formed simultaneously with incoherent signals.},
	journal = {Nature Communications},
	author = {Yang, Hao and Rutte, R. N. and Jones, L. and Simson, M. and Sagawa, R. and Ryll, H. and Huth, M. and Pennycook, T. and Green, Malcolm L. H. and Soltau, H. and Kondo, Yushi and Davis, B. G. and Nellist, P.},
	year = {2016},
	pages = {null},
}

@article{hachtel_sub-angstrom_2018,
	title = {Sub-Ångstrom electric field measurements on a universal detector in a scanning transmission electron microscope},
	volume = {4},
	url = {https://www.semanticscholar.org/paper/a4e08e571384c573ed1577dc8b65b269dd7f3077},
	doi = {10.1186/s40679-018-0059-4},
	abstract = {S2 TL;DR: DPC spatial resolution is demonstrated almost reaching the information limit of a 100 keV electron beam, which opens up new opportunities to understand and design functional materials and devices that involve lattice and charge coupling at nano- and atomic-scales.},
	journal = {Advanced Structural and Chemical Imaging},
	author = {Hachtel, J. and Idrobo, J. and Chi, M.},
	year = {2018},
	pages = {null},
}

@article{chang_ptychographic_2020,
	title = {Ptychographic atomic electron tomography: {Towards} three-dimensional imaging of individual light atoms in materials},
	volume = {102},
	url = {https://www.semanticscholar.org/paper/a77d90cda64d3a7973688c08a8c76f77e7813cfa},
	doi = {10.1103/physrevb.102.174101},
	abstract = {Through numerical simulations, we demonstrate the combination of ptychography and atomic electron tomography as an effective method for low dose imaging of individual low-Z atoms in three dimensions. After generating noisy diffraction patterns with multislice simulations of an aberration-corrected scanning transmission electron microscope through a 5-nm zinc-oxide nanoparticle, we have achieved three-dimensional (3D) imaging of individual zinc and oxygen atoms and their defects by performing tomography on ptychographic projections. The methodology has also been simulated in 2D materials, resolving individual sulfur atoms in vertical {\textbackslash}mathrmW{\textbackslash}mathrmS\_2/{\textbackslash}mathrmWS{\textbackslash}mathrme\_2 van der Waals heterostructure with a low total electron dose where annular-dark-field images fail to resolve. We envision that the development of this method could be instrumental in studying the precise 3D atomic structures of radiation sensitive systems and low-Z atomic structures such as 2D heterostructures, catalysts, functional oxides, and glasses.},
	journal = {Physical Review B},
	author = {Chang, Dillan J. and Kim, Dennis S. and Rana, Arjun and Tian, Xuezeng and Zhou, Jihan and Ercius, P. and Miao, J.},
	year = {2020},
	pages = {174101},
}

@article{pelz_real-time_2021,
	title = {Real-{Time} {Interactive} {4D}-{STEM} {Phase}-{Contrast} {Imaging} {From} {Electron} {Event} {Representation} {Data}: {Less} computation with the right representation},
	volume = {39},
	url = {https://www.semanticscholar.org/paper/9a51c6cbaeb72d79cf0f262e7057d4f656a0bfe2},
	doi = {10.1017/S1431927621001288},
	abstract = {The arrival of direct electron detectors (DEDs) with high frame rates in the field of scanning transmission electron microscopy (TEM) has enabled many experimental techniques that require collection of a full diffraction pattern at each scan position, a field which is subsumed under the name four-dimensional scanning transmission electron microscopy (4D-STEM). DED frame rates approaching 100 kHz require data transmission rates and data storage capabilities that exceed those of the commonly available computing infrastructures. Current commercial DEDs allow the user to make compromises in pixel bit depth, detector binning, or windowing to reduce the per-frame file size and allow higher frame rates. This change in detector specifications requires decisions to be made before data acquisition that may reduce or lose information that could have been advantageous during data analysis.},
	journal = {IEEE Signal Processing Magazine},
	author = {Pelz, Philipp M. and Johnson, I. and Ophus, C. and Ercius, P. and Scott, M.},
	year = {2021},
	pages = {25--31},
}

@article{brown_three-dimensional_2022,
	title = {A {Three}-{Dimensional} {Reconstruction} {Algorithm} for {Scanning} {Transmission} {Electron} {Microscopy} {Data} from a {Single} {Sample} {Orientation}},
	volume = {28},
	url = {https://www.semanticscholar.org/paper/5bc598a5f88b67b6dea6afea4953596379d271f3},
	doi = {10.1017/S1431927622012090},
	abstract = {Abstract Increasing interest in three-dimensional nanostructures adds impetus to electron microscopy techniques capable of imaging at or below the nanoscale in three dimensions. We present a reconstruction algorithm that takes as input a focal series of four-dimensional scanning transmission electron microscopy (4D-STEM) data. We apply the approach to a lead iridate, Pb₂Ir₂O₇, and yttrium-stabilized zirconia, Y$_{\textrm{0.095}}$Zr$_{\textrm{0.905}}$O₂, heterostructure from data acquired with the specimen in a single plan-view orientation, with the epitaxial layers stacked along the beam direction. We demonstrate that Pb–Ir atomic columns are visible in the uppermost layers of the reconstructed volume. We compare this approach to the alternative techniques of depth sectioning using differential phase contrast scanning transmission electron microscopy (DPC-STEM) and multislice ptychographic reconstruction.},
	journal = {Microscopy and Microanalysis},
	author = {Brown, H. and Pelz, Philipp M. and Hsu, S. and Zhang, Zimeng and Ramesh, Ramamoorthy and Inzani, Katherine and Sheridan, Evan and Griffin, S. and Schloz, M. and Pekin, T. and Koch, C. and Findlay, S. and Allen, L. and Scott, M. and Ophus, C. and Ciston, J.},
	year = {2022},
	pages = {1632 -- 1640},
}

@article{gao_electron_2017,
	title = {Electron ptychographic microscopy for three-dimensional imaging},
	volume = {8},
	url = {https://www.semanticscholar.org/paper/b4e5bcf25c588b620960798ac5b0b9fb69c16e56},
	doi = {10.1038/s41467-017-00150-1},
	abstract = {S2 TL;DR: Three-dimensional ptychographic imaging with electrons has remained a challenge because, unlike X-rays, electrons are easily scattered by atoms, but multi-slice methods to electrons in the multiple scattering regime are extended, paving the way to nanometer-scale 3D structure determination with electrons.},
	journal = {Nature Communications},
	author = {Gao, Si and Wang, Peng and Zhang, Fucai and Martinez, G. and Nellist, P. and Pan, Xiaoqing and Kirkland, A.},
	year = {2017},
	pages = {null},
}

@article{mao_multi-convergence-angle_2024,
	title = {Multi-{Convergence}-{Angle} {Ptychography} with {Simultaneous} {Strong} {Contrast} and {High} {Resolution}},
	url = {https://www.semanticscholar.org/paper/0d5020e504e6ee51bea37c3c4a6769f032e8677e},
	abstract = {Advances in bioimaging methods and hardware facilities have revolutionised the determination of numerous biological structures at atomic or near-atomic resolution. Among these developments, electron ptychography has recently attracted considerable attention because of its superior resolution, remarkable sensitivity to light elements, and high electron dose efficiency. Here, we introduce an innovative approach called multi-convergence-angle (MCA) ptychography, which can simultaneously enhance both contrast and resolution with continuous information transfer across a wide spectrum of spatial frequency. Our work provides feasibility of future applications of MCA-ptychography in providing high-quality two-dimensional images as input to three-dimensional reconstruction methods, thereby facilitating more accurate determination of biological structures.},
	author = {Mao, Wei and Zhang, Weiyang and Huang, Chen and Zhou, Liqi and Kim, Judy S. and Gao, Si and Lei, Yu and Wu, Xiaopeng and Hu, Yiming and Pei, Xudong and Fang, Weina and Liu, Xiaoguo and Song, Jingdong and Fan, Chunhai and Nie, Yuefeng and Kirkland, A. and Wang, Peng},
	year = {2024},
}

@article{strauch_live_2021,
	title = {Live {Processing} of {Momentum}-{Resolved} {STEM} {Data} for {First} {Moment} {Imaging} and {Ptychography}},
	volume = {27},
	url = {https://www.semanticscholar.org/paper/ad11b03da478972ef646d16b5bb62bc721d4fe1e},
	doi = {10.1017/S1431927621012423},
	abstract = {Abstract A reformulated implementation of single-sideband ptychography enables analysis and display of live detector data streams in 4D scanning transmission electron microscopy (STEM) using the LiberTEM open-source platform. This is combined with live first moment and further virtual STEM detector analysis. Processing of both real experimental and simulated data shows the characteristics of this method when data are processed progressively, as opposed to the usual offline processing of a complete data set. In particular, the single-sideband method is compared with other techniques such as the enhanced ptychographic engine in order to ascertain its capability for structural imaging at increased specimen thickness. Qualitatively interpretable live results are obtained also if the sample is moved, or magnification is changed during the analysis. This allows live optimization of instrument as well as specimen parameters during the analysis. The methodology is especially expected to improve contrast- and dose-efficient in situ imaging of weakly scattering specimens, where fast live feedback during the experiment is required.},
	journal = {Microscopy and Microanalysis},
	author = {Strauch, A. and Weber, D. and Clausen, Alexander and Lesnichaia, Anastasiia and Bangun, Arya and März, B. and Lyu, Feng Jiao and Chen, Qing and Rosenauer, A. and Dunin‐Borkowski, R. and Müller-Caspary, K.},
	year = {2021},
	pages = {1078 -- 1092},
}

@article{you_magnetic_2023,
	title = {Magnetic phase imaging using {Lorentz} near-field electron ptychography},
	volume = {null},
	url = {https://www.semanticscholar.org/paper/2397cf2fea7ff8aa2bf3808126b8d4aaf8320a51},
	doi = {10.22443/rms.mmc2023.175},
	abstract = {Over the past few years, the combination of diffuser and near-field electron ptychography has drawn more attention by its ability to recover large field of view with few diffraction patterns. In this paper, we purpose a novel design and implementation of amplitude diffuser. The amplitude diffuser introduces structures to the illumination while reducing the inelastic scattering. And the amplitude diffuser is implemented at the condenser lens aperture, allowing us to vary the illumination size under the same microscope setup. We demonstrate the reconstruction results under both conventional Transmission Electron Microscopy (TEM) mode as well as Lorentz mode.},
	journal = {Proceedings of the Microscience Microscopy Congress 2023 incorporating EMAG 2023},
	author = {You, Sheng and Lu, Peng-Han and Kov'acs, A. and Schachinger, T. and Allars, Frederick and Dunin‐Borkowski, R. and Maiden, A.},
	year = {2023},
	pages = {null},
}

@article{song_atomic_2019,
	title = {Atomic {Resolution} {Defocused} {Electron} {Ptychography} at {Low} {Dose} with a {Fast}, {Direct} {Electron} {Detector}},
	volume = {9},
	url = {https://www.semanticscholar.org/paper/1fb08c995bb41d086523115efde7052d93cff3b0},
	doi = {10.1038/s41598-019-40413-z},
	abstract = {null},
	journal = {Scientific Reports},
	author = {Song, Jiamei and Allen, C. and Gao, Si and Huang, Chen and Sawada, H. and Pan, Xiaoqing and Warner, J. and Wang, Peng and Kirkland, A.},
	year = {2019},
	pages = {null},
}

@article{pelz_reconstructing_2020,
	title = {Reconstructing the {Scattering} {Matrix} from {Scanning} {Electron} {Diffraction} {Measurements} {Alone}},
	url = {https://www.semanticscholar.org/paper/4b9fa7b499cf31ac6c3663dc552d9309ec25008c},
	doi = {10.1103/PhysRevResearch.3.023159},
	abstract = {Three-dimensional phase contrast imaging of multiply-scattering samples in X-ray and electron microscopy is extremely challenging, due to small numerical apertures, the unavailability of wavefront shaping optics, and the highly nonlinear inversion required from intensity-only measurements. In this work, we present a new algorithm using the scattering matrix formalism to solve the scattering from a non-crystalline medium from scanning diffraction measurements, and recover the illumination aberrations. Our method will enable 3D imaging and materials characterization at high resolution for a wide range of materials.},
	journal = {arXiv: Computational Physics},
	author = {Pelz, Philipp M. and Brown, H. and Ciston, J. and Findlay, S. and Zhang, Yaqian and Scott, M. and Ophus, C.},
	year = {2020},
	pages = {null},
}

@article{song_hollow_2018,
	title = {Hollow {Electron} {Ptychographic} {Diffractive} {Imaging}.},
	volume = {121 14},
	url = {https://www.semanticscholar.org/paper/0a1d964b23b5e6073f9872444acbedc4a731e34d},
	doi = {10.1103/PhysRevLett.121.146101},
	abstract = {We report a method for quantitative phase recovery and simultaneous electron energy loss spectroscopy analysis using ptychographic reconstruction of a data set of "hollow" diffraction patterns. This has the potential for recovering both structural and chemical information at atomic resolution with a new generation of detectors.},
	journal = {Physical review letters},
	author = {Song, Biying and Ding, Zhiyuan and Allen, C. and Sawada, H. and Zhang, Fucai and Pan, Xiaoqing and Warner, J. and Kirkland, A. and Wang, Peng},
	year = {2018},
	pages = {146101},
}

@article{ophus_four-dimensional_2019,
	title = {Four-{Dimensional} {Scanning} {Transmission} {Electron} {Microscopy} ({4D}-{STEM}): {From} {Scanning} {Nanodiffraction} to {Ptychography} and {Beyond}},
	volume = {25},
	url = {https://www.semanticscholar.org/paper/a2d30ed583d968d7e7b22c6d7568d196234d2d13},
	doi = {10.1017/S1431927619000497},
	abstract = {Abstract Scanning transmission electron microscopy (STEM) is widely used for imaging, diffraction, and spectroscopy of materials down to atomic resolution. Recent advances in detector technology and computational methods have enabled many experiments that record a full image of the STEM probe for many probe positions, either in diffraction space or real space. In this paper, we review the use of these four-dimensional STEM experiments for virtual diffraction imaging, phase, orientation and strain mapping, measurements of medium-range order, thickness and tilt of samples, and phase contrast imaging methods, including differential phase contrast, ptychography, and others.},
	journal = {Microscopy and Microanalysis},
	author = {Ophus, C.},
	year = {2019},
	pages = {563 -- 582},
	file = {Ophus - 2019 - Four-Dimensional Scanning Transmission Electron Mi.pdf:files/1224/Ophus - 2019 - Four-Dimensional Scanning Transmission Electron Mi.pdf:application/pdf},
}

@article{blackburn_practical_2020,
	title = {Practical {Implementation} of {High}-{Resolution} {Electron} {Ptychography} and {Comparison} with {Off}-{Axis} {Electron} {Holography}.},
	volume = {null},
	url = {https://www.semanticscholar.org/paper/e8a292e1648a4380c1890d57d047234fa93ec541},
	doi = {10.1093/jmicro/dfaa055},
	abstract = {Ptychography is a coherent diffractive imaging technique that can determine how an electron wave is transmitted through an object by probing it in many small overlapping regions and processing the diffraction data obtained at each point. The resulting electron transmission model describes both phase and amplitude changes to the electron wave. Ptychography has been adopted in transmission electron microscopy (TEM) in recent years following advances in high speed direct electron detectors and computer algorithms which now make the technique suitable for practical applications. Its ability to retrieve quantitative phase information at high spatial resolution makes it a plausible alternative or complement to electron holography. Furthermore, unlike off-axis electron holography, it can provide phase information without an electron bi-prism assembly or the requirement of a minimally structured region adjacent to the region of interest in the object. However, it does require a well calibrated scanning TEM and a well-managed workflow to manage the calibration, data acquisition and reconstruction process to yield a practical technique. Here we detail this workflow and highlight how this is greatly assisted by acquisition management software. Through experimental data and modelling we also explore the similarities and differences between high resolution ptychography and electron holography. Both techniques show a dependence of the recovered phase on the crystalline orientation of the material which is attributable to dynamical scattering. However, the exact nature of the variation differs reflecting fundamental expectations, but nonetheless equally useful information is obtained from electron holography and the ptychographically determined object transmission function.},
	journal = {Microscopy},
	author = {Blackburn, A. and McLeod, R.},
	year = {2020},
	pages = {null},
}

@article{cao_automatic_2022,
	title = {Automatic parameter selection for electron ptychography via {Bayesian} optimization},
	volume = {12},
	url = {https://www.semanticscholar.org/paper/3dafb2063802ef2639a447056fd556c6de80054c},
	doi = {10.1038/s41598-022-16041-5},
	abstract = {null},
	journal = {Scientific Reports},
	author = {Cao, Michael C. and Chen, Zhen and Jiang, Yi and Han, Yimo},
	year = {2022},
	pages = {null},
}

@article{brown_single-projection_2020,
	title = {A single-projection three-dimensional reconstruction algorithm for scanning transmission electron microscopy data},
	url = {https://www.semanticscholar.org/paper/fb4d57893731c6b93c5793882f135e4e72ed3e1a},
	abstract = {Increasing interest in three-dimensional nanostructures adds impetus to electron microscopy techniques capable of imaging at or below the nanoscale in three dimensions. We present a reconstruction algorithm that takes as input a focal series of four-dimensional scanning transmission electron microscopy (4D-STEM) data. We apply the approach to a lead iridate, Pb₂Ir₂O₇, and yttrium-stabilized zirconia,Y$_{\textrm{0.095}}$Zr$_{\textrm{0.905}}$O₂ , heterostructure from data acquired with the specimen in a single plan-view orientation, with the epitaxial layers stacked along the beam direction. We demonstrate that Pb-Ir atomic columns are visible in the uppermost layers of the reconstructed volume. We compare this approach to the alternative techniques of depth sectioning using differential phase contrast scanning transmission electron microscopy (DPC-STEM) and multislice ptychographic reconstruction.},
	author = {Brown, H. and Pelz, Philipp M. and Hsu, S. and Zhang, Zimeng and Ramesh, R. and Inzani, Katherine and Sheridan, Evan and Griffin, S. and Schloz, M. and Pekin, T. and Koch, C. and Findlay, S. and Allen, L. and Scott, M. and Ophus, C. and Ciston, J.},
	year = {2020},
}

@article{findlay_scattering_2021,
	title = {Scattering {Matrix} {Determination} in {Crystalline} {Materials} from {4D} {Scanning} {Transmission} {Electron} {Microscopy} at a {Single} {Defocus} {Value}},
	volume = {27},
	url = {https://www.semanticscholar.org/paper/22e3ae2ce6736fbc7e10924e672edcc16c456eb9},
	doi = {10.1017/S1431927621000490},
	abstract = {Abstract Recent work has revived interest in the scattering matrix formulation of electron scattering in transmission electron microscopy as a stepping stone toward atomic-resolution structure determination in the presence of multiple scattering. We discuss ways of visualizing the scattering matrix that make its properties clear. Through a simulation-based case study incorporating shot noise, we shown how regularizing on this continuity enables the scattering matrix to be reconstructed from 4D scanning transmission electron microscopy (STEM) measurements from a single defocus value. Intriguingly, for crystalline samples, this process also yields the sample thickness to nanometer accuracy with no a priori knowledge about the sample structure. The reconstruction quality is gauged by using the reconstructed scattering matrix to simulate STEM images at defocus values different from that of the data from which it was reconstructed.},
	journal = {Microscopy and Microanalysis},
	author = {Findlay, S. and Brown, H. and Pelz, Philipp M. and Ophus, C. and Ciston, J. and Allen, L.},
	year = {2021},
	pages = {744 -- 757},
}

@article{ding_three-dimensional_2022,
	title = {Three-dimensional electron ptychography of organic–inorganic hybrid nanostructures},
	volume = {13},
	url = {https://www.semanticscholar.org/paper/9e588f998b168441d8cd9c89817ede01d4b6900e},
	doi = {10.1038/s41467-022-32548-x},
	abstract = {null},
	journal = {Nature Communications},
	author = {Ding, Zhiyuan and Gao, Si and Fang, Weina and Huang, Chen and Zhou, Liqi and Pei, Xudong and Liu, Xiaoguo and Pan, Xiaoqing and Fan, Chunhai and Kirkland, A. and Wang, Peng},
	year = {2022},
	pages = {null},
}

@article{rodenburg_phase_2004,
	title = {A phase retrieval algorithm for shifting illumination},
	volume = {85},
	issn = {0003-6951, 1077-3118},
	url = {https://pubs.aip.org/apl/article/85/20/4795/329473/A-phase-retrieval-algorithm-for-shifting},
	doi = {10.1063/1.1823034},
	abstract = {We propose a method of iterative phase retrieval that uses measured intensities in the diffraction plane to solve the phase problem in a way that bypasses the problem of lens aberration, leading to greatly improved spatial resolution. This method is stable, easy to implement experimentally, and can be used to view a large area of the specimen when that is desired.},
	language = {en},
	number = {20},
	urldate = {2024-09-25},
	journal = {Applied Physics Letters},
	author = {Rodenburg, J. M. and Faulkner, H. M. L.},
	month = nov,
	year = {2004},
	pages = {4795--4797},
	file = {Rodenburg and Faulkner - 2004 - A phase retrieval algorithm for shifting illuminat.pdf:files/603/Rodenburg and Faulkner - 2004 - A phase retrieval algorithm for shifting illuminat.pdf:application/pdf},
}

@article{satta_formation_2021,
	title = {Formation {Mechanisms} and {Phase} {Stability} of {Solid}-{State} {Grown} {CsPbI3} {Perovskites}},
	volume = {11},
	copyright = {http://creativecommons.org/licenses/by/3.0/},
	issn = {2079-4991},
	url = {https://www.mdpi.com/2079-4991/11/7/1823},
	doi = {10.3390/nano11071823},
	abstract = {CsPbI3 inorganic perovskite is synthesized by a solvent-free, solid-state reaction, and its structural and optical properties can be deeply investigated using a multi-technique approach. X-ray Diffraction (XRD) and Raman measurements, optical absorption, steady-time and time-resolved luminescence, as well as High-Resolution Transmission Electron Microscopy (HRTEM) imaging, were exploited to understand phase evolution as a function of synthesis time length. Nanoparticles with multiple, well-defined crystalline domains of different crystalline phases were observed, usually surrounded by a thin, amorphous/out-of-axis shell. By increasing the synthesis time length, in addition to the pure α phase, which was rapidly converted into the δ phase at room temperature, a secondary phase, Cs4PbI6, was observed, together with the 715 nm-emitting γ phase.},
	language = {en},
	number = {7},
	urldate = {2024-09-27},
	journal = {Nanomaterials},
	publisher = {Multidisciplinary Digital Publishing Institute},
	author = {Satta, Jessica and Casu, Alberto and Chiriu, Daniele and Carbonaro, Carlo Maria and Stagi, Luigi and Ricci, Pier Carlo},
	month = jul,
	year = {2021},
	note = {Number: 7},
	keywords = {CsPbI$_{\textrm{3}}$, inorganic lead halide perovskites, phase stability, solid-state synthesis},
	pages = {1823},
	file = {Full Text PDF:files/605/Satta et al. - 2021 - Formation Mechanisms and Phase Stability of Solid-.pdf:application/pdf},
}

@article{bryce_chemical_2022,
	title = {Chemical durability and degradation mechanisms of {CsPbI3} as a potential host phase for cesium and iodine sequestration},
	volume = {12},
	issn = {2046-2069},
	url = {https://pubs.rsc.org/en/content/articlelanding/2022/ra/d2ra01259f},
	doi = {10.1039/D2RA01259F},
	abstract = {Effective nuclear waste management of radioactive cesium and off-gas iodine from complex waste streams of used fuels is essential for the sustainable development of advanced nuclear fuel cycles. Once cesium and iodine are separated from their respective waste streams, host phases are required to immobilize them into a durable waste form matrix for long-term disposition. The inorganic metal halide perovskite, CsPbI3, has a unique crystal structure capable of incorporating both cesium and iodine simultaneously. Exposure to groundwater in geological repositories is a long-term concern for waste forms, as this may cause corrosion and decrease the waste form's ability to retain radionuclides. In this study, we explore the potential of CsPbI3 perovskite as a promising host phase to incorporate Cs and I, and investigate its chemical durability and degradation mechanisms in an aqueous environment. CsPbI3 was synthesized through a solution-based method and was consolidated into dense pellets by spark plasma sintering. The chemical durability of the CsPbI3 pellets was evaluated by static leaching tests in deionized water at different temperatures of 25, 58, and 90 °C. The elemental release mechanisms and surface alteration of the monolithic CsPbI3 pellets were investigated. Both I and Cs displayed a non-congruent leaching behavior and faster release rates as compared to Pb, particularly at longer leaching durations and higher temperatures. At the initial leaching stage, a PbI2 alteration layer formed on the surface of the pellet due to the rapid release of Cs and I, followed by the formation of a PbI(OH) alteration layer. The activation energies for both dissolution and diffusion controlled mechanisms were determined to be 44.90 kJ mol−1 and 45.40 kJ mol−1 for Pb, 27.10 kJ mol−1 and 40.82 kJ mol−1 for I and 24.27 kJ mol−1 and 23.86 kJ mol−1 for Cs, respectively. These results show a clear decrease in activation energies from Pb to I and Cs, suggesting a preferential release of I and Cs. The solution-based synthesis of CsPbI3 as a host phase for Cs and I and the fundamental understanding of the chemical durability and degradation behavior will be useful for further exploring its application for immobilizing iodine and cesium into final durable waste forms for long-term geological disposition.},
	language = {en},
	number = {20},
	urldate = {2024-09-27},
	journal = {RSC Advances},
	publisher = {The Royal Society of Chemistry},
	author = {Bryce, Keith and Yang, Kun and Wang, Yachun and Lian, Jie},
	month = apr,
	year = {2022},
	pages = {12242--12252},
	file = {Full Text PDF:files/607/Bryce et al. - 2022 - Chemical durability and degradation mechanisms of .pdf:application/pdf},
}

@article{teunissen_additivity_2023,
	title = {Additivity of {Atomic} {Strain} {Fields} as a {Tool} to {Strain}-{Engineering} {Phase}-{Stabilized} {CsPbI3} {Perovskites}},
	volume = {127},
	issn = {1932-7447},
	url = {https://doi.org/10.1021/acs.jpcc.3c05770},
	doi = {10.1021/acs.jpcc.3c05770},
	abstract = {CsPbI3 is a promising perovskite material for photovoltaic applications in its photoactive perovskite or black phase. However, the material degrades to a photovoltaically inactive or yellow phase at room temperature. Various mitigation strategies are currently being developed to increase the lifetime of the black phase, many of which rely on inducing strains in the material that hinder the black-to-yellow phase transition. Physical insight into how these strategies exactly induce strain as well as knowledge of the spatial extent over which these strains impact the material is crucial to optimize these approaches but is still lacking. Herein, we combine machine learning potential-based molecular dynamics simulations with our in silico strain engineering approach to accurately quantify strained large-scale atomic structures on a nanosecond time scale. To this end, we first model the strain fields introduced by atomic substitutions as they form the most elementary strain sources. We demonstrate that the magnitude of the induced strain fields decays exponentially with the distance from the strain source, following a decay rate that is largely independent of the specific substitution. Second, we show that the total strain field induced by multiple strain sources can be predicted to an excellent approximation by summing the strain fields of each individual source. Finally, through a case study, we illustrate how this additive character allows us to explain how complex strain fields, induced by spatially extended strain sources, can be predicted by adequately combining the strain fields caused by local strain sources. Hence, the strain additivity proposed here can be adopted to further our insight into the complex strain behavior in perovskites and to design strain from the atomic level onward to enhance their sought-after phase stability.},
	number = {48},
	urldate = {2024-09-27},
	journal = {The Journal of Physical Chemistry C},
	publisher = {American Chemical Society},
	author = {Teunissen, Johannes L. and Braeckevelt, Tom and Skvortsova, Irina and Guo, Jinhui and Pradhan, Bapi and Debroye, Elke and Roeffaers, Maarten B. J. and Hofkens, Johan and Van Aert, Sandra and Bals, Sara and Rogge, Sven M. J. and Van Speybroeck, Veronique},
	month = dec,
	year = {2023},
	pages = {23400--23411},
	file = {Full Text PDF:files/611/Teunissen et al. - 2023 - Additivity of Atomic Strain Fields as a Tool to St.pdf:application/pdf},
}

@article{muller_reaction-diffusion_2024,
	title = {Reaction-diffusion study of electron-beam-induced contamination growth},
	volume = {264},
	issn = {03043991},
	url = {https://linkinghub.elsevier.com/retrieve/pii/S0304399124000743},
	doi = {10.1016/j.ultramic.2024.113995},
	abstract = {A time-dependent reaction-diffusion model was elaborated to better understand the dynamical growth of contamination on surfaces illuminated by an electron beam. The goal of this work was to fully describe the flow of hydrocarbon molecules, denoted as contaminants, and their polymerization in the irradiated area with the number of parameters reduced to a minimum necessary. It was considered that the diffusion process of con­ taminants is driven by the gradient of their surface density generated by the impact of a circular homogeneous electron beam. The contribution of the residual gas atmosphere in the instrument was described by the tendency to reestablish the initial equilibrium surface density of contaminants before irradiation. The four unknown pa­ rameters of the model, the electron interaction cross-section, the diffusion coefficient, the initial surface density of contaminants, and the frequency of the supply of contaminants from the residual gas atmosphere were determined by comparing the modeled contamination growth with experimental results. The experiments were designed such that the influence of the single parameters could be unequivocally separated. To follow the dynamical evolution of the system and to generate time-resolved distinct experimental data, successive contamination measurements were performed at short time intervals up to 20 min. The local height and shape of the grown contamination were quantified by evaluating high-angle annular dark-field (HAADF) scanningtransmission- electron-microcopy (STEM) image intensities and corresponding Monte-Carlo simulations. Our model also applies to nonhomogeneous initial conditions like the reduced local surface density of contaminants after previous beam-showering. The dynamic analyses of this process might provide hints regarding the relative size of the contaminant molecules and also indicate some measures for the reduction of contamination growth.},
	language = {en},
	urldate = {2024-09-30},
	journal = {Ultramicroscopy},
	author = {Müller, Erich and Adrion, Katharina and Hugenschmidt, Milena and Gerthsen, Dagmar},
	month = oct,
	year = {2024},
	pages = {113995},
	file = {Müller et al. - 2024 - Reaction-diffusion study of electron-beam-induced .pdf:files/649/Müller et al. - 2024 - Reaction-diffusion study of electron-beam-induced .pdf:application/pdf},
}

@article{pelz_low-dose_2017,
	title = {Low-dose cryo electron ptychography via non-convex {Bayesian} optimization},
	volume = {7},
	issn = {2045-2322},
	doi = {10.1038/s41598-017-07488-y},
	abstract = {Electron ptychography has seen a recent surge of interest for phase sensitive imaging at atomic or near-atomic resolution. However, applications are so far mainly limited to radiation-hard samples, because the required doses are too high for imaging biological samples at high resolution. We propose the use of non-convex Bayesian optimization to overcome this problem, and show via numerical simulations that the dose required for successful reconstruction can be reduced by two orders of magnitude compared to previous experiments. As an important application we suggest to use this method for imaging single biological macromolecules at cryogenic temperatures and demonstrate 2D single-particle reconstructions from simulated data with a resolution up to 5.4 Å at a dose of 20e - /Å2. When averaging over only 30 low-dose datasets, a 2D resolution around 3.5 Å is possible for macromolecular complexes even below 100 kDa. With its independence from the microscope transfer function, direct recovery of phase contrast, and better scaling of signal-to-noise ratio, low-dose cryo electron ptychography may become a promising alternative to Zernike phase-contrast microscopy.},
	language = {eng},
	number = {1},
	journal = {Scientific Reports},
	author = {Pelz, Philipp Michael and Qiu, Wen Xuan and Bücker, Robert and Kassier, Günther and Miller, R. J. Dwayne},
	month = aug,
	year = {2017},
	pages = {9883},
	file = {Full Text:files/666/Pelz et al. - 2017 - Low-dose cryo electron ptychography via non-convex.pdf:application/pdf},
}

@article{lozano_low-dose_2018,
	title = {Low-{Dose} {Aberration}-{Free} {Imaging} of {Li}-{Rich} {Cathode} {Materials} at {Various} {States} of {Charge} {Using} {Electron} {Ptychography}},
	volume = {18},
	issn = {1530-6984},
	url = {https://doi.org/10.1021/acs.nanolett.8b02718},
	doi = {10.1021/acs.nanolett.8b02718},
	abstract = {Imaging the complete atomic structure of materials, including light elements, with minimal beam-induced damage of the sample is a long-standing challenge in electron microscopy. Annular bright-field scanning transmission electron microscopy is often used to image elements with low atomic numbers, but due to its low efficiency and high sensitivity to precise imaging parameters it comes at the price of potentially significant beam damage. In this paper, we show that electron ptychography is a powerful technique to retrieve reconstructed phase images that provide the full structure of beam-sensitive materials containing light and heavy elements. Due to its much higher efficiency, we can reduce the beam currents used down to the subpicoampere range. Electron ptychography also allows residual lens aberrations to be corrected at the postprocessing stage, which avoids the need for fine-tuning of the probe that would result in further beam damage and provides aberration-free reconstructed phase images. We have used electron ptychography to obtain structural information from aberration-free reconstructed phase images in the technologically relevant lithium-rich transition metal oxides at different states of charge. We can unambiguously determine the position of the lithium and oxygen atomic columns while amorphization of the surface, formation of beam-induced surface reconstruction layers, or migration of transition metals to the alkali layers are drastically reduced.},
	number = {11},
	urldate = {2024-10-01},
	journal = {Nano Letters},
	publisher = {American Chemical Society},
	author = {Lozano, Juan G. and Martinez, Gerardo T. and Jin, Liyu and Nellist, Peter D. and Bruce, Peter G.},
	month = nov,
	year = {2018},
	pages = {6850--6855},
	file = {Full Text PDF:files/674/Lozano et al. - 2018 - Low-Dose Aberration-Free Imaging of Li-Rich Cathod.pdf:application/pdf},
}

@article{thibault_maximum-likelihood_2012,
	title = {Maximum-likelihood refinement for coherent diffractive imaging},
	volume = {14},
	issn = {1367-2630},
	url = {https://iopscience.iop.org/article/10.1088/1367-2630/14/6/063004},
	doi = {10.1088/1367-2630/14/6/063004},
	number = {6},
	urldate = {2024-10-01},
	journal = {New Journal of Physics},
	author = {Thibault, P and Guizar-Sicairos, M},
	month = jun,
	year = {2012},
	pages = {063004},
	file = {Accepted Version:files/676/Thibault and Guizar-Sicairos - 2012 - Maximum-likelihood refinement for coherent diffrac.pdf:application/pdf},
}

@article{hegerl_dynamische_1970,
	title = {Dynamische {Theorie} der {Kristallstrukturanalyse} durch {Elektronenbeugung} im inhomogenen {Primärstrahlwellenfeld}},
	volume = {74},
	copyright = {http://onlinelibrary.wiley.com/termsAndConditions\#vor},
	issn = {0005-9021, 0005-9021},
	url = {https://onlinelibrary.wiley.com/doi/10.1002/bbpc.19700741112},
	doi = {10.1002/bbpc.19700741112},
	abstract = {Abstract
            
              Vor einiger Zeit wurde ein neues Prinzip zur Registrierung der Gesamtinformation (Amplituden und Phasen) in einem Beugungsbild angegeben, welches nicht – wie die Holographie – Interferenzen der Streuwellen mit
              einer
              Referenzwelle voraussetzt. Von wesentlicher Bedeutung sind hierbei Interferenzen
              benachbarter
              Streuwellen, die entstehen, wenn die Objektfunktion
              Q(x,y)
              mit einer verallgemeinerten Primärwellenfunktion
              p(x,y)
              multipliziert wird. Im Fourierraum (Beugungsdiagramm) entstehen dabei Faltungen der Fouriertransformierten dieser Funktionen; durch geeignete Formung von
              p(x,y)
              läßt sich erreichen, daß die erwähnten, zur Phasenbestimmung ausnutzbaren Interferenzen benachbarter Streustrahlen resultieren. Zum Unterschied zur Holographie soll dieses Verfahren “Ptychographie” (ξ = Falte) genannt werden. Das Verfahren ist für periodische und nichtperiodische Strukturen anwendbar; die einfachsten und übersichtlichsten Beziehungen ergeben sich für Kreuzgitter. In der vorliegenden Arbeit wird die Theorie für Raumgitter mit und ohne Berücksichtigung der dynamischen Theorie erweitert. An einem praktischen Beispiel werden die auftretenden Effekte demonstriert.
            
          , 
            
              Some time ago a new principle was proposed for the registration of the complete information (amplitudes and phases) in a diffraction diagram, which does not –as does Holography – require the interference of the scattered waves with a single reference wave. The basis of the principle lies in the interference of neighbouring scattered waves which result when the object function
              Q(x,y)
              is multiplied by a generalized primary wave function
              p(x,y)
              in Fourier space (diffraction diagram) this is a convolution of the Fourier transforms of these functions. The above mentioned interferences necessary for the phase determination can be obtained by suitable choice of the shape of
              p(x,y)
              . To distinguish it from holography this procedure is designated “ptychography” (ξ = fold). The procedure is applicable to periodic and aperiodic structures. The relationships are simplest for plane lattices. In this paper the theory is extended to space lattices both with and without consideration of the dynamic theory. The resulting effects are demonstrated using a practical example.},
	language = {en},
	number = {11},
	urldate = {2024-10-01},
	journal = {Berichte der Bunsengesellschaft für physikalische Chemie},
	author = {Hegerl, R. and Hoppe, W.},
	month = nov,
	year = {1970},
	pages = {1148--1154},
}

@article{maiden_ptychographic_2012,
	title = {Ptychographic transmission microscopy in three dimensions using a multi-slice approach},
	volume = {29},
	copyright = {https://doi.org/10.1364/OA\_License\_v1\#VOR},
	issn = {1084-7529, 1520-8532},
	url = {https://opg.optica.org/abstract.cfm?URI=josaa-29-8-1606},
	doi = {10.1364/JOSAA.29.001606},
	language = {en},
	number = {8},
	urldate = {2024-10-01},
	journal = {Journal of the Optical Society of America A},
	author = {Maiden, A. M. and Humphry, M. J. and Rodenburg, J. M.},
	month = aug,
	year = {2012},
	pages = {1606},
	file = {Maiden et al. - 2012 - Ptychographic transmission microscopy in three dim.pdf:files/679/Maiden et al. - 2012 - Ptychographic transmission microscopy in three dim.pdf:application/pdf},
}

@article{jannis_aan_nodate,
	title = {aan de {Universiteit} {Antwerpen}, te verdedigen door},
	language = {en},
	author = {Jannis, Daen},
	file = {Jannis - aan de Universiteit Antwerpen, te verdedigen door.pdf:files/680/Jannis - aan de Universiteit Antwerpen, te verdedigen door.pdf:application/pdf},
}

@article{winkler_direct_2020,
	title = {Direct measurement of electrostatic potentials at the atomic scale: {A} conceptual comparison between electron holography and scanning transmission electron microscopy},
	volume = {210},
	issn = {03043991},
	shorttitle = {Direct measurement of electrostatic potentials at the atomic scale},
	url = {https://linkinghub.elsevier.com/retrieve/pii/S0304399119303092},
	doi = {10.1016/j.ultramic.2019.112926},
	abstract = {Off-axis electron holography and first moment STEM are sensitive to electromagnetic potentials or fields, respectively. In this work, we investigate in what sense the results obtained from both techniques are equivalent and work out the major differences. The analysis is focused on electrostatic (Coulomb) potentials at atomic spatial resolution. It is shown that the probe-forming/objective aperture strongly affects the reconstructed electrostatic potentials and that, as a result of the different illumination setups, dynamical diffraction effects show a specific response with increasing specimen thickness. It is shown that thermal diffuse scattering is negligible for a wide range of specimen thicknesses, when evaluating the first moment of diffraction patterns.},
	language = {en},
	urldate = {2024-10-01},
	journal = {Ultramicroscopy},
	author = {Winkler, Florian and Barthel, Juri and Dunin-Borkowski, Rafal E. and Müller-Caspary, Knut},
	month = mar,
	year = {2020},
	pages = {112926},
	file = {Winkler et al. - 2020 - Direct measurement of electrostatic potentials at .pdf:files/683/Winkler et al. - 2020 - Direct measurement of electrostatic potentials at .pdf:application/pdf},
}

@book{williams_transmission_2009,
	address = {New York},
	edition = {2. ed},
	title = {Transmission electron microscopy: a textbook for materials science},
	isbn = {978-0-387-76500-6},
	shorttitle = {Transmission electron microscopy},
	language = {en},
	publisher = {Springer},
	author = {Williams, David B. and Carter, C. Barry},
	year = {2009},
	file = {Williams and Carter - 2009 - Transmission electron microscopy a textbook for m.pdf:files/687/Williams and Carter - 2009 - Transmission electron microscopy a textbook for m.pdf:application/pdf},
}

@incollection{carter_imaging_2016,
	address = {Cham},
	title = {Imaging in the {STEM}},
	copyright = {http://www.springer.com/tdm},
	isbn = {978-3-319-26649-7 978-3-319-26651-0},
	url = {http://link.springer.com/10.1007/978-3-319-26651-0_11},
	doi = {10.1007/978-3-319-26651-0_11},
	language = {en},
	urldate = {2024-10-01},
	booktitle = {Transmission {Electron} {Microscopy}},
	publisher = {Springer International Publishing},
	author = {Pennycook, Stephen J.},
	editor = {Carter, C. Barry and Williams, David B.},
	year = {2016},
	pages = {283--342},
	file = {Pennycook - 2016 - Imaging in the STEM.pdf:files/688/Pennycook - 2016 - Imaging in the STEM.pdf:application/pdf},
}

@article{madsen_abtem_2021,
	title = {The {abTEM} code: transmission electron microscopy from first principles},
	volume = {1},
	issn = {2732-5121},
	shorttitle = {The {abTEM} code},
	url = {https://open-research-europe.ec.europa.eu/articles/1-24/v1},
	doi = {10.12688/openreseurope.13015.1},
	abstract = {Simulation of transmission electron microscopy (TEM) images or diffraction patterns is often required to interpret experimental data. Since nuclear cores dominate electron scattering, the scattering potential is typically described using the independent atom model, which completely neglects valence bonding and its effect on the transmitting electrons. As instrumentation has advanced, new measurements have revealed subtle details of the scattering potential that were previously not accessible to experiment.
            
            We have created an open-source simulation code designed to meet these demands by integrating the ability to calculate the potential via density functional theory (DFT) with a flexible modular software design. abTEM can simulate most standard imaging modes and incorporates the latest algorithmic developments. The development of new techniques requires a program that is accessible to domain experts without extensive programming experience. abTEM is written purely in Python and designed for easy modification and extension.
            
            The effective use of modern open-source libraries makes the performance of abTEM highly competitive with existing optimized codes on both CPUs and GPUs and allows us to leverage an extensive ecosystem of libraries, such as the Atomic Simulation Environment and the DFT code GPAW. abTEM is designed to work in an interactive Python notebook, creating a seamless and reproducible workflow from defining an atomic structure, calculating molecular dynamics (MD) and electrostatic potentials, to the analysis of results, all in a single, easy-to-read document. 
            
            This article provides ongoing documentation of abTEM development. In this first version, we show use cases for hexagonal boron nitride, where valence bonding can be detected, a 4D-STEM simulation of molybdenum disulfide including ptychographic phase reconstruction, a comparison of MD and frozen phonon modeling for convergent-beam electron diffraction of a 2.6-million-atom silicon system, and a performance comparison of our fast implementation of the PRISM algorithm for a decahedral 20000-atom gold nanoparticle.},
	language = {en},
	urldate = {2024-10-01},
	journal = {Open Research Europe},
	author = {Madsen, Jacob and Susi, Toma},
	month = mar,
	year = {2021},
	pages = {24},
}

@article{chen_electron_2021,
	title = {Electron ptychography achieves atomic-resolution limits set by lattice vibrations},
	volume = {372},
	issn = {0036-8075, 1095-9203},
	url = {https://www.science.org/doi/10.1126/science.abg2533},
	doi = {10.1126/science.abg2533},
	abstract = {Transmission electron microscopes use electrons with wavelengths of a few picometers, potentially capable of imaging individual atoms in solids at a resolution ultimately set by the intrinsic size of an atom. Unfortunately, due to imperfections in the imaging lenses and multiple scattering of electrons in the sample, the image resolution reached is 3 to 10 times worse. Here, by inversely solving the multiple scattering problem and overcoming the aberrations of the electron probe using electron ptychography to recover a linear phase response in thick samples, we demonstrate an instrumental blurring of under 20 picometers. The widths of atomic columns in the measured electrostatic potential are now no longer limited by the imaging system, but instead by the thermal fluctuations of the atoms. We also demonstrate that electron ptychography can potentially reach a sub-nanometer depth resolution and locate embedded atomic dopants in all three dimensions with only a single projection measurement.},
	language = {en},
	number = {6544},
	urldate = {2024-10-01},
	journal = {Science},
	author = {Chen, Zhen and Jiang, Yi and Shao, Yu-Tsun and Holtz, Megan E. and Odstrčil, Michal and Guizar-Sicairos, Manuel and Hanke, Isabelle and Ganschow, Steffen and Schlom, Darrell G. and Muller, David A.},
	month = may,
	year = {2021},
	pages = {826--831},
	file = {Chen et al. - 2021 - Electron ptychography achieves atomic-resolution l.pdf:files/694/Chen et al. - 2021 - Electron ptychography achieves atomic-resolution l.pdf:application/pdf},
}

@article{xiang_review_2021,
	title = {A review on the stability of inorganic metal halide perovskites: challenges and opportunities for stable solar cells},
	volume = {14},
	issn = {1754-5692, 1754-5706},
	shorttitle = {A review on the stability of inorganic metal halide perovskites},
	url = {https://xlink.rsc.org/?DOI=D1EE00157D},
	doi = {10.1039/D1EE00157D},
	abstract = {The composition, light, moisture and oxygen all affect the stability of metal halide inorganic perovskites, whose degradation mechanisms are significantly different from those of hybrid perovskites.
          , 
            Inorganic perovskite based solar cells (PSCs) have been receiving unprecedented attention worldwide in the past several years due to their higher intrinsic stability towards high temperatures and high theoretical power conversion efficiencies. Since a photovoltaic performance of 20.37\% has been achieved for inorganic PSCs recently, the operational stability of these devices has become the major bottleneck which impedes their commercialization. The high thermal stability associated with inorganic perovskites comes along with poorer phase stability compared to their hybrid counterparts and therefore needs thorough understanding. Lattice strain and vacancies within the perovskite crystals are found to be the origin of these phase instability issues. This review summarizes the progress in stability research on inorganic perovskites. Specifically, the degradation mechanisms of inorganic perovskites towards temperature, moisture and oxygen are summarized and discussed. Solutions for tackling these stability issues are reviewed and an outlook on further strategies is provided.},
	language = {en},
	number = {4},
	urldate = {2024-10-03},
	journal = {Energy \& Environmental Science},
	author = {Xiang, Wanchun and Liu, Shengzhong (Frank) and Tress, Wolfgang},
	year = {2021},
	pages = {2090--2113},
}

@article{peng_observation_2022,
	title = {Observation of formation and local structures of metal-organic layers via complementary electron microscopy techniques},
	volume = {13},
	issn = {2041-1723},
	url = {https://www.nature.com/articles/s41467-022-32330-z},
	doi = {10.1038/s41467-022-32330-z},
	abstract = {Abstract
            Metal-organic layers (MOLs) are highly attractive for application in catalysis, separation, sensing and biomedicine, owing to their tunable framework structure. However, it is challenging to obtain comprehensive information about the formation and local structures of MOLs using standard electron microscopy methods due to serious damage under electron beam irradiation. Here, we investigate the growth processes and local structures of MOLs utilizing a combination of liquid-phase transmission electron microscopy, cryogenic electron microscopy and electron ptychography. Our results show a multistep formation process, where precursor clusters first form in solution, then they are complexed with ligands to form non-crystalline solids, followed by the arrangement of the cluster-ligand complex into crystalline sheets, with additional possible growth by the addition of clusters to surface edges. Moreover, high-resolution imaging allows us to identify missing clusters, dislocations, loop and flat surface terminations and ligand connectors in the MOLs. Our observations provide insights into controllable MOL crystal morphology, defect engineering, and surface modification, thus assisting novel MOL design and synthesis.},
	language = {en},
	number = {1},
	urldate = {2024-10-03},
	journal = {Nature Communications},
	author = {Peng, Xinxing and Pelz, Philipp M. and Zhang, Qiubo and Chen, Peican and Cao, Lingyun and Zhang, Yaqian and Liao, Hong-Gang and Zheng, Haimei and Wang, Cheng and Sun, Shi-Gang and Scott, Mary C.},
	month = sep,
	year = {2022},
	pages = {5197},
	file = {Full Text:files/703/Peng et al. - 2022 - Observation of formation and local structures of m.pdf:application/pdf},
}

@article{hoppe_beugung_1969,
	title = {Beugung im inhomogenen {Primärstrahlwellenfeld}. {III}. {Amplituden}- und {Phasenbestimmung} bei unperiodischen {Objekten}},
	volume = {25},
	copyright = {http://journals.iucr.org/services/copyrightpolicy.html},
	issn = {0567-7394},
	url = {https://scripts.iucr.org/cgi-bin/paper?S0567739469001069},
	doi = {10.1107/S0567739469001069},
	number = {4},
	urldate = {2024-10-03},
	journal = {Acta Crystallographica Section A},
	author = {Hoppe, W.},
	month = jul,
	year = {1969},
	pages = {508--514},
}

@article{hoppe_beugung_1969-1,
	title = {Beugung in inhomogenen {Primärstrahlenwellenfeld}. {II}. {Lichtoptische} {Analogieversuche} zur {Phasenmessung} von {Gitterinterferenzen}},
	volume = {25},
	copyright = {http://journals.iucr.org/services/copyrightpolicy.html},
	issn = {0567-7394},
	url = {https://scripts.iucr.org/cgi-bin/paper?S0567739469001057},
	doi = {10.1107/S0567739469001057},
	number = {4},
	urldate = {2024-10-03},
	journal = {Acta Crystallographica Section A},
	author = {Hoppe, W. and Strube, G.},
	month = jul,
	year = {1969},
	pages = {502--507},
}

@article{hoppe_beugung_1969-2,
	title = {Beugung im inhomogenen {Primärstrahlwellenfeld}. {I}. {Prinzip} einer {Phasenmessung} von {Elektronenbeungungsinterferenzen}},
	volume = {25},
	copyright = {http://journals.iucr.org/services/copyrightpolicy.html},
	issn = {0567-7394},
	url = {https://scripts.iucr.org/cgi-bin/paper?S0567739469001045},
	doi = {10.1107/S0567739469001045},
	number = {4},
	urldate = {2024-10-03},
	journal = {Acta Crystallographica Section A},
	author = {Hoppe, W.},
	month = jul,
	year = {1969},
	pages = {495--501},
}

@article{rodenburg_experimental_1993,
	title = {Experimental tests on double-resolution coherent imaging via {STEM}},
	volume = {48},
	issn = {0304-3991},
	url = {https://www.sciencedirect.com/science/article/pii/0304399193901057},
	doi = {10.1016/0304-3991(93)90105-7},
	abstract = {Experimental data collected in a scanning transmission electron microscope (STEM) is used to obtain an image of the specimen at twice the conventional resolution determined by the size of the objective aperture. The measured data also provide a robust estimate of the quality of the reconstruction, determining such variables as specimen drift, defocus in the objective lens, source coherence and mechanical instability.},
	number = {3},
	urldate = {2024-10-03},
	journal = {Ultramicroscopy},
	author = {Rodenburg, J. M. and McCallum, B. C. and Nellist, P. D.},
	month = mar,
	year = {1993},
	pages = {304--314},
	file = {Rodenburg et al. - 1993 - Experimental tests on double-resolution coherent i.pdf:files/711/Rodenburg et al. - 1993 - Experimental tests on double-resolution coherent i.pdf:application/pdf;ScienceDirect Snapshot:files/710/0304399193901057.html:text/html},
}

@article{yang_electron_2017,
	title = {Electron ptychographic phase imaging of light elements in crystalline materials using {Wigner} distribution deconvolution},
	volume = {180},
	issn = {03043991},
	url = {https://linkinghub.elsevier.com/retrieve/pii/S0304399117300773},
	doi = {10.1016/j.ultramic.2017.02.006},
	language = {en},
	urldate = {2024-10-03},
	journal = {Ultramicroscopy},
	author = {Yang, Hao and MacLaren, Ian and Jones, Lewys and Martinez, Gerardo T. and Simson, Martin and Huth, Martin and Ryll, Henning and Soltau, Heike and Sagawa, Ryusuke and Kondo, Yukihito and Ophus, Colin and Ercius, Peter and Jin, Lei and Kovács, András and Nellist, Peter D.},
	month = sep,
	year = {2017},
	pages = {173--179},
	file = {Full Text:files/716/Yang et al. - 2017 - Electron ptychographic phase imaging of light elem.pdf:application/pdf},
}

@article{egerton_mechanisms_2012,
	title = {Mechanisms of radiation damage in beam‐sensitive specimens, for {TEM} accelerating voltages between 10 and 300 {kV}},
	volume = {75},
	copyright = {http://onlinelibrary.wiley.com/termsAndConditions\#vor},
	issn = {1059-910X, 1097-0029},
	url = {https://analyticalsciencejournals.onlinelibrary.wiley.com/doi/10.1002/jemt.22099},
	doi = {10.1002/jemt.22099},
	abstract = {Abstract
            
              Ionization damage (radiolysis) and knock‐on displacement are compared in terms of scattering cross section and stopping power, for thin organic specimens exposed to the electrons in a TEM. Based on stopping power, which includes secondary processes, radiolysis is found to be predominant for all incident energies (10–300 keV), even in materials containing hydrogen. For conducting inorganic specimens, knock‐on displacement is the only damage mechanism but an electron dose exceeding 1000 C cm
              −2
              is usually required. Ways of experimentally determining the damage mechanism (with a view to minimizing damage) are discussed. Microsc. Res. Tech., 2012. © 2012 Wiley Periodicals, Inc.},
	language = {en},
	number = {11},
	urldate = {2024-10-03},
	journal = {Microscopy Research and Technique},
	author = {Egerton, R.F.},
	month = nov,
	year = {2012},
	pages = {1550--1556},
	file = {Full Text:files/758/Egerton - 2012 - Mechanisms of radiation damage in beam‐sensitive s.pdf:application/pdf},
}

@article{egerton_control_2013,
	title = {Control of radiation damage in the {TEM}},
	volume = {127},
	issn = {03043991},
	url = {https://linkinghub.elsevier.com/retrieve/pii/S0304399112001763},
	doi = {10.1016/j.ultramic.2012.07.006},
	language = {en},
	urldate = {2024-10-03},
	journal = {Ultramicroscopy},
	author = {Egerton, R.F.},
	month = apr,
	year = {2013},
	pages = {100--108},
	file = {Egerton - 2013 - Control of radiation damage in the TEM.pdf:files/801/Egerton - 2013 - Control of radiation damage in the TEM.pdf:application/pdf},
}

@article{egerton_radiation_2019,
	title = {Radiation damage to organic and inorganic specimens in the {TEM}},
	volume = {119},
	issn = {09684328},
	url = {https://linkinghub.elsevier.com/retrieve/pii/S0968432818304359},
	doi = {10.1016/j.micron.2019.01.005},
	language = {en},
	urldate = {2024-10-03},
	journal = {Micron},
	author = {Egerton, R.F.},
	month = apr,
	year = {2019},
	pages = {72--87},
	file = {Egerton - 2019 - Radiation damage to organic and inorganic specimen.pdf:files/802/Egerton - 2019 - Radiation damage to organic and inorganic specimen.pdf:application/pdf},
}

@article{smets_multi-wavelength_2024,
	title = {Multi-wavelength {Raman} microscopy of nickel-based electron transport in cable bacteria},
	volume = {15},
	issn = {1664-302X},
	url = {https://www.frontiersin.org/articles/10.3389/fmicb.2024.1208033/full},
	doi = {10.3389/fmicb.2024.1208033},
	abstract = {Cable bacteria embed a network of conductive protein fibers in their cell envelope that efficiently guides electron transport over distances spanning up to several centimeters. This form of long-distance electron transport is unique in biology and is mediated by a metalloprotein with a sulfur-coordinated nickel (Ni) cofactor. However, the molecular structure of this cofactor remains presently unknown. Here, we applied multi-wavelength Raman microscopy to identify cell compounds linked to the unique cable bacterium physiology, combined with stable isotope labeling, and orientation-dependent and ultralow-frequency Raman microscopy to gain insight into the structure and organization of this novel Ni-cofactor. Raman spectra of native cable bacterium filaments reveal vibrational modes originating from cytochromes, polyphosphate granules, proteins, as well as the Ni-cofactor. After selective extraction of the conductive fiber network from the cell envelope, the Raman spectrum becomes simpler, and primarily retains vibrational modes associated with the Ni-cofactor. These Ni-cofactor modes exhibit intense Raman scattering as well as a strong orientation-dependent response. The signal intensity is particularly elevated when the polarization of incident laser light is parallel to the direction of the conductive fibers. This orientation dependence allows to selectively identify the modes that are associated with the Ni-cofactor. We identified 13 such modes, some of which display strong Raman signals across the entire range of applied wavelengths (405–1,064 nm). Assignment of vibrational modes, supported by stable isotope labeling, suggest that the structure of the Ni-cofactor shares a resemblance with that of nickel bis(1,2-dithiolene) complexes. Overall, our results indicate that cable bacteria have evolved a unique cofactor structure that does not resemble any of the known Ni-cofactors in biology.},
	urldate = {2024-10-03},
	journal = {Frontiers in Microbiology},
	author = {Smets, Bent and Boschker, Henricus T. S. and Wetherington, Maxwell T. and Lelong, Gérald and Hidalgo-Martinez, Silvia and Polerecky, Lubos and Nuyts, Gert and De Wael, Karolien and Meysman, Filip J. R.},
	month = mar,
	year = {2024},
	pages = {1208033},
	file = {Full Text:files/725/Smets et al. - 2024 - Multi-wavelength Raman microscopy of nickel-based .pdf:application/pdf},
}

@article{schrenker_investigation_2024,
	title = {Investigation of the {Octahedral} {Network} {Structure} in {Formamidinium} {Lead} {Bromide} {Nanocrystals} by {Low}-{Dose} {Scanning} {Transmission} {Electron} {Microscopy}},
	volume = {24},
	copyright = {https://doi.org/10.15223/policy-029},
	issn = {1530-6984, 1530-6992},
	url = {https://pubs.acs.org/doi/10.1021/acs.nanolett.4c02811},
	doi = {10.1021/acs.nanolett.4c02811},
	language = {en},
	number = {35},
	urldate = {2024-10-07},
	journal = {Nano Letters},
	author = {Schrenker, Nadine J. and Braeckevelt, Tom and De Backer, Annick and Livakas, Nikolaos and Yu, Chu-Ping and Friedrich, Thomas and Roeffaers, Maarten B. J. and Hofkens, Johan and Verbeeck, Johan and Manna, Liberato and Van Speybroeck, Veronique and Van Aert, Sandra and Bals, Sara},
	month = sep,
	year = {2024},
	pages = {10936--10942},
	file = {Full Text:files/751/Schrenker et al. - 2024 - Investigation of the Octahedral Network Structure .pdf:application/pdf},
}

@article{rothmann_structural_2018,
	title = {Structural and {Chemical} {Changes} to {CH} $_{\textrm{3}}$ {NH} $_{\textrm{3}}$ {PbI} $_{\textrm{3}}$ {Induced} by {Electron} and {Gallium} {Ion} {Beams}},
	volume = {30},
	issn = {0935-9648, 1521-4095},
	url = {https://onlinelibrary.wiley.com/doi/10.1002/adma.201800629},
	doi = {10.1002/adma.201800629},
	abstract = {Abstract
            
              Organic–inorganic hybrid perovskites, such as CH
              3
              NH
              3
              PbI
              3,
              have shown highly promising photovoltaic performance. Electron microscopy (EM) is a powerful tool for studying the crystallography, morphology, interfaces, lattice defects, composition, and charge carrier collection and recombination properties at the nanoscale. Here, the sensitivity of CH
              3
              NH
              3
              PbI
              3
              to electron beam irradiation is examined. CH
              3
              NH
              3
              PbI
              3
              undergoes continuous structural and compositional changes with increasing electron dose, with the total dose, rather than dose rate, being the key operative parameter. Importantly, the first structural change is subtle and easily missed and occurs after an electron dose significantly smaller than that typically applied in conventional EM techniques. The electron dose conditions under which these structural changes occur are identified. With appropriate dose‐minimization techniques, electron diffraction patterns can be obtained from pristine material consistent with the tetragonal CH
              3
              NH
              3
              PbI
              3
              phases determined by X‐ray diffraction. Radiation damage incurred at liquid nitrogen temperatures and using Ga
              +
              irradiation in a focused ion beam instrument are also examined. Finally, some simple guidelines for how to minimize electron‐beam‐induced artifacts when using EM to study hybrid perovskite materials are provided.},
	language = {en},
	number = {25},
	urldate = {2024-10-07},
	journal = {Advanced Materials},
	author = {Rothmann, Mathias Uller and Li, Wei and Zhu, Ye and Liu, Amelia and Ku, Zhiliang and Bach, Udo and Etheridge, Joanne and Cheng, Yi‐Bing},
	month = jun,
	year = {2018},
	pages = {1800629},
	file = {Accepted Version:files/753/Rothmann et al. - 2018 - Structural and Chemical Changes to CH 3.pdf:application/pdf},
}

@article{klein-kedem_effects_2016,
	title = {Effects of {Light} and {Electron} {Beam} {Irradiation} on {Halide} {Perovskites} and {Their} {Solar} {Cells}},
	volume = {49},
	issn = {0001-4842, 1520-4898},
	url = {https://pubs.acs.org/doi/10.1021/acs.accounts.5b00469},
	doi = {10.1021/acs.accounts.5b00469},
	language = {en},
	number = {2},
	urldate = {2024-10-14},
	journal = {Accounts of Chemical Research},
	author = {Klein-Kedem, Nir and Cahen, David and Hodes, Gary},
	month = feb,
	year = {2016},
	pages = {347--354},
}

@article{banhart_irradiation_1999,
	title = {Irradiation effects in carbon nanostructures},
	volume = {62},
	issn = {0034-4885, 1361-6633},
	url = {https://iopscience.iop.org/article/10.1088/0034-4885/62/8/201},
	doi = {10.1088/0034-4885/62/8/201},
	number = {8},
	urldate = {2024-10-14},
	journal = {Reports on Progress in Physics},
	author = {Banhart, Florian},
	month = aug,
	year = {1999},
	pages = {1181--1221},
}

@article{ugurlu_radiolysis_2011,
	title = {Radiolysis to knock-on damage transition in zeolites under electron beam irradiation},
	volume = {83},
	copyright = {http://link.aps.org/licenses/aps-default-license},
	issn = {1098-0121, 1550-235X},
	url = {https://link.aps.org/doi/10.1103/PhysRevB.83.113408},
	doi = {10.1103/PhysRevB.83.113408},
	language = {en},
	number = {11},
	urldate = {2024-10-15},
	journal = {Physical Review B},
	author = {Ugurlu, O. and Haus, J. and Gunawan, A. A. and Thomas, M. G. and Maheshwari, S. and Tsapatsis, M. and Mkhoyan, K. A.},
	month = mar,
	year = {2011},
	pages = {113408},
	file = {Accepted Version:files/788/Ugurlu et al. - 2011 - Radiolysis to knock-on damage transition in zeolit.pdf:application/pdf},
}

@article{guerrero_scaling-up_2024,
	title = {Scaling-{Up} {Microwave}-{Assisted} {Synthesis} of {Highly} {Defective} {Pd}@{UiO}-66-{NH} $_{\textrm{2}}$ {Catalysts} for {Selective} {Olefin} {Hydrogenation} under {Ambient} {Conditions}},
	copyright = {https://creativecommons.org/licenses/by/4.0/},
	issn = {1944-8244, 1944-8252},
	url = {https://pubs.acs.org/doi/10.1021/acsami.4c03106},
	doi = {10.1021/acsami.4c03106},
	language = {en},
	urldate = {2024-10-15},
	journal = {ACS Applied Materials \& Interfaces},
	author = {Guerrero, Raúl M. and Lemir, Ignacio D. and Carrasco, Sergio and Fernández-Ruiz, Carlos and Kavak, Safiyye and Pizarro, Patricia and Serrano, David P. and Bals, Sara and Horcajada, Patricia and Pérez, Yolanda},
	month = apr,
	year = {2024},
	pages = {acsami.4c03106},
	file = {Full Text:files/790/Guerrero et al. - 2024 - Scaling-Up Microwave-Assisted Synthesis of Highly .pdf:application/pdf},
}

@article{zhou_low-dose_2020,
	title = {Low-dose phase retrieval of biological specimens using cryo-electron ptychography},
	volume = {11},
	copyright = {2020 The Author(s)},
	issn = {2041-1723},
	url = {https://www.nature.com/articles/s41467-020-16391-6},
	doi = {10.1038/s41467-020-16391-6},
	abstract = {Cryo-electron microscopy is an essential tool for high-resolution structural studies of biological systems. This method relies on the use of phase contrast imaging at high defocus to improve information transfer at low spatial frequencies at the expense of higher spatial frequencies. Here we demonstrate that electron ptychography can recover the phase of the specimen with continuous information transfer across a wide range of the spatial frequency spectrum, with improved transfer at lower spatial frequencies, and as such is more efficient for phase recovery than conventional phase contrast imaging. We further show that the method can be used to study frozen-hydrated specimens of rotavirus double-layered particles and HIV-1 virus-like particles under low-dose conditions (5.7 e/Å2) and heterogeneous objects in an Adenovirus-infected cell over large fields of view (1.14 × 1.14 μm), thus making it suitable for studies of many biologically important structures.},
	language = {en},
	number = {1},
	urldate = {2024-10-15},
	journal = {Nature Communications},
	publisher = {Nature Publishing Group},
	author = {Zhou, Liqi and Song, Jingdong and Kim, Judy S. and Pei, Xudong and Huang, Chen and Boyce, Mark and Mendonça, Luiza and Clare, Daniel and Siebert, Alistair and Allen, Christopher S. and Liberti, Emanuela and Stuart, David and Pan, Xiaoqing and Nellist, Peter D. and Zhang, Peijun and Kirkland, Angus I. and Wang, Peng},
	month = jun,
	year = {2020},
	keywords = {Cryoelectron microscopy, Transmission electron microscopy},
	pages = {2773},
	file = {Full Text PDF:files/798/Zhou et al. - 2020 - Low-dose phase retrieval of biological specimens u.pdf:application/pdf},
}

@article{maiden_improved_2009,
	title = {An improved ptychographical phase retrieval algorithm for diffractive imaging},
	volume = {109},
	copyright = {https://www.elsevier.com/tdm/userlicense/1.0/},
	issn = {03043991},
	url = {https://linkinghub.elsevier.com/retrieve/pii/S0304399109001284},
	doi = {10.1016/j.ultramic.2009.05.012},
	abstract = {The ptychographical iterative engine (or PIE) is a recently developed phase retrieval algorithm that employs a series of diffraction patterns recorded as a known illumination function is translated to a set of overlapping positions relative to a target sample. The technique has been demonstrated successfully at optical and X-ray wavelengths and has been shown to be robust to detector noise and to converge considerably faster than support-based phase retrieval methods. In this paper, the PIE is extended so that the requirement for an accurate model of the illumination function is removed.},
	language = {en},
	number = {10},
	urldate = {2024-10-17},
	journal = {Ultramicroscopy},
	author = {Maiden, Andrew M. and Rodenburg, John M.},
	month = sep,
	year = {2009},
	pages = {1256--1262},
	file = {Maiden and Rodenburg - 2009 - An improved ptychographical phase retrieval algori.pdf:files/815/Maiden and Rodenburg - 2009 - An improved ptychographical phase retrieval algori.pdf:application/pdf},
}

@article{boschker_efficient_2021,
	title = {Efficient long-range conduction in cable bacteria through nickel protein wires},
	volume = {12},
	issn = {2041-1723},
	url = {https://www.nature.com/articles/s41467-021-24312-4},
	doi = {10.1038/s41467-021-24312-4},
	abstract = {Abstract
            Filamentous cable bacteria display long-range electron transport, generating electrical currents over centimeter distances through a highly ordered network of fibers embedded in their cell envelope. The conductivity of these periplasmic wires is exceptionally high for a biological material, but their chemical structure and underlying electron transport mechanism remain unresolved. Here, we combine high-resolution microscopy, spectroscopy, and chemical imaging on individual cable bacterium filaments to demonstrate that the periplasmic wires consist of a conductive protein core surrounded by an insulating protein shell layer. The core proteins contain a sulfur-ligated nickel cofactor, and conductivity decreases when nickel is oxidized or selectively removed. The involvement of nickel as the active metal in biological conduction is remarkable, and suggests a hitherto unknown form of electron transport that enables efficient conduction in centimeter-long protein structures.},
	language = {en},
	number = {1},
	urldate = {2024-10-17},
	journal = {Nature Communications},
	author = {Boschker, Henricus T. S. and Cook, Perran L. M. and Polerecky, Lubos and Eachambadi, Raghavendran Thiruvallur and Lozano, Helena and Hidalgo-Martinez, Silvia and Khalenkow, Dmitry and Spampinato, Valentina and Claes, Nathalie and Kundu, Paromita and Wang, Da and Bals, Sara and Sand, Karina K. and Cavezza, Francesca and Hauffman, Tom and Bjerg, Jesper Tataru and Skirtach, Andre G. and Kochan, Kamila and McKee, Merrilyn and Wood, Bayden and Bedolla, Diana and Gianoncelli, Alessandra and Geerlings, Nicole M. J. and Van Gerven, Nani and Remaut, Han and Geelhoed, Jeanine S. and Millan-Solsona, Ruben and Fumagalli, Laura and Nielsen, Lars Peter and Franquet, Alexis and Manca, Jean V. and Gomila, Gabriel and Meysman, Filip J. R.},
	month = jun,
	year = {2021},
	pages = {3996},
	file = {Boschker et al. - 2021 - Efficient long-range conduction in cable bacteria .pdf:files/817/Boschker et al. - 2021 - Efficient long-range conduction in cable bacteria .pdf:application/pdf},
}

@article{chennit_assessing_2024,
	title = {Assessing {Ptychographic} {Methods} for {Maximum} {Low} {Dose} {Performance}},
	volume = {129},
	copyright = {© The Authors, published by EDP Sciences, 2024},
	issn = {2117-4458},
	url = {https://www.bio-conferences.org/articles/bioconf/abs/2024/48/bioconf_emc2024_04034/bioconf_emc2024_04034.html},
	doi = {10.1051/bioconf/202412904034},
	abstract = {BIO Web of Conferences, open access proceedings in biology, life sciences and health},
	language = {en},
	urldate = {2024-10-25},
	journal = {BIO Web of Conferences},
	publisher = {EDP Sciences},
	author = {Chennit, Tamazouzt and Hofer, Christoph and Yuan, Biao and Li, Songge and Maiden, Andrew and Pennycook, Timothy},
	year = {2024},
	pages = {04034},
	file = {Full Text PDF:files/820/Chennit et al. - 2024 - Assessing Ptychographic Methods for Maximum Low Do.pdf:application/pdf},
}

@article{maiden_further_2017,
	title = {Further improvements to the ptychographical iterative engine},
	volume = {4},
	issn = {2334-2536},
	url = {https://opg.optica.org/optica/abstract.cfm?uri=optica-4-7-736},
	doi = {10.1364/OPTICA.4.000736},
	abstract = {Ptychography is a form of phase imaging that uses iterative algorithms to reconstruct an image of a specimen from a series of diffraction patterns. It is swiftly developing into a mainstream technique, with a growing list of applications across a range of imaging modalities. As the field has advanced, numerous reconstruction algorithms have been proposed, yet the early approaches have not seen major improvement and remain popular. In this paper, we revisit the first such algorithm, the ptychographical iterative engine (PIE), and show how a simple revision and powerful extension can deliver an order of magnitude speed increase and handle difficult data sets where the original version fails completely.},
	language = {EN},
	number = {7},
	urldate = {2024-10-28},
	journal = {Optica},
	publisher = {Optica Publishing Group},
	author = {Maiden, Andrew and Johnson, Daniel and Li, Peng},
	month = jul,
	year = {2017},
	keywords = {Ptychography, Machine learning, Neural networks, Phase imaging, Three dimensional imaging, X-ray imaging},
	pages = {736--745},
	file = {Accepted Version:files/822/Maiden et al. - 2017 - Further improvements to the ptychographical iterat.pdf:application/pdf},
}

@article{kucukoglu_low-dose_2024,
	title = {Low-dose cryo-electron ptychography of proteins at sub-nanometer resolution},
	volume = {15},
	copyright = {2024 The Author(s)},
	issn = {2041-1723},
	url = {https://www.nature.com/articles/s41467-024-52403-5},
	doi = {10.1038/s41467-024-52403-5},
	abstract = {Cryo-transmission electron microscopy (cryo-EM) of frozen hydrated specimens is an efficient method for the structural analysis of purified biological molecules. However, cryo-EM and cryo-electron tomography are limited by the low signal-to-noise ratio (SNR) of recorded images, making detection of smaller particles challenging. For dose-resilient samples often studied in the physical sciences, electron ptychography – a coherent diffractive imaging technique using 4D scanning transmission electron microscopy (4D-STEM) – has recently demonstrated excellent SNR and resolution down to tens of picometers for thin specimens imaged at room temperature. Here we apply 4D-STEM and ptychographic data analysis to frozen hydrated proteins, reaching sub-nanometer resolution 3D reconstructions. We employ low-dose cryo-EM with an aberration-corrected, convergent electron beam to collect 4D-STEM data for our reconstructions. The high frame rate of the electron detector allows us to record large datasets of electron diffraction patterns with substantial overlaps between the interaction volumes of adjacent scan positions, from which the scattering potentials of the samples are iteratively reconstructed. The reconstructed micrographs show strong SNR enabling the reconstruction of the structure of apoferritin protein at up to 5.8 Å resolution. We also show structural analysis of the Phi92 capsid and sheath, tobacco mosaic virus, and bacteriorhodopsin at slightly lower resolutions.},
	language = {en},
	number = {1},
	urldate = {2024-10-28},
	journal = {Nature Communications},
	publisher = {Nature Publishing Group},
	author = {Küçükoğlu, Berk and Mohammed, Inayathulla and Guerrero-Ferreira, Ricardo C. and Ribet, Stephanie M. and Varnavides, Georgios and Leidl, Max Leo and Lau, Kelvin and Nazarov, Sergey and Myasnikov, Alexander and Kube, Massimo and Radecke, Julika and Sachse, Carsten and Müller-Caspary, Knut and Ophus, Colin and Stahlberg, Henning},
	month = sep,
	year = {2024},
	keywords = {Cryoelectron microscopy, Transmission electron microscopy, Biochemistry, Biophysics},
	pages = {8062},
	file = {Full Text PDF:files/826/Küçükoğlu et al. - 2024 - Low-dose cryo-electron ptychography of proteins at.pdf:application/pdf},
}

@article{chen_imaging_2020,
	title = {Imaging {Beam}‐{Sensitive} {Materials} by {Electron} {Microscopy}},
	volume = {32},
	issn = {0935-9648, 1521-4095},
	url = {https://onlinelibrary.wiley.com/doi/10.1002/adma.201907619},
	doi = {10.1002/adma.201907619},
	abstract = {Abstract
            Electron microscopy allows the extraction of multidimensional spatiotemporally correlated structural information of diverse materials down to atomic resolution, which is essential for figuring out their structure–property relationships. Unfortunately, the high‐energy electrons that carry this important information can cause damage by modulating the structures of the materials. This has become a significant problem concerning the recent boost in materials science applications of a wide range of beam‐sensitive materials, including metal–organic frameworks, covalent–organic frameworks, organic–inorganic hybrid materials, 2D materials, and zeolites. To this end, developing electron microscopy techniques that minimize the electron beam damage for the extraction of intrinsic structural information turns out to be a compelling but challenging need. This article provides a comprehensive review on the revolutionary strategies toward the electron microscopic imaging of beam‐sensitive materials and associated materials science discoveries, based on the principles of electron–matter interaction and mechanisms of electron beam damage. Finally, perspectives and future trends in this field are put forward.},
	language = {en},
	number = {16},
	urldate = {2024-10-31},
	journal = {Advanced Materials},
	author = {Chen, Qiaoli and Dwyer, Christian and Sheng, Guan and Zhu, Chongzhi and Li, Xiaonian and Zheng, Changlin and Zhu, Yihan},
	month = apr,
	year = {2020},
	pages = {1907619},
	file = {Accepted Version:files/828/Chen et al. - 2020 - Imaging Beam‐Sensitive Materials by Electron Micro.pdf:application/pdf},
}

@phdthesis{oleary_development_2020,
	type = {http://purl.org/dc/dcmitype/{Text}},
	title = {The development and applications of {STEM} ptychography using direct electron detectors},
	url = {https://ora.ox.ac.uk/objects/uuid:a2f937d8-8b6b-4b7d-b0b4-feeb9d28964c},
	abstract = {{\textless}p{\textgreater}Since the introduction of direct electron detectors to scanning transmission electron microscopy (STEM), electron ptychography - a technique which utilises the interference in diffraction patterns to reconstruct the sample-induced phase changes of a transmitted electron wave - has significantly extended the capabilities of electron microscopy. However, a number of limitations to electron ptychography exist, namely the poor contrast transfer for low and high spatial frequencies to the phase reconstruction, and the relatively slow detector speeds used to acquire the ptychographic data ({\textasciitilde} 1,000 fps). In this thesis, a number of strategies are introduced to further improve the robustness and dose-efficiency of focused-probe electron ptychography (FPP), after which several applications of FPP techniques are demonstrated.{\textless}/p{\textgreater} {\textless}p{\textgreater}Firstly, the contrast transfer properties of single side-band (SSB) ptychography are experimentally measured from an amorphous carbon sample in order to determine the optimal experimental parameters for ptychography. It is demonstrated that the probe convergence semi-angle can be used to tune the phase-contrast transfer function (PCTF) for each experiment, such that the relevant sample information is transferred with high contrast. Furthermore, careful consideration of the noise in the ptychographic data can provide an enhanced PCTF which broadens the transfer window in the image plane. These strategies are combined with a 1-bit fast (12,500 fps) acquisition scheme to enable the atomic-resolution phase reconstruction of a beam-sensitive zeolite sample using a low electron dose of 1.0 x 10{\textasciicircum}(5) e nm{\textasciicircum}(-2). By implementing these experimental and analytical strategies, the efficiency of FPP techniques can be signifificantly improved.{\textless}/p{\textgreater} {\textless}p{\textgreater}At the end of this thesis, several experimental challenges common to STEM are overcome using electron ptychography. Firstly, the precision of phase reconstructions are improved considerably by increasing the electron dose via multi-frame image acquisition, hence avoiding the slow-scan instabilities inherent to long STEM acquisition times. Furthermore, three-dimensional analysis of an unknown graphene defect is performed using a single ptychographic data set. Finally, electron ptychography is used to visualise oxygen vacancies in uranium dioxide for the first time. {\textless}/p{\textgreater}},
	language = {English},
	urldate = {2024-11-25},
	school = {University of Oxford},
	author = {O'Leary, C.},
	year = {2020},
}

@misc{noauthor_wasp_nodate,
	title = {{WASP}: weighted average of sequential projections for ptychographic phase retrieval},
	url = {https://opg.optica.org/oe/fulltext.cfm?uri=oe-32-12-21327&id=551196},
	urldate = {2024-12-19},
}

@article{wen_simultaneous_2019,
	title = {Simultaneous {Identification} of {Low} and {High} {Atomic} {Number} {Atoms} in {Monolayer} {2D} {Materials} {Using} {4D} {Scanning} {Transmission} {Electron} {Microscopy}},
	volume = {19},
	issn = {1530-6984},
	url = {https://doi.org/10.1021/acs.nanolett.9b02717},
	doi = {10.1021/acs.nanolett.9b02717},
	abstract = {Simultaneous imaging of individual low and high atomic number atoms using annular dark field scanning transmission electron microscopy (ADF-STEM) is often challenging due to substantial differences in their scattering cross sections. This often leads to contrast from only the high atomic number species when imaged using ADF-STEM such as the Mo and 2S sites in monolayer MoS2 crystals, without detection of lighter atoms such as C, O, or N. Here, we show that by capturing an array of convergent beam electron diffraction patterns using a 2D pixelated electron detector (2D-PED) in a 4D STEM geometry enables identification of individual low and high atomic number atoms in 2D materials by multicomponent imaging. We have used ptychographic phase reconstructions, combined with angular dependent ADF-STEM reconstructions, to image light elements at lateral (nanopores) and vertical interfaces (surface dopants) within 2D monolayer MoS2. Differential phase contrast imaging (Div(DPC)) using quadrant segmentation of the 2D pixelated direct electron detector data not only qualitatively matches the ptychographic phase reconstructions in both resolution and contrast but also offers the additional potential for real time display. Using 4D-STEM, we have identified surface adatoms on MoS2 monolayers and have separated atomic columns with similar total atomic number into their relative combinations of low and high atomic number elements. These results demonstrate the rich information present in the data obtained during 4D-STEM imaging of ultrathin 2D materials and the ability of this approach to extract unique insights beyond conventional imaging.},
	number = {9},
	urldate = {2024-12-19},
	journal = {Nano Letters},
	publisher = {American Chemical Society},
	author = {Wen, Yi and Ophus, Colin and Allen, Christopher S. and Fang, Shiang and Chen, Jun and Kaxiras, Efthimios and Kirkland, Angus I. and Warner, Jamie H.},
	month = sep,
	year = {2019},
	pages = {6482--6491},
	file = {Full Text PDF:files/907/Wen et al. - 2019 - Simultaneous Identification of Low and High Atomic.pdf:application/pdf},
}

@article{lazic_single-particle_2022,
	title = {Single-particle cryo-{EM} structures from {iDPC}–{STEM} at near-atomic resolution},
	volume = {19},
	copyright = {2022 The Author(s)},
	issn = {1548-7105},
	url = {https://www.nature.com/articles/s41592-022-01586-0},
	doi = {10.1038/s41592-022-01586-0},
	abstract = {In electron cryomicroscopy (cryo-EM), molecular images of vitrified biological samples are obtained by conventional transmission microscopy (CTEM) using large underfocuses and subsequently computationally combined into a high-resolution three-dimensional structure. Here, we apply scanning transmission electron microscopy (STEM) using the integrated differential phase contrast mode also known as iDPC–STEM to two cryo-EM test specimens, keyhole limpet hemocyanin (KLH) and tobacco mosaic virus (TMV). The micrographs show complete contrast transfer to high resolution and enable the cryo-EM structure determination for KLH at 6.5 Å resolution, as well as for TMV at 3.5 Å resolution using single-particle reconstruction methods, which share identical features with maps obtained by CTEM of a previously acquired same-sized TMV data set. These data show that STEM imaging in general, and in particular the iDPC–STEM approach, can be applied to vitrified single-particle specimens to determine near-atomic resolution cryo-EM structures of biological macromolecules.},
	language = {en},
	number = {9},
	urldate = {2024-12-19},
	journal = {Nature Methods},
	publisher = {Nature Publishing Group},
	author = {Lazić, Ivan and Wirix, Maarten and Leidl, Max Leo and de Haas, Felix and Mann, Daniel and Beckers, Maximilian and Pechnikova, Evgeniya V. and Müller-Caspary, Knut and Egoavil, Ricardo and Bosch, Eric G. T. and Sachse, Carsten},
	month = sep,
	year = {2022},
	keywords = {Cryoelectron microscopy, Cryoelectron tomography},
	pages = {1126--1136},
	file = {Full Text PDF:files/909/Lazić et al. - 2022 - Single-particle cryo-EM structures from iDPC–STEM .pdf:application/pdf},
}

@article{liu_direct_2020,
	title = {Direct {Imaging} of {Atomically} {Dispersed} {Molybdenum} that {Enables} {Location} of {Aluminum} in the {Framework} of {Zeolite} {ZSM}-5},
	volume = {59},
	copyright = {© 2020 Wiley-VCH Verlag GmbH \& Co. KGaA, Weinheim},
	issn = {1521-3773},
	url = {https://onlinelibrary.wiley.com/doi/abs/10.1002/anie.201909834},
	doi = {10.1002/anie.201909834},
	abstract = {Integrated differential phase-contrast scanning transmission electron microscopy (iDPC-STEM) is capable of directly probing guest molecules in zeolites, owing to its sufficient and interpretable image contrast for both heavy and light elements under low-dose conditions. This unique ability is demonstrated by imaging volatile organic compounds adsorbed in zeolite Silicalite-1; iDPC-STEM was then used to investigate molybdenum supported on various zeolites including Silicalite-1, ZSM-5, and mordenite. Isolated single-Mo clusters were observed in the micropores of ZSM-5, demonstrating the crucial role of framework Al in driving Mo atomically dispersed into the micropores. Importantly, the specific one-to-one Mo-Al interaction makes it possible to locate Al atoms, that is, catalytic active sites, in the ZSM-5 framework from the images, according to the positions of Mo atoms in the micropores.},
	language = {en},
	number = {2},
	urldate = {2024-12-19},
	journal = {Angewandte Chemie International Edition},
	author = {Liu, Lingmei and Wang, Ning and Zhu, Chongzhi and Liu, Xiaona and Zhu, Yihan and Guo, Peng and Alfilfil, Lujain and Dong, Xinglong and Zhang, Daliang and Han, Yu},
	year = {2020},
	note = {\_eprint: https://onlinelibrary.wiley.com/doi/pdf/10.1002/anie.201909834},
	keywords = {electron microscopy, heterogeneous catalysis, host–guest systems, zeolites},
	pages = {819--825},
	file = {Full Text PDF:files/913/Liu et al. - 2020 - Direct Imaging of Atomically Dispersed Molybdenum .pdf:application/pdf;Snapshot:files/914/anie.html:text/html},
}

@article{dong_atomic-level_2023,
	title = {Atomic-{Level} {Imaging} of {Zeolite} {Local} {Structures} {Using} {Electron} {Ptychography}},
	volume = {145},
	issn = {0002-7863},
	url = {https://doi.org/10.1021/jacs.2c12673},
	doi = {10.1021/jacs.2c12673},
	abstract = {Zeolites are among the most important heterogeneous catalysts, widely employed in separation reaction, fine chemical production, and petroleum refining. Through rational design of the frameworks, zeolites with versatile functions can be synthesized. Local imaging of zeolite structures at the atomic scale, including the basic framework atoms (Si, Al, and O) and extra-framework cations, is necessary to understand the structure–function relationship of zeolites. Herein, we implemented electron ptychography into direct imaging of local structures of two zeolites, Na-LTA and ZSM-5. Not only all the framework atoms but also extra-framework Na+ cations with only 1/4 occupation probabilities in Na-LTA were directly observed. Local structures of ZSM-5 zeolites having guest molecules among channels with different orientations were also unraveled using different reconstruction algorithms. The approach presented here provides a new way to locally image zeolites structure, and it is expected to be an essential key for further studying and tuning zeolites active sites at the atomic level.},
	number = {12},
	urldate = {2024-12-19},
	journal = {Journal of the American Chemical Society},
	publisher = {American Chemical Society},
	author = {Dong, Zhuoya and Zhang, Enci and Jiang, Yilan and Zhang, Qing and Mayoral, Alvaro and Jiang, Huaidong and Ma, Yanhang},
	month = mar,
	year = {2023},
	pages = {6628--6632},
	file = {Full Text PDF:files/916/Dong et al. - 2023 - Atomic-Level Imaging of Zeolite Local Structures U.pdf:application/pdf},
}

@article{hao_atomic-scale_2023,
	title = {Atomic-scale imaging of polyvinyl alcohol crystallinity using electron ptychography},
	volume = {284},
	issn = {0032-3861},
	url = {https://www.sciencedirect.com/science/article/pii/S0032386123006353},
	doi = {10.1016/j.polymer.2023.126305},
	abstract = {Polyvinyl alcohol (PVA) is considered to have great potential in medical, pharmaceutical, and packaging applications because of its outstanding biocompatibility, water solubility, low density and relatively low cost. PVA crystallinity, central to the materials properties, has been studied by X-ray diffraction, but two possible crystal structures are mooted. Electron microscopic techniques can potentially image PVA at high resolution. Still, it is challenging for conventional electron microscopies because of the relatively low crystallinity of PVA, its severe beam sensitivity, and the poor contrast of light elements. Electron ptychography makes use of a 4D STEM dataset comprising the intensity in the STEM detector plane recorded as a function of each probe position and has lower sample damage and better phase-contrast compared to traditional techniques. Here, we use electron ptychography to image PVA crystallinity. The reconstructed images, which show good agreement in the unit cell dimension with X-ray diffraction data, can show how the atoms order in the materials, however, deviations from previous models derived from X-ray diffraction are observed. To interpret the data, we propose a series of changes based on previous models to formulate a description of PVA crystal structure. Simulated results from this new model accord well with the experimental images. This study manages to image both carbon and oxygen atoms in PVA, which has not previously been achieved by any conventional method. The results are expected to bring a new and deeper understanding of PVA crystal structure, and illustrate the opportunity presented by this approach for directly imaging molecular order in polymer crystals.},
	urldate = {2024-12-19},
	journal = {Polymer},
	author = {Hao, Botao and Ding, Zhiyuan and Tao, Xudong and Nellist, Peter D. and Assender, Hazel E.},
	month = oct,
	year = {2023},
	keywords = {4D STEM, Electron ptychography, Scanning transmission electron microscopy, Polyvinyl alcohol},
	pages = {126305},
	file = {Full Text:files/918/Hao et al. - 2023 - Atomic-scale imaging of polyvinyl alcohol crystall.pdf:application/pdf;ScienceDirect Snapshot:files/919/S0032386123006353.html:text/html},
}

@misc{noauthor_situ_nodate,
	title = {In situ imaging of the atomic phase transition dynamics in metal halide perovskites {\textbar} {Nature} {Communications}},
	url = {https://www.nature.com/articles/s41467-023-42999-5},
	urldate = {2024-12-19},
	file = {In situ imaging of the atomic phase transition dynamics in metal halide perovskites | Nature Communications:files/921/s41467-023-42999-5.html:text/html},
}

@misc{noauthor_direct_nodate,
	title = {Direct {Imaging} of {Tunable} {Crystal} {Surface} {Structures} of {MOF} {MIL}-101 {Using} {High}-{Resolution} {Electron} {Microscopy} {\textbar} {Journal} of the {American} {Chemical} {Society}},
	url = {https://pubs.acs.org/doi/10.1021/jacs.9b04896},
	urldate = {2024-12-19},
	file = {Direct Imaging of Tunable Crystal Surface Structures of MOF MIL-101 Using High-Resolution Electron Microscopy | Journal of the American Chemical Society:files/925/jacs.html:text/html},
}

@article{king_damage_1987,
	title = {Damage effects of high energy electrons on metals},
	volume = {23},
	issn = {0304-3991},
	url = {https://www.sciencedirect.com/science/article/pii/0304399187902452},
	doi = {10.1016/0304-3991(87)90245-2},
	abstract = {In this paper, we present results obtained both experimentally, using in-situ electrical resistivity measurements in the high voltage electron microscope, and theoretically, using molecular dynamics computer simulations. These data give a detailed picture of the mechanisms of atomic displacement in Cu and make it possible to quantitatively predict the production rate of point defects given the incident electron beam energy and direction relative to the crystalline axes. Damage effects are found to be important at energies as low as 100–400 keV. We also review the work that has been carried out on other metals and the effect of temperature on the damage production process.},
	number = {3},
	urldate = {2024-12-19},
	journal = {Ultramicroscopy},
	author = {King, Wayne E. and Benedek, R. and Merkle, K. L. and Meshii, M.},
	month = jan,
	year = {1987},
	pages = {345--353},
	file = {ScienceDirect Snapshot:files/933/0304399187902452.html:text/html},
}

@article{hobbs_radiation_1994,
	title = {Radiation effects in ceramics},
	volume = {216},
	issn = {0022-3115},
	url = {https://www.sciencedirect.com/science/article/pii/0022311594900175},
	doi = {10.1016/0022-3115(94)90017-5},
	abstract = {Ceramics represent a large class of solids with a wide spectrum of applicability, whose structures range from simple to complex, whose bonding runs from highly ionic to almost entirely covalent and, in some cases, partially metallic, and whose band structures yield wide-gap insulators, narrow-gap semiconductors or even superconductors. These solids exhibit responses to irradiation which are more complex than those for metals. In ceramic materials, atomic displacements can be produced by direct momentum transfer to often more than one distinguishable sublattice, and in some cases radiolytically by electronic excitations, and result in point defects which are in general not simple. Radiation-induced defect interaction, accumulation and aggregation modes differ significantly from those found in metals. Amorphization is a frequent option in response to high-density defect perturbation and is strongly related to structural topology. These fundamental responses to irradiation result in significant changes to important applicable properties, such as strength, toughness, electrical and thermal conductivities, dielectric response and optical behavior. The understanding of such phenomena is less well-understood than the simple responses of metals but is being increasingly driven by critical applications in fusion energy production, nuclear waste disposal and optical communications.},
	urldate = {2024-12-19},
	journal = {Journal of Nuclear Materials},
	author = {Hobbs, Linn W. and Clinard, Frank W. and Zinkle, Steven J. and Ewing, Rodney C.},
	month = oct,
	year = {1994},
	pages = {291--321},
	file = {ScienceDirect Snapshot:files/937/0022311594900175.html:text/html},
}

@article{egerton_dose_2021,
	title = {Dose measurement in the {TEM} and {STEM}},
	volume = {229},
	issn = {0304-3991},
	url = {https://www.sciencedirect.com/science/article/pii/S0304399121001455},
	doi = {10.1016/j.ultramic.2021.113363},
	abstract = {Practical aspects of dosimetry are considered, including the measurement of electron-beam current and current density. Complications that arise in the case of a focused probe or a STEM image are discussed and solutions proposed. Advantages of expressing the radiation dose in Grays are listed and a simple formula given for converting electron fluence to Gray units, based on a near constancy of the stopping power per atomic electron. Comparisons with stopping-power calculations and EELS measurements suggest that this formula is accurate to within 5\%. Based on the stopping power formula, a new way of measuring the local mass-thickness of light-element specimens is proposed. The average energy loss per inelastic collision is shown to be higher than previous expectations.},
	urldate = {2024-12-19},
	journal = {Ultramicroscopy},
	author = {Egerton, R. F.},
	month = oct,
	year = {2021},
	keywords = {STEM, dosimetry, Gray units, radiation damage, stopping power},
	pages = {113363},
	file = {ScienceDirect Snapshot:files/940/S0304399121001455.html:text/html},
}

@article{egerton_limits_2007,
	title = {Limits to the spatial, energy and momentum resolution of electron energy-loss spectroscopy},
	volume = {107},
	issn = {0304-3991},
	url = {https://www.sciencedirect.com/science/article/pii/S0304399106002245},
	doi = {10.1016/j.ultramic.2006.11.005},
	abstract = {We discuss various factors that determine the performance of electron energy-loss spectroscopy (EELS) and energy-filtered (EFTEM) imaging in a transmission electron microscope. Some of these factors are instrumental and have undergone substantial improvement in recent years, including the development of electron monochromators and aberration correctors. Others, such as radiation damage, delocalization of inelastic scattering and beam broadening in the specimen, derive from basic physics and are likely to remain as limitations. To aid the experimentalist, analytical expressions are given for beam broadening, delocalization length, energy broadening due to core-hole and excited-electron lifetimes, and for the momentum resolution in angle-resolved EELS.},
	number = {8},
	urldate = {2024-12-19},
	journal = {Ultramicroscopy},
	author = {Egerton, R. F.},
	month = aug,
	year = {2007},
	keywords = {TEM, EELS, Electron scattering, Fourier optics, Radiation effects},
	pages = {575--586},
}

@article{rez_coherent_2021,
	series = {80th {Birthdays} of {Colin} {Humphreys} and {Knut} {Urban}, 75th {Birthdays} of {Wolfgang} {Baumeister} and {John} {Spence} - {PICO} 2021 – {Sixth} {Conference} on {Frontiers} of {Aberration} {Corrected} {Electron} {Microscopy}},
	title = {Coherent and incoherent imaging of biological specimens with electrons and {X}-rays},
	volume = {231},
	issn = {0304-3991},
	url = {https://www.sciencedirect.com/science/article/pii/S0304399121000887},
	doi = {10.1016/j.ultramic.2021.113301},
	abstract = {Since radiation damage is proportional to fluence, radiation damage limits the spatial resolution of biological structures determined by either X-ray or electron scattering. If only elastic scattering is used for structural information then electrons are superior as the ratio of elastic to inelastic scattering is higher than for X-rays. For soft X-rays in the water window below the O K edge photoabsorption contrast might be better than elastic scattering for distinguishing different biological materials. Phase contrast elastic scattering is most effective in the hard X-ray region up to about 10 keV. Radiation damage limits spatial resolution for most X-ray imaging to 10-20 nm. Local molar concentrations of Na,K and Ca ions can be determined at somewhat lower spatial resolutions using relevant absorption edges. At higher energies resolution res is only limited by the fluence available from the light source, since energy deposition is small.},
	urldate = {2024-12-19},
	journal = {Ultramicroscopy},
	author = {Rez, Peter},
	month = dec,
	year = {2021},
	pages = {113301},
}

@article{egerton_voxel_2024,
	title = {Voxel dose-limited resolution for thick beam-sensitive specimens imaged in a {TEM} or {STEM}},
	volume = {177},
	issn = {0968-4328},
	url = {https://www.sciencedirect.com/science/article/pii/S0968432823001749},
	doi = {10.1016/j.micron.2023.103576},
	abstract = {The resolution limit imposed by radiation damage is quantified in terms of a voxel dose-limited resolution (DLR), applicable to small features within a thick specimen. An analytical formula for this DLR is derived and applied to bright-field mass-thickness contrast from organic (polymer or biological) specimens of thickness between 400 nm and 20 µm. For a permissible dose of 330 MGy (typical of frozen-hydrated tissue), the TEM or STEM image resolution is determined by radiation damage rather than by lens aberrations or beam-broadening effects, which can be restricted by use of a small angle-limiting aperture. DLR is improved by a up to factor of 2 by increasing the primary-electron energy from 300 keV to 3 MeV, or by up to a factor of 3 by heavy-metal staining. For stained samples, a higher electron fluence allows better resolution but the improvement is modest because the voxel DLR is proportional to the 1/4 power of electron dose. The relevance of voxel and columnar DLR is discussed, for both thick and thin samples.},
	urldate = {2024-12-19},
	journal = {Micron},
	author = {Egerton, R. F.},
	month = feb,
	year = {2024},
	keywords = {STEM, TEM, Brightfield microscopy, Dose-limited resolution, Electron tomography, Radiation damage},
	pages = {103576},
}

@article{mcmullan_electron_2007,
	title = {Electron imaging with {Medipix2} hybrid pixel detector},
	volume = {107},
	issn = {0304-3991},
	url = {https://www.sciencedirect.com/science/article/pii/S0304399106001963},
	doi = {10.1016/j.ultramic.2006.10.005},
	abstract = {The electron imaging performance of Medipix2 is described. Medipix2 is a hybrid pixel detector composed of two layers. It has a sensor layer and a layer of readout electronics, in which each 55μm×55μm pixel has upper and lower energy discrimination and MHz rate counting. The sensor layer consists of a 300μm slab of pixellated monolithic silicon and this is bonded to the readout chip. Experimental measurement of the detective quantum efficiency, DQE(0) at 120keV shows that it can reach ∼85\% independent of electron exposure, since the detector has zero noise, and the DQE(Nyquist) can reach ∼35\% of that expected for a perfect detector (4/π2). Experimental measurement of the modulation transfer function (MTF) at Nyquist resolution for 120keV electrons using a 60keV lower energy threshold, yields a value that is 50\% of that expected for a perfect detector (2/π). Finally, Monte Carlo simulations of electron tracks and energy deposited in adjacent pixels have been performed and used to calculate expected values for the MTF and DQE as a function of the threshold energy. The good agreement between theory and experiment allows suggestions for further improvements to be made with confidence. The present detector is already very useful for experiments that require a high DQE at very low doses.},
	number = {4},
	urldate = {2024-12-19},
	journal = {Ultramicroscopy},
	author = {McMullan, G. and Cattermole, D. M. and Chen, S. and Henderson, R. and Llopart, X. and Summerfield, C. and Tlustos, L. and Faruqi, A. R.},
	month = apr,
	year = {2007},
	keywords = {Electron microscopy, Hybrid pixel detectors, Imaging detectors},
	pages = {401--413},
}

@article{ballabriga_medipix3_2011,
	series = {11th {International} {Workshop} on {Radiation} {Imaging} {Detectors} ({IWORID})},
	title = {Medipix3: {A} 64   k pixel detector readout chip working in single photon counting mode with improved spectrometric performance},
	volume = {633},
	issn = {0168-9002},
	shorttitle = {Medipix3},
	url = {https://www.sciencedirect.com/science/article/pii/S0168900210012982},
	doi = {10.1016/j.nima.2010.06.108},
	abstract = {Medipix3 is a 256×256 channel hybrid pixel detector readout chip working in a single photon counting mode with a new inter-pixel architecture, which aims to improve the energy resolution in pixelated detectors by mitigating the effects of charge sharing between channels. Charges are summed in all 2×2 pixel clusters on the chip and a given hit is allocated locally to the pixel summing circuit with the biggest total charge on an event-by-event basis. Each pixel contains also two 12-bit binary counters with programmable depth and overflow control. The chip is configurable such that either the dimensions of each detector pixel match those of one readout pixel or detector pixels are four times greater in area than the readout pixels. In the latter case, event-by-event summing is still possible between the larger pixels. Each pixel has around 1600 transistors and the analog static power consumption is below 15μW in the charge summing mode and 9μW in the single pixel mode. The chip has been built in an 8-metal 0.13μm CMOS technology. This paper describes the chip from the pixel to the periphery and first electrical results are summarized.},
	urldate = {2024-12-19},
	journal = {Nuclear Instruments and Methods in Physics Research Section A: Accelerators, Spectrometers, Detectors and Associated Equipment},
	author = {Ballabriga, R. and Campbell, M. and Heijne, E. and Llopart, X. and Tlustos, L. and Wong, W.},
	month = may,
	year = {2011},
	keywords = {Charge sharing, Charge summing, CMOS, Medipix, Photon counting, Pixel, X-rays},
	pages = {S15--S18},
}

@article{llopart_timepix_2007,
	series = {{VCI} 2007},
	title = {Timepix, a 65k programmable pixel readout chip for arrival time, energy and/or photon counting measurements},
	volume = {581},
	issn = {0168-9002},
	url = {https://www.sciencedirect.com/science/article/pii/S0168900207017020},
	doi = {10.1016/j.nima.2007.08.079},
	abstract = {A novel approach for the readout of a TPC at the future linear collider is to use a CMOS pixel detector combined with some kind of gas gain grid. A first test using the photon counting chip Medipix2 with GEM or Micromegas demonstrated the feasibility of such an approach. Although this experiment demonstrated that single primary electrons could be detected the chip did not provide information on the arrival time of the electron in the sensitive gas volume nor did it give any indication of the quantity of charge detected. The Timepix chip uses an external clock with a frequency of up to 100MHz as a time reference. Each pixel contains a preamplifier, a discriminator with hysteresis and 4-bit DAC for threshold adjustment, synchronization logic and a 14-bit counter with overflow control. Moreover, each pixel can be independently configured in one of four different modes: masked mode: pixel is off, counting mode: 1-count for each signal over threshold, TOT mode: the counter is incremented continuously as long as the signal is above threshold, and arrival time mode: the counter is incremented continuously from the time the first hit arrives until the end of the shutter. The chip resembles very much the Medipix2 chip physically and can be readout using slightly modified versions of the various existing systems. This paper presents the main features of the new design, electrical measurements and some first images.},
	number = {1},
	urldate = {2024-12-19},
	journal = {Nuclear Instruments and Methods in Physics Research Section A: Accelerators, Spectrometers, Detectors and Associated Equipment},
	author = {Llopart, X. and Ballabriga, R. and Campbell, M. and Tlustos, L. and Wong, W.},
	month = oct,
	year = {2007},
	keywords = {CMOS, Medipix, Photon counting, Pixel, Arrival time, Micro-pattern gas detectors},
	pages = {485--494},
	file = {ScienceDirect Snapshot:files/946/S0168900207017020.html:text/html},
}

@article{poikela_timepix3_2014,
	title = {Timepix3: a {65K} channel hybrid pixel readout chip with simultaneous {ToA}/{ToT} and sparse readout},
	volume = {9},
	issn = {1748-0221},
	shorttitle = {Timepix3},
	url = {https://dx.doi.org/10.1088/1748-0221/9/05/C05013},
	doi = {10.1088/1748-0221/9/05/C05013},
	abstract = {The Timepix3, hybrid pixel detector (HPD) readout chip, a successor to the Timepix {\textbackslash}citetimepix2007 chip, can record time-of-arrival (ToA) and time-over-threshold (ToT) simultaneously in each pixel. ToA information is recorded in a 14-bit register at 40 MHz and can be refined by a further 4 bits with a nominal resolution of 1.5625 ns (640 MHz). ToT is recorded in a 10-bit overflow controlled counter at 40 MHz. Pixels can be programmed to record 14 bits of integral ToT and 10 bits of event counting, both at 40 MHz. The chip is designed in 130 nm CMOS and contains 256 × 256 pixel channels (55 × 55 μm2). The chip, which has more than 170 M transistors, has been conceived as a general-purpose readout chip for HPDs used in a wide range of applications. Common requirements of these applications are operation without a trigger signal, and sparse readout where only pixels containing event information are read out. A new architecture has been designed for sparse readout and can achieve a throughput of up to 40 Mhits/s/cm2. The flexible architecture offers readout schemes ranging from serial (one link) readout (40 Mbps) to faster parallel (up to 8 links) readout of 5.12 Gbps. In the ToA/ToT operation mode, readout is simultaneous with data acquisition thus keeping pixels sensitive at all times. The pixel matrix is formed by super pixel (SP) structures of 2 × 4 pixels. This optimizes resources by sharing the pixel readout logic which transports data from SPs to End-of-Column (EoC) using a 2-phase handshake protocol. To reduce power consumption in applications with a low duty cycle, an on-chip power pulsing scheme has been implemented. The logic switches bias currents of the analog front-ends in a sequential manner, and all front-ends can be switched in 800 ns. The digital design uses a mixture of commercial and custom standard cell libraries and was verified using Open Verification Methodology (OVM) and commercial timing analysis tools. The analog front-end and a voltage-controlled oscillator for 1.5625 ns timing resolution have been designed using full custom techniques.},
	language = {en},
	number = {05},
	urldate = {2024-12-19},
	journal = {Journal of Instrumentation},
	author = {Poikela, T. and Plosila, J. and Westerlund, T. and Campbell, M. and Gaspari, M. De and Llopart, X. and Gromov, V. and Kluit, R. and Beuzekom, M. van and Zappon, F. and Zivkovic, V. and Brezina, C. and Desch, K. and Fu, Y. and Kruth, A.},
	month = may,
	year = {2014},
	pages = {C05013},
	file = {IOP Full Text PDF:files/943/Poikela et al. - 2014 - Timepix3 a 65K channel hybrid pixel readout chip .pdf:application/pdf},
}

@article{tate_high_2016,
	title = {High {Dynamic} {Range} {Pixel} {Array} {Detector} for {Scanning} {Transmission} {Electron} {Microscopy}},
	volume = {22},
	issn = {1431-9276},
	url = {https://doi.org/10.1017/S1431927615015664},
	doi = {10.1017/S1431927615015664},
	abstract = {We describe a hybrid pixel array detector (electron microscope pixel array detector, or EMPAD) adapted for use in electron microscope applications, especially as a universal detector for scanning transmission electron microscopy. The 128×128 pixel detector consists of a 500 µm thick silicon diode array bump-bonded pixel-by-pixel to an application-specific integrated circuit. The in-pixel circuitry provides a 1,000,000:1 dynamic range within a single frame, allowing the direct electron beam to be imaged while still maintaining single electron sensitivity. A 1.1 kHz framing rate enables rapid data collection and minimizes sample drift distortions while scanning. By capturing the entire unsaturated diffraction pattern in scanning mode, one can simultaneously capture bright field, dark field, and phase contrast information, as well as being able to analyze the full scattering distribution, allowing true center of mass imaging. The scattering is recorded on an absolute scale, so that information such as local sample thickness can be directly determined. This paper describes the detector architecture, data acquisition system, and preliminary results from experiments with 80–200 keV electron beams.},
	number = {1},
	urldate = {2024-12-19},
	journal = {Microscopy and Microanalysis},
	author = {Tate, Mark W and Purohit, Prafull and Chamberlain, Darol and Nguyen, Kayla X and Hovden, Robert and Chang, Celesta S and Deb, Pratiti and Turgut, Emrah and Heron, John T and Schlom, Darrell G and Ralph, Daniel C and Fuchs, Gregory D and Shanks, Katherine S and Philipp, Hugh T and Muller, David A and Gruner, Sol M},
	month = feb,
	year = {2016},
	pages = {237--249},
	file = {Full Text PDF:files/947/Tate et al. - 2016 - High Dynamic Range Pixel Array Detector for Scanni.pdf:application/pdf},
}

@article{nellist_beyond_1994,
	title = {Beyond the conventional information limit: the relevant coherence function},
	volume = {54},
	issn = {0304-3991},
	shorttitle = {Beyond the conventional information limit},
	url = {https://www.sciencedirect.com/science/article/pii/0304399194900922},
	doi = {10.1016/0304-3991(94)90092-2},
	abstract = {The theory of partial coherence functions as applied to a super-resolution reconstruction algorithm is developed in detail, taking into account the phase gradients across the aperture function, the source and detector sizes, and temporal coherence. Experimental examples demonstrate the main properties of the relevant coherence envelopes, and why they need not limit the total band-pass of the microscope. It is demonstrated that finite detector size in the STEM configuration can facilitate a simpler reconstruction algorithm by creating a virtual objective aperture which obviates the need for a physical objective aperture. These results are also relevant to the question of uniquely deconvolving shadow images.},
	number = {1},
	urldate = {2024-12-19},
	journal = {Ultramicroscopy},
	author = {Nellist, P. D. and Rodenburg, J. M.},
	month = may,
	year = {1994},
	pages = {61--74},
	file = {ScienceDirect Snapshot:files/955/0304399194900922.html:text/html},
}

@article{godard_noise_2012,
	title = {Noise models for low counting rate coherent diffraction imaging},
	volume = {20},
	copyright = {© 2012 Optical Society of America},
	issn = {1094-4087},
	url = {https://opg.optica.org/oe/abstract.cfm?uri=oe-20-23-25914},
	doi = {10.1364/OE.20.025914},
	abstract = {Coherent diffraction imaging (CDI) is a lens-less microscopy method that extracts the complex-valued exit field from intensity measurements alone. It is of particular importance for microscopy imaging with diffraction set-ups where high quality lenses are not available. The inversion scheme allowing the phase retrieval is based on the use of an iterative algorithm. In this work, we address the question of the choice of the iterative process in the case of data corrupted by photon or electron shot noise. Several noise models are presented and further used within two inversion strategies, the ordered subset and the scaled gradient. Based on analytical and numerical analysis together with Monte-Carlo studies, we show that any physical interpretations drawn from a CDI iterative technique require a detailed understanding of the relationship between the noise model and the used inversion method. We observe that iterative algorithms often assume implicitly a noise model. For low counting rates, each noise model behaves differently. Moreover, the used optimization strategy introduces its own artefacts. Based on this analysis, we develop a hybrid strategy which works efficiently in the absence of an informed initial guess. Our work emphasises issues which should be considered carefully when inverting experimental data.},
	language = {EN},
	number = {23},
	urldate = {2024-12-19},
	journal = {Optics Express},
	publisher = {Optica Publishing Group},
	author = {Godard, Pierre and Allain, Marc and Chamard, Virginie and Rodenburg, John},
	month = nov,
	year = {2012},
	keywords = {Phase retrieval, Spatial resolution, CCD cameras, Phase shift, Shot noise, Zone plates},
	pages = {25914--25934},
	file = {Full Text:files/959/Godard et al. - 2012 - Noise models for low counting rate coherent diffra.pdf:application/pdf},
}

@article{li_direct_2019,
	title = {Direct {Imaging} of {Tunable} {Crystal} {Surface} {Structures} of {MOF} {MIL}-101 {Using} {High}-{Resolution} {Electron} {Microscopy}},
	volume = {141},
	issn = {0002-7863},
	url = {https://doi.org/10.1021/jacs.9b04896},
	doi = {10.1021/jacs.9b04896},
	abstract = {Metal–organic frameworks (MOFs) are often synthesized using various additives to modulate the crystallization. Here, we report the direct imaging of the crystal surface of MOF MIL-101 synthesized with different additives, using low-dose high-resolution transmission electron microscopy (HRTEM), and identify three distinct surface structures, at subunit cell resolution. We find that the mesoporous cages at the outermost surface of MIL-101 can be opened up by vacuum heating treatment at different temperatures, depending on the MIL-101 samples. We monitor the structural evolution of MIL-101 upon vacuum heating, using in situ X-ray diffraction, and find the results to be in good agreement with HRTEM observations, which leads us to speculate that additives have an influence not only on the surface structure but also on the stability of framework. In addition, we observe solid–solid phase transformation from MIL-101 to MIL-53 taking place in the sample synthesized with hydrofluoric acid.},
	number = {30},
	urldate = {2024-12-19},
	journal = {Journal of the American Chemical Society},
	publisher = {American Chemical Society},
	author = {Li, Xinghua and Wang, Jianjian and Liu, Xin and Liu, Lingmei and Cha, Dongkyu and Zheng, Xinliang and Yousef, Ali A. and Song, Kepeng and Zhu, Yihan and Zhang, Daliang and Han, Yu},
	month = jul,
	year = {2019},
	pages = {12021--12028},
	file = {Full Text PDF:files/961/Li et al. - 2019 - Direct Imaging of Tunable Crystal Surface Structur.pdf:application/pdf},
}

@misc{robert_benchmarking_2025,
	title = {Benchmarking analytical electron ptychography methods for the low-dose imaging of beam-sensitive materials},
	url = {http://arxiv.org/abs/2501.08874},
	doi = {10.48550/arXiv.2501.08874},
	abstract = {This publication presents an investigation of the performance of different analytical electron ptychography methods for low-dose imaging. In particular, benchmarking is performed for two modelobjects, monolayer MoS2 and apoferritin, by means of multislice simulations. Specific attention is given to cases where the individual diffraction patterns remain sparse. After a first rigorous introduction to the theoretical foundations of the methods, an implementation based on the scan-frequency partitioning of calculation steps is described, permitting a significant reduction of memory needs and high sampling flexibility. By analyzing the role of contrast transfer and illumination conditions, this work provides insights into the trade-off between resolution, signal-to-noise ratio and probe focus, as is necessary for the optimization of practical experiments. Furthermore, important differences between the different methods are demonstrated. Overall, the results obtained for the two model-objects demonstrate that analytical ptychography is an attractive option for the low-dose imaging of beam-sensitive materials.},
	language = {en},
	urldate = {2025-01-16},
	publisher = {arXiv},
	author = {Robert, Hoelen L. Lalandec and Leidl, Max Leo and Müller-Caspary, Knut and Verbeeck, Jo},
	month = jan,
	year = {2025},
	note = {arXiv:2501.08874 [physics]},
	keywords = {Physics - Applied Physics},
	file = {Robert et al. - 2025 - Benchmarking analytical electron ptychography meth.pdf:files/987/Robert et al. - 2025 - Benchmarking analytical electron ptychography meth.pdf:application/pdf},
}

@article{melnyk_connections_2022,
	title = {On connections between {Amplitude} {Flow} and {Error} {Reduction} for phase retrieval and ptychography},
	volume = {20},
	issn = {2730-5724},
	url = {https://doi.org/10.1007/s43670-022-00035-5},
	doi = {10.1007/s43670-022-00035-5},
	abstract = {In this paper, we consider two iterative algorithms for the phase retrieval problem: the well-known Error Reduction method and the Amplitude Flow algorithm, which performs minimization of the amplitude-based squared loss via the gradient descent. We show that Error Reduction can be interpreted as a scaled gradient method applied to minimize the same amplitude-based squared loss, which allows to establish its convergence properties. Moreover, we show that for a class of measurement scenarios, such as ptychography, both methods have the same computational complexity and sometimes even coincide.},
	language = {en},
	number = {2},
	urldate = {2025-01-23},
	journal = {Sampling Theory, Signal Processing, and Data Analysis},
	author = {Melnyk, Oleh},
	month = sep,
	year = {2022},
	keywords = {Ptychography, Phase retrieval, 47J25, 47J26, 78A46, 78M50, 90C26, Amplitude Flow, Error Reduction},
	pages = {16},
	file = {Full Text PDF:files/992/Melnyk - 2022 - On connections between Amplitude Flow and Error Re.pdf:application/pdf},
}

@article{oleary_phase_2020,
	title = {Phase reconstruction using fast binary {4D} {STEM} data},
	volume = {116},
	issn = {0003-6951},
	url = {https://doi.org/10.1063/1.5143213},
	doi = {10.1063/1.5143213},
	abstract = {We report the application of focused probe ptychography using binary 4D datasets obtained using scanning transmission electron microscopy (STEM). Modern fast pixelated detectors have enabled imaging of individual convergent beam electron diffraction patterns in a STEM raster scan at frame rates in the range of 1000–8000 Hz using conventional counting modes. Changing the bit depth of a counting detector, such that only values of 0 or 1 can be recorded at each pixel, allows one to decrease the dwell time and increase the frame rate to 12.5 kHz, reducing the electron exposure of the sample for a given beam current. Atomically resolved phase contrast of an aluminosilicate zeolite (ZSM-5) is observed from sparse diffraction patterns with isolated individual electrons, demonstrating the potential of binary ptychography as a low-dose 4D STEM technique.},
	number = {12},
	urldate = {2025-02-03},
	journal = {Applied Physics Letters},
	author = {O'Leary, C. M. and Allen, C. S. and Huang, C. and Kim, J. S. and Liberti, E. and Nellist, P. D. and Kirkland, A. I.},
	month = mar,
	year = {2020},
	pages = {124101},
	file = {Full Text:files/1027/O'Leary et al. - 2020 - Phase reconstruction using fast binary 4D STEM dat.pdf:application/pdf;Snapshot:files/1028/Phase-reconstruction-using-fast-binary-4D-STEM.html:text/html},
}

@article{thibault_reconstructing_2013,
	title = {Reconstructing state mixtures from diffraction measurements},
	volume = {494},
	copyright = {2013 Springer Nature Limited},
	issn = {1476-4687},
	url = {https://www.nature.com/articles/nature11806},
	doi = {10.1038/nature11806},
	abstract = {An imaging technique has been developed to characterize state mixtures caused by partial coherence and fluctuations in dynamical systems.},
	language = {en},
	number = {7435},
	urldate = {2025-02-13},
	journal = {Nature},
	publisher = {Nature Publishing Group},
	author = {Thibault, Pierre and Menzel, Andreas},
	month = feb,
	year = {2013},
	keywords = {X-rays, Applied physics, Scientific data},
	pages = {68--71},
	file = {Full Text PDF:files/1045/Thibault and Menzel - 2013 - Reconstructing state mixtures from diffraction mea.pdf:application/pdf},
}

@misc{noauthor_overlap_nodate,
	title = {On {Overlap} {Ratio} in {Defocused} {Electron} {Ptychography}},
	url = {https://arxiv.org/html/2502.00762v1},
	urldate = {2025-02-17},
	file = {On Overlap Ratio in Defocused Electron Ptychograph.pdf:files/1051/On Overlap Ratio in Defocused Electron Ptychograph.pdf:application/pdf;On Overlap Ratio in Defocused Electron Ptychography:files/1047/2502.html:text/html},
}

@misc{zhao_electron_2025,
	title = {Electron {Fourier} ptychography for phase reconstruction},
	url = {http://arxiv.org/abs/2502.08342},
	doi = {10.48550/arXiv.2502.08342},
	abstract = {Phase reconstruction is important in transmission electron microscopy for structural studies. We describe electron Fourier ptychography and its application to phase reconstruction of both radiation-resistant and beam-sensitive materials. We demonstrate that the phase of the exit wave can be reconstructed at high resolution using a modified iterative phase retrieval algorithm with data collected using an alternative optical geometry. This method achieves a spatial resolution of 0.63 nm at a fluence of \$4.5 {\textbackslash}times 10{\textasciicircum}2 {\textbackslash}, e{\textasciicircum}-/{\textbackslash}text\{nm\}{\textasciicircum}2\$, as validated on Cry11Aa protein crystals under cryogenic conditions. Notably, this method requires no additional hardware modifications, is straightforward to implement, and can be seamlessly integrated with existing data collection software, providing a broadly accessible approach for structural studies.},
	urldate = {2025-02-17},
	publisher = {arXiv},
	author = {Zhao, Jingjing and Huang, Chen and Mostaed, Ali and Moshtaghpour, Amirafshar and Parkhurst, James M. and Lobato, Ivan and Gallagher-Jones, Marcus and Kim, Judy S. and Boyce, Mark and Stuart, David and Andreeva, Elena A. and Colletier, Jacques-Philippe and Kirkland, Angus I.},
	month = feb,
	year = {2025},
	note = {arXiv:2502.08342 [physics]},
	keywords = {Condensed Matter - Materials Science, Physics - Applied Physics},
	file = {Preprint PDF:files/1049/Zhao et al. - 2025 - Electron Fourier ptychography for phase reconstruc.pdf:application/pdf;Snapshot:files/1050/2502.html:text/html},
}

@misc{chennit_investigating_2025,
	title = {Investigating the convergence properties of iterative ptychography for atomic-resolution low-dose imaging},
	url = {http://arxiv.org/abs/2507.06756},
	doi = {10.48550/arXiv.2507.06756},
	abstract = {This study investigates the convergence properties of a collection of iterative electron ptychography methods, under low electron doses (\${\textless}\$ 10\${\textasciicircum}3\$ \$e{\textasciicircum}-/A{\textasciicircum}2\$) and gives particular attention to the impact of the user-defined update strengths. We demonstrate that carefully chosen values for this parameter, ideally smaller than those conventionally met in the literature, are essential for achieving accurate reconstructions of the projected electrostatic potential. Using a 4D dataset of a thin hybrid organic-inorganic formamidinium lead bromide (FAPbBr\$\_\{3\}\$) sample, we show that convergence is in practice achievable only when the update strengths for both the object and probe are relatively small compared to what is found in literature. Additionally we demonstrate that under low electron doses, the reconstructions initial error increases when the update strength coefficients are reduced below a certain threshold emphasizing the existence of critical values beyond which the algorithms are trapped in local minima. These findings highlight the need for carefully optimized reconstruction parameters in iterative ptychography, especially when working with low electron doses, ensuring both effective convergence and correctness of the result.},
	urldate = {2025-09-07},
	publisher = {arXiv},
	author = {Chennit, Tamazouzt and Li, Songge and Robert, Hoelen L. Lalandec and Hofer, Christoph and Schrenker, Nadine J. and Manna, Liberato and Bals, Sara and Pennycook, Timothy J. and Verbeeck, Jo},
	month = jul,
	year = {2025},
	note = {arXiv:2507.06756 [cond-mat]},
	keywords = {Condensed Matter - Materials Science, Physics - Applied Physics, Physics - Instrumentation and Detectors},
	file = {Preprint PDF:files/1221/Chennit et al. - 2025 - Investigating the convergence properties of iterat.pdf:application/pdf;Snapshot:files/1220/2507.html:text/html},
}

@article{maiden_superresolution_2011,
	title = {Superresolution imaging via ptychography},
	volume = {28},
	copyright = {https://doi.org/10.1364/OA\_License\_v1\#VOR},
	issn = {1084-7529, 1520-8532},
	url = {https://opg.optica.org/abstract.cfm?URI=josaa-28-4-604},
	doi = {10.1364/JOSAA.28.000604},
	language = {en},
	number = {4},
	urldate = {2025-09-16},
	journal = {Journal of the Optical Society of America A},
	author = {Maiden, Andrew M. and Humphry, Martin J. and Zhang, Fucai and Rodenburg, John M.},
	month = apr,
	year = {2011},
	pages = {604},
	file = {Maiden et al. - 2011 - Superresolution imaging via ptychography.pdf:files/1226/Maiden et al. - 2011 - Superresolution imaging via ptychography.pdf:application/pdf},
}

@article{fannjiang_absolute_2012,
	title = {Absolute {Uniqueness} of {Phase} {Retrieval} with {Random} {Illumination}},
	volume = {28},
	issn = {0266-5611, 1361-6420},
	url = {http://arxiv.org/abs/1110.5097},
	doi = {10.1088/0266-5611/28/7/075008},
	abstract = {Random illumination is proposed to enforce absolute uniqueness and resolve all types of ambiguity, trivial or nontrivial, in phase retrieval. Almost sure irreducibility is proved for any complex-valued object whose support set has rank ≥ 2. While the new irreducibility result can be viewed as a probabilistic version of the classical result by Bruck, Sodin and Hayes, it provides a novel perspective and an eﬀective method for phase retrieval. In particular, almost sure uniqueness, up to a global phase, is proved for complex-valued objects under general two-point conditions. Under a tight sector constraint absolute uniqueness is proved to hold with probability exponentially close to unity as the object sparsity increases. Under a magnitude constraint with random amplitude illumination, uniqueness modulo global phase is proved to hold with probability exponentially close to unity as object sparsity increases. For general complex-valued objects without any constraint, almost sure uniqueness up to global phase is established with two sets of Fourier magnitude data under two independent illuminations. Numerical experiments suggest that random illumination essentially alleviates most, if not all, numerical problems commonly associated with the standard phasing algorithms.},
	language = {en},
	number = {7},
	urldate = {2025-09-16},
	journal = {Inverse Problems},
	author = {Fannjiang, Albert},
	month = jul,
	year = {2012},
	note = {arXiv:1110.5097 [physics]},
	keywords = {Physics - Optics, Computer Science - Computer Vision and Pattern Recognition, Mathematical Physics, Mathematics - Mathematical Physics},
	pages = {075008},
	file = {Fannjiang - 2012 - Absolute Uniqueness of Phase Retrieval with Random.pdf:files/1229/Fannjiang - 2012 - Absolute Uniqueness of Phase Retrieval with Random.pdf:application/pdf},
}

@misc{wittwer_object_2022,
	title = {Object {Initialization} {For} {Ptychographic} {Scans} {With} {Reduced} {Overlap}},
	url = {http://arxiv.org/abs/2207.13760},
	doi = {10.48550/arXiv.2207.13760},
	abstract = {X-ray ptychography utilizes overlapping illuminations to reconstruct the object’s phase and absorption signal with spatial resolutions much smaller than the focus size. Usually, the illumination overlap is chosen to be between 50 \% and 60 \% in order to ensure high quality reconstructions at reasonable scan times and/or doses. Here, we experimentally demonstrate that ptychographic iteration with object instead of ﬂat initialization allows for a signiﬁcant reduction of the overlap with only a modest loss in reconstruction quality. This approach could prove beneﬁcial for dose sensitive experiments and for rapid feedback overview scans.},
	language = {en},
	urldate = {2025-09-16},
	publisher = {arXiv},
	author = {Wittwer, Felix and Modregger, Peter},
	month = jul,
	year = {2022},
	note = {arXiv:2207.13760 [cond-mat]},
	keywords = {Condensed Matter - Materials Science, Physics - Optics, Physics - Medical Physics},
	file = {Wittwer and Modregger - 2022 - Object Initialization For Ptychographic Scans With.pdf:files/1230/Wittwer and Modregger - 2022 - Object Initialization For Ptychographic Scans With.pdf:application/pdf},
}

@incollection{hawkes_ptychography_2019,
	address = {Cham},
	title = {Ptychography},
	isbn = {978-3-030-00068-4 978-3-030-00069-1},
	url = {http://link.springer.com/10.1007/978-3-030-00069-1_17},
	doi = {10.1007/978-3-030-00069-1_17},
	abstract = {Ptychography is a computational imaging technique. A detector records an extensive data set consisting of many inference patterns obtained as an object is displaced to various positions relative to an illumination field. A computer algorithm of some type is then used to invert this data into an image. It has three key advantages: it does not depend upon a good-quality lens, or indeed on using any lens at all; it can obtain the image wave in phase as well as in intensity; and it can self-calibrate in the sense that errors that arise in the experimental set-up can be accounted for and their effects removed. Its transfer function is in theory perfect, with resolution being wavelength-limited. Although the main concepts of ptychography were developed many years ago, it has only recently (over the last ten years) become widely adopted. This chapter surveys visible light, X-ray, electron, and EUV ptychography as applied to microscopic imaging. It describes the principal experimental arrangements used at these various wavelengths. It reviews the most common inversion algorithms that are nowadays employed, giving examples of meta code to implement these. It describes, for those new to the field, how to avoid the most common pitfalls in obtaining good quality reconstructions. It also discusses more advanced techniques such as modal decomposition and strategies to cope with 3D multiple scattering.},
	language = {en},
	urldate = {2025-09-16},
	booktitle = {Springer {Handbook} of {Microscopy}},
	publisher = {Springer International Publishing},
	author = {Rodenburg, John and Maiden, Andrew},
	editor = {Hawkes, Peter W. and Spence, John C. H.},
	year = {2019},
	note = {Series Title: Springer Handbooks},
	pages = {819--904},
	file = {Rodenburg and Maiden - 2019 - Ptychography.pdf:files/1231/Rodenburg and Maiden - 2019 - Ptychography.pdf:application/pdf},
}

@article{dalfonso_dose-dependent_2016,
	title = {Dose-dependent high-resolution electron ptychography},
	volume = {119},
	issn = {0021-8979},
	url = {https://doi.org/10.1063/1.4941269},
	doi = {10.1063/1.4941269},
	abstract = {Recent reports of electron ptychography at atomic resolution have ushered in a new era of coherent diffractive imaging in the context of electron microscopy. We report and discuss electron ptychography under variable electron dose conditions, exploring the prospects of an approach which has considerable potential for imaging where low dose is needed.},
	number = {5},
	urldate = {2025-09-16},
	journal = {Journal of Applied Physics},
	author = {D'Alfonso, A. J. and Allen, L. J. and Sawada, H. and Kirkland, A. I.},
	month = feb,
	year = {2016},
	pages = {054302},
	file = {Snapshot:files/1239/Dose-dependent-high-resolution-electron.html:text/html},
}

@article{dalfonso_deterministic_2014,
	title = {Deterministic electron ptychography at atomic resolution},
	volume = {89},
	url = {https://link.aps.org/doi/10.1103/PhysRevB.89.064101},
	doi = {10.1103/PhysRevB.89.064101},
	abstract = {We present a fast deterministic approach to the ptychographic reconstruction of the transmission function of a specimen at atomic resolution. The method is demonstrated using a data set obtained from a cerium dioxide nanoparticle using an aberration-corrected electron microscope and is compared to established approaches to ptychography. The method is based on the solution of an overdetermined set of linear equations, and is computationally efficient and robust to measurement noise. The set of linear equations is efficiently solved using the conjugate gradient least squares method implemented using fast Fourier transforms.},
	number = {6},
	urldate = {2025-09-16},
	journal = {Physical Review B},
	publisher = {American Physical Society},
	author = {D'Alfonso, A. J. and Morgan, A. J. and Yan, A. W. C. and Wang, P. and Sawada, H. and Kirkland, A. I. and Allen, L. J.},
	month = feb,
	year = {2014},
	pages = {064101},
}

@article{ilett_studying_2024,
	title = {Studying crystallisation processes using electron microscopy: {The} importance of sample preparation},
	volume = {295},
	issn = {0022-2720, 1365-2818},
	shorttitle = {Studying crystallisation processes using electron microscopy},
	url = {https://onlinelibrary.wiley.com/doi/10.1111/jmi.13300},
	doi = {10.1111/jmi.13300},
	abstract = {We present a comparison of common electron microscopy sample preparation methods for studying crystallisation processes from solution using both scanning and transmission electron microscopy (SEM and TEM). We focus on two widely studied inorganic systems: calcium sulphate, gypsum (CaSO4⋅2H2O) and calcium carbonate (CaCO3). We find significant differences in crystallisation kinetics and polymorph selection between the different sample preparation methods, which indicate that drying and chemical quenching can induce severe artefacts that are capable of masking the true native state of the crystallising solution. Overall, these results highlight the importance of cryogenic (cryo)quenching crystallising solutions and the use of full cryo-TEM as the most reliable method for studying the early stages of crystallisation.},
	language = {en},
	number = {3},
	urldate = {2025-10-03},
	journal = {Journal of Microscopy},
	author = {Ilett, Martha and Afzali, Maryam and Abdulkarim, Bilal and Aslam, Zabeada and Foster, Stephanie and Burgos‐Ruiz, Miguel and Kim, Yi‐Yeoun and Meldrum, Fiona C. and Drummond‐Brydson, Rik M.},
	month = sep,
	year = {2024},
	pages = {243--256},
	file = {Ilett et al. - 2024 - Studying crystallisation processes using electron .pdf:files/1241/Ilett et al. - 2024 - Studying crystallisation processes using electron .pdf:application/pdf},
}

@article{li_4d-stem_2022,
	title = {{4D}-{STEM} {Ptychography} for {Electron}-{Beam}-{Sensitive} {Materials}},
	volume = {8},
	copyright = {https://creativecommons.org/licenses/by/4.0/},
	issn = {2374-7943, 2374-7951},
	url = {https://pubs.acs.org/doi/10.1021/acscentsci.2c01137},
	doi = {10.1021/acscentsci.2c01137},
	abstract = {Recent advances in high-speed pixelated electron detectors have substantially facilitated the implementation of four-dimensional scanning transmission electron microscopy (4D-STEM). A critical application of 4D-STEM is electron ptychography, which reveals the atomic structure of a specimen by reconstructing its transmission function from redundant convergent-beam electron diffraction patterns. Although 4D-STEM ptychography offers many advantages over conventional imaging modes, this emerging technique has not been fully applied to materials highly sensitive to electron beams. In this Outlook, we introduce the fundamentals of 4D-STEM ptychography, focusing on data collection and processing methods, and present the current applications of 4D-STEM ptychography in various materials. Next, we discuss the potential advantages of imaging electron-beam-sensitive materials using 4D-STEM ptychography and explore its feasibility by performing simulations and experiments on a zeolite material. The preliminary results demonstrate that, at the low electron dose required to preserve the zeolite structure, 4D-STEM ptychography can reliably provide higher resolution and greater tolerance to the specimen thickness and probe defocus as compared to existing imaging techniques. In the final section, we discuss the challenges and possible strategies to further reduce the electron dose for 4D-STEM ptychography. If successful, it will be a game-changer for imaging extremely sensitive materials, such as metal−organic frameworks, hybrid halide perovskites, and supramolecular crystals.},
	language = {en},
	number = {12},
	urldate = {2025-10-03},
	journal = {ACS Central Science},
	author = {Li, Guanxing and Zhang, Hui and Han, Yu},
	month = dec,
	year = {2022},
	pages = {1579--1588},
	file = {Li et al. - 2022 - 4D-STEM Ptychography for Electron-Beam-Sensitive M.pdf:files/1243/Li et al. - 2022 - 4D-STEM Ptychography for Electron-Beam-Sensitive M.pdf:application/pdf},
}

@misc{meysman_hierarchical_2025,
	title = {A hierarchical nickel organic framework confers high conductivity over long distances in cable bacteria},
	copyright = {© 2025, Posted by Cold Spring Harbor Laboratory. This pre-print is available under a Creative Commons License (Attribution-NonCommercial-NoDerivs 4.0 International), CC BY-NC-ND 4.0, as described at http://creativecommons.org/licenses/by-nc-nd/4.0/},
	url = {https://www.biorxiv.org/content/10.1101/2025.10.10.681601v1},
	doi = {10.1101/2025.10.10.681601},
	abstract = {Multi-cellular cable bacteria have evolved a unique machinery that efficiently transports electrons across centimetre-scale distances. Currents flow through a parallel network of periplasmic fibres, which display an extraordinary conductivity for a biological material. However, the conduction mechanism remains elusive as the molecular structure of the fibres has not been resolved. Here, we demonstrate that each fibre embeds a bundle of intertwined nanoribbons, which are built from Nickel Bis(Dithiolene) (NiBiD) repeat units that are formed by linking nickel centres with ethenetetrathiolate ligands. The planar and conjugated NiBiD complexes are aligned and stacked to form an elongated supramolecular coordination network, thus explaining the observed organo-metal electronic properties of the fibres. Our results hence demonstrate that biology is capable of producing extensive metal organic frameworks. These structures enable highly conductive one-dimensional conduits, ensuring efficient charge transport over macroscale distances, thus providing a novel design principle for bio-based, sustainable organo-electronic materials.},
	language = {en},
	urldate = {2025-10-14},
	publisher = {bioRxiv},
	author = {Meysman, Filip J. R. and Smets, Bent and Martinez, Silvia Hidalgo and Claes, Nathalie and Schroeder, Bob C. and Geelhoed, Jeanine S. and Liu, Yun and Choyikutty, Jiji and Chennit, Tamazouzt and Bodson, Thijs and Collauto, Alberto and Roessler, Maxie M. and Karpov, Dmitry and Bohic, Sylvain and Aramini, Matteo and Hayama, Shusaku and Wetherington, Maxwell and Zwijnenburg, Martijn A. and Pankratova, Galina and Pintelon, Isabel and Timmermans, Jean-Pierre and Nuyts, Gert and Wael, Karolien De and Bals, Sara and Verbeeck, Johan and Remaut, Han and Boschker, Henricus T. S.},
	month = oct,
	year = {2025},
	note = {ISSN: 2692-8205
Pages: 2025.10.10.681601
Section: New Results},
	file = {Full Text PDF:files/1248/Meysman et al. - 2025 - A hierarchical nickel organic framework confers hi.pdf:application/pdf},
}

@article{sun_challenge_2025,
	title = {The challenge of imaging electron-beam sensitive {LiCoO2} cathode at atomic scale and a ptychography solution},
	volume = {111},
	issn = {2095-4956},
	url = {https://www.sciencedirect.com/science/article/pii/S2095495625005893},
	doi = {10.1016/j.jechem.2025.07.037},
	urldate = {2025-11-13},
	journal = {Journal of Energy Chemistry},
	author = {Sun, Kang and Sha, Haozhi and Cui, Jizhe and Zhang, Jingxi and Liu, Zehui and Dong, Yanhao and Yu, Rong},
	month = dec,
	year = {2025},
	pages = {61--66},
	file = {ScienceDirect Snapshot:files/1250/S2095495625005893.html:text/html;Sun et al. - 2025 - The challenge of imaging electron-beam sensitive L.pdf:files/1251/Sun et al. - 2025 - The challenge of imaging electron-beam sensitive L.pdf:application/pdf},
}

@article{wang_ptychographic_2025,
	title = {Ptychographic {Observation} of {Lithium} {Atoms} in the {Irradiation}-{Sensitive} {Garnet}-{Type} {Solid} {Electrolyte} at {Sub}-{Angstrom} {Resolution}},
	volume = {147},
	issn = {0002-7863},
	url = {https://doi.org/10.1021/jacs.5c03627},
	doi = {10.1021/jacs.5c03627},
	abstract = {Garnet-type solid electrolyte cubic Li7La3Zr2O12 (c-LLZO) emerges as a promising candidate for establishing reliable and high-performance lithium-ion batteries. Its extreme sensitivity to electron irradiation poses a significant challenge in atomically resolving its structure using conventional transmission electron microscopy (TEM) techniques. We demonstrate that the combination of low-dose four-dimensional scanning TEM (4D-STEM) with multislice ptychographic retrieval methodology manages to achieve a sub-Angstrom resolution and the direct visualization of lithium atoms within c-LLZO. The distribution of lithium in depth direction is also obtained. This work provides atomic-scale insights into the distribution of light elements within irradiation-sensitive dense crystals, paving the way for investigating the microstructure–property relationship.},
	number = {21},
	urldate = {2025-11-13},
	journal = {Journal of the American Chemical Society},
	publisher = {American Chemical Society},
	author = {Wang, Zeyu and Hu, Xiangchen and Zhang, Yue and Wu, Xiaoyan and Shi, Hongsheng and Liu, Wei and Yu, Yi},
	month = may,
	year = {2025},
	pages = {18025--18032},
	file = {Full Text PDF:files/1300/Wang et al. - 2025 - Ptychographic Observation of Lithium Atoms in the .pdf:application/pdf},
}

@article{noauthor_direct_2022,
	title = {Direct imaging of oxygen shifts associated with the oxygen redox of {Li}-rich layered oxides},
	volume = {6},
	issn = {2542-4351},
	url = {https://www.sciencedirect.com/science/article/pii/S2542435122001465},
	doi = {10.1016/j.joule.2022.04.008},
	abstract = {Li-rich metal oxides, such as Li1.2Ni0.13Mn0.54Co0.13O2, can deliver high specific capacities because of the redox of lattice O2− in addition to the c…},
	language = {en-US},
	number = {5},
	urldate = {2025-11-13},
	journal = {Joule},
	publisher = {Cell Press},
	month = may,
	year = {2022},
	pages = {1049--1065},
	file = {Snapshot:files/1302/S2542435122001465.html:text/html},
}

@article{song_direct_2022,
	title = {Direct imaging of oxygen shifts associated with the oxygen redox of {Li}-rich layered oxides},
	volume = {6},
	issn = {2542-4351},
	url = {https://www.sciencedirect.com/science/article/pii/S2542435122001465},
	doi = {10.1016/j.joule.2022.04.008},
	abstract = {Li-rich metal oxides, such as Li1.2Ni0.13Mn0.54Co0.13O2, can deliver high specific capacities because of the redox of lattice O2− in addition to the cations. Observing oxygen distortions is key to understand the redox process. Electron ptychography is a phase-reconstruction method in 4D scanning transmission electron microscopy, providing atomic-resolution phase images with high signal-to-noise ratio and dose efficiency. Herein, we use electron ptychography to image the oxygen shift in Li1.2Ni0.13Mn0.54Co0.13O2 during the first cycle. The picometer-scale precision measurement shows distinct oxygen shifts in the bulk and surface after charging and compares with various theoretical anionic redox models. The shift after discharging is not seen to recover in the bulk accounting for voltage hysteresis; however, it recovers close to the surface, although with a phase change. We suggest that Li1.2Ni0.13Mn0.54Co0.13O2 proceeds distinct oxygen redox in the bulk and surface. The altered oxygen sublattice after first cycle potentially explains the changed voltage profiles of following cycles.},
	number = {5},
	urldate = {2025-11-13},
	journal = {Joule},
	author = {Song, Weixin and Pérez-Osorio, Miguel A. and Marie, John-Joseph and Liberti, Emanuela and Luo, Xiaonan and O’Leary, Colum and House, Robert A. and Bruce, Peter G. and Nellist, Peter D.},
	month = may,
	year = {2022},
	keywords = {electron ptychography, Li-rich layered oxides, LiNiMnCoO, oxygen distortion, oxygen redox},
	pages = {1049--1065},
	file = {Full Text:files/1305/Song et al. - 2022 - Direct imaging of oxygen shifts associated with th.pdf:application/pdf;ScienceDirect Snapshot:files/1304/S2542435122001465.html:text/html},
}

@article{liang_nanocomposite_2018,
	title = {Nanocomposite {Materials} for the {Sodium}–{Ion} {Battery}: {A} {Review}},
	volume = {14},
	issn = {1613-6810},
	shorttitle = {Nanocomposite {Materials} for the {Sodium}–{Ion} {Battery}},
	url = {https://onlinelibrary.wiley.com/doi/full/10.1002/smll.201702514},
	doi = {10.1002/smll.201702514},
	abstract = {Abstract Clean energy has become an important topic in recent decades because of the serious global issues related to the development of energy, such as environmental contamination, and the intermittence of the traditional energy sources. Creating new battery-related energy storage facilities is an urgent subject for human beings to address and for solutions for the future. Compared with lithium-based batteries, sodium?ion batteries have become the new focal point in the competition for clean energy solutions and have more potential for commercialization due to the huge natural abundance of sodium. Nevertheless, sodium?ion batteries still exhibit some challenges, like inferior electrochemical performance caused by the bigger ionic size of Na+ ions, the detrimental volume expansion, and the low conductivity of the active materials. To solve these issues, nanocomposites have recently been applied as a new class of electrodes to enhance the electrochemical performance in sodium batteries based on advantages that include the size effect, high stability, and excellent conductivity. In this Review, the recent development of nanocomposite materials applied in sodium?ion batteries is summarized, and the existing challenges and the potential solutions are presented.},
	number = {5},
	urldate = {2025-11-14},
	journal = {Small},
	publisher = {John Wiley \& Sons, Ltd},
	author = {Liang, Yaru and Lai, Wei-Hong and Miao, Zongcheng and Chou, Shu-Lei},
	month = feb,
	year = {2018},
	keywords = {anodes, cathodes, nanocomposites, sodium-ion batteries},
	pages = {1702514},
}

@article{nurohmah_sodium-ion_2022,
	title = {Sodium-ion battery from sea salt: a review},
	volume = {11},
	issn = {2194-1467},
	shorttitle = {Sodium-ion battery from sea salt},
	url = {https://doi.org/10.1007/s40243-022-00208-1},
	doi = {10.1007/s40243-022-00208-1},
	abstract = {The electrical energy storage is important right now, because it is influenced by increasing human energy needs, and the battery is a storage energy that is being developed simultaneously. Furthermore, it is planned to switch the lithium-ion batteries with the sodium-ion batteries and the abundance of the sodium element and its economical price compared to lithium is the main point. The main components anode and cathode have significant effect on the sodium battery performance. This review briefly describes the components of the sodium battery, including the anode, cathode, electrolyte, binder, and separator, and the sources of sodium raw material is the most important in material synthesis or installation. Sea salt or NaCl has potential ability as a raw material for sodium battery cathodes, and the usage of sea salt in the cathode synthesis process reduces production costs, because the salt is very abundant and environmentally friendly as well. When a cathode using a source of Na2CO3, which was synthesized independently from NaCl can save about 16.66\% after being calculated and anode with sodium metal when synthesized independently with NaCl can save about 98\% after being calculated, because sodium metal is classified as expensive matter.},
	language = {en},
	number = {1},
	urldate = {2025-11-14},
	journal = {Materials for Renewable and Sustainable Energy},
	author = {Nurohmah, Anisa Raditya and Nisa, Shofirul Sholikhatun and Stulasti, Khikmah Nur Rikhy and Yudha, Cornelius Satria and Suci, Windhu Griyasti and Aliwarga, Kiwi and Widiyandari, Hendri and Purwanto, Agus},
	month = apr,
	year = {2022},
	keywords = {Battery, Energy storage, Sea salt, Sodium-ion},
	pages = {71--89},
	file = {Full Text PDF:files/1312/Nurohmah et al. - 2022 - Sodium-ion battery from sea salt a review.pdf:application/pdf},
}

@article{schafer_multiscale_2024,
	title = {Multiscale {Investigation} of {Sodium}-{Ion} {Battery} {Anodes}: {Analytical} {Techniques} and {Applications}},
	volume = {14},
	issn = {1614-6840},
	shorttitle = {Multiscale {Investigation} of {Sodium}-{Ion} {Battery} {Anodes}},
	url = {https://onlinelibrary.wiley.com/doi/abs/10.1002/aenm.202302830},
	doi = {10.1002/aenm.202302830},
	abstract = {The anode/electrolyte interface behavior, and by extension, the overall cell performance of sodium-ion batteries is determined by a complex interaction of processes that occur at all components of the electrochemical cell across a wide range of size- and timescales. Single-scale studies may provide incomplete insights, as they cannot capture the full picture of this complex and intertwined behavior. Broad, multiscale studies are essential to elucidate these processes. Within this perspectives article, several analytical and theoretical techniques are introduced, and described how they can be combined to provide a more complete and comprehensive understanding of sodium-ion battery (SIB) performance throughout its lifetime, with a special focus on the interfaces of hard carbon anodes. These methods target various length- and time scales, ranging from micro to nano, from cell level to atomistic structures, and account for a broad spectrum of physical and (electro)chemical characteristics. Specifically, how mass spectrometric, microscopic, spectroscopic, electrochemical, thermodynamic, and physical methods can be employed to obtain the various types of information required to understand battery behavior will be explored. Ways are then discussed how these methods can be coupled together in order to elucidate the multiscale phenomena at the anode interface and develop a holistic understanding of their relationship to overall sodium-ion battery function.},
	language = {en},
	number = {15},
	urldate = {2025-11-14},
	journal = {Advanced Energy Materials},
	author = {Schäfer, David and Hankins, Kie and Allion, Michelle and Krewer, Ulrike and Karcher, Franziska and Derr, Laurin and Schuster, Rolf and Maibach, Julia and Mück, Stefan and Kramer, Dominik and Mönig, Reiner and Jeschull, Fabian and Daboss, Sven and Philipp, Tom and Neusser, Gregor and Romer, Jan and Palanisamy, Krishnaveni and Kranz, Christine and Buchner, Florian and Behm, R. Jürgen and Ahmadian, Ali and Kübel, Christian and Mohammad, Irshad and Samoson, Ago and Witter, Raiker and Smarsly, Bernd and Rohnke, Marcus},
	year = {2024},
	note = {\_eprint: https://advanced.onlinelibrary.wiley.com/doi/pdf/10.1002/aenm.202302830},
	keywords = {hard carbon, multiscale, SEI, sodium-ion battery, solid electrolyte interphase},
	pages = {2302830},
	file = {Full Text PDF:files/1356/Schäfer et al. - 2024 - Multiscale Investigation of Sodium-Ion Battery Ano.pdf:application/pdf},
}

@article{zulueta_unraveling_2024,
	title = {Unraveling fundamental characteristics of {Na2Mg3Cl8} as a solid-state electrolyte for {Na}-ion batteries},
	volume = {14},
	issn = {2046-2069},
	url = {https://pubs.rsc.org/en/content/articlelanding/2024/ra/d4ra06490a},
	doi = {10.1039/D4RA06490A},
	abstract = {In this theoretical study, we harnessed advanced atomistic computations to unravel several features of Na2Mg3Cl8, an unexplored but promising chloride compound for solid-state electrolytes in Na-batteries. First, Na2Mg3Cl8 exhibits an insulating behavior, characterized by an energy gap of ∼5 eV, arising from the hybridization of [NaCl] trigonal prismatic and [MgCl6] octahedral units. Second, the compound possesses mechanical stability and ductility, which render it suitable for practical fabrication. Improved electrolyte/electrode contact can reduce resistance and enhance battery performance. The electrochemical performance of Na2Mg3Cl8 involves an open cell voltage of 1.2 V and a theoretical capacity of 133 mA h g−1. Finally, its transport characteristics include low activation energy for diffusion and conduction as well as a remarkable room-temperature conductivity of 1.26 mS cm−1, comparable to those of current superionic conductors.},
	language = {en},
	number = {45},
	urldate = {2025-11-14},
	journal = {RSC Advances},
	publisher = {The Royal Society of Chemistry},
	author = {Zulueta, Yohandys A. and Fernández-Gamboa, Jose R. and Phung, Thi Viet Bac and Pham-Ho, My Phuong and Nguyen, Minh Tho},
	month = oct,
	year = {2024},
	pages = {33619--33628},
	file = {Full Text PDF:files/1358/Zulueta et al. - 2024 - Unraveling fundamental characteristics of Na2Mg3Cl.pdf:application/pdf},
}

@article{xu_unravelling_2025,
	title = {Unravelling nonclassical beam damage mechanisms in metal-organic frameworks by low-dose electron microscopy},
	volume = {16},
	copyright = {2024 The Author(s)},
	issn = {2041-1723},
	url = {https://www.nature.com/articles/s41467-024-55632-w},
	doi = {10.1038/s41467-024-55632-w},
	abstract = {Recent advances in direct electron detectors and low-dose imaging techniques have opened up captivating possibilities for real-space visualization of radiation-induced structural dynamics. This has significantly contributed to our understanding of electron-beam radiation damage in materials, serving as the foundation for modern electron microscopy. In light of these developments, the exploration of more precise and specific beam damage mechanisms, along with the development of associated descriptive models, has expanded the theoretical framework of radiation damage beyond classical mechanisms. We unravel, in this work, the nonclassical beam damage mechanisms of an open-framework material, i.e. UiO-66(Hf) metal-organic framework, by integrating low-dose electron microscopy and ab initio simulations of radiation induced structural dynamics. The physical origins of radiation damage phenomena, spanning across multiple scales including morphological, lattice, and molecular levels, have been unequivocally unveiled. Based on these observations, potential alternative mechanisms including reversible radiolysis and radiolysis-enhanced knock-on displacement are proposed, which account for their respective dynamic crystalline-to-amorphous interconversion and site-specific ligand knockout events occurring during continuous beam radiation. The current study propels the fundamental understanding of beam damage mechanisms from dynamic and correlated perspectives. Moreover, it fuels technical innovations, such as low-dose ultrafast electron microscopy, enabling imaging of beam-sensitive materials with uncompromised spatial resolution.},
	language = {en},
	number = {1},
	urldate = {2025-11-18},
	journal = {Nature Communications},
	publisher = {Nature Publishing Group},
	author = {Xu, Xiaoqiu and Xia, Liwei and Zheng, Changlin and Liu, Yikuan and Yu, Dongyang and Li, Jingjing and Zhong, Shigui and Li, Cuiyu and Song, Huijun and Liu, Yunzhou and Sun, Tulai and Li, Yonghe and Han, Yu and Zhao, Jia and Lin, Qiang and Li, Xiaonian and Zhu, Yihan},
	month = jan,
	year = {2025},
	keywords = {Coordination chemistry, Metal–organic frameworks},
	pages = {261},
	file = {Full Text PDF:files/1360/Xu et al. - 2025 - Unravelling nonclassical beam damage mechanisms in.pdf:application/pdf},
}

@article{li_atomically_2025,
	title = {Atomically resolved imaging of radiation-sensitive metal-organic frameworks via electron ptychography},
	volume = {16},
	copyright = {2025 The Author(s)},
	issn = {2041-1723},
	url = {https://www.nature.com/articles/s41467-025-56215-z},
	doi = {10.1038/s41467-025-56215-z},
	abstract = {Electron ptychography, recognized as an ideal technique for low-dose imaging, consistently achieves deep sub-angstrom resolution at electron doses of several thousand electrons per square angstrom (e−/Å2) or higher. Despite its proven efficacy, the application of electron ptychography at even lower doses—necessary for materials highly sensitive to electron beams—raises questions regarding its feasibility and the attainable resolution under such stringent conditions. Herein, we demonstrate the implementation of near-atomic-resolution ( {\textasciitilde} 2 Å) electron ptychography reconstruction at electron doses as low as {\textasciitilde}100 e−/Å2, for metal-organic frameworks (MOFs), which are known for their extreme sensitivity. The reconstructed images clearly resolve organic linkers, metal clusters, and even atomic columns within these clusters, while unravelling various local structural features in MOFs, including missing linkers, extra clusters, and surface termination modes. By combining the findings from simulations and experiments, we have identified that employing a small convergence semi-angle during data acquisition is crucial for effective iterative ptychographic reconstruction under such low-dose conditions. This important insight advances our understanding of the rapidly evolving electron ptychography technique and provides a novel approach to high-resolution imaging of various sensitive materials.},
	language = {en},
	number = {1},
	urldate = {2025-11-18},
	journal = {Nature Communications},
	publisher = {Nature Publishing Group},
	author = {Li, Guanxing and Xu, Ming and Tang, Wen-Qi and Liu, Ying and Chen, Cailing and Zhang, Daliang and Liu, Lingmei and Ning, Shoucong and Zhang, Hui and Gu, Zhi-Yuan and Lai, Zhiping and Muller, David A. and Han, Yu},
	month = jan,
	year = {2025},
	keywords = {Imaging techniques, Coordination chemistry, Organic–inorganic nanostructures},
	pages = {914},
	file = {Full Text PDF:files/1380/Li et al. - 2025 - Atomically resolved imaging of radiation-sensitive.pdf:application/pdf},
}

@article{basavappa_role_1994,
	title = {Role and mechanism of the maturation cleavage of {VP0} in poliovirus assembly: {Structure} of the empty capsid assembly intermediate at 2.9 Å resolution},
	volume = {3},
	copyright = {http://onlinelibrary.wiley.com/termsAndConditions\#vor},
	issn = {0961-8368, 1469-896X},
	shorttitle = {Role and mechanism of the maturation cleavage of {VP0} in poliovirus assembly},
	url = {https://onlinelibrary.wiley.com/doi/10.1002/pro.5560031005},
	doi = {10.1002/pro.5560031005},
	abstract = {Abstract
            The crystal structure of the P1/Mahoney poliovirus empty capsid has been determined at 2.9 Å resolution. The empty capsids differ from mature virions in that they lack the viral RNA and have yet to undergo a stabilizing maturation cleavage of VPO to yield the mature capsid proteins VP4 and VP2. The outer surface and the bulk of the protein shell are very similar to those of the mature virion. The major differences between the 2 structures are focused in a network formed by the N‐terminal extensions of the capsid proteins on the inner surface of the shell. In the empty capsids, the entire N‐terminal extension of VP1, as well as portions corresponding to VP4 and the N‐terminal extension of VP2, are disordered, and many stabilizing interactions that are present in the mature virion are missing. In the empty capsid, the VP1 scissile bond is located some 20 Å away from the positions in the mature virion of the termini generated by VP0 cleavage. The scissile bond is located on the rim of a trefoilshaped depression in the inner surface of the shell that is highly reminiscent of an RNA binding site in bean pod mottle virus. The structure suggests plausible (and ultimately testable) models for the initiation of encapsidation, for the RNA‐dependent autocatalytic cleavage of VP0, and for the role of the cleavage in establishing the ordered N‐terminal network and in generating stable virions.},
	language = {en},
	number = {10},
	urldate = {2025-12-19},
	journal = {Protein Science},
	author = {Basavappa, R. and Filman, D.J. and Syed, R. and Flore, O. and Icenogle, J.P. and Hogle, J.M.},
	month = oct,
	year = {1994},
	pages = {1651--1669},
	file = {Full Text:files/1448/Basavappa et al. - 1994 - Role and mechanism of the maturation cleavage of V.pdf:application/pdf},
}

@article{liu_universal_2026,
	title = {A {Universal} {FIB} {Approach} for {Contamination}- and {Damage}-{Free} {Plan}-{View} {TEM} {Lamellae} {Using} {NaCl} {Sacrificial} {Layers}},
	issn = {0304-3991},
	url = {https://www.sciencedirect.com/science/article/pii/S0304399126000124},
	doi = {10.1016/j.ultramic.2026.114319},
	abstract = {The preparation of high-quality plan-view transmission electron microscopy (TEM) lamellae is essential for investigating the in-plane properties of thin films. However, current focused ion beam (FIB) techniques are limited by ion-beam damage, surface contamination, and time-consuming workflows. Here, we introduce NaCl microcrystals as a sacrificial protective layer, which effectively shields the surface from ion irradiation and can be completely removed by simple dissolution in water, leaving a pristine surface. Building on this, we established a universal and streamlined FIB workflow for plan-view lamellae fabrication from thin films that eliminates the need for conventional Pt/C deposition and avoids custom hardware, relying solely on standard commercial components. Using a classic metal multilayer and an ultrathin epitaxial oxide film as representative model systems, we demonstrate that the prepared plan-view lamellae exhibit large uniform areas, preserved film structures, and contamination-free surfaces, enabling reliable surface-sensitive TEM analyses. This time-efficient and user-friendly approach offers a powerful solution for the contamination- and damage-free preparation of plan-view TEM lamellae across diverse thin-film systems, paving the way for in-depth investigations of their in-plane properties.},
	urldate = {2026-01-13},
	journal = {Ultramicroscopy},
	author = {Liu, Chen and Xu, Jingkai and Wang, Qingxiao and Guo, Tianchao and Chen, Maolin and Zheng, Dongxing and Alshareef, Husam N. and Zhang, Xixiang},
	month = jan,
	year = {2026},
	keywords = {Contamination-free, Damage-free, Focused ion beam (FIB), NaCl protective layer, Plan-view TEM lamella preparation, Thin films},
	pages = {114319},
	file = {PDF:files/1451/Liu et al. - 2026 - A Universal FIB Approach for Contamination- and Damage-Free Plan-View TEM Lamellae Using NaCl Sacrif.pdf:application/pdf;ScienceDirect Snapshot:files/1450/S0304399126000124.html:text/html},
}

@article{li_atomic-scale_2023,
	title = {Atomic-scale probing of short-range order and its impact on electrochemical properties in cation-disordered oxide cathodes},
	volume = {14},
	copyright = {2023 The Author(s)},
	issn = {2041-1723},
	url = {https://www.nature.com/articles/s41467-023-43356-2},
	doi = {10.1038/s41467-023-43356-2},
	abstract = {Chemical short-range-order has been widely noticed to dictate the electrochemical properties of Li-excess cation-disordered rocksalt oxides, a class of cathode based on earth abundant elements for next-generation high-energy-density batteries. Existence of short-range-order is normally evidenced by a diffused intensity pattern in reciprocal space, however, derivation of local atomic arrangements of short-range-order in real space is hardly possible. Here, by a combination of aberration-corrected scanning transmission electron microscopy, electron diffraction, and cluster-expansion Monte Carlo simulations, we reveal the short-range-order is a convolution of three basic types: tetrahedron, octahedron, and cube. We discover that short-range-order directly correlates with Li percolation channels, which correspondingly affects Li transport behavior. We further demonstrate that short-range-order can be effectively manipulated by anion doping or post-synthesis thermal treatment, creating new avenues for tailoring the electrochemical properties. Our results provide fundamental insights for decoding the complex relationship between local chemical ordering and properties of crystalline compounds.},
	language = {en},
	number = {1},
	urldate = {2026-01-15},
	journal = {Nature Communications},
	publisher = {Nature Publishing Group},
	author = {Li, Linze and Ouyang, Bin and Lun, Zhengyan and Huo, Haoyan and Chen, Dongchang and Yue, Yuan and Ophus, Colin and Tong, Wei and Chen, Guoying and Ceder, Gerbrand and Wang, Chongmin},
	month = nov,
	year = {2023},
	keywords = {Batteries},
	pages = {7448},
	file = {Full Text PDF:files/1453/Li et al. - 2023 - Atomic-scale probing of short-range order and its impact on electrochemical properties in cation-dis.pdf:application/pdf},
}

@misc{noauthor_cation-disordered_nodate,
	title = {Cation-disordered rocksalt-type high-entropy cathodes for {Li}-ion batteries {\textbar} {Nature} {Materials}},
	url = {https://www.nature.com/articles/s41563-020-00816-0},
	urldate = {2026-01-15},
}

@article{park_highthroughput_nodate,
	title = {High‐{Throughput} {Synthesis} of {Mn}‐{Based} {Disordered} {Rock}‐{Salt} {Li}‐{Ion} {Cathodes} with {Improved} {Rate} {Capability} via {Rapid} {Joule}‐{Heating}},
	url = {https://advanced.onlinelibrary.wiley.com/doi/10.1002/aenm.202503496},
	doi = {10.1002/aenm.202503496},
	abstract = {Rapid and energy-efficient Joule-heating synthesis enables the formation of high-performance Mn-based disordered rock-salt (DRX) cathodes through multiscale structural optimization. A case study on L...},
	language = {en},
	urldate = {2026-01-15},
	author = {Park, Sang-Wook and Kim, Hojoon and Han, Sangwook and Kim, Kyoung Sun and Song, You-Yeob and Lee, Hyun-Gi and Jang, Wonyoung and Lee, Dae-Hyung and Bae, Seongjae and Kim, Dongwoo and Park, Jungwoo and Park, Inchul and Nam, Sang-Cheol and Kim, Hyungsub and Lee, Jinhyuk and Kang, Kisuk and Seo, Dong-Hwa},
}

@article{f_carbon_2019,
	title = {Carbon doping of {WS2} monolayers: {Bandgap} reduction and p-type doping transport},
	shorttitle = {Carbon doping of {WS2} monolayers},
	url = {https://pubmed.ncbi.nlm.nih.gov/31139746/?utm_source=chatgpt.com},
	abstract = {Chemical doping constitutes an effective route to alter the electronic, chemical, and optical properties of two-dimensional transition metal dichalcogenides (2D-TMDs). We used a plasma-assisted method to introduce carbon-hydrogen (CH) units into WS$_{\textrm{2}}$ monolayers. We found CH-groups to be th …},
	language = {en},
	urldate = {2026-01-15},
	journal = {PubMed},
	author = {F, Zhang and Y, Lu and Ds, Schulman and T, Zhang and K, Fujisawa and Z, Lin and Y, Lei and Al, Elias and S, Das and Sb, Sinnott and M, Terrones},
	year = {2019},
	file = {Snapshot:files/1470/31139746.html:text/html},
}

@misc{noauthor_zotero_nodate,
	title = {Zotero {\textbar} {Your} personal research assistant},
	url = {https://www.zotero.org/download/},
	urldate = {2026-04-21},
	file = {Zotero | Your personal research assistant:files/1480/download.html:text/html},
}

@article{faruqi_direct_2018,
	series = {Radiation {Imaging} {Techniques} and {Applications}},
	title = {Direct imaging detectors for electron microscopy},
	volume = {878},
	issn = {0168-9002},
	url = {https://www.sciencedirect.com/science/article/pii/S0168900217307787},
	doi = {10.1016/j.nima.2017.07.037},
	abstract = {Electronic detectors used for imaging in electron microscopy are reviewed in this paper. Much of the detector technology is based on the developments in microelectronics, which have allowed the design of direct detectors with fine pixels, fast readout and which are sufficiently radiation hard for practical use. Detectors included in this review are hybrid pixel detectors, monolithic active pixel sensors based on CMOS technology and pnCCDs, which share one important feature: they are all direct imaging detectors, relying on directly converting energy in a semiconductor. Traditional methods of recording images in the electron microscope such as film and CCDs, are mentioned briefly along with a more detailed description of direct electronic detectors. Many applications benefit from the use of direct electron detectors and a few examples are mentioned in the text. In recent years one of the most dramatic advances in structural biology has been in the deployment of the new backthinned CMOS direct detectors to attain near-atomic resolution molecular structures with electron cryo-microscopy (cryo-EM). The development of direct detectors, along with a number of other parallel advances, has seen a very significant amount of new information being recorded in the images, which was not previously possible—and this forms the main emphasis of the review.},
	urldate = {2026-04-23},
	journal = {Nuclear Instruments and Methods in Physics Research Section A: Accelerators, Spectrometers, Detectors and Associated Equipment},
	author = {Faruqi, A. R. and McMullan, G.},
	month = jan,
	year = {2018},
	keywords = {Imaging detectors, CMOS detectors, Direct detectors, Electron cryo-microscopy, Electron Microscopy},
	pages = {180--190},
	file = {ScienceDirect Snapshot:files/1482/S0168900217307787.html:text/html},
}

@article{yuan_atomically_2025,
	title = {Atomically resolved edges and defects in lead halide perovskites},
	volume = {647},
	copyright = {2025 The Author(s), under exclusive licence to Springer Nature Limited},
	issn = {1476-4687},
	url = {https://www.nature.com/articles/s41586-025-09693-6},
	doi = {10.1038/s41586-025-09693-6},
	abstract = {Although edges and defects constitute only a small fraction of crystalline materials, they exert an outsized impact on a material′s properties. Organic–inorganic halide perovskites are promising next-generation semiconductor materials with superior cost effectiveness and interesting optoelectronic properties1–3. However, clear images of their edges have remained challenging to obtain owing to their extreme sensitivity4,5. Using truly high-speed ultralow-dose four-dimensional scanning transmission electron microscopy with dose fractionation, we perform ptychography at, to our knowledge, the lowest-dose atomic resolution to date, revealing not only the detailed atomic structure of the edges of a halide perovskite but also their structural dynamics. A majority methylammonium (MA) and iodine (I) edge termination is observed in methylammonium lead iodide (MAPbI3), and the damage rate of its edges and internal defects is found to depend on the concentration and type of vacancies present, with a preponderance of I vacancies in particular correlating with higher rates of damage.},
	language = {en},
	number = {8089},
	urldate = {2026-04-29},
	journal = {Nature},
	publisher = {Nature Publishing Group},
	author = {Yuan, Biao and Wang, Zeyu and Zhang, Shuchen and Hofer, Christoph and Gao, Chuang and Chennit, Tamazouzt and Shi, Hongsheng and Wu, Xiaoyan and Han, Yu and Dou, Letian and Yu, Yi and Pennycook, Timothy J.},
	month = nov,
	year = {2025},
	keywords = {Transmission electron microscopy, Imaging techniques, Solar cells, Phase-contrast microscopy},
	pages = {364--368},
	file = {Full Text PDF:files/1484/Yuan et al. - 2025 - Atomically resolved edges and defects in lead halide perovskites.pdf:application/pdf},
}

@misc{sandholt_designing_2026,
	title = {Designing dislocation-driven polar vortex networks in twisted perovskites},
	url = {http://arxiv.org/abs/2603.27272},
	doi = {10.48550/arXiv.2603.27272},
	abstract = {Twisting two atomic layers produces a geometric moire pattern, but bonding-induced interfacial reconstruction fundamentally transforms this into an ordered dislocation network - a distinction obscured in weakly-bonded van der Waals systems. Although in-plane topological vortex nanostructures arising from twisting-induced lateral strain modulation have been linked to periodic moire patterns in freestanding perovskite layers and 2D bilayers, their coupling to the interfacial dislocation network in twisted layers remains unresolved. Here we demonstrate that twisting freestanding SrTiO3 layers undergo interfacial reconstruction into a network of screw dislocations, accompanied by the emergence of in-plane topological vortices. Unlike in previous reports, these vortices are associated with the periodicity of the dislocation network rather than with geometric moire patterns. Four-dimensional scanning transmission electron microscopy (4D-STEM) reveals long-range ordered vortex-antivortex arrays with nearly continuous polarisation rotation. A machine-learning interatomic potential, trained on first-principles calculations, together with phase-field modelling, confirms that competing strains within the dislocation network stabilize polar vortex-antivortex pairs and drive the emergence of an electronic superlattice with a well-defined periodicity. Our results establish twist-controlled dislocation networks as a new and versatile route to designing local polar and electronic structures in oxide materials.},
	urldate = {2026-04-29},
	publisher = {arXiv},
	author = {Sandholt, William and Gauquelin, Nicolas and Mangeri, John and Dollekamp, Edwin and Panchal, Gyanendra and Chennit, Tamazouzt and Backer, Annick De and Annys, Arno and Vitaliti, Nikolas and Insinga, Andrea Roberto and Hansen, Jonas Mejlby and Mandal, Rajesh and Rodrigues, Davi R. and Aert, Sandra van and Wurster, Katja I. and Bhowmik, Arghya and Castelli, Ivano E. and Simonsen, Søren B. and Jespersen, Thomas S. and James, Richard D. and Jalan, Bharat and Verbeeck, Jo and Lastra, Juan Maria Garcia and Pryds, Nini},
	month = mar,
	year = {2026},
	note = {arXiv:2603.27272 [cond-mat]},
	keywords = {Condensed Matter - Materials Science},
	file = {Preprint PDF:files/1488/Sandholt et al. - 2026 - Designing dislocation-driven polar vortex networks in twisted perovskites.pdf:application/pdf;Snapshot:files/1487/2603.html:text/html},
}

@article{wang_ptychography_2025,
	title = {Ptychography at all wavelengths},
	volume = {5},
	issn = {2662-8449},
	url = {https://www.nature.com/articles/s43586-025-00438-3},
	doi = {10.1038/s43586-025-00438-3},
	language = {en},
	number = {1},
	urldate = {2026-04-29},
	journal = {Nature Reviews Methods Primers},
	author = {Wang, Ruihai and Zhao, Qianhao and Loetgering, Lars and Allars, Frederick and Hong, Zhixuan and Pennycook, Timothy J. and Horstmeyer, Roarke and Rodenburg, John and Maiden, Andrew and Zheng, Guoan},
	month = oct,
	year = {2025},
	pages = {68},
}

@article{miao_computational_2025,
	title = {Computational microscopy with coherent diffractive imaging and ptychography},
	volume = {637},
	issn = {0028-0836, 1476-4687},
	url = {https://www.nature.com/articles/s41586-024-08278-z},
	doi = {10.1038/s41586-024-08278-z},
	language = {en},
	number = {8045},
	urldate = {2026-04-29},
	journal = {Nature},
	author = {Miao, Jianwei},
	month = jan,
	year = {2025},
	pages = {281--295},
}

@article{mccallum_two-dimensional_1992,
	title = {Two-dimensional demonstration of {Wigner} phase-retrieval microscopy in the {STEM} configuration},
	volume = {45},
	copyright = {https://www.elsevier.com/tdm/userlicense/1.0/},
	issn = {03043991},
	url = {https://linkinghub.elsevier.com/retrieve/pii/030439919290149E},
	doi = {10.1016/0304-3991(92)90149-E},
	language = {en},
	number = {3-4},
	urldate = {2026-04-29},
	journal = {Ultramicroscopy},
	author = {McCallum, B.C and Rodenburg, J.M},
	month = nov,
	year = {1992},
	pages = {371--380},
}

@article{faulkner_movable_2004,
	title = {Movable {Aperture} {Lensless} {Transmission} {Microscopy}: {A} {Novel} {Phase} {Retrieval} {Algorithm}},
	volume = {93},
	copyright = {http://link.aps.org/licenses/aps-default-license},
	issn = {0031-9007, 1079-7114},
	shorttitle = {Movable {Aperture} {Lensless} {Transmission} {Microscopy}},
	url = {https://link.aps.org/doi/10.1103/PhysRevLett.93.023903},
	doi = {10.1103/PhysRevLett.93.023903},
	language = {en},
	number = {2},
	urldate = {2026-04-29},
	journal = {Physical Review Letters},
	author = {Faulkner, H. M. L. and Rodenburg, J. M.},
	month = jul,
	year = {2004},
	pages = {023903},
}

@article{guizar-sicairos_phase_2008,
	title = {Phase retrieval with transverse translation diversity: a nonlinear optimization approach},
	volume = {16},
	copyright = {https://doi.org/10.1364/OA\_License\_v1\#VOR-OA},
	issn = {1094-4087},
	shorttitle = {Phase retrieval with transverse translation diversity},
	url = {https://opg.optica.org/oe/abstract.cfm?uri=oe-16-10-7264},
	doi = {10.1364/OE.16.007264},
	language = {en},
	number = {10},
	urldate = {2026-04-29},
	journal = {Optics Express},
	author = {Guizar-Sicairos, Manuel and Fienup, James R.},
	month = may,
	year = {2008},
	pages = {7264},
}

@article{muller_strain_2012,
	title = {Strain {Measurement} in {Semiconductor} {Heterostructures} by {Scanning} {Transmission} {Electron} {Microscopy}},
	volume = {18},
	copyright = {https://www.cambridge.org/core/terms},
	issn = {1431-9276, 1435-8115},
	url = {https://academic.oup.com/mam/article/18/5/995/6932247},
	doi = {10.1017/S1431927612001274},
	abstract = {Abstract
            
              This article deals with the measurement of strain in semiconductor heterostructures from convergent beam electron diffraction patterns. In particular, three different algorithms in the field of (circular) pattern recognition are presented that are able to detect diffracted disc positions accurately, from which the strain in growth direction is calculated. Although the three approaches are very different as one is based on edge detection, one on rotational averages, and one on cross correlation with masks, it is found that identical strain profiles result for an In
              
                x
              
              Ga
              
                1−
                x
              
              N
              
                y
              
              As
              
                1−
                y
              
              /GaAs heterostructure consisting of five compressively and tensile strained layers. We achieve a precision of strain measurements of 7–9·10
              −4
              and a spatial resolution of 0.5–0.7 nm over the whole width of the layer stack which was 350 nm. Being already very applicable to strain measurements in contemporary nanostructures, we additionally suggest future hardware and software designs optimized for fast and direct acquisition of strain distributions, motivated by the present studies.},
	language = {en},
	number = {5},
	urldate = {2026-04-29},
	journal = {Microscopy and Microanalysis},
	author = {Müller, Knut and Rosenauer, Andreas and Schowalter, Marco and Zweck, Josef and Fritz, Rafael and Volz, Kerstin},
	month = oct,
	year = {2012},
	pages = {995--1009},
}

@article{plackett_merlin_2013,
	title = {Merlin: a fast versatile readout system for {Medipix3}},
	volume = {8},
	issn = {1748-0221},
	shorttitle = {Merlin},
	url = {https://iopscience.iop.org/article/10.1088/1748-0221/8/01/C01038},
	doi = {10.1088/1748-0221/8/01/C01038},
	number = {01},
	urldate = {2026-04-29},
	journal = {Journal of Instrumentation},
	author = {Plackett, R and Horswell, I and Gimenez, E N and Marchal, J and Omar, D and Tartoni, N},
	month = jan,
	year = {2013},
	pages = {C01038--C01038},
}

@article{ryll_pnccd-based_2016,
	title = {A {pnCCD}-based, fast direct single electron imaging camera for {TEM} and {STEM}},
	volume = {11},
	copyright = {http://iopscience.iop.org/info/page/text-and-data-mining},
	issn = {1748-0221},
	url = {https://iopscience.iop.org/article/10.1088/1748-0221/11/04/P04006},
	doi = {10.1088/1748-0221/11/04/P04006},
	number = {04},
	urldate = {2026-04-29},
	journal = {Journal of Instrumentation},
	author = {Ryll, H. and Simson, M. and Hartmann, R. and Holl, P. and Huth, M. and Ihle, S. and Kondo, Y. and Kotula, P. and Liebel, A. and Müller-Caspary, K. and Rosenauer, A. and Sagawa, R. and Schmidt, J. and Soltau, H. and Strüder, L.},
	month = apr,
	year = {2016},
	pages = {P04006--P04006},
}

@article{philipp_very-high_2022,
	title = {Very-{High} {Dynamic} {Range}, 10,000 {Frames}/{Second} {Pixel} {Array} {Detector} for {Electron} {Microscopy}},
	volume = {28},
	issn = {1431-9276, 1435-8115},
	url = {https://academic.oup.com/mam/article/28/2/425/6889399},
	doi = {10.1017/S1431927622000174},
	abstract = {Precision and accuracy of quantitative scanning transmission electron microscopy (STEM) methods such as ptychography, and the mapping of electric, magnetic, and strain fields depend on the dose. Reasonable acquisition time requires high beam current and the ability to quantitatively detect both large and minute changes in signal. A new hybrid pixel array detector (PAD), the second-generation Electron Microscope Pixel Array Detector (EMPAD-G2), addresses this challenge by advancing the technology of a previous generation PAD, the EMPAD. The EMPAD-G2 images continuously at a frame-rates up to 10 kHz with a dynamic range that spans from low-noise detection of single electrons to electron beam currents exceeding 180 pA per pixel, even at electron energies of 300 keV. The EMPAD-G2 enables rapid collection of high-quality STEM data that simultaneously contain full diffraction information from unsaturated bright-field disks to usable Kikuchi bands and higher-order Laue zones. Test results from 80 to 300 keV are presented, as are first experimental results demonstrating ptychographic reconstructions, strain and polarization maps. We introduce a new information metric, the maximum usable imaging speed (MUIS), to identify when a detector becomes electron-starved, saturated or its pixel count is mismatched with the beam current.},
	language = {en},
	number = {2},
	urldate = {2026-04-29},
	journal = {Microscopy and Microanalysis},
	author = {Philipp, Hugh T. and Tate, Mark W. and Shanks, Katherine S. and Mele, Luigi and Peemen, Maurice and Dona, Pleun and Hartong, Reinout and Van Veen, Gerard and Shao, Yu-Tsun and Chen, Zhen and Thom-Levy, Julia and Muller, David A. and Gruner, Sol M.},
	month = apr,
	year = {2022},
	pages = {425--440},
	file = {Full Text PDF:files/1656/Philipp et al. - 2022 - Very-High Dynamic Range, 10,000 FramesSecond Pixel Array Detector for Electron Microscopy.pdf:application/pdf},
}

@article{zambon_kite_2023,
	title = {{KITE}: {High} frame rate, high count rate pixelated electron counting {ASIC} for {4D} {STEM} applications featuring high-{Z} sensor},
	volume = {1048},
	issn = {01689002},
	shorttitle = {{KITE}},
	url = {https://linkinghub.elsevier.com/retrieve/pii/S0168900222011809},
	doi = {10.1016/j.nima.2022.167888},
	language = {en},
	urldate = {2026-04-29},
	journal = {Nuclear Instruments and Methods in Physics Research Section A: Accelerators, Spectrometers, Detectors and Associated Equipment},
	author = {Zambon, P. and Bottinelli, S. and Schnyder, R. and Musarra, D. and Boye, D. and Dudina, A. and Lehmann, N. and De Carlo, S. and Rissi, M. and Schulze-Briese, C. and Meffert, M. and Campanini, M. and Erni, R. and Piazza, L.},
	month = mar,
	year = {2023},
	pages = {167888},
}

@article{ercius_4d_2024,
	title = {The {4D} {Camera}: {An} 87 {kHz} {Direct} {Electron} {Detector} for {Scanning}/{Transmission} {Electron} {Microscopy}},
	volume = {30},
	copyright = {https://creativecommons.org/licenses/by-nc/4.0/},
	issn = {1431-9276, 1435-8115},
	shorttitle = {The {4D} {Camera}},
	url = {https://academic.oup.com/mam/article/30/5/903/7762045},
	doi = {10.1093/mam/ozae086},
	abstract = {Abstract
            We describe the development, operation, and application of the 4D Camera—a 576 by 576 pixel active pixel sensor for scanning/transmission electron microscopy which operates at 87,000 Hz. The detector generates data at ∼480 Gbit/s which is captured by dedicated receiver computers with a parallelized software infrastructure that has been implemented to process the resulting 10–700 Gigabyte-sized raw datasets. The back illuminated detector provides the ability to detect single electron events at accelerating voltages from 30 to 300 kV. Through electron counting, the resulting sparse data sets are reduced in size by 10--300× compared to the raw data, and open-source sparsity-based processing algorithms offer rapid data analysis. The high frame rate allows for large and complex scanning diffraction experiments to be accomplished with typical scanning transmission electron microscopy scanning parameters.},
	language = {en},
	number = {5},
	urldate = {2026-04-29},
	journal = {Microscopy and Microanalysis},
	author = {Ercius, Peter and Johnson, Ian J and Pelz, Philipp and Savitzky, Benjamin H and Hughes, Lauren and Brown, Hamish G and Zeltmann, Steven E and Hsu, Shang-Lin and Pedroso, Cassio C S and Cohen, Bruce E and Ramesh, Ramamoorthy and Paul, David and Joseph, John M and Stezelberger, Thorsten and Czarnik, Cory and Lent, Matthew and Fong, Erin and Ciston, Jim and Scott, Mary C and Ophus, Colin and Minor, Andrew M and Denes, Peter},
	month = nov,
	year = {2024},
	pages = {903--912},
	file = {Full Text:files/1657/Ercius et al. - 2024 - The 4D Camera An 87 kHz Direct Electron Detector for ScanningTransmission Electron Microscopy.pdf:application/pdf},
}

@article{yang_4d_2015,
	title = {{4D} {STEM}: {High} efficiency phase contrast imaging using a fast pixelated detector},
	volume = {644},
	copyright = {http://iopscience.iop.org/info/page/text-and-data-mining},
	issn = {1742-6588, 1742-6596},
	shorttitle = {{4D} {STEM}},
	url = {https://iopscience.iop.org/article/10.1088/1742-6596/644/1/012032},
	doi = {10.1088/1742-6596/644/1/012032},
	urldate = {2026-04-29},
	journal = {Journal of Physics: Conference Series},
	author = {Yang, H and Jones, L and Ryll, H and Simson, M and Soltau, H and Kondo, Y and Sagawa, R and Banba, H and MacLaren, I and Nellist, P D},
	month = oct,
	year = {2015},
	pages = {012032},
	file = {Full Text:files/1658/Yang et al. - 2015 - 4D STEM High efficiency phase contrast imaging using a fast pixelated detector.pdf:application/pdf},
}

@article{muller-caspary_comparison_2019,
	title = {Comparison of first moment {STEM} with conventional differential phase contrast and the dependence on electron dose},
	volume = {203},
	issn = {03043991},
	url = {https://linkinghub.elsevier.com/retrieve/pii/S0304399118302730},
	doi = {10.1016/j.ultramic.2018.12.018},
	language = {en},
	urldate = {2026-04-29},
	journal = {Ultramicroscopy},
	author = {Müller-Caspary, Knut and Krause, Florian F. and Winkler, Florian and Béché, Armand and Verbeeck, Johan and Van Aert, Sandra and Rosenauer, Andreas},
	month = aug,
	year = {2019},
	pages = {95--104},
}

@article{strauch_live_2021-1,
	title = {Live {Processing} of {Momentum}-{Resolved} {STEM} {Data} for {First} {Moment} {Imaging} and {Ptychography}},
	volume = {27},
	issn = {1431-9276, 1435-8115},
	url = {https://academic.oup.com/mam/article/27/5/1078/6888074},
	doi = {10.1017/S1431927621012423},
	abstract = {A reformulated implementation of single-sideband ptychography enables analysis and display of live detector data streams in 4D scanning transmission electron microscopy (STEM) using the LiberTEM open-source platform. This is combined with live first moment and further virtual STEM detector analysis. Processing of both real experimental and simulated data shows the characteristics of this method when data are processed progressively, as opposed to the usual offline processing of a complete data set. In particular, the single-sideband method is compared with other techniques such as the enhanced ptychographic engine in order to ascertain its capability for structural imaging at increased specimen thickness. Qualitatively interpretable live results are obtained also if the sample is moved, or magnification is changed during the analysis. This allows live optimization of instrument as well as specimen parameters during the analysis. The methodology is especially expected to improve contrast- and dose-efficient
              in situ
              imaging of weakly scattering specimens, where fast live feedback during the experiment is required.},
	language = {en},
	number = {5},
	urldate = {2026-04-29},
	journal = {Microscopy and Microanalysis},
	author = {Strauch, Achim and Weber, Dieter and Clausen, Alexander and Lesnichaia, Anastasiia and Bangun, Arya and März, Benjamin and Lyu, Feng Jiao and Chen, Qing and Rosenauer, Andreas and Dunin-Borkowski, Rafal and Müller-Caspary, Knut},
	month = oct,
	year = {2021},
	pages = {1078--1092},
	file = {Full Text PDF:files/1659/Strauch et al. - 2021 - Live Processing of Momentum-Resolved STEM Data for First Moment Imaging and Ptychography.pdf:application/pdf},
}

@article{yu_real-time_2022,
	title = {Real-{Time} {Integration} {Center} of {Mass} ({riCOM}) {Reconstruction} for {4D} {STEM}},
	volume = {28},
	copyright = {https://creativecommons.org/licenses/by/4.0/},
	issn = {1435-8115, 1431-9276},
	url = {https://academic.oup.com/mam/article/28/5/1526/6995606},
	doi = {10.1017/S1431927622000617},
	abstract = {Abstract
            A real-time image reconstruction method for scanning transmission electron microscopy (STEM) is proposed. With an algorithm requiring only the center of mass of the diffraction pattern at one probe position at a time, it is able to update the resulting image each time a new probe position is visited without storing any intermediate diffraction patterns. The results show clear features at high spatial frequency, such as atomic column positions. It is also demonstrated that some common post-processing methods, such as band-pass filtering, can be directly integrated in the real-time processing flow. Compared with other reconstruction methods, the proposed method produces high-quality reconstructions with good noise robustness at extremely low memory and computational requirements. An efficient, interactive open source implementation of the concept is further presented, which is compatible with frame-based, as well as event-based camera/file types. This method provides the attractive feature of immediate feedback that microscope operators have become used to, for example, conventional high-angle annular dark field STEM imaging, allowing for rapid decision-making and fine-tuning to obtain the best possible images for beam-sensitive samples at the lowest possible dose.},
	language = {en},
	number = {5},
	urldate = {2026-04-29},
	journal = {Microscopy and Microanalysis},
	author = {Yu, Chu-Ping and Friedrich, Thomas and Jannis, Daen and Van Aert, Sandra and Verbeeck, Johan},
	month = oct,
	year = {2022},
	pages = {1526--1537},
	file = {Full Text:files/1660/Yu et al. - 2022 - Real-Time Integration Center of Mass (riCOM) Reconstruction for 4D STEM.pdf:application/pdf},
}

@article{bangun_wigner_2023,
	title = {Wigner {Distribution} {Deconvolution} {Adaptation} for {Live} {Ptychography} {Reconstruction}},
	volume = {29},
	copyright = {https://creativecommons.org/licenses/by/4.0/},
	issn = {1431-9276, 1435-8115},
	url = {https://academic.oup.com/mam/article/29/3/994/7072677},
	doi = {10.1093/micmic/ozad021},
	abstract = {Abstract
            We propose a modification of Wigner distribution deconvolution (WDD) to support live processing ptychography. Live processing allows to reconstruct and display the specimen transmission function gradually while diffraction patterns are acquired. For this purpose, we reformulate WDD and apply a dimensionality reduction technique that reduces memory consumption and increases processing speed. We show numerically that this approach maintains the reconstruction quality of specimen transfer functions as well as reduces computational complexity during acquisition processes. Although we only present the reconstruction for scanning transmission electron microscopy datasets, in general, the live processing algorithm we present in this paper can be applied to real-time ptychographic reconstruction for different fields of application.},
	language = {en},
	number = {3},
	urldate = {2026-04-29},
	journal = {Microscopy and Microanalysis},
	author = {Bangun, Arya and Baumeister, Paul F and Clausen, Alexander and Weber, Dieter and Dunin-Borkowski, Rafal E},
	month = jun,
	year = {2023},
	pages = {994--1008},
	file = {Full Text:files/1661/Bangun et al. - 2023 - Wigner Distribution Deconvolution Adaptation for Live Ptychography Reconstruction.pdf:application/pdf},
}

@article{weber_live_2024,
	title = {Live {Iterative} {Ptychography}},
	volume = {30},
	copyright = {https://creativecommons.org/licenses/by/4.0/},
	issn = {1431-9276, 1435-8115},
	url = {https://academic.oup.com/mam/article/30/1/103/7611447},
	doi = {10.1093/mam/ozae004},
	abstract = {Abstract
            We demonstrate live-updating ptychographic reconstruction with the extended ptychographical iterative engine, an iterative ptychography method, during ongoing data acquisition. The reconstruction starts with a small subset of the total data, and as the acquisition proceeds the data used for reconstruction are extended. This creates a live-updating view of object and illumination that allows monitoring the ongoing experiment and adjusting parameters with quick turn around. This is particularly advantageous for long-running acquisitions. We show that such a gradual reconstruction yields interpretable results already with a small subset of the data. We show simulated live processing with various scan patterns, parallelized reconstruction, and real-world live processing at the hard X-ray ptychographic nanoanalytical microscope PtyNAMi at the PETRA III beamline.},
	language = {en},
	number = {1},
	urldate = {2026-04-29},
	journal = {Microscopy and Microanalysis},
	author = {Weber, Dieter and Ehrig, Simeon and Schropp, Andreas and Clausen, Alexander and Achilles, Silvio and Hoffmann, Nico and Bussmann, Michael and Dunin-Borkowski, Rafal E and Schroer, Christian G},
	month = mar,
	year = {2024},
	pages = {103--117},
	file = {Full Text PDF:files/1662/Weber et al. - 2024 - Live Iterative Ptychography.pdf:application/pdf},
}

@article{frojdh_timepix3_2015,
	title = {Timepix3: first measurements and characterization of a hybrid-pixel detector working in event driven mode},
	volume = {10},
	copyright = {http://iopscience.iop.org/info/page/text-and-data-mining},
	issn = {1748-0221},
	shorttitle = {Timepix3},
	url = {https://iopscience.iop.org/article/10.1088/1748-0221/10/01/C01039},
	doi = {10.1088/1748-0221/10/01/C01039},
	number = {01},
	urldate = {2026-04-29},
	journal = {Journal of Instrumentation},
	author = {Frojdh, E and Campbell, M and Gaspari, M De and Kulis, S and Llopart, X and Poikela, T and Tlustos, L},
	month = jan,
	year = {2015},
	pages = {C01039--C01039},
}

@article{llopart_timepix4_2022,
	title = {Timepix4, a large area pixel detector readout chip which can be tiled on 4 sides providing sub-200 ps timestamp binning},
	volume = {17},
	issn = {1748-0221},
	url = {https://iopscience.iop.org/article/10.1088/1748-0221/17/01/C01044},
	doi = {10.1088/1748-0221/17/01/C01044},
	abstract = {Abstract
            
              Timepix4 is a 24.7 × 30.0 mm
              2
              hybrid pixel detector readout ASIC which has been designed to permit detector tiling on 4 sides. It consists of 448 × 512 pixels which can be bump bonded to a sensor with square pixels at a pitch of 55 µm. Like its predecessor, Timepix3, it can operate in data driven mode sending out information (Time of Arrival, ToA and Time over Threshold, ToT) only when a pixel has a hit above a pre-defined and programmable threshold. In this mode hits can be tagged to a time bin of {\textless}200 ps and Timepix4 can record hits correctly at incoming rates of ∼3.6 MHz/mm
              2
              /s. In photon counting (or frame-based) mode it can count incoming hits at rates of up to 5 GHz/mm
              2
              /s. In both modes data is output via between 2 and 16 serializers each running at a programmable data bandwidth of between 40 Mbps and 10 Gbps. The specifications, architecture and circuit implementation are described along with first electrical measurements and measurements with radioactive sources. In photon counting mode X-ray images have been taken at a threshold of 650 e
              −
              (with {\textless}10 masked pixels). In data driven mode images were taken of ToA/ToT data using a
              90
              Sr source at a threshold of 800 e
              −
              (with ∼120 masked pixels).},
	number = {01},
	urldate = {2026-04-29},
	journal = {Journal of Instrumentation},
	author = {Llopart, X. and Alozy, J. and Ballabriga, R. and Campbell, M. and Casanova, R. and Gromov, V. and Heijne, E.H.M. and Poikela, T. and Santin, E. and Sriskaran, V. and Tlustos, L. and Vitkovskiy, A.},
	month = jan,
	year = {2022},
	pages = {C01044},
	file = {Full Text:files/1663/Llopart et al. - 2022 - Timepix4, a large area pixel detector readout chip which can be tiled on 4 sides providing sub-200 p.pdf:application/pdf},
}

@article{auad_time_2024,
	title = {Time calibration studies for the {Timepix3} hybrid pixel detector in electron microscopy},
	volume = {257},
	issn = {03043991},
	url = {https://linkinghub.elsevier.com/retrieve/pii/S0304399123002061},
	doi = {10.1016/j.ultramic.2023.113889},
	language = {en},
	urldate = {2026-04-29},
	journal = {Ultramicroscopy},
	author = {Auad, Yves and Baaboura, Jassem and Blazit, Jean-Denis and Tencé, Marcel and Stéphan, Odile and Kociak, Mathieu and Tizei, Luiz H.G.},
	month = mar,
	year = {2024},
	pages = {113889},
}

@article{kuttruff_real-time_2024,
	title = {Real-time electron clustering in an event-driven hybrid pixel detector},
	volume = {255},
	issn = {03043991},
	url = {https://linkinghub.elsevier.com/retrieve/pii/S030439912300181X},
	doi = {10.1016/j.ultramic.2023.113864},
	language = {en},
	urldate = {2026-04-29},
	journal = {Ultramicroscopy},
	author = {Kuttruff, J. and Holder, J. and Meng, Y. and Baum, P.},
	month = jan,
	year = {2024},
	pages = {113864},
}

@article{jannis_event_2022-1,
	title = {Event driven {4D} {STEM} acquisition with a {Timepix3} detector: {Microsecond} dwell time and faster scans for high precision and low dose applications},
	volume = {233},
	issn = {03043991},
	shorttitle = {Event driven {4D} {STEM} acquisition with a {Timepix3} detector},
	url = {https://linkinghub.elsevier.com/retrieve/pii/S0304399121001996},
	doi = {10.1016/j.ultramic.2021.113423},
	language = {en},
	urldate = {2026-04-29},
	journal = {Ultramicroscopy},
	author = {Jannis, D. and Hofer, C. and Gao, C. and Xie, X. and Béché, A. and Pennycook, T.J. and Verbeeck, J.},
	month = mar,
	year = {2022},
	pages = {113423},
}

@article{annys_removing_2025,
	title = {Removing constraints of {4D}-{STEM} with a framework for event-driven acquisition and processing},
	volume = {277},
	issn = {03043991},
	url = {https://linkinghub.elsevier.com/retrieve/pii/S0304399125001044},
	doi = {10.1016/j.ultramic.2025.114206},
	language = {en},
	urldate = {2026-04-29},
	journal = {Ultramicroscopy},
	author = {Annys, Arno and Robert, Hoelen L. Lalandec and Gholam, Saleh and Hadermann, Joke and Verbeeck, Jo},
	month = nov,
	year = {2025},
	pages = {114206},
}

@article{pfeiffer_x-ray_2018,
	title = {X-ray ptychography},
	volume = {12},
	issn = {1749-4885, 1749-4893},
	url = {https://www.nature.com/articles/s41566-017-0072-5},
	doi = {10.1038/s41566-017-0072-5},
	language = {en},
	number = {1},
	urldate = {2026-04-29},
	journal = {Nature Photonics},
	author = {Pfeiffer, Franz},
	month = jan,
	year = {2018},
	pages = {9--17},
}

@article{pennycook_high_2019-1,
	title = {High dose efficiency atomic resolution imaging via electron ptychography},
	volume = {196},
	issn = {03043991},
	url = {https://linkinghub.elsevier.com/retrieve/pii/S0304399118302316},
	doi = {10.1016/j.ultramic.2018.10.005},
	language = {en},
	urldate = {2026-04-29},
	journal = {Ultramicroscopy},
	author = {Pennycook, Timothy J. and Martinez, Gerardo T. and Nellist, Peter D. and Meyer, Jannik C.},
	month = jan,
	year = {2019},
	pages = {131--135},
}

@article{oleary_contrast_2021,
	title = {Contrast transfer and noise considerations in focused-probe electron ptychography},
	volume = {221},
	issn = {03043991},
	url = {https://linkinghub.elsevier.com/retrieve/pii/S0304399120303314},
	doi = {10.1016/j.ultramic.2020.113189},
	language = {en},
	urldate = {2026-04-29},
	journal = {Ultramicroscopy},
	author = {O’Leary, Colum M. and Martinez, Gerardo T. and Liberti, Emanuela and Humphry, Martin J. and Kirkland, Angus I. and Nellist, Peter D.},
	month = feb,
	year = {2021},
	pages = {113189},
}

@article{jilek_simulation_2025,
	title = {Simulation {Study} of {Low}-{Dose} {4D}-{STEM} {Phase} {Contrast} {Techniques} at the {Nanoscale} in {SEM}},
	volume = {15},
	issn = {2079-4991},
	url = {https://www.mdpi.com/2079-4991/15/1/70},
	doi = {10.3390/nano15010070},
	abstract = {Phase contrast imaging is well-suited for studying weakly scattering samples. Its strength lies in its ability to measure how the phase of the electron beam is affected by the sample, even when other imaging techniques yield low contrast. In this study, we explore via simulations two phase contrast techniques: integrated center of mass (iCOM) and ptychography, specifically using the extended ptychographical iterative engine (ePIE). We simulate the four-dimensional scanning transmission electron microscopy (4D-STEM) datasets for specific parameters corresponding to a scanning electron microscope (SEM) with an immersive objective and a given pixelated detector. The performance of these phase contrast techniques is analyzed using a contrast transfer function. Simulated datasets from a sample consisting of graphene sheets and carbon nanotubes are used for iCOM and ePIE reconstructions for two aperture sizes and two electron doses. We highlight the influence of aperture size, showing that for a smaller aperture, the radiation dose is spent mostly on larger sample features, which may aid in imaging sensitive samples while minimizing radiation damage.},
	language = {en},
	number = {1},
	urldate = {2026-04-29},
	journal = {Nanomaterials},
	author = {Jílek, Zvonimír and Radlička, Tomáš and Krzyžánek, Vladislav},
	month = jan,
	year = {2025},
	pages = {70},
}

@article{lalandec_robert_benchmarking_2025,
	title = {Benchmarking analytical electron ptychography methods for the low-dose imaging of beam-sensitive materials},
	volume = {100},
	copyright = {https://creativecommons.org/licenses/by/4.0},
	issn = {1286-0042, 1286-0050},
	url = {https://www.epjap.org/10.1051/epjap/2025018},
	doi = {10.1051/epjap/2025018},
	abstract = {This publication presents an investigation of the performance of different analytical electron ptychography methods for low-dose imaging. In particular, benchmarking is performed for two model-objects, monolayer MoS
              2
              and apoferritin, by means of multislice simulations. Specific attention is given to cases where the individual diffraction patterns remain sparse. After a first rigorous introduction to the theoretical foundations of the methods, an implementation based on the scan-frequency partitioning of calculation steps is described, permitting a significant reduction of memory needs and high sampling flexibility. By analyzing the role of contrast transfer and illumination conditions, this work provides insights into the trade-off between resolution, signal-to-noise ratio and probe focus, as is necessary for the optimization of practical experiments. Furthermore, important differences between the different methods are demonstrated. Overall, the results obtained for the two model-objects demonstrate that analytical ptychography is an attractive option for the low-dose imaging of beam-sensitive materials.},
	urldate = {2026-04-29},
	journal = {The European Physical Journal Applied Physics},
	author = {Lalandec Robert, Hoelen L. and Leidl, Max Leo and Müller-Caspary, Knut and Verbeeck, Jo},
	year = {2025},
	pages = {20},
}

@article{dearg_stability_2025,
	title = {Stability of electron ptychography at low electron dose},
	volume = {300},
	issn = {0022-2720, 1365-2818},
	url = {https://onlinelibrary.wiley.com/doi/10.1111/jmi.70011},
	doi = {10.1111/jmi.70011},
	abstract = {Abstract
            Electron ptychography provides a promising avenue towards dose‐efficient, high‐resolution materials characterisation. Prior work demonstrates the feasibility of this approach, but an overarching view on the reliability of ptychographic images in low‐dose scenarios is required. Here, we address this limitation with a systematic study of image clarity across dose, thickness and convergence semi‐angle, on a range of materials science specimens. With the now widespread adoption of 4D‐STEM and ptychographic imaging, the establishment of the practical parameter space in which one can anticipate a reliably interpretable phase image is urgently needed. In some cases, our parameter space exploration confirms high‐resolution imaging at doses of 200 Å.
          , 
            LAY DESCRIPTION
            Electron ptychography is an increasingly popular imaging technique, valued for its ability to image at much lower doses than conventional STEM imaging methods. In this manuscript we quantify the gains possible via ptychography in a comparison of methods (SSB, WDD, ePIE, iCoM), across different materials (GaN, STO, ZSM‐5) over thickness and dose series with both simulated and experimental data analysed.
            The phase imaging methods used include direct (non‐iterative) ptychography methods: SSB, WDD; iterative ptychography method (ePIE) and a related non‐ptychographic phase retrieval approach (iCoM).
            The samples studied include Gallium Nitride (previously studied in related papers), Strontium Titanate (a classic test material used in the electron microscopy community for its stable crystal structure) and Zeolite Socony Mobil–5 (ZSM‐5) a particularly beam‐sensitive material on which there is interesting prior experimental results in the literature.
            Through a series of simulations and experimental studies, we are able to characterise the behaviour of these imaging techniques across an important range of experimentally relevant conditions ‐ which we hope will be useful to colleagues in their future experimental design processes.},
	language = {en},
	number = {2},
	urldate = {2026-04-29},
	journal = {Journal of Microscopy},
	author = {Dearg, M. and Michaelides, N. and Gilbert, J. and Ding, Z. and Aslam, Z. and Hopkinson, D. G. and Allen, C. S. and Clark, L.},
	month = nov,
	year = {2025},
	pages = {217--226},
}

@article{yang_simultaneous_2016-1,
	title = {Simultaneous atomic-resolution electron ptychography and {Z}-contrast imaging of light and heavy elements in complex nanostructures},
	volume = {7},
	issn = {2041-1723},
	url = {https://www.nature.com/articles/ncomms12532},
	doi = {10.1038/ncomms12532},
	abstract = {Abstract
            
              The aberration-corrected scanning transmission electron microscope (STEM) has emerged as a key tool for atomic resolution characterization of materials, allowing the use of imaging modes such as
              Z
              -contrast and spectroscopic mapping. The STEM has not been regarded as optimal for the phase-contrast imaging necessary for efficient imaging of light materials. Here, recent developments in fast electron detectors and data processing capability is shown to enable electron ptychography, to extend the capability of the STEM by allowing quantitative phase images to be formed simultaneously with incoherent signals. We demonstrate this capability as a practical tool for imaging complex structures containing light and heavy elements, and use it to solve the structure of a beam-sensitive carbon nanostructure. The contrast of the phase image contrast is maximized through the post-acquisition correction of lens aberrations. The compensation of defocus aberrations is also used for the measurement of three-dimensional sample information through post-acquisition optical sectioning.},
	language = {en},
	number = {1},
	urldate = {2026-04-29},
	journal = {Nature Communications},
	author = {Yang, H. and Rutte, R. N. and Jones, L. and Simson, M. and Sagawa, R. and Ryll, H. and Huth, M. and Pennycook, T. J. and Green, M.L.H. and Soltau, H. and Kondo, Y. and Davis, B. G. and Nellist, P. D.},
	month = aug,
	year = {2016},
	pages = {12532},
}

@article{leidl_dynamical_2023,
	title = {Dynamical scattering in ice-embedded proteins in conventional and scanning transmission electron microscopy},
	volume = {10},
	issn = {2052-2525},
	url = {https://journals.iucr.org/paper?S2052252523004505},
	doi = {10.1107/S2052252523004505},
	abstract = {Structure determination of biological macromolecules using cryogenic electron microscopy is based on applying the phase object (PO) assumption and the weak phase object (WPO) approximation to reconstruct the 3D potential density of the molecule. To enhance the understanding of image formation of protein complexes embedded in glass-like ice in a transmission electron microscope, this study addresses multiple scattering in tobacco mosaic virus (TMV) specimens. This includes the propagation inside the molecule while also accounting for the effect of structural noise. The atoms in biological macromolecules are light but are distributed over several nanometres. Commonly, PO and WPO approximations are used in most simulations and reconstruction models. Therefore, dynamical multislice simulations of TMV specimens embedded in glass-like ice were performed based on fully atomistic molecular-dynamics simulations. In the first part, the impact of multiple scattering is studied using different numbers of slices. In the second part, different sample thicknesses of the ice-embedded TMV are considered in terms of additional ice layers. It is found that single-slice models yield full frequency transfer up to a resolution of 2.5 Å, followed by attenuation up to 1.4 Å. Three slices are sufficient to reach an information transfer up to 1.0 Å. In the third part, ptychographic reconstructions based on scanning transmission electron microscopy (STEM) and single-slice models are compared with conventional TEM simulations. The ptychographic reconstructions do not need the deliberate introduction of aberrations, are capable of post-acquisition aberration correction and promise benefits for information transfer, especially at resolutions beyond 1.8 Å.},
	number = {4},
	urldate = {2026-04-29},
	journal = {IUCrJ},
	author = {Leidl, Max Leo and Sachse, Carsten and Müller-Caspary, Knut},
	month = jul,
	year = {2023},
	pages = {475--486},
}

@article{lozano_low-dose_2018-1,
	title = {Low-{Dose} {Aberration}-{Free} {Imaging} of {Li}-{Rich} {Cathode} {Materials} at {Various} {States} of {Charge} {Using} {Electron} {Ptychography}},
	volume = {18},
	copyright = {http://pubs.acs.org/page/policy/authorchoice\_ccby\_termsofuse.html},
	issn = {1530-6984, 1530-6992},
	url = {https://pubs.acs.org/doi/10.1021/acs.nanolett.8b02718},
	doi = {10.1021/acs.nanolett.8b02718},
	language = {en},
	number = {11},
	urldate = {2026-04-29},
	journal = {Nano Letters},
	author = {Lozano, Juan G. and Martinez, Gerardo T. and Jin, Liyu and Nellist, Peter D. and Bruce, Peter G.},
	month = nov,
	year = {2018},
	pages = {6850--6855},
}

@article{song_direct_2022-1,
	title = {Direct imaging of oxygen shifts associated with the oxygen redox of {Li}-rich layered oxides},
	volume = {6},
	issn = {25424351},
	url = {https://linkinghub.elsevier.com/retrieve/pii/S2542435122001465},
	doi = {10.1016/j.joule.2022.04.008},
	language = {en},
	number = {5},
	urldate = {2026-04-29},
	journal = {Joule},
	author = {Song, Weixin and Pérez-Osorio, Miguel A. and Marie, John-Joseph and Liberti, Emanuela and Luo, Xiaonan and O’Leary, Colum and House, Robert A. and Bruce, Peter G. and Nellist, Peter D.},
	month = may,
	year = {2022},
	pages = {1049--1065},
}

@article{song_visualization_2024,
	title = {Visualization of {Tetrahedral} {Li} in the {Alkali} {Layers} of {Li}-{Rich} {Layered} {Metal} {Oxides}},
	volume = {146},
	copyright = {https://creativecommons.org/licenses/by/4.0/},
	issn = {0002-7863, 1520-5126},
	url = {https://pubs.acs.org/doi/10.1021/jacs.4c05556},
	doi = {10.1021/jacs.4c05556},
	language = {en},
	number = {34},
	urldate = {2026-04-29},
	journal = {Journal of the American Chemical Society},
	author = {Song, Weixin and Pérez-Osorio, Miguel A. and Chen, Jun and Ding, Zhiyuan and Marie, John-Joseph and Juelsholt, Mikkel and House, Robert A. and Bruce, Peter G. and Nellist, Peter D.},
	month = aug,
	year = {2024},
	pages = {23814--23824},
}

@article{sha_ptychographic_2023,
	title = {Ptychographic measurements of varying size and shape along zeolite channels},
	volume = {9},
	issn = {2375-2548},
	url = {https://www.science.org/doi/10.1126/sciadv.adf1151},
	doi = {10.1126/sciadv.adf1151},
	abstract = {Sub-angstrom resolution imaging of porous materials like zeolites is important to reveal their structure-property relationships involved in ion exchange, molecule adsorption and separation, and catalysis. Using multislice electron ptychography, we successfully measured the atomic structure of zeolite at sub-angstrom lateral resolution for 100-nanometer-thick samples. Both lateral and depth deformations of the straight channels are mapped, showing the three-dimensional structural inhomogeneity and flexibility. Since most zeolites in industrial applications are usually tens to hundreds of nanometers thick, the sub-angstrom resolution imaging and accurate measurements of depth-dependent local structures with electron ptychography at low-dose condition will find wide applications in porous materials close to their industrially relevant conditions.
          , 
            Three-dimensional deformation of zeolite channels is imaged at sub angstrom resolution with electron ptychography.},
	language = {en},
	number = {11},
	urldate = {2026-04-29},
	journal = {Science Advances},
	author = {Sha, Haozhi and Cui, Jizhe and Li, Jialu and Zhang, Yuxuan and Yang, Wenfeng and Li, Yadong and Yu, Rong},
	month = mar,
	year = {2023},
	pages = {eadf1151},
}

@article{zhang_three-dimensional_2023,
	title = {Three-dimensional inhomogeneity of zeolite structure and composition revealed by electron ptychography},
	volume = {380},
	issn = {0036-8075, 1095-9203},
	url = {https://www.science.org/doi/10.1126/science.adg3183},
	doi = {10.1126/science.adg3183},
	abstract = {Structural and compositional inhomogeneity is common in zeolites and considerably affects their properties. Thickness-limited lateral resolution, lack of depth resolution, and electron dose-constrained focusing limit local structural studies of zeolites in conventional transmission electron microscopy (TEM). We demonstrate that a multislice ptychography method based on four-dimensional scanning TEM (4D-STEM) data can overcome these limitations. Images obtained from a {\textasciitilde}40-nanometer-thick MFI zeolite exhibited a lateral resolution of {\textasciitilde}0.85 angstrom that enabled the identification of individual framework oxygen (O) atoms and the precise determination of the orientations of adsorbed molecules. Furthermore, a depth resolution of {\textasciitilde}6.6 nanometers allowed probing of the three-dimensional distribution of O vacancies, as well as the phase boundaries in intergrown MFI and MEL zeolites. The 4D-STEM ptychography can be generally applied to other materials with similar high electron-beam sensitivity.
          , 
            Editor’s summary
            
              Zeolite structures are prone to structural and compositional inhomogeneities that can cause batch-to-batch variations in applications. However, imaging these variations is difficult in transmission electron microscopy (TEM) because zeolites are prone to electron beam damage at the doses needed for atomic resolution. Zhang
              et al
              . found that electron ptychography based on low-dose four-dimensional scanning TEM data could achieve subangstrom resolution. The authors resolved individual oxygen atom columns in various zeolites with specimen thicknesses of up to 40 nanometers and mapped the distribution of oxygen vacancies throughout a zeolite. Complex intergrown structures between different zeolite phases were also imaged. —Phil Szuromi
            
          , 
            Low-electron-dose ptychography reveals zeolite local structures such as oxygen vacancies and intergrowth phase boundaries.},
	language = {en},
	number = {6645},
	urldate = {2026-04-29},
	journal = {Science},
	author = {Zhang, Hui and Li, Guanxing and Zhang, Jiaxing and Zhang, Daliang and Chen, Zhen and Liu, Xiaona and Guo, Peng and Zhu, Yihan and Chen, Cailing and Liu, Lingmei and Guo, Xinwen and Han, Yu},
	month = may,
	year = {2023},
	pages = {633--638},
}

@article{dong_atomic-level_2023-1,
	title = {Atomic-{Level} {Imaging} of {Zeolite} {Local} {Structures} {Using} {Electron} {Ptychography}},
	volume = {145},
	copyright = {https://doi.org/10.15223/policy-029},
	issn = {0002-7863, 1520-5126},
	url = {https://pubs.acs.org/doi/10.1021/jacs.2c12673},
	doi = {10.1021/jacs.2c12673},
	language = {en},
	number = {12},
	urldate = {2026-04-29},
	journal = {Journal of the American Chemical Society},
	author = {Dong, Zhuoya and Zhang, Enci and Jiang, Yilan and Zhang, Qing and Mayoral, Alvaro and Jiang, Huaidong and Ma, Yanhang},
	month = mar,
	year = {2023},
	pages = {6628--6632},
}

@article{mitsuishi_direct_2023,
	title = {Direct observation of {Cu} in high-silica chabazite zeolite by electron ptychography using {Wigner} distribution deconvolution},
	volume = {13},
	issn = {2045-2322},
	url = {https://www.nature.com/articles/s41598-023-27452-3},
	doi = {10.1038/s41598-023-27452-3},
	abstract = {Abstract
            Direct observation of Cu in Cu-chabazite (CHA) zeolite has been achieved by electron ptychography using the Wigner distribution deconvolution. The imaging properties of ptychographically reconstructed images were evaluated by comparing the intensities of six-membered-ring columns of the zeolite with and without Cu using simulated ptychography images. It was concluded that although false contrast may appear at Cu-free columns for some acquisition conditions, ptychography can discriminate columns with and without Cu. Experimental observation of CHA with and without Cu was performed. Images obtained from the Cu-containing sample showed contrast at the six-membered-rings, while no contrast was observed for the Cu-free sample. The results show that ptychography is a promising technique for visualizing the atomic structures of beam-sensitive materials.},
	language = {en},
	number = {1},
	urldate = {2026-04-29},
	journal = {Scientific Reports},
	author = {Mitsuishi, Kazutaka and Nakazawa, Katsuaki and Sagawa, Ryusuke and Shimizu, Masahiko and Matsumoto, Hajime and Shima, Hisashi and Takewaki, Takahiko},
	month = jan,
	year = {2023},
	pages = {316},
}

@article{chen_direct_2024,
	title = {Direct observation of single-atom defects in monolayer two-dimensional materials by using electron ptychography at 200 {kV} acceleration voltage},
	volume = {14},
	issn = {2045-2322},
	url = {https://www.nature.com/articles/s41598-023-50784-z},
	doi = {10.1038/s41598-023-50784-z},
	abstract = {Abstract
            Electron ptychography has emerged as a popular technology for high-resolution imaging by combining the high coherence of electron sources with the ultra-fast scanning electron coil. However, the limitations of conventional pixelated detectors, including poor dynamic range and slow data readout speeds, have posed restrictions in the past on conducting electron ptychography experiments. We used the Gatan STELA pixelated detector to capture sequential diffraction data of monolayer two-dimensional (2D) materials for ptychographic reconstruction. By using the pixelated detector and electron ptychography, we demonstrate the observation of the radiation damage at atomic resolution in Transition Metal Dichalcogenides (TMDs).},
	language = {en},
	number = {1},
	urldate = {2026-04-29},
	journal = {Scientific Reports},
	author = {Chen, Ying and Chou, Tzu-Chieh and Fang, Ching-Hsing and Lu, Cheng-Yi and Hsiao, Chien-Nan and Hsu, Wei-Ting and Chen, Chien-Chun},
	month = jan,
	year = {2024},
	pages = {277},
}

@article{loh_electron_2025,
	title = {Electron {Ptychography} for {Atom}-by-{Atom} {Quantification} of {1D} {Defect} {Complexes} in {Monolayer} {MoS}$_{\textrm{2}}$},
	volume = {19},
	copyright = {https://doi.org/10.15223/policy-029},
	issn = {1936-0851, 1936-086X},
	url = {https://pubs.acs.org/doi/10.1021/acsnano.4c14988},
	doi = {10.1021/acsnano.4c14988},
	language = {en},
	number = {6},
	urldate = {2026-04-29},
	journal = {ACS Nano},
	author = {Loh, Leyi and Ning, Shoucong and Kieczka, Daria and Chen, Yuan and Yang, Jianmin and Wang, Zhe and Pennycook, Stephen J. and Eda, Goki and Shluger, Alexander L. and Bosman, Michel},
	month = feb,
	year = {2025},
	pages = {6195--6208},
}

@article{hofer_detecting_2025,
	title = {Detecting charge transfer at defects in {2D} materials with electron ptychography},
	volume = {300},
	issn = {0022-2720, 1365-2818},
	url = {https://onlinelibrary.wiley.com/doi/10.1111/jmi.13404},
	doi = {10.1111/jmi.13404},
	abstract = {Abstract
            
              Electronic charge transfer at the atomic scale can reveal fundamental information about chemical bonding, but is far more challenging to directly image than the atomic structure. The charge density is dominated by the atomic nuclei, with bonding causing only a small perturbation. Thus detecting any change due to bonding requires a higher level of sensitivity than imaging structure and the overall charge density. Here we achieve the sensitivity required to detect charge transfer in both pristine and defected monolayer WS
              2
              using the high dose efficiency of electron ptychography and its ability to correct for lens aberrations. Excellent agreement is achieved with first‐principles image simulations including where thermal diffuse scattering is explicitly modelled via finite‐temperature molecular dynamics based on density functional theory. The focused‐probe ptychography configuration we use also provides the important ability to concurrently collect the annular dark‐field signal, which can be unambiguously interpreted in terms of the atomic structure and chemical identity of the atoms, independently of the charge transfer. Our results demonstrate both the power of ptychographic reconstructions and the importance of quantitatively accurate simulations to aid their interpretation.},
	language = {en},
	number = {2},
	urldate = {2026-04-29},
	journal = {Journal of Microscopy},
	author = {Hofer, Christoph and Madsen, Jacob and Susi, Toma and Pennycook, Timothy J.},
	month = nov,
	year = {2025},
	pages = {156--166},
}

@article{cheng_primer_2015,
	title = {A {Primer} to {Single}-{Particle} {Cryo}-{Electron} {Microscopy}},
	volume = {161},
	issn = {00928674},
	url = {https://linkinghub.elsevier.com/retrieve/pii/S0092867415003700},
	doi = {10.1016/j.cell.2015.03.050},
	language = {en},
	number = {3},
	urldate = {2026-04-29},
	journal = {Cell},
	author = {Cheng, Yifan and Grigorieff, Nikolaus and Penczek, Pawel A. and Walz, Thomas},
	month = apr,
	year = {2015},
	pages = {438--449},
}

@article{nakane_single-particle_2020,
	title = {Single-particle cryo-{EM} at atomic resolution},
	volume = {587},
	issn = {0028-0836, 1476-4687},
	url = {https://www.nature.com/articles/s41586-020-2829-0},
	doi = {10.1038/s41586-020-2829-0},
	language = {en},
	number = {7832},
	urldate = {2026-04-29},
	journal = {Nature},
	author = {Nakane, Takanori and Kotecha, Abhay and Sente, Andrija and McMullan, Greg and Masiulis, Simonas and Brown, Patricia M. G. E. and Grigoras, Ioana T. and Malinauskaite, Lina and Malinauskas, Tomas and Miehling, Jonas and Uchański, Tomasz and Yu, Lingbo and Karia, Dimple and Pechnikova, Evgeniya V. and De Jong, Erwin and Keizer, Jeroen and Bischoff, Maarten and McCormack, Jamie and Tiemeijer, Peter and Hardwick, Steven W. and Chirgadze, Dimitri Y. and Murshudov, Garib and Aricescu, A. Radu and Scheres, Sjors H. W.},
	month = nov,
	year = {2020},
	pages = {152--156},
}

@article{odstrcil_iterative_2018,
	title = {Iterative least-squares solver for generalized maximum-likelihood ptychography},
	volume = {26},
	issn = {1094-4087},
	url = {https://opg.optica.org/abstract.cfm?URI=oe-26-3-3108},
	doi = {10.1364/OE.26.003108},
	language = {en},
	number = {3},
	urldate = {2026-04-29},
	journal = {Optics Express},
	author = {Odstrčil, Michal and Menzel, Andreas and Guizar-Sicairos, Manuel},
	month = feb,
	year = {2018},
	pages = {3108},
}

@article{schloz_overcoming_2020,
	title = {Overcoming information reduced data and experimentally uncertain parameters in ptychography with regularized optimization},
	volume = {28},
	issn = {1094-4087},
	url = {https://opg.optica.org/abstract.cfm?URI=oe-28-19-28306},
	doi = {10.1364/OE.396925},
	abstract = {The overdetermination of the mathematical problem underlying ptychography is reduced by a host of experimentally more desirable settings. Furthermore, reconstruction of the sample-induced phase shift is typically limited by uncertainty in the experimental parameters and finite sample thicknesses. Presented is a conjugate gradient descent algorithm, regularized optimization for ptychography (ROP), that recovers the partially known experimental parameters along with the phase shift, improves resolution by incorporating the multislice formalism to treat finite sample thicknesses, and includes regularization in the optimization process, thus achieving reliable results from noisy data with severely reduced and underdetermined information.},
	language = {en},
	number = {19},
	urldate = {2026-04-29},
	journal = {Optics Express},
	author = {Schloz, Marcel and Pekin, Thomas Christopher and Chen, Zhen and Van Den Broek, Wouter and Muller, David Anthony and Koch, Christoph Tobias},
	month = sep,
	year = {2020},
	pages = {28306},
}

@article{marchesini_sharp_2016,
	title = {\textit{{SHARP}} : a distributed {GPU}-based ptychographic solver},
	volume = {49},
	issn = {1600-5767},
	shorttitle = {\textit{{SHARP}}},
	url = {https://journals.iucr.org/paper?S1600576716008074},
	doi = {10.1107/S1600576716008074},
	abstract = {Ever brighter light sources, fast parallel detectors and advances in phase retrieval methods have made ptychography a practical and popular imaging technique. Compared to previous techniques, ptychography provides superior robustness and resolution at the expense of more advanced and time-consuming data analysis. By taking advantage of massively parallel architectures, high-throughput processing can expedite this analysis and provide microscopists with immediate feedback. These advances allow real-time imaging at wavelength-limited resolution, coupled with a large field of view. This article describes a set of algorithmic and computational methodologies used at the Advanced Light Source and US Department of Energy light sources. These are packaged as a CUDA-based software environment named
              SHARP
              (http://camera.lbl.gov/sharp), aimed at providing state-of-the-art high-throughput ptychography reconstructions for the coming era of diffraction-limited light sources.},
	number = {4},
	urldate = {2026-04-29},
	journal = {Journal of Applied Crystallography},
	author = {Marchesini, Stefano and Krishnan, Hari and Daurer, Benedikt J. and Shapiro, David A. and Perciano, Talita and Sethian, James A. and Maia, Filipe R. N. C.},
	month = aug,
	year = {2016},
	pages = {1245--1252},
}

@article{wakonig_ptychoshelves_2020,
	title = {\textit{{PtychoShelves}} , a versatile high-level framework for high-performance analysis of ptychographic data},
	volume = {53},
	issn = {1600-5767},
	url = {https://journals.iucr.org/paper?S1600576720001776},
	doi = {10.1107/S1600576720001776},
	abstract = {Over the past decade, ptychography has been proven to be a robust tool for non-destructive high-resolution quantitative electron, X-ray and optical microscopy. It allows for quantitative reconstruction of the specimen's transmissivity, as well as recovery of the illuminating wavefront. Additionally, various algorithms have been developed to account for systematic errors and improved convergence. With fast ptychographic microscopes and more advanced algorithms, both the complexity of the reconstruction task and the data volume increase significantly.
              PtychoShelves
              is a software package which combines high-level modularity for easy and fast changes to the data-processing pipeline, and high-performance computing on CPUs and GPUs.},
	number = {2},
	urldate = {2026-04-29},
	journal = {Journal of Applied Crystallography},
	author = {Wakonig, Klaus and Stadler, Hans-Christian and Odstrčil, Michal and Tsai, Esther H. R. and Diaz, Ana and Holler, Mirko and Usov, Ivan and Raabe, Jörg and Menzel, Andreas and Guizar-Sicairos, Manuel},
	month = apr,
	year = {2020},
	pages = {574--586},
}

@article{yu_scalable_2022,
	title = {Scalable and accurate multi-{GPU}-based image reconstruction of large-scale ptychography data},
	volume = {12},
	issn = {2045-2322},
	url = {https://www.nature.com/articles/s41598-022-09430-3},
	doi = {10.1038/s41598-022-09430-3},
	abstract = {Abstract
            
              While the advances in synchrotron light sources, together with the development of focusing optics and detectors, allow nanoscale ptychographic imaging of materials and biological specimens, the corresponding experiments can yield terabyte-scale volumes of data that can impose a heavy burden on the computing platform. Although graphics processing units (GPUs) provide high performance for such large-scale ptychography datasets, a single GPU is typically insufficient for analysis and reconstruction. Several works have considered leveraging multiple GPUs to accelerate the ptychographic reconstruction. However, most of these works utilize only the Message Passing Interface to handle the communications between GPUs. This approach poses inefficiency for a hardware configuration that has multiple GPUs in a single node, especially while reconstructing a single large projection, since it provides no optimizations to handle the heterogeneous GPU interconnections containing both low-speed (e.g., PCIe) and high-speed links (e.g., NVLink). In this paper, we provide an optimized intranode multi-GPU implementation that can efficiently solve large-scale ptychographic reconstruction problems. We focus on the maximum likelihood reconstruction problem using a conjugate gradient (CG) method for the solution and propose a novel hybrid parallelization model to address the performance bottlenecks in the CG solver. Accordingly, we have developed a tool, called PtyGer (
              Pty
              chographic
              G
              PU(multipl
              e
              )-based
              r
              econstruction), implementing our hybrid parallelization model design. A comprehensive evaluation verifies that PtyGer can fully preserve the original algorithm’s accuracy while achieving outstanding intranode GPU scalability.},
	language = {en},
	number = {1},
	urldate = {2026-04-29},
	journal = {Scientific Reports},
	author = {Yu, Xiaodong and Nikitin, Viktor and Ching, Daniel J. and Aslan, Selin and Gürsoy, Doğa and Biçer, Tekin},
	month = mar,
	year = {2022},
	pages = {5334},
}

@inproceedings{wang_image_2022,
	address = {Dallas, TX, USA},
	title = {Image {Gradient} {Decomposition} for {Parallel} and {Memory}-{Efficient} {Ptychographic} {Reconstruction}},
	copyright = {https://doi.org/10.15223/policy-029},
	isbn = {978-1-6654-5444-5},
	url = {https://ieeexplore.ieee.org/document/10045785/},
	doi = {10.1109/SC41404.2022.00013},
	urldate = {2026-04-29},
	booktitle = {{SC22}: {International} {Conference} for {High} {Performance} {Computing}, {Networking}, {Storage} and {Analysis}},
	publisher = {IEEE},
	author = {Wang, Xiao and Tsaris, Aristeidis and Mukherjee, Debangshu and Wahib, Mohamed and Chen, Peng and Oxley, Mark and Ovchinnikova, Olga and Hinkle, Jacob},
	month = nov,
	year = {2022},
	pages = {1--13},
}

@article{skoupy_ptychoscopy_2025,
	title = {Ptychoscopy: a user friendly experimental design tool for ptychography},
	volume = {15},
	issn = {2045-2322},
	shorttitle = {Ptychoscopy},
	url = {https://www.nature.com/articles/s41598-025-09871-6},
	doi = {10.1038/s41598-025-09871-6},
	abstract = {Abstract
            
              Electron ptychography is a rapidly growing diffractive imaging technique providing superior image contrast, high resolution, efficient dose usage and versatile imaging conditions. Originally coming from the physical sciences, it has also become a method of interest in the life sciences. Regardless of the scientific field, the successful ptychographic reconstruction relies on various combinations of adjustable experimental parameters, along with the choice of the reconstruction algorithm. In this work we present
              ptychoScopy
              , a Python-based tool developed to simplify the selection of these parameters and aid successful experimental design. Using a
              
                \$\${\textbackslash}hbox \{SmB\}\_\{6\}\$\$
              
              sample, we show the influence of key parameters such as probe convergence angle, defocus and electron dose on the reconstruction quality using direct and iterative reconstruction algorithms. We investigate the influence of real and reciprocal space sampling and their influence on the speed and quality of ptychographic reconstructions. PtychoScopy simplifies experimental design and guides the researchers across diverse scientific fields in setting up successful ptychographic experiments.},
	language = {en},
	number = {1},
	urldate = {2026-04-29},
	journal = {Scientific Reports},
	author = {Skoupy, Radim and Müller, Elisabeth and Pennycook, Timothy J. and Guizar-Sicairos, Manuel and Fabbri, Emiliana and Poghosyan, Emiliya},
	month = jul,
	year = {2025},
	pages = {24959},
}

@article{mukherjee_roadmap_2022,
	title = {A {Roadmap} for {Edge} {Computing} {Enabled} {Automated} {Multidimensional} {Transmission} {Electron} {Microscopy}},
	volume = {30},
	copyright = {https://academic.oup.com/journals/pages/open\_access/funder\_policies/chorus/standard\_publication\_model},
	issn = {2150-3583, 1551-9295},
	url = {https://academic.oup.com/mt/article/30/6/10/6995490},
	doi = {10.1017/S1551929522001286},
	abstract = {Abstract:
            The advent of modern, high-speed electron detectors has made the collection of multidimensional hyperspectral transmission electron microscopy datasets, such as 4D-STEM, a routine. However, many microscopists find such experiments daunting since analysis, collection, long-term storage, and networking of such datasets remain challenging. Some common issues are their large and unwieldy size that often are several gigabytes, non-standardized data analysis routines, and a lack of clarity about the computing and network resources needed to utilize the electron microscope. The existing computing and networking bottlenecks introduce significant penalties in each step of these experiments, and thus, real-time analysis-driven automated experimentation for multidimensional TEM is challenging. One solution is to integrate microscopy with edge computing, where moderately powerful computational hardware performs the preliminary analysis before handing off the heavier computation to high-performance computing (HPC) systems. Here we trace the roots of computation in modern electron microscopy, demonstrate deep learning experiments running on an edge system, and discuss the networking requirements for tying together microscopes, edge computers, and HPC systems.},
	language = {en},
	number = {6},
	urldate = {2026-04-29},
	journal = {Microscopy Today},
	author = {Mukherjee, Debangshu and Roccapriore, Kevin M and Al-Najjar, Anees and Ghosh, Ayana and Hinkle, Jacob D and Lupini, Andrew R and Vasudevan, Rama K and Kalinin, Sergei V and Ovchinnikova, Olga S and Ziatdinov, Maxim A and Rao, Nageswara S},
	month = nov,
	year = {2022},
	pages = {10--19},
}

@article{welborn_streaming_2025,
	title = {Streaming {Large}-{Scale} {Microscopy} {Data} to a {Supercomputing} {Facility}},
	volume = {31},
	copyright = {https://creativecommons.org/licenses/by/4.0/},
	issn = {1431-9276, 1435-8115},
	url = {https://academic.oup.com/mam/article/doi/10.1093/mam/ozae109/7900426},
	doi = {10.1093/mam/ozae109},
	abstract = {Abstract
            Data management is a critical component of modern experimental workflows. As data generation rates increase, transferring data from acquisition servers to processing servers via conventional file-based methods is becoming increasingly impractical. The 4D Camera at the National Center for Electron Microscopy generates data at a nominal rate of 480 Gbit s−1 (87,000 frames s−1), producing a 700 GB dataset in 15 s. To address the challenges associated with storing and processing such quantities of data, we developed a streaming workflow that utilizes a high-speed network to connect the 4D Camera’s data acquisition system to supercomputing nodes at the National Energy Research Scientific Computing Center, bypassing intermediate file storage entirely. In this work, we demonstrate the effectiveness of our streaming pipeline in a production setting through an hour-long experiment that generated over 10 TB of raw data, yielding high-quality datasets suitable for advanced analyses. Additionally, we compare the efficacy of this streaming workflow against the conventional file-transfer workflow by conducting a postmortem analysis on historical data from experiments performed by real users. Our findings show that the streaming workflow significantly improves data turnaround time, enables real-time decision-making, and minimizes the potential for human error by eliminating manual user interactions.},
	language = {en},
	number = {1},
	urldate = {2026-04-29},
	journal = {Microscopy and Microanalysis},
	author = {Welborn, Samuel S and Harris, Chris and Ribet, Stephanie M and Varnavides, Georgios and Ophus, Colin and Enders, Bjoern and Ercius, Peter},
	month = feb,
	year = {2025},
	pages = {ozae109},
}

@article{maiden_further_2017-1,
	title = {Further improvements to the ptychographical iterative engine},
	volume = {4},
	issn = {2334-2536},
	url = {https://opg.optica.org/abstract.cfm?URI=optica-4-7-736},
	doi = {10.1364/OPTICA.4.000736},
	language = {en},
	number = {7},
	urldate = {2026-04-29},
	journal = {Optica},
	author = {Maiden, Andrew and Johnson, Daniel and Li, Peng},
	month = jul,
	year = {2017},
	pages = {736},
}

@article{bangun_inverse_2022,
	title = {Inverse {Multislice} {Ptychography} by {Layer}-{Wise} {Optimisation} and {Sparse} {Matrix} {Decomposition}},
	volume = {8},
	copyright = {https://creativecommons.org/licenses/by/4.0/legalcode},
	issn = {2333-9403, 2334-0118, 2573-0436},
	url = {https://ieeexplore.ieee.org/document/9936607/},
	doi = {10.1109/TCI.2022.3218993},
	urldate = {2026-04-29},
	journal = {IEEE Transactions on Computational Imaging},
	author = {Bangun, Arya and Melnyk, Oleh and Marz, Benjamin and Diederichs, Benedikt and Clausen, Alexander and Weber, Dieter and Filbir, Frank and Muller-Caspary, Knut},
	year = {2022},
	pages = {996--1011},
}

@article{leidl_influence_2024,
	title = {Influence of loss function and electron dose on ptychography of {2D} materials using the {Wirtinger} flow},
	volume = {185},
	issn = {09684328},
	url = {https://linkinghub.elsevier.com/retrieve/pii/S0968432824001057},
	doi = {10.1016/j.micron.2024.103688},
	language = {en},
	urldate = {2026-04-29},
	journal = {Micron},
	author = {Leidl, Max Leo and Diederichs, Benedikt and Sachse, Carsten and Müller-Caspary, Knut},
	month = oct,
	year = {2024},
	pages = {103688},
}

@article{muller_atomic_2014,
	title = {Atomic electric fields revealed by a quantum mechanical approach to electron picodiffraction},
	volume = {5},
	issn = {2041-1723},
	url = {https://www.nature.com/articles/ncomms6653},
	doi = {10.1038/ncomms6653},
	abstract = {Abstract
            
              By focusing electrons on probes with a diameter of 50 pm, aberration-corrected scanning transmission electron microscopy (STEM) is currently crossing the border to probing subatomic details. A major challenge is the measurement of atomic electric fields using differential phase contrast (DPC) microscopy, traditionally exploiting the concept of a field-induced shift of diffraction patterns. Here we present a simplified quantum theoretical interpretation of DPC. This enables us to calculate the momentum transferred to the STEM probe from diffracted intensities recorded on a pixel array instead of conventional segmented bright-field detectors. The methodical development yielding atomic electric field, charge and electron density is performed using simulations for binary GaN as an ideal model system. We then present a detailed experimental study of SrTiO
              3
              yielding atomic electric fields, validated by comprehensive simulations. With this interpretation and upgraded instrumentation, STEM is capable of quantifying atomic electric fields and high-contrast imaging of light atoms.},
	language = {en},
	number = {1},
	urldate = {2026-04-29},
	journal = {Nature Communications},
	author = {Müller, Knut and Krause, Florian F. and Béché, Armand and Schowalter, Marco and Galioit, Vincent and Löffler, Stefan and Verbeeck, Johan and Zweck, Josef and Schattschneider, Peter and Rosenauer, Andreas},
	month = dec,
	year = {2014},
	pages = {5653},
}

@article{lazic_phase_2016,
	title = {Phase contrast {STEM} for thin samples: {Integrated} differential phase contrast},
	volume = {160},
	issn = {03043991},
	shorttitle = {Phase contrast {STEM} for thin samples},
	url = {https://linkinghub.elsevier.com/retrieve/pii/S0304399115300449},
	doi = {10.1016/j.ultramic.2015.10.011},
	language = {en},
	urldate = {2026-04-29},
	journal = {Ultramicroscopy},
	author = {Lazić, Ivan and Bosch, Eric G.T. and Lazar, Sorin},
	month = jan,
	year = {2016},
	pages = {265--280},
}

@article{yucelen_phase_2018,
	title = {Phase contrast scanning transmission electron microscopy imaging of light and heavy atoms at the limit of contrast and resolution},
	volume = {8},
	issn = {2045-2322},
	url = {https://www.nature.com/articles/s41598-018-20377-2},
	doi = {10.1038/s41598-018-20377-2},
	abstract = {Abstract
            
              Using state of the art scanning transmission electron microscopy (STEM) it is nowadays possible to directly image single atomic columns at sub-Å resolution. In standard (high angle) annular dark field STEM ((HA)ADF-STEM), however, light elements are usually invisible when imaged together with heavier elements in one image. Here we demonstrate the capability of the recently introduced Integrated Differential Phase Contrast STEM (iDPC-STEM) technique to image both light and heavy atoms in a thin sample at sub-Å resolution. We use the technique to resolve both the Gallium and Nitrogen dumbbells in a GaN crystal in [

                  \$\$\{{\textbackslash}bf\{10\}\}{\textbackslash}bar\{\{{\textbackslash}bf\{1\}\}\}\{{\textbackslash}bf\{1\}\}\$\$
                  
                    10
                    
                      1
                      ¯
                    
                    1

              ] orientation, which each have a separation of only 63 pm. Reaching this ultimate resolution even for light elements is possible due to the fact that iDPC-STEM is a direct phase imaging technique that allows fine-tuning the microscope while imaging. Apart from this qualitative imaging result, we also demonstrate a quantitative match of ratios of the measured intensities with theoretical predictions based on simulations.},
	language = {en},
	number = {1},
	urldate = {2026-04-29},
	journal = {Scientific Reports},
	author = {Yücelen, Emrah and Lazić, Ivan and Bosch, Eric G. T.},
	month = feb,
	year = {2018},
	pages = {2676},
}

@article{bates_sub-angstrom_1989,
	title = {Sub-ångström transmission microscopy: {A} fourier transform algorithm for microdiffraction plane intensity information},
	volume = {31},
	copyright = {https://www.elsevier.com/tdm/userlicense/1.0/},
	issn = {03043991},
	shorttitle = {Sub-ångström transmission microscopy},
	url = {https://linkinghub.elsevier.com/retrieve/pii/0304399189900521},
	doi = {10.1016/0304-3991(89)90052-1},
	language = {en},
	number = {3},
	urldate = {2026-04-29},
	journal = {Ultramicroscopy},
	author = {Bates, R.H.T. and Rodenburg, J.M.},
	month = nov,
	year = {1989},
	pages = {303--307},
}

@article{rodenburg_theory_1992,
	title = {The theory of super-resolution electron microscopy via {Wigner}-distribution deconvolution},
	volume = {339},
	copyright = {https://royalsociety.org/journals/ethics-policies/data-sharing-mining/},
	issn = {0962-8428, 2054-0299},
	url = {https://royalsocietypublishing.org/rsta/article/339/1655/521/47763/The-theory-of-super-resolution-electron-microscopy},
	doi = {10.1098/rsta.1992.0050},
	abstract = {Abstract
            The theory of deconvolving the microdiffraction data-set available in a scanning transmission electron microscope or, equivalently, the set of all bright- and dark-field images available in a conventional transmission electron microscope to obtain super- resolution micrographs (which are not limited by the transfer function of the objective lens) is developed and described with reference to holography and other phase-retrieval schemes. By the use of a Wigner distribution, influences of the instrument function can be entirely separated from the information pertaining to the specimen. The final solution yields an unambiguous estimate of the complex value of the specimen function at a resolution which in theory is only limited by the electron wavelength. The faithfulness of the image processing is shown to be not seriously affected by specimen thickness or partial coherence in the illuminating beam. The inversion procedure is remarkably noise insensitive, implying that it should result in a robust and practicable experimental technique, though one that will require very large computing facilities.},
	language = {en},
	number = {1655},
	urldate = {2026-04-29},
	journal = {Philosophical Transactions of the Royal Society of London. Series A: Physical and Engineering Sciences},
	author = {Rodenburg, J. M. and Bates, R. H. T.},
	month = jun,
	year = {1992},
	pages = {521--553},
}

@article{li_ptychographic_2014,
	title = {Ptychographic inversion via {Wigner} distribution deconvolution: {Noise} suppression and probe design},
	volume = {147},
	issn = {03043991},
	shorttitle = {Ptychographic inversion via {Wigner} distribution deconvolution},
	url = {https://linkinghub.elsevier.com/retrieve/pii/S0304399114001272},
	doi = {10.1016/j.ultramic.2014.07.004},
	language = {en},
	urldate = {2026-04-29},
	journal = {Ultramicroscopy},
	author = {Li, Peng and Edo, Tega B. and Rodenburg, John M.},
	month = dec,
	year = {2014},
	pages = {106--113},
}

@article{rodenburg_experimental_1993-1,
	title = {Experimental tests on double-resolution coherent imaging via {STEM}},
	volume = {48},
	copyright = {https://www.elsevier.com/tdm/userlicense/1.0/},
	issn = {03043991},
	url = {https://linkinghub.elsevier.com/retrieve/pii/0304399193901057},
	doi = {10.1016/0304-3991(93)90105-7},
	language = {en},
	number = {3},
	urldate = {2026-04-29},
	journal = {Ultramicroscopy},
	author = {Rodenburg, J.M. and McCallum, B.C. and Nellist, P.D.},
	month = mar,
	year = {1993},
	pages = {304--314},
}

@article{pennycook_efficient_2015-1,
	title = {Efficient phase contrast imaging in {STEM} using a pixelated detector. {Part} 1: {Experimental} demonstration at atomic resolution},
	volume = {151},
	issn = {03043991},
	shorttitle = {Efficient phase contrast imaging in {STEM} using a pixelated detector. {Part} 1},
	url = {https://linkinghub.elsevier.com/retrieve/pii/S0304399114001934},
	doi = {10.1016/j.ultramic.2014.09.013},
	language = {en},
	urldate = {2026-04-29},
	journal = {Ultramicroscopy},
	author = {Pennycook, Timothy J. and Lupini, Andrew R. and Yang, Hao and Murfitt, Matthew F. and Jones, Lewys and Nellist, Peter D.},
	month = apr,
	year = {2015},
	pages = {160--167},
}

@article{yang_efficient_2015,
	title = {Efficient phase contrast imaging in {STEM} using a pixelated detector. {Part} {II}: {Optimisation} of imaging conditions},
	volume = {151},
	issn = {03043991},
	shorttitle = {Efficient phase contrast imaging in {STEM} using a pixelated detector. {Part} {II}},
	url = {https://linkinghub.elsevier.com/retrieve/pii/S0304399114002058},
	doi = {10.1016/j.ultramic.2014.10.013},
	language = {en},
	urldate = {2026-04-29},
	journal = {Ultramicroscopy},
	author = {Yang, Hao and Pennycook, Timothy J. and Nellist, Peter D.},
	month = apr,
	year = {2015},
	pages = {232--239},
}

@article{guo_electron-event_2020,
	title = {Electron-event representation data enable efficient {cryoEM} file storage with full preservation of spatial and temporal resolution},
	volume = {7},
	issn = {2052-2525},
	url = {https://journals.iucr.org/paper?S205225252000929X},
	doi = {10.1107/S205225252000929X},
	abstract = {Direct detector device (DDD) cameras have revolutionized electron cryomicroscopy (cryoEM) with their high detective quantum efficiency (DQE) and output of movie data. A high ratio of camera frame rate (frames per second) to camera exposure rate (electrons per pixel per second) allows electron counting, which further improves the DQE and enables the recording of super-resolution information. Movie output also allows the correction of specimen movement and compensation for radiation damage. However, these movies come at the cost of producing large volumes of data. It is common practice to sum groups of successive camera frames to reduce the final frame rate, and therefore the file size, to one suitable for storage and image processing. This reduction in the temporal resolution of the camera requires decisions to be made during data acquisition that may result in the loss of information that could have been advantageous during image analysis. Here, experimental analysis of a new electron-event representation (EER) data format for electron-counting DDD movies is presented, which is enabled by new hardware developed by Thermo Fisher Scientific for their Falcon DDD cameras. This format enables the recording of DDD movies at the raw camera frame rate without sacrificing either spatial or temporal resolution. Experimental data demonstrate that the method retains super-resolution information and allows the correction of specimen movement at the physical frame rate of the camera while maintaining manageable file sizes. The EER format will enable the development of new methods that can utilize the full spatial and temporal resolution of DDD cameras.},
	number = {5},
	urldate = {2026-04-29},
	journal = {IUCrJ},
	author = {Guo, Hui and Franken, Erik and Deng, Yuchen and Benlekbir, Samir and Singla Lezcano, Garbi and Janssen, Bart and Yu, Lingbo and Ripstein, Zev A. and Tan, Yong Zi and Rubinstein, John L.},
	month = sep,
	year = {2020},
	pages = {860--869},
}

@article{datta_data_2021,
	title = {A data reduction and compression description for high throughput time-resolved electron microscopy},
	volume = {12},
	issn = {2041-1723},
	url = {https://www.nature.com/articles/s41467-020-20694-z},
	doi = {10.1038/s41467-020-20694-z},
	abstract = {Abstract
            Fast, direct electron detectors have significantly improved the spatio-temporal resolution of electron microscopy movies. Preserving both spatial and temporal resolution in extended observations, however, requires storing prohibitively large amounts of data. Here, we describe an efficient and flexible data reduction and compression scheme (ReCoDe) that retains both spatial and temporal resolution by preserving individual electron events. Running ReCoDe on a workstation we demonstrate on-the-fly reduction and compression of raw data streaming off a detector at 3 GB/s, for hours of uninterrupted data collection. The output was 100-fold smaller than the raw data and saved directly onto network-attached storage drives over a 10 GbE connection. We discuss calibration techniques that support electron detection and counting (e.g., estimate electron backscattering rates, false positive rates, and data compressibility), and novel data analysis methods enabled by ReCoDe (e.g., recalibration of data post acquisition, and accurate estimation of coincidence loss).},
	language = {en},
	number = {1},
	urldate = {2026-04-29},
	journal = {Nature Communications},
	author = {Datta, Abhik and Ng, Kian Fong and Balakrishnan, Deepan and Ding, Melissa and Chee, See Wee and Ban, Yvonne and Shi, Jian and Loh, N. Duane},
	month = jan,
	year = {2021},
	pages = {664},
}

@article{cao_theory_2018,
	title = {Theory and practice of electron diffraction from single atoms and extended objects using an {EMPAD}},
	volume = {67},
	copyright = {https://academic.oup.com/journals/pages/about\_us/legal/notices},
	issn = {2050-5698, 2050-5701},
	url = {https://academic.oup.com/jmicro/article/67/suppl_1/i150/4835603},
	doi = {10.1093/jmicro/dfx123},
	language = {en},
	number = {suppl\_1},
	urldate = {2026-04-29},
	journal = {Microscopy},
	author = {Cao, Michael C and Han, Yimo and Chen, Zhen and Jiang, Yi and Nguyen, Kayla X and Turgut, Emrah and Fuchs, Gregory D and Muller, David A},
	month = mar,
	year = {2018},
	pages = {i150--i161},
}

@article{aharonov_significance_1959,
	title = {Significance of {Electromagnetic} {Potentials} in the {Quantum} {Theory}},
	volume = {115},
	copyright = {http://link.aps.org/licenses/aps-default-license},
	issn = {0031-899X},
	url = {https://link.aps.org/doi/10.1103/PhysRev.115.485},
	doi = {10.1103/PhysRev.115.485},
	language = {en},
	number = {3},
	urldate = {2026-04-29},
	journal = {Physical Review},
	author = {Aharonov, Y. and Bohm, D.},
	month = aug,
	year = {1959},
	pages = {485--491},
}

@article{cederquist_phase-retrieval_1987,
	title = {Phase-retrieval error: a lower bound},
	volume = {4},
	copyright = {https://doi.org/10.1364/OA\_License\_v1\#VOR},
	issn = {1084-7529, 1520-8532},
	shorttitle = {Phase-retrieval error},
	url = {https://opg.optica.org/abstract.cfm?URI=josaa-4-9-1788},
	doi = {10.1364/JOSAA.4.001788},
	language = {en},
	number = {9},
	urldate = {2026-04-29},
	journal = {Journal of the Optical Society of America A},
	author = {Cederquist, J. N. and Wackerman, C. C.},
	month = sep,
	year = {1987},
	pages = {1788},
}

@article{luczka_master_1991,
	title = {A master equation for quantum systems driven by {Poisson} white noise},
	volume = {24},
	issn = {0305-4470, 1361-6447},
	url = {https://iopscience.iop.org/article/10.1088/0305-4470/24/17/010},
	doi = {10.1088/0305-4470/24/17/010},
	number = {17},
	urldate = {2026-04-29},
	journal = {Journal of Physics A: Mathematical and General},
	author = {Luczka, J and Niemiec, M},
	month = sep,
	year = {1991},
	pages = {L1021--L1024},
}

@article{drenth_problem_1975,
	title = {The {Problem} of {Phase} {Retrieval} in {Light} and {Electron} {Microscopy} of {Strong} {Objects}},
	volume = {22},
	issn = {0030-3909},
	url = {https://www.tandfonline.com/doi/full/10.1080/713819083},
	doi = {10.1080/713819083},
	language = {en},
	number = {7},
	urldate = {2026-04-29},
	journal = {Optica Acta: International Journal of Optics},
	author = {Drenth, A.J.J. and Huiser, A.M.J. and Ferwerda, H.A.},
	month = jul,
	year = {1975},
	pages = {615--628},
}

@article{van_schayck_sub-pixel_2020,
	title = {Sub-pixel electron detection using a convolutional neural network},
	volume = {218},
	issn = {03043991},
	url = {https://linkinghub.elsevier.com/retrieve/pii/S0304399120302424},
	doi = {10.1016/j.ultramic.2020.113091},
	language = {en},
	urldate = {2026-04-29},
	journal = {Ultramicroscopy},
	author = {Van Schayck, J. Paul and Van Genderen, Eric and Maddox, Erik and Roussel, Lucas and Boulanger, Hugo and Fröjdh, Erik and Abrahams, Jan-Pieter and Peters, Peter J. and Ravelli, Raimond B.G.},
	month = nov,
	year = {2020},
	pages = {113091},
}

@article{seki_theoretical_2018,
	title = {Theoretical framework of statistical noise in scanning transmission electron microscopy},
	volume = {193},
	issn = {03043991},
	url = {https://linkinghub.elsevier.com/retrieve/pii/S0304399118300603},
	doi = {10.1016/j.ultramic.2018.06.014},
	language = {en},
	urldate = {2026-04-29},
	journal = {Ultramicroscopy},
	author = {Seki, Takehito and Ikuhara, Yuichi and Shibata, Naoya},
	month = oct,
	year = {2018},
	pages = {118--125},
}

@article{bian_fourier_2016,
	title = {Fourier ptychographic reconstruction using {Poisson} maximum likelihood and truncated {Wirtinger} gradient},
	volume = {6},
	issn = {2045-2322},
	url = {https://www.nature.com/articles/srep27384},
	doi = {10.1038/srep27384},
	abstract = {Abstract
            Fourier ptychographic microscopy (FPM) is a novel computational coherent imaging technique for high space-bandwidth product imaging. Mathematically, Fourier ptychographic (FP) reconstruction can be implemented as a phase retrieval optimization process, in which we only obtain low resolution intensity images corresponding to the sub-bands of the sample’s high resolution (HR) spatial spectrum and aim to retrieve the complex HR spectrum. In real setups, the measurements always suffer from various degenerations such as Gaussian noise, Poisson noise, speckle noise and pupil location error, which would largely degrade the reconstruction. To efficiently address these degenerations, we propose a novel FP reconstruction method under a gradient descent optimization framework in this paper. The technique utilizes Poisson maximum likelihood for better signal modeling and truncated Wirtinger gradient for effective error removal. Results on both simulated data and real data captured using our laser-illuminated FPM setup show that the proposed method outperforms other state-of-the-art algorithms. Also, we have released our source code for non-commercial use.},
	language = {en},
	number = {1},
	urldate = {2026-04-29},
	journal = {Scientific Reports},
	author = {Bian, Liheng and Suo, Jinli and Chung, Jaebum and Ou, Xiaoze and Yang, Changhuei and Chen, Feng and Dai, Qionghai},
	month = jun,
	year = {2016},
	pages = {27384},
}

@article{seifert_maximum-likelihood_2023,
	title = {Maximum-likelihood estimation in ptychography in the presence of {Poisson}–{Gaussian} noise statistics},
	volume = {48},
	issn = {0146-9592, 1539-4794},
	url = {https://opg.optica.org/abstract.cfm?URI=ol-48-22-6027},
	doi = {10.1364/OL.502344},
	abstract = {Optical measurements often exhibit mixed Poisson–Gaussian noise statistics, which hampers the image quality, particularly under low signal-to-noise ratio (SNR) conditions. Computational imaging falls short in such situations when solely Poissonian noise statistics are assumed. In response to this challenge, we define a loss function that explicitly incorporates this mixed noise nature. By using a maximum-likelihood estimation, we devise a practical method to account for a camera readout noise in gradient-based ptychography optimization. Our results, based on both experimental and numerical data, demonstrate that this approach outperforms the conventional one, enabling enhanced image reconstruction quality under challenging noise conditions through a straightforward methodological adjustment.},
	language = {en},
	number = {22},
	urldate = {2026-04-29},
	journal = {Optics Letters},
	author = {Seifert, Jacob and Shao, Yifeng and Van Dam, Rens and Bouchet, Dorian and Van Leeuwen, Tristan and Mosk, Allard P.},
	month = nov,
	year = {2023},
	pages = {6027},
}

@article{fujiwara_relativistic_1961,
	title = {Relativistic {Dynamical} {Theory} of {Electron} {Diffraction}},
	volume = {16},
	issn = {0031-9015, 1347-4073},
	url = {https://journals.jps.jp/doi/10.1143/JPSJ.16.2226},
	doi = {10.1143/JPSJ.16.2226},
	language = {en},
	number = {11},
	urldate = {2026-04-29},
	journal = {Journal of the Physical Society of Japan},
	author = {Fujiwara, Kunio},
	month = nov,
	year = {1961},
	pages = {2226--2238},
}

@article{cowley_electron_1972,
	title = {Electron {Microscope} {Image} {Contrast} for {Thin} {Crystal}},
	volume = {27},
	copyright = {http://creativecommons.org/licenses/by-nc-nd/3.0/},
	issn = {1865-7109, 0932-0784},
	url = {https://www.degruyter.com/document/doi/10.1515/zna-1972-0312/html},
	doi = {10.1515/zna-1972-0312},
	abstract = {Abstract
            
              High resolution electron microscope images showing the detailed distribution of metal atoms within the unit cells of complex oxide structures have been recorded recently and as a first approximation may be interpreted as amplitude-object images if obtained with the degree of defocus corresponding to the "optimum-defocus condition" for the phase-contrast imaging of thin phase objects. Detailed observations of images of Ti
              2
              Nb
              10
              O
              29
              crystals having thicknesses of the order of 100 Å reveal that the thin phase-object approximation, which assumes that only small phase-shifts are involved, is inadequate to explain some features of the image intensities including the variation of contrast with crystal thickness. A very aproximate treatment of the phase contrast due to defocussing of phase objects having large phase shifts is evolved and shown to give a qualitativity correct account of the observations. The variation of image contrast with tilt away from a principle orientation is discussed. From the symmetry of the image contrast it is deduced that the symmetry of the crystal structure as derived from X-ray diffraction studies can not be correct.},
	language = {en},
	number = {3},
	urldate = {2026-04-29},
	journal = {Zeitschrift für Naturforschung A},
	author = {Cowley, J. M. and Iijima, Sumio},
	month = mar,
	year = {1972},
	pages = {445--451},
}

@article{chen_lorentz_2022,
	title = {Lorentz electron ptychography for imaging magnetic textures beyond the diffraction limit},
	volume = {17},
	issn = {1748-3387, 1748-3395},
	url = {https://www.nature.com/articles/s41565-022-01224-y},
	doi = {10.1038/s41565-022-01224-y},
	language = {en},
	number = {11},
	urldate = {2026-04-29},
	journal = {Nature Nanotechnology},
	author = {Chen, Zhen and Turgut, Emrah and Jiang, Yi and Nguyen, Kayla X. and Stolt, Matthew J. and Jin, Song and Ralph, Daniel C. and Fuchs, Gregory D. and Muller, David A.},
	month = nov,
	year = {2022},
	pages = {1165--1170},
}

@article{cui_antiferromagnetic_2024,
	title = {Antiferromagnetic imaging via ptychographic phase retrieval},
	volume = {69},
	issn = {20959273},
	url = {https://linkinghub.elsevier.com/retrieve/pii/S2095927323009167},
	doi = {10.1016/j.scib.2023.12.044},
	language = {en},
	number = {4},
	urldate = {2026-04-29},
	journal = {Science Bulletin},
	author = {Cui, Jizhe and Sha, Haozhi and Yang, Wenfeng and Yu, Rong},
	month = feb,
	year = {2024},
	pages = {466--472},
}

@article{robert_dynamical_2022,
	title = {Dynamical diffraction of high-energy electrons investigated by focal series momentum-resolved scanning transmission electron microscopy at atomic resolution},
	volume = {233},
	issn = {03043991},
	url = {https://linkinghub.elsevier.com/retrieve/pii/S030439912100200X},
	doi = {10.1016/j.ultramic.2021.113425},
	language = {en},
	urldate = {2026-04-29},
	journal = {Ultramicroscopy},
	author = {Robert, H.L. and Lobato, I. and Lyu, F.J. and Chen, Q. and Van Aert, S. and Van Dyck, D. and Müller-Caspary, K.},
	month = mar,
	year = {2022},
	pages = {113425},
}

@article{bunk_influence_2008,
	title = {Influence of the overlap parameter on the convergence of the ptychographical iterative engine},
	volume = {108},
	copyright = {https://www.elsevier.com/tdm/userlicense/1.0/},
	issn = {03043991},
	url = {https://linkinghub.elsevier.com/retrieve/pii/S0304399107001969},
	doi = {10.1016/j.ultramic.2007.08.003},
	language = {en},
	number = {5},
	urldate = {2026-04-29},
	journal = {Ultramicroscopy},
	author = {Bunk, Oliver and Dierolf, Martin and Kynde, Søren and Johnson, Ian and Marti, Othmar and Pfeiffer, Franz},
	month = apr,
	year = {2008},
	pages = {481--487},
}

@article{humphreys_absorption_1968,
	title = {Absorption parameters in electron diffraction theory},
	volume = {18},
	issn = {0031-8086},
	url = {https://www.tandfonline.com/doi/full/10.1080/14786436808227313},
	doi = {10.1080/14786436808227313},
	language = {en},
	number = {151},
	urldate = {2026-04-29},
	journal = {The Philosophical Magazine: A Journal of Theoretical Experimental and Applied Physics},
	author = {Humphreys, C. J. and Hirsch, P. B.},
	month = jul,
	year = {1968},
	pages = {115--122},
}

@article{beyer_influence_2020,
	title = {Influence of plasmon excitations on atomic-resolution quantitative {4D} scanning transmission electron microscopy},
	volume = {10},
	issn = {2045-2322},
	url = {https://www.nature.com/articles/s41598-020-74434-w},
	doi = {10.1038/s41598-020-74434-w},
	abstract = {Abstract
            
              Scanning transmission electron microscopy (STEM) allows to gain quantitative information on the atomic-scale structure and composition of materials, satisfying one of todays major needs in the development of novel nanoscale devices. The aim of this study is to quantify the impact of inelastic, i.e. plasmon excitations (PE), on the angular dependence of STEM intensities and answer the question whether these excitations are responsible for a drastic mismatch between experiments and contemporary image simulations observed at scattering angles below

                  \$\${\textbackslash}sim \$\$
                  
                    ∼

              40 mrad. For the two materials silicon and platinum, the angular dependencies of elastic and inelastic scattering are investigated. We utilize energy filtering in two complementary microscopes, which are representative for the systems used for quantitative STEM, to form position-averaged diffraction patterns as well as atomically resolved 4D STEM data sets for different energy ranges. The resulting five-dimensional data are used to elucidate the distinct features in real and momentum space for different energy losses. We find different angular distributions for the elastic and inelastic scattering, resulting in an increased low-angle intensity (

                  \$\${\textbackslash}sim \$\$
                  
                    ∼

              10–40 mrad). The ratio of inelastic/elastic scattering increases with rising sample thickness, while the general shape of the angular dependency is maintained. Moreover, the ratio increases with the distance to an atomic column in the low-angle regime. Since PE are usually neglected in image simulations, consequently the experimental intensity is underestimated at these angles, which especially affects bright field or low-angle annular dark field imaging. The high-angle regime, however, is unaffected. In addition, we find negligible impact of inelastic scattering on first-moment imaging in momentum-resolved STEM, which is important for STEM techniques to measure internal electric fields in functional nanostructures. To resolve the discrepancies between experiment and simulation, we present an adopted simulation scheme including PE. This study highlights the necessity to take into account PE to achieve quantitative agreement between simulation and experiment. Besides solving the fundamental question of missing physics in established simulations, this finally allows for the quantitative evaluation of low-angle scattering, which contains valuable information about the material investigated.},
	language = {en},
	number = {1},
	urldate = {2026-04-29},
	journal = {Scientific Reports},
	author = {Beyer, Andreas and Krause, Florian F. and Robert, Hoel L. and Firoozabadi, Saleh and Grieb, Tim and Kükelhan, Pirmin and Heimes, Damien and Schowalter, Marco and Müller-Caspary, Knut and Rosenauer, Andreas and Volz, Kerstin},
	month = oct,
	year = {2020},
	pages = {17890},
}

@article{barthel_role_2021,
	title = {Role of ionization in imaging and spectroscopy utilizing fast electrons that have excited phonons},
	volume = {104},
	issn = {2469-9950, 2469-9969},
	url = {https://link.aps.org/doi/10.1103/PhysRevB.104.104108},
	doi = {10.1103/PhysRevB.104.104108},
	language = {en},
	number = {10},
	urldate = {2026-04-29},
	journal = {Physical Review B},
	author = {Barthel, Juri and Allen, Leslie J.},
	month = sep,
	year = {2021},
	pages = {104108},
}

@article{robert_contribution_2022,
	title = {Contribution of multiple plasmon scattering in low-angle electron diffraction investigated by energy-filtered atomically resolved {4D}-{STEM}},
	volume = {121},
	issn = {0003-6951, 1077-3118},
	url = {https://pubs.aip.org/apl/article/121/21/213502/2834682/Contribution-of-multiple-plasmon-scattering-in-low},
	doi = {10.1063/5.0129692},
	abstract = {We report the influence of multiple plasmon losses on the dynamical diffraction of high-energy electrons, in a scanning transmission electron microscopy (STEM) study. Using an experimental setup enabling energy-filtered momentum-resolved STEM, it is shown that the successive excitation of up to five plasmons within the imaged material results in a subsequent and significant redistribution of low-angle intensity in diffraction space. An empirical approach, based on the convolution with a Lorentzian kernel, is shown to reliably model this redistribution in dependence of the energy-loss. Our study demonstrates that both the significant impact of inelastic scattering in low-angle diffraction at elevated specimen thickness and a rather straightforward model can be applied to mimic multiple plasmon scattering, which otherwise is currently not within reach for multislice simulations due to computational complexity.},
	language = {en},
	number = {21},
	urldate = {2026-04-29},
	journal = {Applied Physics Letters},
	author = {Robert, H. L. and Diederichs, B. and Müller-Caspary, K.},
	month = nov,
	year = {2022},
	pages = {213502},
}

@article{van_dyck_is_2009,
	title = {Is the frozen phonon model adequate to describe inelastic phonon scattering?},
	volume = {109},
	copyright = {https://www.elsevier.com/tdm/userlicense/1.0/},
	issn = {03043991},
	url = {https://linkinghub.elsevier.com/retrieve/pii/S0304399109000023},
	doi = {10.1016/j.ultramic.2009.01.001},
	language = {en},
	number = {6},
	urldate = {2026-04-29},
	journal = {Ultramicroscopy},
	author = {Van Dyck, D.},
	month = may,
	year = {2009},
	pages = {677--682},
}

@article{ruskin_quantitative_2013,
	title = {Quantitative characterization of electron detectors for transmission electron microscopy},
	volume = {184},
	issn = {10478477},
	url = {https://linkinghub.elsevier.com/retrieve/pii/S1047847713002815},
	doi = {10.1016/j.jsb.2013.10.016},
	language = {en},
	number = {3},
	urldate = {2026-04-29},
	journal = {Journal of Structural Biology},
	author = {Ruskin, Adrian I. and Yu, Zhiheng and Grigorieff, Nikolaus},
	month = dec,
	year = {2013},
	pages = {385--393},
}

@article{capitani_practical_2006,
	title = {A practical method to detect and correct for lens distortion in the {TEM}},
	volume = {106},
	copyright = {https://www.elsevier.com/tdm/userlicense/1.0/},
	issn = {03043991},
	url = {https://linkinghub.elsevier.com/retrieve/pii/S0304399105001087},
	doi = {10.1016/j.ultramic.2005.06.003},
	language = {en},
	number = {2},
	urldate = {2026-04-29},
	journal = {Ultramicroscopy},
	author = {Capitani, Gian Carlo and Oleynikov, Peter and Hovmöller, Sven and Mellini, Marcello},
	month = jan,
	year = {2006},
	pages = {66--74},
}

@article{li_integrated_2022,
	title = {Integrated {Differential} {Phase} {Contrast} ({IDPC})-{STEM} {Utilizing} a {Multi}-{Sector} {Detector} for {Imaging} {Thick} {Samples}},
	volume = {28},
	issn = {1431-9276, 1435-8115},
	url = {https://academic.oup.com/mam/article/28/3/611/6889424},
	doi = {10.1017/S1431927622000289},
	abstract = {Abstract
            
              The integrated differential phase contrast (IDPC) method is useful for generating the potential map of a thin sample. We evaluate theoretically the potential of IDPC imaging for thick samples by varying the focus at different sample thicknesses. Our calculations show that high defocus values result in enhanced anisotropy of the contrast transfer function (CTF) and uninterpretable images, if a quadrant detector is applied. We further show that applying a multi-sector detector can result in an almost isotropic CTF. By sector number-dependent calculations for both
              C
              
                c
              
              /
              C
              3
              -corrected and
              C
              3
              -corrected scanning transmission electron microscopy (STEM), we show that the increase of detector sectors not only removes the anisotropy of the CTF, but also improves image contrast and resolution. For a proof-of-principle IDPC-STEM (uncorrected) experiment, we realize the functionality of a 12-sector detector from a physical quadrant detector and demonstrate the improvement in contrast and resolution on the example of InGaN/GaN quantum well structure.},
	language = {en},
	number = {3},
	urldate = {2026-04-29},
	journal = {Microscopy and Microanalysis},
	author = {Li, Zhongbo and Biskupek, Johannes and Kaiser, Ute and Rose, Harald},
	month = jun,
	year = {2022},
	pages = {611--621},
}

@article{grieb_gan_2024,
	title = {{GaN} atomic electric fields from a segmented {STEM} detector: {Experiment} and simulation},
	volume = {295},
	issn = {0022-2720, 1365-2818},
	shorttitle = {{GaN} atomic electric fields from a segmented {STEM} detector},
	url = {https://onlinelibrary.wiley.com/doi/10.1111/jmi.13276},
	doi = {10.1111/jmi.13276},
	abstract = {Summary
            Atomic electric fields in a thin GaN sample are measured with the centre‐of‐mass approach in 4D‐scanning transmission electron microscopy (4D‐STEM) using a 12‐segmented STEM detector in a Spectra 300 microscope. The electric fields, charge density and potential are compared to simulations and an experimental measurement using a pixelated 4D‐STEM detector. The segmented detector benefits from a high recording speed, which enables measurements at low radiation doses. However, there is measurement uncertainty due to the limited number of segments analysed in this study.},
	language = {en},
	number = {2},
	urldate = {2026-04-29},
	journal = {Journal of Microscopy},
	author = {Grieb, Tim and Krause, Florian F. and Mehrtens, Thorsten and Mahr, Christoph and Gerken, Beeke and Schowalter, Marco and Freitag, Bert and Rosenauer, Andreas},
	month = aug,
	year = {2024},
	pages = {140--146},
}

@article{pollath_differential_2021,
	title = {The differential phase contrast uncertainty relation: {Connection} between electron dose and field resolution},
	volume = {228},
	issn = {03043991},
	shorttitle = {The differential phase contrast uncertainty relation},
	url = {https://linkinghub.elsevier.com/retrieve/pii/S030439912100125X},
	doi = {10.1016/j.ultramic.2021.113342},
	language = {en},
	urldate = {2026-04-29},
	journal = {Ultramicroscopy},
	author = {Pöllath, Simon and Schwarzhuber, Felix and Zweck, Josef},
	month = sep,
	year = {2021},
	pages = {113342},
}

@article{lorenzen_imaging_2024,
	title = {Imaging built-in electric fields and light matter by {Fourier}-precession {TEM}},
	volume = {14},
	issn = {2045-2322},
	url = {https://www.nature.com/articles/s41598-024-51423-x},
	doi = {10.1038/s41598-024-51423-x},
	abstract = {Abstract
            
              We report the precise measurement of electric fields in nanostructures, and high-contrast imaging of soft matter at ultralow electron doses by transmission electron microscopy (TEM). In particular, a versatile method based on the theorem of reciprocity is introduced to enable differential phase contrast imaging and ptychography in conventional, plane-wave illumination TEM. This is realised by a series of TEM images acquired under different tilts, thereby introducing the sampling rate in reciprocal space as a tuneable parameter, in contrast to momentum-resolved scanning techniques. First, the electric field of a
              p–n
              junction in GaAs is imaged. Second, low-dose, in-focus ptychographic and DPC characterisation of Kagome pores in weakly scattering covalent organic frameworks is demonstrated by using a precessing electron beam in combination with a direct electron detector. The approach offers utmost flexibility to record relevant spatial frequencies selectively, while acquisition times and dose requirements are significantly reduced compared to the 4D-STEM counterpart.},
	language = {en},
	number = {1},
	urldate = {2026-04-29},
	journal = {Scientific Reports},
	author = {Lorenzen, Tizian and März, Benjamin and Xue, Tianhao and Beyer, Andreas and Volz, Kerstin and Bein, Thomas and Müller-Caspary, Knut},
	month = jan,
	year = {2024},
	pages = {1320},
}

@article{cowley_image_1969,
	title = {{IMAGE} {CONTRAST} {IN} {A} {TRANSMISSION} {SCANNING} {ELECTRON} {MICROSCOPE}},
	volume = {15},
	issn = {0003-6951, 1077-3118},
	url = {https://pubs.aip.org/apl/article/15/2/58/41737/IMAGE-CONTRAST-IN-A-TRANSMISSION-SCANNING-ELECTRON},
	doi = {10.1063/1.1652901},
	abstract = {The appearance of phase-contrast, Fresnel fringes and various forms of diffraction contrast in images produced by transmission scanning electron microscopes can be understood simply by invoking the principle of reciprocity to equate the imaging conditions to those relevant to a conventional electron microscope.},
	language = {en},
	number = {2},
	urldate = {2026-04-29},
	journal = {Applied Physics Letters},
	author = {Cowley, J. M.},
	month = jul,
	year = {1969},
	pages = {58--59},
}

@article{krause_reciprocity_2017,
	title = {Reciprocity relations in transmission electron microscopy: {A} rigorous derivation},
	volume = {92},
	issn = {09684328},
	shorttitle = {Reciprocity relations in transmission electron microscopy},
	url = {https://linkinghub.elsevier.com/retrieve/pii/S0968432816301421},
	doi = {10.1016/j.micron.2016.09.007},
	language = {en},
	urldate = {2026-04-29},
	journal = {Micron},
	author = {Krause, Florian F. and Rosenauer, Andreas},
	month = jan,
	year = {2017},
	pages = {1--5},
}

@article{cowley_scattering_1957,
	title = {The scattering of electrons by atoms and crystals. {I}. {A} new theoretical approach},
	volume = {10},
	copyright = {http://journals.iucr.org/services/copyrightpolicy.html},
	issn = {0365-110X},
	url = {https://journals.iucr.org/paper?S0365110X57002194},
	doi = {10.1107/S0365110X57002194},
	number = {10},
	urldate = {2026-04-29},
	journal = {Acta Crystallographica},
	author = {Cowley, J. M. and Moodie, A. F.},
	month = oct,
	year = {1957},
	pages = {609--619},
}

@article{goodman_numerical_1974,
	title = {Numerical evaluations of \textit{{N}} -beam wave functions in electron scattering by the multi-slice method},
	volume = {30},
	copyright = {http://journals.iucr.org/services/copyrightpolicy.html},
	issn = {0567-7394},
	url = {https://journals.iucr.org/paper?S056773947400057X},
	doi = {10.1107/S056773947400057X},
	number = {2},
	urldate = {2026-04-29},
	journal = {Acta Crystallographica Section A},
	author = {Goodman, P. and Moodie, A. F.},
	month = mar,
	year = {1974},
	pages = {280--290},
}

@article{ishizuka_new_1977,
	title = {A new theoretical and practical approach to the multislice method},
	volume = {33},
	copyright = {http://journals.iucr.org/services/copyrightpolicy.html},
	issn = {0567-7394},
	url = {https://journals.iucr.org/paper?S0567739477001879},
	doi = {10.1107/S0567739477001879},
	number = {5},
	urldate = {2026-04-29},
	journal = {Acta Crystallographica Section A},
	author = {Ishizuka, K. and Uyeda, N.},
	month = sep,
	year = {1977},
	pages = {740--749},
}

@article{butan_cryo-electron_2014,
	title = {Cryo-{Electron} {Microscopy} {Reconstruction} {Shows} {Poliovirus} {135S} {Particles} {Poised} for {Membrane} {Interaction} and {RNA} {Release}},
	volume = {88},
	issn = {0022-538X, 1098-5514},
	url = {https://journals.asm.org/doi/10.1128/JVI.01949-13},
	doi = {10.1128/JVI.01949-13},
	abstract = {ABSTRACT
            During infection, binding of mature poliovirus to cell surface receptors induces an irreversible expansion of the capsid, to form an infectious cell-entry intermediate particle that sediments at 135S. In these expanded virions, the major capsid proteins (VP1 to VP3) adopt an altered icosahedral arrangement to open holes in the capsid at 2-fold and quasi-3-fold axes, and internal polypeptides VP4 and the N terminus of VP1, which can bind membranes, become externalized. Cryo-electron microscopy images for 117,330 particles were collected using Leginon and reconstructed using FREALIGN. Improved rigid-body positioning of major capsid proteins established reliably which polypeptide segments become disordered or rearranged. The virus-to-135S transition includes expansion of 4\%, rearrangements of the GH loops of VP3 and VP1, and disordering of C-terminal extensions of VP1 and VP2. The N terminus of VP1 rearranges to become externalized near its quasi-3-fold exit, binds to rearranged GH loops of VP3 and VP1, and attaches to the top surface of VP2. These details improve our understanding of subsequent stages of infection, including endocytosis and RNA transfer into the cytoplasm.},
	language = {en},
	number = {3},
	urldate = {2026-04-29},
	journal = {Journal of Virology},
	author = {Butan, Carmen and Filman, David J. and Hogle, James M.},
	month = feb,
	year = {2014},
	pages = {1758--1770},
}

@article{dimova_measurement_2025,
	title = {Measurement of the resolution of the {Timepix4} detector for 100 {keV} and 200 {keV} electrons for transmission electron microscopy},
	volume = {1075},
	issn = {01689002},
	url = {https://linkinghub.elsevier.com/retrieve/pii/S0168900225001366},
	doi = {10.1016/j.nima.2025.170335},
	language = {en},
	urldate = {2026-04-29},
	journal = {Nuclear Instruments and Methods in Physics Research Section A: Accelerators, Spectrometers, Detectors and Associated Equipment},
	author = {Dimova, N. and Plackett, R. and Weatherill, D. and Wood, D. and O’Ryan, L. and Crevatin, G. and Barnard, J.S. and Gallagher-Jones, M. and Hynds, D. and Goldsbrough, R. and Shipsey, I. and Bortoletto, D. and Kirkland, A.},
	month = jun,
	year = {2025},
	pages = {170335},
}

@article{li_improving_2025,
	title = {Improving the low-dose performance of aberration correction in single sideband ptychography},
	volume = {277},
	issn = {03043991},
	url = {https://linkinghub.elsevier.com/retrieve/pii/S0304399125001238},
	doi = {10.1016/j.ultramic.2025.114225},
	language = {en},
	urldate = {2026-04-29},
	journal = {Ultramicroscopy},
	author = {Li, Songge and Gauquelin, Nicolas and Lalandec Robert, Hoelen L. and Annys, Arno and Gao, Chuang and Hofer, Christoph and Pennycook, Timothy J. and Verbeeck, Jo},
	month = nov,
	year = {2025},
	pages = {114225},
}

@article{thibault_probe_2009,
	title = {Probe retrieval in ptychographic coherent diffractive imaging},
	volume = {109},
	issn = {03043991},
	url = {https://linkinghub.elsevier.com/retrieve/pii/S0304399108003458},
	doi = {10.1016/j.ultramic.2008.12.011},
	language = {en},
	number = {4},
	urldate = {2026-04-29},
	journal = {Ultramicroscopy},
	author = {Thibault, Pierre and Dierolf, Martin and Bunk, Oliver and Menzel, Andreas and Pfeiffer, Franz},
	month = mar,
	year = {2009},
	pages = {338--343},
}

@article{humphry_ptychographic_2012,
	title = {Ptychographic electron microscopy using high-angle dark-field scattering for sub-nanometre resolution imaging},
	volume = {3},
	issn = {2041-1723},
	url = {https://www.nature.com/articles/ncomms1733},
	doi = {10.1038/ncomms1733},
	language = {en},
	number = {1},
	urldate = {2026-04-29},
	journal = {Nature Communications},
	author = {Humphry, M.J. and Kraus, B. and Hurst, A.C. and Maiden, A.M. and Rodenburg, J.M.},
	month = mar,
	year = {2012},
	pages = {730},
}

@article{sayre_implications_1952,
	title = {Some implications of a theorem due to {Shannon}},
	volume = {5},
	copyright = {http://journals.iucr.org/services/copyrightpolicy.html},
	issn = {0365-110X},
	url = {https://journals.iucr.org/paper?S0365110X52002276},
	doi = {10.1107/S0365110X52002276},
	number = {6},
	urldate = {2026-04-29},
	journal = {Acta Crystallographica},
	author = {Sayre, D.},
	month = nov,
	year = {1952},
	pages = {843--843},
}

@article{gerchberg_super-resolution_1974,
	title = {Super-resolution through {Error} {Energy} {Reduction}},
	volume = {21},
	issn = {0030-3909},
	url = {https://www.tandfonline.com/doi/full/10.1080/713818946},
	doi = {10.1080/713818946},
	language = {en},
	number = {9},
	urldate = {2026-04-29},
	journal = {Optica Acta: International Journal of Optics},
	author = {Gerchberg, R.W.},
	month = sep,
	year = {1974},
	pages = {709--720},
}

@article{close_towards_2015,
	title = {Towards quantitative, atomic-resolution reconstruction of the electrostatic potential via differential phase contrast using electrons},
	volume = {159},
	issn = {03043991},
	url = {https://linkinghub.elsevier.com/retrieve/pii/S0304399115300255},
	doi = {10.1016/j.ultramic.2015.09.002},
	language = {en},
	urldate = {2026-04-29},
	journal = {Ultramicroscopy},
	author = {Close, R. and Chen, Z. and Shibata, N. and Findlay, S.D.},
	month = dec,
	year = {2015},
	pages = {124--137},
}

@article{addiego_thickness_2020,
	title = {Thickness and defocus dependence of inter-atomic electric fields measured by scanning diffraction},
	volume = {208},
	issn = {03043991},
	url = {https://linkinghub.elsevier.com/retrieve/pii/S0304399119301093},
	doi = {10.1016/j.ultramic.2019.112850},
	language = {en},
	urldate = {2026-04-29},
	journal = {Ultramicroscopy},
	author = {Addiego, Christopher and Gao, Wenpei and Pan, Xiaoqing},
	month = jan,
	year = {2020},
	pages = {112850},
}

@article{burger_influence_2020,
	title = {Influence of lens aberrations, specimen thickness and tilt on differential phase contrast {STEM} images},
	volume = {219},
	issn = {03043991},
	url = {https://linkinghub.elsevier.com/retrieve/pii/S0304399120302692},
	doi = {10.1016/j.ultramic.2020.113118},
	language = {en},
	urldate = {2026-04-29},
	journal = {Ultramicroscopy},
	author = {Bürger, Julius and Riedl, Thomas and Lindner, Jörg K.N.},
	month = dec,
	year = {2020},
	pages = {113118},
}

@article{liang_optimizing_2023,
	title = {Optimizing experimental parameters of integrated differential phase contrast ({iDPC}) for atomic resolution imaging},
	volume = {246},
	issn = {03043991},
	url = {https://linkinghub.elsevier.com/retrieve/pii/S0304399123000037},
	doi = {10.1016/j.ultramic.2023.113686},
	language = {en},
	urldate = {2026-04-29},
	journal = {Ultramicroscopy},
	author = {Liang, Zhiyao and Song, Dongsheng and Ge, Binghui},
	month = apr,
	year = {2023},
	pages = {113686},
}

@article{clark_effect_2023,
	title = {The {Effect} of {Dynamical} {Scattering} on {Single}-plane {Phase} {Retrieval} in {Electron} {Ptychography}},
	volume = {29},
	copyright = {https://creativecommons.org/licenses/by-nc/4.0/},
	issn = {1431-9276, 1435-8115},
	url = {https://academic.oup.com/mam/article/29/1/384/6948181},
	doi = {10.1093/micmic/ozac022},
	abstract = {Abstract
            Segmented and pixelated detectors on scanning transmission electron microscopes enable the complex specimen transmission function to be reconstructed. Imaging the transmission function is key to interpreting the electric and magnetic properties of the specimen, and as such four-dimensional scanning transmission electron microscopy (4D-STEM) imaging techniques are crucial for our understanding of functional materials. Many of the algorithms used in the reconstruction of the transmission function rely on the multiplicative approximation and the (weak) phase object approximation, which are not valid for many materials, particularly at high resolution. Herein, we study the breakdown of simple phase imaging in thicker samples. We demonstrate the behavior of integrated center of mass imaging, single-side band ptychography, and Wigner distribution deconvolution over a thickness series of simulated GaN 4D-STEM datasets. We further give guidance as to the optimal focal conditions for obtaining a more interpretable dataset using these algorithms.},
	language = {en},
	number = {1},
	urldate = {2026-04-29},
	journal = {Microscopy and Microanalysis},
	author = {Clark, Laura and Martinez, Gerardo T and O’Leary, Colum M and Yang, Hao and Ding, Zhiyuan and Petersen, Timothy C and Findlay, Scott D and Nellist, Peter D},
	month = feb,
	year = {2023},
	pages = {384--394},
}

@article{shahmoradian_three-dimensional_2017,
	title = {Three-{Dimensional} {Imaging} of {Biological} {Tissue} by {Cryo} {X}-{Ray} {Ptychography}},
	volume = {7},
	issn = {2045-2322},
	url = {https://www.nature.com/articles/s41598-017-05587-4},
	doi = {10.1038/s41598-017-05587-4},
	abstract = {Abstract
            High-throughput three-dimensional cryogenic imaging of thick biological specimens is valuable for identifying biologically- or pathologically-relevant features of interest, especially for subsequent correlative studies. Unfortunately, high-resolution imaging techniques at cryogenic conditions often require sample reduction through sequential physical milling or sectioning for sufficient penetration to generate each image of the 3-D stack. This study represents the first demonstration of using ptychographic hard X-ray tomography at cryogenic temperatures for imaging thick biological tissue in a chemically-fixed, frozen-hydrated state without heavy metal staining and organic solvents. Applied to mammalian brain, this label-free cryogenic imaging method allows visualization of myelinated axons and sub-cellular features such as age-related pigmented cellular inclusions at a spatial resolution of {\textasciitilde}100 nanometers and thicknesses approaching 100 microns. Because our approach does not require dehydration, staining or reduction of the sample, we introduce the possibility for subsequent analysis of the same tissue using orthogonal approaches that are expected to yield direct complementary insight to the biological features of interest.},
	language = {en},
	number = {1},
	urldate = {2026-04-29},
	journal = {Scientific Reports},
	author = {Shahmoradian, S. H. and Tsai, E. H. R. and Diaz, A. and Guizar-Sicairos, M. and Raabe, J. and Spycher, L. and Britschgi, M. and Ruf, A. and Stahlberg, H. and Holler, M.},
	month = jul,
	year = {2017},
	pages = {6291},
}

@article{wang_optical_2023,
	title = {Optical ptychography for biomedical imaging: recent progress and future directions [{Invited}]},
	volume = {14},
	issn = {2156-7085, 2156-7085},
	shorttitle = {Optical ptychography for biomedical imaging},
	url = {https://opg.optica.org/abstract.cfm?URI=boe-14-2-489},
	doi = {10.1364/BOE.480685},
	abstract = {Ptychography is an enabling microscopy technique for both fundamental
					and applied sciences. In the past decade, it has become an
					indispensable imaging tool in most X-ray synchrotrons and national
					laboratories worldwide. However, ptychography’s limited
					resolution and throughput in the visible light regime have prevented
					its wide adoption in biomedical research. Recent developments in this
					technique have resolved these issues and offer turnkey solutions for
					high-throughput optical imaging with minimum hardware modifications.
					The demonstrated imaging throughput is now greater than that of a
					high-end whole slide scanner. In this review, we discuss the basic
					principle of ptychography and summarize the main milestones of its
					development. Different ptychographic implementations are categorized
					into four groups based on their lensless/lens-based configurations and
					coded-illumination/coded-detection operations. We also highlight the
					related biomedical applications, including digital pathology, drug
					screening, urinalysis, blood analysis, cytometric analysis, rare cell
					screening, cell culture monitoring, cell and tissue imaging in 2D and
					3D, polarimetric analysis, among others. Ptychography for
					high-throughput optical imaging, currently in its early stages, will
					continue to improve in performance and expand in its applications. We
					conclude this review article by pointing out several directions for
					its future development.},
	language = {en},
	number = {2},
	urldate = {2026-04-29},
	journal = {Biomedical Optics Express},
	author = {Wang, Tianbo and Jiang, Shaowei and Song, Pengming and Wang, Ruihai and Yang, Liming and Zhang, Terrance and Zheng, Guoan},
	month = feb,
	year = {2023},
	pages = {489},
}

@article{kunze_xray_2025,
	title = {X‐ray imaging methods for multiscale characterization of batteries},
	volume = {46},
	issn = {1229-5949, 1229-5949},
	url = {https://onlinelibrary.wiley.com/doi/10.1002/bkcs.70009},
	doi = {10.1002/bkcs.70009},
	abstract = {Abstract
            X‐ray imaging is transforming battery research by delivering multiscale insights into structure, composition, and chemistry, spanning micrometer to nanometer scales. As indispensable components of energy storage, electric vehicles, and mobile devices, batteries face significant challenges due to their intricate electrochemical processes, hierarchical architectures, and inaccessible components such as buried interfaces and air‐sensitive materials. These complexities demand advanced techniques capable of uncovering localized effects and addressing the heterogeneity that often obscures critical phenomena. Advanced x‐ray imaging techniques are rising to meet these demands, bridging knowledge gaps by providing detailed visualization of battery processes in ways that conventional methods cannot. Operando x‐ray imaging, in particular, captures real‐time changes during battery cycling; enabling researchers to observe dynamic processes and material transformations that are otherwise inaccessible. This approach overcomes the limitations of traditional destructive post‐mortem analyses by offering a non‐invasive, real‐time window into battery operation. Recent advances in spatial resolution, computational power, and data analysis tools have made x‐ray imaging increasingly accessible and effective for studying critical phenomena, such as buried interfaces, phase transitions, and surface‐bulk differences. This work introduces four advanced x‐ray imaging techniques, outlining their principles, capabilities, and contributions to battery research. These methods illuminate key processes, advance our understanding of battery behavior, and guide targeted performance improvements. Finally, we explore the future of x‐ray imaging as a mainstream tool for addressing the pressing challenges in energy storage systems, emphasizing its pivotal role in guiding innovations for next‐generation batteries.},
	language = {en},
	number = {4},
	urldate = {2026-04-29},
	journal = {Bulletin of the Korean Chemical Society},
	author = {Kunze, Sebastian and Nam, Chihyun and Kim, Hwiho and Chung, Jinkyu and Hong, Eunki and Song, Jaejung and Choi, Hanbi and Lim, Jongwoo},
	month = apr,
	year = {2025},
	pages = {360--380},
}

@article{donnelly_high-resolution_2016,
	title = {High-resolution hard x-ray magnetic imaging with dichroic ptychography},
	volume = {94},
	copyright = {http://link.aps.org/licenses/aps-default-license},
	issn = {2469-9950, 2469-9969},
	url = {https://link.aps.org/doi/10.1103/PhysRevB.94.064421},
	doi = {10.1103/PhysRevB.94.064421},
	language = {en},
	number = {6},
	urldate = {2026-04-29},
	journal = {Physical Review B},
	author = {Donnelly, Claire and Scagnoli, Valerio and Guizar-Sicairos, Manuel and Holler, Mirko and Wilhelm, Fabrice and Guillou, Francois and Rogalev, Andrei and Detlefs, Carsten and Menzel, Andreas and Raabe, Jörg and Heyderman, Laura J.},
	month = aug,
	year = {2016},
	pages = {064421},
}

@article{kp_electron_2025,
	title = {Electron ptychography reveals a ferroelectricity dominated by anion displacements},
	volume = {24},
	issn = {1476-1122, 1476-4660},
	url = {https://www.nature.com/articles/s41563-025-02205-x},
	doi = {10.1038/s41563-025-02205-x},
	language = {en},
	number = {9},
	urldate = {2026-04-29},
	journal = {Nature Materials},
	author = {Kp, Harikrishnan and Xu, Ruijuan and Patel, Kinnary and Crust, Kevin J. and Khandelwal, Aarushi and Zhang, Chenyu and Prosandeev, Sergey and Zhou, Hua and Shao, Yu-Tsun and Bellaiche, Laurent and Hwang, Harold Y. and Muller, David A.},
	month = sep,
	year = {2025},
	pages = {1433--1440},
}

@article{butcher_imaging_2025,
	title = {Imaging ferroelectric domains with soft-x-ray ptychography at the oxygen {K}-edge},
	volume = {23},
	issn = {2331-7019},
	url = {https://link.aps.org/doi/10.1103/PhysRevApplied.23.L011002},
	doi = {10.1103/PhysRevApplied.23.L011002},
	abstract = {The ferroelectric domain structure of a freestanding

                    Bi
                    Fe
                    O
                  
                  3

              film was visualized by ptychographic dichroic imaging with linearly polarized x-rays at the

                    O

              K-edge around 530 eV. The dichroic contrast is maximized at the energy of the hybridization of the

                    O

              2p state and the
              
                Fe
              
              3d orbitals, which is split by the octahedral crystal field of the perovskite structure. The microscopy images thus obtained complement the ptychographic imaging of the antiferromagnetic contribution at the
              
                Fe

                    L
                  
                  3

              -edge. The approach can be extended to the separation of different ferroic contributions in other multiferroic oxides.},
	language = {en},
	number = {1},
	urldate = {2026-04-29},
	journal = {Physical Review Applied},
	author = {Butcher, Tim A. and Phillips, Nicholas W. and Wei, Chia-Chun and Chang, Shih-Chao and Beinik, Igor and Thånell, Karina and Yang, Jan-Chi and Huang, Shih-Wen and Raabe, Jörg and Finizio, Simone},
	month = jan,
	year = {2025},
	pages = {L011002},
}

@article{black_spherical_1957,
	title = {Spherical aberration and the information content of optical images},
	volume = {239},
	copyright = {https://royalsociety.org/journals/ethics-policies/data-sharing-mining/},
	issn = {0080-4630, 2053-9169},
	url = {https://royalsocietypublishing.org/rspa/article/239/1219/522/10004/Spherical-aberration-and-the-information-content},
	doi = {10.1098/rspa.1957.0059},
	abstract = {Abstract
            The main purpose of the paper is to investigate the effect of spherical aberration on the information content of low-contrast photographic images with specified spread and noise characteristics. In § 1 an outline is given of the basic ideas and the relevant formalism. In § 2 the response functions of monochromats with spherical aberration and defocusing are considered, and their computed values in selected special cases are displayed in figures 3 to 5. A digression is made in § 3 in order to discuss, with the help of these values, a point of topical interest, namely, the variation of best focus with the line frequency of a sinusoidal test object. In § 4 the notion of the equivalent receiving surface of a low-contrast photographic process is introduced and two equivalent receiving surfaces are defined, with the help of some experimental results of Higgins \& Jones (1952), which correspond to model photographic processes used with low-contrast objects. In § 5 the mean information density in the images of an aberration-free monochromat is calculated for two model emulsions, at selected noise levels, over a range of focal settings. In both models, the correctly focused images of a random object set are found to contain about one bit per Airy disk when the signal-to-noise ratio is 100 and the effect of defocusing are similar in the two cases. The effect of spherical aberration on information density at different focal settings is then examined in the second model. It appears that, for amounts of fourth-power aberration up to two fringes, the informationally best focus is approximately midway between paraxial and marginal foci, and that the acceptance of a 20\% drop in information content corresponds to a focal tolerance of approximately ± ½ fringe.},
	language = {en},
	number = {1219},
	urldate = {2026-04-29},
	journal = {Proceedings of the Royal Society of London. Series A. Mathematical and Physical Sciences},
	author = {Black, G. and Linfoot, E. H.},
	month = apr,
	year = {1957},
	pages = {522--540},
}

@misc{robert_guided_2026,
	title = {Guided progressive reconstructive imaging: a new quantization-based framework for low-dose, high-throughput and real-time analytical ptychography},
	shorttitle = {Guided progressive reconstructive imaging},
	url = {http://arxiv.org/abs/2512.17561},
	doi = {10.48550/arXiv.2512.17561},
	abstract = {By profiting from recent developments in detector technologies, making it possible to access a stream of detection events with few-ns time resolutions, a new ptychographic workflow is established. This methodological framework, referred to as guided progressive reconstructive imaging, relies on a quantization-based description of the acquired intensity, through an elementary derivation. Established direct phase retrieval solutions, such as the Wigner distribution deconvolution approach, can then be adapted to a continuous treatment of received counts, with no need for a dense data representation. Consequently, the result is obtained in the form of a progressively improving estimate, while providing immediate user feedback thanks to a remarkable processing speed, able to surpass the acquisition bandwidth. This fast measurement is enabled by the cumulative usage of a pre-calculated library of kernel-limited guide functions, compiling count-wise contributions as a function of the triggered detector pixel. Hence, the reconstruction offers the same advantages of direct phase retrieval methods, in particular a high dose-efficiency and the absence of complex convergence dynamics, with much less stringent restrictions on the field of view than is typical in current alternatives. Its implementation is also significantly more straightforward and flexible. Overall, this work constitutes a major evolution in the state-of-the-art, facilitating repeatable and low-dose experiments with high accessibility, and being applicable to electron-based imaging, X-ray diffraction and optical microscopy.},
	language = {en},
	urldate = {2026-04-29},
	publisher = {arXiv},
	author = {Robert, Hoelen L. Lalandec and Annys, Arno and Chennit, Tamazouzt and Verbeeck, Jo},
	month = mar,
	year = {2026},
	note = {arXiv:2512.17561 [physics]},
	keywords = {Physics - Applied Physics},
	file = {PDF:files/1664/Robert et al. - 2026 - Guided progressive reconstructive imaging a new quantization-based framework for low-dose, high-thr.pdf:application/pdf},
}

@misc{sandholt_designing_2026-1,
	title = {Designing dislocation-driven polar vortex networks in twisted perovskites},
	url = {http://arxiv.org/abs/2603.27272},
	doi = {10.48550/arXiv.2603.27272},
	abstract = {Twisting two atomic layers produces a geometric moire pattern, but bonding-induced interfacial reconstruction fundamentally transforms this into an ordered dislocation network - a distinction obscured in weakly-bonded van der Waals systems. Although in-plane topological vortex nanostructures arising from twisting-induced lateral strain modulation have been linked to periodic moire patterns in freestanding perovskite layers and 2D bilayers, their coupling to the interfacial dislocation network in twisted layers remains unresolved. Here we demonstrate that twisting freestanding SrTiO3 layers undergo interfacial reconstruction into a network of screw dislocations, accompanied by the emergence of in-plane topological vortices. Unlike in previous reports, these vortices are associated with the periodicity of the dislocation network rather than with geometric moire patterns. Four-dimensional scanning transmission electron microscopy (4D-STEM) reveals long-range ordered vortex-antivortex arrays with nearly continuous polarisation rotation. A machine-learning interatomic potential, trained on first-principles calculations, together with phase-field modelling, confirms that competing strains within the dislocation network stabilize polar vortex-antivortex pairs and drive the emergence of an electronic superlattice with a well-defined periodicity. Our results establish twist-controlled dislocation networks as a new and versatile route to designing local polar and electronic structures in oxide materials.},
	urldate = {2026-06-18},
	publisher = {arXiv},
	author = {Sandholt, William and Gauquelin, Nicolas and Mangeri, John and Dollekamp, Edwin and Panchal, Gyanendra and Chennit, Tamazouzt and Backer, Annick De and Annys, Arno and Vitaliti, Nikolas and Insinga, Andrea Roberto and Hansen, Jonas Mejlby and Mandal, Rajesh and Rodrigues, Davi R. and Aert, Sandra van and Wurster, Katja I. and Bhowmik, Arghya and Castelli, Ivano E. and Simonsen, Søren B. and Jespersen, Thomas S. and James, Richard D. and Jalan, Bharat and Verbeeck, Jo and Lastra, Juan Maria Garcia and Pryds, Nini},
	month = mar,
	year = {2026},
	note = {arXiv:2603.27272 [cond-mat.mtrl-sci]},
	keywords = {Condensed Matter - Materials Science},
	file = {Preprint PDF:files/1681/Sandholt et al. - 2026 - Designing dislocation-driven polar vortex networks in twisted perovskites.pdf:application/pdf;Snapshot:files/1682/2603.html:text/html},
}

@article{palacio_ultra-thin_nodate,
	title = {Ultra-thin {NaCl} films as protective layers for graphene},
	url = {https://pubs.rsc.org/en/content/articlelanding/2019/nr/c9nr03970h},
	doi = {10.1039/C9NR03970H},
	abstract = {The ageing of graphene is an important issue that limits its technological applications. Capping layers are a good option for circumventing this problem. In this work, we propose the use of ultra-thin NaCl films as easily-removable protective layers. We have carried out a detailed characterization of the NaC},
	language = {en},
	urldate = {2026-06-30},
	publisher = {The Royal Society of Chemistry},
	author = {Palacio, Irene and Lauwaet, Koen and Vázquez, Luis and Palomares, Francisco Javier and González-Herrero, Héctor and Martínez, José Ignacio and Aballe, Lucía and Foerster, Michael and García-Hernández, Mar and Martín-Gago, José Ángel},
	file = {Full Text PDF:files/1685/Palacio et al. - Ultra-thin NaCl films as protective layers for graphene.pdf:application/pdf},
}

@article{lehnert_quasi-two-dimensional_2021,
	title = {Quasi-two-dimensional {NaCl} crystals encapsulated between graphene sheets and their decomposition under an electron beam},
	volume = {13},
	issn = {2040-3364, 2040-3372},
	url = {https://xlink.rsc.org/?DOI=D1NR04792B},
	doi = {10.1039/D1NR04792B},
	abstract = {Graphene encapsulation was used for systematic
              in situ
              investigations of the decomposition processes of encapsulated quasi-2-dimensional sodium chloride (NaCl) crystals in a transmission electron microscope (TEM).
            
          , 
            
              Quasi-two-dimensional (2D) sodium chloride (NaCl) crystals of various lateral sizes between graphene sheets were manufactured
              via
              supersaturation from a saline solution. Aberration-corrected transmission electron microscopy was used for systematic
              in situ
              investigations of the crystals and their decomposition under an 80 kV electron beam. Counterintuitively, bigger clusters were found to disintegrate faster under electron irradiation, but in general no correlation between crystal sizes and electron doses at which the crystals decompose was found. As for the destruction process, an abrupt decomposition of the crystals was observed, which can be described by a logistic decay function. Density-functional theory molecular dynamics simulations provide insights into the destruction mechanism, and indicate that even without account for ionization and electron excitations, free-standing NaCl crystals must quickly disintegrate due to the ballistic displacement of atoms from their surface and edges during imaging. However, graphene sheets mitigate damage development by stopping the displaced atoms and enable the immediate recombination of defects at the surface of the crystal. At the same time, once a hole in graphene appears, the displaced atoms escape, giving rise to the quick destruction of the crystal. Our results provide quantitative data on the stability of encapsulated quasi 2D NaCl crystals under electron irradiation and allow the conclusion that only high-quality graphene is suitable for protecting ionic crystals from beam damage in electron microscopy studies.},
	language = {en},
	number = {46},
	urldate = {2026-06-30},
	journal = {Nanoscale},
	author = {Lehnert, Tibor and Kretschmer, Silvan and Bräuer, Fredrik and Krasheninnikov, Arkady V. and Kaiser, Ute},
	year = {2021},
	pages = {19626--19633},
}

@article{t_atomistic_2012,
	title = {Atomistic description of electron beam damage in nitrogen-doped graphene and single-walled carbon nanotubes},
	url = {https://pubmed.ncbi.nlm.nih.gov/23009666/},
	abstract = {By combining ab initio simulations with state-of-the-art electron microscopy and electron energy loss spectroscopy, we study the mechanism of electron beam damage in nitrogen-doped graphene and carbon nanotubes. Our results show that the incorporation of nitrogen atoms results in noticeable knock-on …},
	language = {fr},
	urldate = {2026-06-30},
	journal = {PubMed},
	author = {T, Susi and J, Kotakoski and R, Arenal and S, Kurasch and H, Jiang and V, Skakalova and O, Stephan and Av, Krasheninnikov and Ei, Kauppinen and U, Kaiser and Jc, Meyer},
	year = {2012},
	file = {Snapshot:files/1690/23009666.html:text/html},
}

@article{elibol_atomic_2018,
	title = {Atomic {Structure} of {Intrinsic} and {Electron}-{Irradiation}-{Induced} {Defects} in {MoTe}$_{\textrm{2}}$},
	volume = {30},
	copyright = {http://pubs.acs.org/page/policy/authorchoice\_ccby\_termsofuse.html},
	issn = {0897-4756, 1520-5002},
	url = {https://pubs.acs.org/doi/10.1021/acs.chemmater.7b03760},
	doi = {10.1021/acs.chemmater.7b03760},
	abstract = {Studying the atomic structure of intrinsic defects in two-dimensional transition-metal dichalcogenides is difficult since they damage quickly under the intense electron irradiation in transmission electron microscopy (TEM). However, this can also lead to insights into the creation of defects and their atom-scale dynamics. We first show that MoTe2 monolayers without protection indeed quickly degrade during scanning TEM (STEM) imaging, and discuss the observed atomic-level dynamics, including a transformation from the 1H phase into 1T′, 3-fold rotationally symmetric defects, and the migration of line defects between two 1H grains with a 60° misorientation. We then analyze the atomic structure of MoTe2 encapsulated between two graphene sheets to mitigate damage, finding the as-prepared material to contain an unexpectedly large concentration of defects. These include similar point defects (or quantum dots, QDs) as those created in the nonencapsulated material and two different types of line defects (or quantum wires, QWs) that can be transformed from one to the other under electron irradiation. Our density functional theory simulations indicate that the QDs and QWs embedded in MoTe2 introduce new midgap states into the semiconducting material and may thus be used to control its electronic and optical properties. Finally, the edge of the encapsulated material appears amorphous, possibly due to the pressure caused by the encapsulation.},
	language = {en},
	number = {4},
	urldate = {2026-06-30},
	journal = {Chemistry of Materials},
	author = {Elibol, Kenan and Susi, Toma and Argentero, Giacomo and Reza Ahmadpour Monazam, Mohammad and Pennycook, Timothy J. and Meyer, Jannik C. and Kotakoski, Jani},
	month = feb,
	year = {2018},
	pages = {1230--1238},
	file = {Full Text:files/1694/Elibol et al. - 2018 - Atomic Structure of Intrinsic and Electron-Irradiation-Induced Defects in MoTe2.pdf:application/pdf},
}

@article{ilett_analysis_2020,
	title = {Analysis of complex, beam-sensitive materials by transmission electron microscopy and associated techniques},
	volume = {378},
	issn = {1364-503X},
	url = {https://doi.org/10.1098/rsta.2019.0601},
	doi = {10.1098/rsta.2019.0601},
	abstract = {We review the use of transmission electron microscopy (TEM) and associated techniques for the analysis of beam-sensitive materials and complex, multiphase systems in-situ or close to their native state. We focus on materials prone to damage by radiolysis and explain that this process cannot be eliminated or switched off, requiring TEM analysis to be done within a dose budget to achieve an optimum dose-limited resolution. We highlight the importance of determining the damage sensitivity of a particular system in terms of characteristic changes that occur on irradiation under both an electron fluence and flux by presenting results from a series of molecular crystals. We discuss the choice of electron beam accelerating voltage and detectors for optimizing resolution and outline the different strategies employed for low-dose microscopy in relation to the damage processes in operation. In particular, we discuss the use of scanning TEM (STEM) techniques for maximizing information content from high-resolution imaging and spectroscopy of minerals and molecular crystals. We suggest how this understanding can then be carried forward for in-situ analysis of samples interacting with liquids and gases, provided any electron beam-induced alteration of a specimen is controlled or used to drive a chosen reaction. Finally, we demonstrate that cryo-TEM of nanoparticle samples snap-frozen in vitreous ice can play a significant role in benchmarking dynamic processes at higher resolution.This article is part of a discussion meeting issue ‘Dynamic in situ microscopy relating structure and function’.},
	number = {2186},
	urldate = {2026-06-30},
	journal = {Philosophical Transactions of the Royal Society A: Mathematical, Physical and Engineering Sciences},
	author = {Ilett, Martha and S'ari, Mark and Freeman, Helen and Aslam, Zabeada and Koniuch, Natalia and Afzali, Maryam and Cattle, James and Hooley, Robert and Roncal-Herrero, Teresa and Collins, Sean M. and Hondow, Nicole and Brown, Andy and Brydson, Rik},
	month = oct,
	year = {2020},
	pages = {20190601},
	file = {Full Text PDF:files/1696/Ilett et al. - 2020 - Analysis of complex, beam-sensitive materials by transmission electron microscopy and associated tec.pdf:application/pdf;Snapshot:files/1697/rsta.2019.html:text/html},
}

@article{a_low_2015,
	title = {Low voltage transmission electron microscopy of graphene},
	url = {https://pubmed.ncbi.nlm.nih.gov/25408379/},
	abstract = {The initial isolation of graphene in 2004 spawned massive interest in this two-dimensional pure sp(2) carbon structure due to its incredible electrical, optical, mechanical, and thermal effects. This in turn led to the rapid development of various characterization tools for graphene. Examples includ …},
	language = {en},
	urldate = {2026-06-30},
	journal = {PubMed},
	author = {A, Bachmatiuk and J, Zhao and Sm, Gorantla and Ig, Martinez and J, Wiedermann and C, Lee and J, Eckert and Mh, Rummeli},
	year = {2015},
	file = {Snapshot:files/1699/25408379.html:text/html},
}

@article{orekhov_room_2025,
	title = {Room temperature electron beam sensitive viscoplastic response of ultra-ductile amorphous olivine films},
	volume = {282},
	issn = {1359-6454},
	url = {https://www.sciencedirect.com/science/article/pii/S1359645424008280},
	doi = {10.1016/j.actamat.2024.120479},
	abstract = {The mechanical properties of amorphous olivine (a-olivine) deformed at room temperature are investigated in situ in a TEM under uniaxial tension using a Push-to-Pull (PTP) device. Thin films of a-olivine were produced by pulsed laser deposition (PLD). With or without electron irradiation, a-olivine films deform plastically, with a gradual transition that makes impossible the determination of a precise threshold. The strength attains values up to 2.5 GPa. The increasing strain-rate in load control results in an apparent softening with stress drop. The fracture strain reaches values close to 30 \% without e-beam irradiation. Under electron illumination at 200 kV, the strength is lower, around 1.7 GPa, while higher elongations close to 36 \% are obtained. Alternating beam-off and beam-on sequences lead to exceptionally large fracture strains equal to 68 \% at 200 kV and 139 \% at 80 kV. EELS measurements were performed to characterize the interaction between the electron beam and a-olivine. At a voltage of 80 kV, radiolysis accompanied by oxygen release dominates whereas at high voltage (300 kV) the interaction is dominated by knock-on type defects. Radiolysis is also the main interaction mechanism at 200 kV with low exposition which corresponds to most of our in situ TEM deformation experiments. To interpret the mechanical data, a simple 1D model has been developed to rationalize the load transfer between the PTP device and the specimen. The strain-rate sensitivity is 6 to 10 times higher when a-olivine is deformed under electron irradiation.},
	urldate = {2026-06-30},
	journal = {Acta Materialia},
	author = {Orekhov, Andrey and Gauquelin, Nicolas and Kermouche, Guillaume and Gomez-Perez, Alejandro and Baral, Paul and Dohmen, Ralf and Coulombier, Michaël and Verbeeck, Johan and Raskin, Jean Pierre and Pardoen, Thomas and Schryvers, Dominique and Lin, Jun and Cordier, Patrick and Idrissi, Hosni},
	month = jan,
	year = {2025},
	keywords = {Transmission electron microscopy, Amorphous olivine, Electron irradiation, Nanomechanical testing},
	pages = {120479},
	file = {ScienceDirect Full Text PDF:files/1702/Orekhov et al. - 2025 - Room temperature electron beam sensitive viscoplastic response of ultra-ductile amorphous olivine fi.pdf:application/pdf;ScienceDirect Snapshot:files/1701/S1359645424008280.html:text/html},
}

@article{meyer_direct_2008,
	title = {Direct {Imaging} of {Lattice} {Atoms} and {Topological} {Defects} in {Graphene} {Membranes}},
	volume = {8},
	issn = {1530-6984},
	url = {https://doi.org/10.1021/nl801386m},
	doi = {10.1021/nl801386m},
	abstract = {We present a transmission electron microscopy investigation of graphene membranes, crystalline foils with a thickness of only 1 atom. By using aberration-correction in combination with a monochromator, 1-Å resolution is achieved at an acceleration voltage of only 80 kV. The low voltage is crucial for the stability of these membranes. As a result, every individual carbon atom in the field of view is detected and resolved. We observe a highly crystalline lattice along with occasional point defects. The formation and annealing of Stone−Wales defects is observed in situ. Multiple five- and seven-membered rings appear exclusively in combinations that avoid dislocations and disclinations, in contrast to previous observations on highly curved (tube- or fullerene-like) graphene surfaces.},
	number = {11},
	urldate = {2026-06-30},
	journal = {Nano Letters},
	publisher = {American Chemical Society},
	author = {Meyer, Jannik C. and Kisielowski, C. and Erni, R. and Rossell, Marta D. and Crommie, M. F. and Zettl, A.},
	month = nov,
	year = {2008},
	pages = {3582--3586},
}

@misc{noauthor_electron-beam-induced_nodate,
	title = {Electron-beam-induced degradation of halide-perovskite-related semiconductor nanomaterials {\textbar} {Chinese} {Optics} {Letters} -- 中国光学期刊网},
	url = {https://www.opticsjournal.net/Articles/OJd6b4e697d237cba8/FullText?alichlgref=https
	urldate = {2026-06-30},
	file = {Electron-beam-induced degradation of halide-perovskite-related semiconductor nanomaterials | Chinese Optics Letters -- 中国光学期刊网:files/1705/FullText.html:text/html},
}

@article{wiktor_transmission_2017,
	title = {Transmission electron microscopy on metal–organic frameworks – a review},
	volume = {5},
	issn = {2050-7488},
	url = {https://doi.org/10.1039/c7ta00194k},
	doi = {10.1039/c7ta00194k},
	abstract = {Versatile materials like metal–organic frameworks require careful characterization. Transmission electron microscopy is a very powerful method that can address a multitude of investigative challenges. In this review we present TEM studies that yielded valuable insights into the investigated MOFs to illustrate the potential of TEM despite the sensitivity of MOFs to the electron beam.},
	number = {29},
	urldate = {2026-06-30},
	journal = {Journal of Materials Chemistry A},
	author = {Wiktor, Christian and Meledina, Maria and Turner, Stuart and Lebedev, Oleg I. and Fischer, Roland A.},
	month = aug,
	year = {2017},
	pages = {14969--14989},
	file = {Snapshot:files/1707/c7ta00194k.html:text/html},
}

@article{glaeser_limitations_1971,
	title = {Limitations to significant information in biological electron microscopy as a result of radiation damage},
	volume = {36},
	issn = {0022-5320},
	url = {https://www.sciencedirect.com/science/article/pii/S0022532071801181},
	doi = {10.1016/S0022-5320(71)80118-1},
	abstract = {Quantitative measurements of radiation damage in crystalline specimens of l-valine, adenosine, and catalase (uranyl acetate stained) have been made by observing the loss of the electron diffraction pattern. Reciprocity of specimen lifetime and current density at the specimen demonstrates the absence of any dose-rate effect, such as specimen heating, as a cause of specimen damage within the range 10−3 to 10−5 amperes/cm2 current density at the specimen. Specimen lifetimes at high voltages are about two and a half times greater than at conventional voltages, and it is shown that this is consistent with the dependence of linear energy loss upon accelerating voltage. The limiting resolution for meaningful observation is considered in terms of the statistics of observation at partticle fluxes that are specified from the specimen lifetime data. The best values are probably not better than 50 Å for l-valine, 20 Å for adenosine, and 15 Å for catalase.},
	number = {3},
	urldate = {2026-06-30},
	journal = {Journal of Ultrastructure Research},
	author = {Glaeser, Robert M.},
	month = aug,
	year = {1971},
	pages = {466--482},
	file = {ScienceDirect Snapshot:files/1709/S0022532071801181.html:text/html},
}

@article{y_unravelling_nodate,
	title = {Unravelling surface and interfacial structures of a metal-organic framework by transmission electron microscopy},
	url = {https://pubmed.ncbi.nlm.nih.gov/28218922/},
	abstract = {Metal-organic frameworks (MOFs) are crystalline porous materials with designable topology, porosity and functionality, having promising applications in gas storage and separation, ion conduction and catalysis. It is challenging to observe MOFs with transmission electron microscopy (TEM) due to the e …},
	language = {en},
	urldate = {2026-06-30},
	journal = {PubMed},
	author = {Y, Zhu and J, Ciston and B, Zheng and X, Miao and C, Czarnik and Y, Pan and R, Sougrat and Z, Lai and Ce, Hsiung and K, Yao and I, Pinnau and M, Pan and Y, Han},
	file = {Snapshot:files/1711/28218922.html:text/html},
}

@article{ekmanis_radiolysis_1984,
	title = {Radiolysis behaviour in alkali halide crystals},
	volume = {1},
	issn = {0168-583X},
	url = {https://www.sciencedirect.com/science/article/pii/0168583X84901113},
	doi = {10.1016/0168-583X(84)90111-3},
	abstract = {It is well established that coagulation of primary radiation defects produces metallic colloidal particles and atomic halogen. These are the final products of the radiolysis process in alkali halide crystals. The primary products of the radiolysis (F- and H-centres) further coagulate into larger clusters under different conditions of radiation and temperature. Radiolysis behaviour at low temperatures and low irradiation dose has recently been described using the Jain-Lidiard theory. While this theory explains certain features of radiolysis as applied to NaCl, the behaviour of this process in other alkali halides has been insufficiently studied. At high dose rates and elevated temperature in nonactivated crystals the radiolysis process can be most precisely described by a mechanism of multiparticle coagulation. This mechanism allows a good qualitative description of the experimental kinetics of colloidal growth to be given.},
	number = {2},
	urldate = {2026-06-30},
	journal = {Nuclear Instruments and Methods in Physics Research Section B: Beam Interactions with Materials and Atoms},
	author = {Ekmanis, Yu.},
	month = feb,
	year = {1984},
	pages = {473--474},
	file = {ScienceDirect Snapshot:files/1713/0168583X84901113.html:text/html},
}

@article{soppe_jain-lidiard_1992,
	title = {Jain-{Lidiard} model calculations of radiation damage in sodium chloride},
	volume = {65},
	issn = {0168-583X},
	url = {https://www.sciencedirect.com/science/article/pii/0168583X92950926},
	doi = {10.1016/0168-583X(92)95092-6},
	abstract = {In this paper numerical results on radiation damage of NaCl, obtained with an improved version of the Jain-Lidiard model are presented. The major radiation damage products are colloidal Na particles and Cl2 molecules, both embedded in the NaCl lattice. The geometry of the metallic Na particles is likely to be fractal. It is shown that the assumption of a fractal dimensionality leads to theoretical damage concentrations which are appreciably smaller than predicted by the original Jain-Lidiard model in which a nonfractal (Euclidean) dimension of the colloidal particles is assumed.},
	number = {1},
	urldate = {2026-06-30},
	journal = {Nuclear Instruments and Methods in Physics Research Section B: Beam Interactions with Materials and Atoms},
	author = {Soppe, W. J.},
	month = mar,
	year = {1992},
	pages = {493--496},
	file = {ScienceDirect Snapshot:files/1715/0168583X92950926.html:text/html},
}

@article{kavak_high-resolution_2025,
	title = {High-resolution electron microscopy imaging of {MOFs} at optimized electron dose},
	volume = {13},
	issn = {2050-7488},
	url = {https://doi.org/10.1039/d4ta06724j},
	doi = {10.1039/d4ta06724j},
	abstract = {Local and high-resolution structural investigation of metal–organic frameworks (MOFs) is essential for understanding the role of defects and incorporated elements. In this paper, we characterize the structure of metalated versions of (Hf)PCN-222(H2) and locate the position of the additional metal atoms. Transmission electron microscopy (TEM) is a powerful technique for this purpose, but MOFs are highly sensitive to the electron beam. To avoid structural alterations, it is therefore crucial to establish the maximum electron dose that can be applied. In this study, we apply a systematic workflow to measure the critical electron dose, enabling the identification of the optimal technique for extracting reliable information about the local structure of MOFs. We examined the electron beam stability of benchmarked (Zr)NU-1000, (Hf)PCN-222(H2) and its metalated versions, (Hf)PCN-222(Fe) and (Hf)PCN-222(Pd), and identified factors influencing the stability under the electron beam. After the threshold for electron dose was established, we applied low-dose, four-dimensional scanning transmission electron microscopy (4D-STEM). We then compared annular bright field (ABF), annular dark field (ADF), and real-time integrated center of mass (riCOM) images that could be extracted from the 4D dataset. The riCOM technique successfully revealed the structure of investigated MOFs with minimal beam-induced alterations and provides insights into local features, including organic linkers and additional metalation elements.},
	number = {6},
	urldate = {2026-06-30},
	journal = {Journal of Materials Chemistry A},
	author = {Kavak, Safiyye and Jannis, Daen and De Backer, Annick and Esteban, Daniel Arenas and Annys, Arno and Carrasco, Sergio and Ferrando-Ferrero, Javier and Guerrero, Raúl M. and Horcajada, Patricia and Verbeeck, Jo and Van Aert, Sandra and Bals, Sara},
	month = feb,
	year = {2025},
	pages = {4281--4291},
	file = {Snapshot:files/1717/d4ta06724j.html:text/html},
}

@article{hollenbach_beam_2026,
	title = {Beam teleportation for precision dose control in scanning transmission electron microscopy},
	volume = {284},
	issn = {0304-3991},
	url = {https://www.sciencedirect.com/science/article/pii/S030439912600063X},
	doi = {10.1016/j.ultramic.2026.114371},
	abstract = {Spatiotemporal dose control utilizing custom, vectorized scan trajectories in scanning transmission electron microscopy can be used to reduce image artifacts, mitigate beam-induced damage, minimize electrostatic charging, and enable the precise manipulation of sample structure. However, the inductance of the magnetic deflectors in the scan system induces an effective inertia of the beam spot, causing unintended exposure between target scan points (“beam dragging”) and reduced dwell time at the target scan points (“beam lagging”). Using a fast electrostatic beam blanker synchronized with the scan controller to blank during inter-pixel transients, it is possible to “teleport” the beam spot between scan points, reducing or eliminating the extra dose and image background associated with deflector settling. We demonstrate this method on randomized scan patterns which are strongly affected by beam spot inertia. We show that background from the beam lagging can be discarded to simultaneously decrease dose and noise. As a result, this work opens up new possibilities for dose-rate-controlled imaging and structural processing via high-fidelity vector scans.},
	urldate = {2026-06-30},
	journal = {Ultramicroscopy},
	author = {Hollenbach, Jonathan D. and Koppell, Stewart A. and Smalley, Darian and Francis, Carter and Levin, Barnaby D. A. and Macy, Juan and Wang, Chih-Feng and Reed, Bryan W. and Taheri, Mitra L.},
	month = jul,
	year = {2026},
	keywords = {Scanning transmission electron microscopy, Beam teleportation, Dose control, Electrostatic blanking, Low dose imaging},
	pages = {114371},
	file = {ScienceDirect Snapshot:files/1719/S030439912600063X.html:text/html},
}

@article{velazco_evaluation_2020,
	title = {Evaluation of different rectangular scan strategies for {STEM} imaging},
	volume = {215},
	issn = {0304-3991},
	url = {https://www.sciencedirect.com/science/article/pii/S0304399120300334},
	doi = {10.1016/j.ultramic.2020.113021},
	abstract = {STEM imaging is typically performed by raster scanning a focused electron probe over a sample. Here we investigate and compare three different scan patterns, making use of a programmable scan engine that allows to arbitrarily set the sequence of probe positions that are consecutively visited on the sample. We compare the typical raster scan with a so-called ‘snake’ pattern where the scan direction is reversed after each row and a novel Hilbert scan pattern that changes scan direction rapidly and provides an homogeneous treatment of both scan directions. We experimentally evaluate the imaging performance on a single crystal test sample by varying dwell time and evaluating behaviour with respect to sample drift. We demonstrate the ability of the Hilbert scan pattern to more faithfully represent the high frequency content of the image in the presence of sample drift. It is also shown that Hilbert scanning provides reduced bias when measuring lattice parameters from the obtained scanned images while maintaining similar precision in both scan directions which is especially important when e.g. performing strain analysis. Compared to raster scanning with flyback correction, both snake and Hilbert scanning benefit from dose reduction as only small probe movement steps occur.},
	urldate = {2026-06-30},
	journal = {Ultramicroscopy},
	author = {Velazco, A. and Nord, M. and Béché, A. and Verbeeck, J.},
	month = aug,
	year = {2020},
	keywords = {Aberration corrected STEM, Drift, Hilbert scan pattern, Precision, Scanning distortions},
	pages = {113021},
	file = {ScienceDirect Snapshot:files/1721/S0304399120300334.html:text/html;Submitted Version:files/1722/Velazco et al. - 2020 - Evaluation of different rectangular scan strategies for STEM imaging.pdf:application/pdf},
}

@article{ede_partial_2020,
	title = {Partial {Scanning} {Transmission} {Electron} {Microscopy} with {Deep} {Learning}},
	volume = {10},
	issn = {2045-2322},
	url = {https://pmc.ncbi.nlm.nih.gov/articles/PMC7239858/},
	doi = {10.1038/s41598-020-65261-0},
	abstract = {Compressed sensing algorithms are used to decrease electron microscope scan time and electron beam exposure with minimal information loss. Following successful applications of deep learning to compressed sensing, we have developed a two-stage multiscale generative adversarial neural network to complete realistic 512 × 512 scanning transmission electron micrographs from spiral, jittered gridlike, and other partial scans. For spiral scans and mean squared error based pre-training, this enables electron beam coverage to be decreased by 17.9× with a 3.8\% test set root mean squared intensity error, and by 87.0× with a 6.2\% error. Our generator networks are trained on partial scans created from a new dataset of 16227 scanning transmission electron micrographs. High performance is achieved with adaptive learning rate clipping of loss spikes and an auxiliary trainer network. Our source code, new dataset, and pre-trained models are publicly available.},
	urldate = {2026-06-30},
	journal = {Scientific Reports},
	author = {Ede, Jeffrey M. and Beanland, Richard},
	month = may,
	year = {2020},
	pages = {8332},
	file = {Full Text:files/1725/Ede and Beanland - 2020 - Partial Scanning Transmission Electron Microscopy with Deep Learning.pdf:application/pdf},
}

@misc{noauthor_interlacing_nodate,
	title = {Interlacing in {Atomic} {Resolution} {Scanning} {Transmission} {Electron} {Microscopy} {\textbar} {Microscopy} and {Microanalysis} {\textbar} {Oxford} {Academic}},
	url = {https://academic.oup.com/mam/article/29/4/1373/7191704},
	urldate = {2026-06-30},
	file = {Interlacing in Atomic Resolution Scanning Transmission Electron Microscopy | Microscopy and Microanalysis | Oxford Academic:files/1727/7191704.html:text/html},
}

@article{pennycook_high-resolution_1991,
	title = {High-resolution \textit{{Z}}-contrast imaging of crystals},
	volume = {37},
	issn = {0304-3991},
	url = {https://www.sciencedirect.com/science/article/pii/030439919190004P},
	doi = {10.1016/0304-3991(91)90004-P},
	abstract = {The use of a high-angle annular detector in a scanning transmission electron microscope is shown to provide incoherent images of crystalline materials with strong compositional sensitivity. How this occurs, even in the presence of strong dynamical diffraction of the low-angle beams, becomes very clear in a Bloch wave description of the imaging, which shows that only tightly bound s-type Bloch states contribute significantly to the image. Interference effects are therefore precluded and the image can be described as a convolution. There are no contrast reversals with thickness or defocus and no Fresnel fringe effects at interfaces. Each atomic column contributes to the image independently of its neighbors until the s-states themselves overlap. With an optimum imaging probe the nature of the convolution can be visualized intuitively to a scale well below the resolution limit. To first order, therefore, each object has only one possible image, and since the same probe is used for all objects, an unknown structure can be interpreted directly. These ideas will be illustrated with images from semiconductors, superconductors, and alloys.},
	number = {1},
	urldate = {2026-07-01},
	journal = {Ultramicroscopy},
	author = {Pennycook, S. J. and Jesson, D. E.},
	month = aug,
	year = {1991},
	pages = {14--38},
	file = {ScienceDirect Snapshot:files/1729/030439919190004P.html:text/html},
}

@article{zhou_defect_2021,
	title = {Defect activity in metal halide perovskites with wide and narrow bandgap},
	volume = {6},
	copyright = {2021 Springer Nature Limited},
	issn = {2058-8437},
	url = {https://www.nature.com/articles/s41578-021-00331-x},
	doi = {10.1038/s41578-021-00331-x},
	abstract = {Metal halide perovskites (MHPs) constitute a rich library of materials with huge potential for disruptive optoelectronic technologies. Their main strength comes from the possibility of easily tuning their bandgap to integrate them in devices with different functionalities — in principle. In reality, this cannot be achieved yet. In fact, whereas defect tolerance can be claimed for MHPs with a bandgap of about 1.6 eV, the model system that is the object of intense investigations, MHPs with lower and higher bandgaps are far from being defect-tolerant. These materials show various forms of instabilities that are mainly driven by strong defect activity. Here we critically assess the most recent advances in elucidating the physical and chemical activity of defects in both high-bandgap and low-bandgap MHPs, while correlating it to performance and stability losses, especially for solar cells, the driving application for these materials. We also provide an overview of the strategies so far implemented to eventually overcome the remaining materials-based and device-based challenges.},
	language = {en},
	number = {11},
	urldate = {2026-08-06},
	journal = {Nature Reviews Materials},
	publisher = {Nature Publishing Group},
	author = {Zhou, Yang and Poli, Isabella and Meggiolaro, Daniele and De Angelis, Filippo and Petrozza, Annamaria},
	month = nov,
	year = {2021},
	keywords = {Solar cells, Electronic devices},
	pages = {986--1002},
}

@article{skvortsova_arm-length-controlled_2026,
	title = {Arm-{Length}-{Controlled} {CsPbBr3} {Nanocrystals} for {Tunable} {Optical} and {Assembly} {Behavior}},
	volume = {38},
	issn = {1521-4095},
	url = {https://onlinelibrary.wiley.com/doi/abs/10.1002/adma.202519211},
	doi = {10.1002/adma.202519211},
	abstract = {Colloidal cesium lead bromide (CsPbBr3) nanocrystals (NCs) are excellent candidates for various photonic and optoelectronic applications due to their bright and stable green emission. Here, we establish arm length control as a central structural parameter that governs both the optical properties of individual CsPbBr3 NCs and their self-assembly behavior. Armed NCs, featuring a cubic core with multiple protruding arms, are synthesized by controlling seed size, concentration, and injection temperature, while arm length is tuned via cesium oleate concentration. Prolonged storage in toluene is shown to lead to a time-dependent morphological evolution from armed NCs to 26-faceted rhombicuboctahedra, with short-armed structures as intermediates. NCs with a longer arm length yield enhanced radiative efficiency, extended photoluminescence (PL) lifetimes, and suppressed blinking, making such NCs suitable for light-emitting devices and quantum photonic applications. In contrast, short-armed NCs exhibit faster recombination, stronger PL intermittency, and increased surface accessibility, which are favorable for sensing and high-speed single-photon emission. The arm length also governs self-assembly behavior, hereby opening new possibilities for applications. Armed NCs form densely 3D-packed assemblies with tunable configurations. This work demonstrates how arm length tuning expands the functional potential of CsPbBr3 NCs by linking morphological control to both optical response and self-assembly characteristics.},
	language = {en},
	number = {23},
	urldate = {2026-08-06},
	journal = {Advanced Materials},
	author = {Skvortsova, Irina and Seth, Sudipta and Zito, Juliette and Girod, Robin and Van Hout, Bob and De Backer, Annick and Stoops, Tom and Abakumov, Sergey and Vlasov, Evgenii and Behera, Tejmani and Van Aert, Sandra and Debroye, Elke and Hofkens, Johan and Bals, Sara},
	year = {2026},
	keywords = {assemblies, blinking, electron tomography, halide perovskites},
	pages = {e19211},
	file = {Full Text PDF:files/1745/Skvortsova et al. - 2026 - Arm-Length-Controlled CsPbBr3 Nanocrystals for Tunable Optical and Assembly Behavior.pdf:application/pdf;Snapshot:files/1744/adma.html:text/html},
}

@article{martani_defect_2023,
	title = {Defect {Engineering} to {Achieve} {Photostable} {Wide} {Bandgap} {Metal} {Halide} {Perovskites}},
	volume = {8},
	issn = {2380-8195},
	url = {https://doi.org/10.1021/acsenergylett.3c00610},
	doi = {10.1021/acsenergylett.3c00610},
	abstract = {Bandgap tuning is
a crucial characteristic of metal-halide perovskites,
with benchmark lead-iodide compounds having a bandgap of 1.6 eV. To increase the bandgap up to 2.0 eV, a straightforward strategy is to partially substitute iodide with bromide in so-called mixed-halide lead perovskites. Such compounds are prone, however, to light-induced halide segregation resulting in bandgap instability, which limits their application in tandem solar cells and a variety of optoelectronic devices. Crystallinity improvement and surface passivation strategies can effectively slow down, but not completely stop, such light-induced instability. Here we identify the defects and the intragap electronic states that trigger the material transformation and bandgap shift. Based on such knowledge, we engineer the perovskite band edge energetics by replacing lead with tin and radically deactivate the photoactivity of such defects. This leads to metal halide perovskites with a photostable bandgap over a wide spectral range and associated solar cells with photostable open circuit voltages.},
	number = {6},
	urldate = {2026-08-06},
	journal = {ACS Energy Letters},
	author = {Martani, Samuele and Zhou, Yang and Poli, Isabella and Aktas, Ece and Meggiolaro, Daniele and Jiménez-López, Jesús and Wong, E Laine and Gregori, Luca and Prato, Mirko and Di Girolamo, Diego and Abate, Antonio and De Angelis, Filippo and Petrozza, Annamaria},
	month = may,
	year = {2023},
	pages = {2801--2808},
	file = {Full Text PDF:files/1747/Martani et al. - 2023 - Defect Engineering to Achieve Photostable Wide Bandgap Metal Halide Perovskites.pdf:application/pdf;Snapshot:files/1748/acsenergylett.html:text/html},
}

@article{cahen_are_2021,
	title = {Are {Defects} in {Lead}-{Halide} {Perovskites} {Healed}, {Tolerated}, or {Both}?},
	volume = {6},
	issn = {2380-8195},
	url = {https://doi.org/10.1021/acsenergylett.1c02027},
	doi = {10.1021/acsenergylett.1c02027},
	number = {11},
	urldate = {2026-08-06},
	journal = {ACS Energy Letters},
	author = {Cahen, David and Kronik, Leeor and Hodes, Gary},
	month = oct,
	year = {2021},
	pages = {4108--4114},
	file = {Full Text PDF:files/1750/Cahen et al. - 2021 - Are Defects in Lead-Halide Perovskites Healed, Tolerated, or Both.pdf:application/pdf;Snapshot:files/1751/acsenergylett.html:text/html},
}

@article{mastrangelo_structural_2026,
	title = {Structural biology of ferritin nanocages},
	volume = {600},
	copyright = {© 2026 The Author(s). FEBS Letters published by John Wiley \& Sons Ltd on behalf of Federation of European Biochemical Societies.},
	issn = {1873-3468},
	url = {https://onlinelibrary.wiley.com/doi/abs/10.1002/1873-3468.70302},
	doi = {10.1002/1873-3468.70302},
	abstract = {Ferritin is a ubiquitous and evolutionarily conserved iron-storage protein that plays a fundamental role in cellular iron homeostasis. By catalyzing the oxidation of ferrous iron and sequestering it as a ferric mineral within a protein nanocage, ferritin prevents toxic accumulation of labile iron and reactive oxygen species that damage proteins, lipids, and DNA. In humans, ferritin assembles into a 24-subunit nearly spherical shell enclosing a central cavity that safely stores thousands of iron atoms. This organized architecture enables ferritin to act as both an efficient iron detoxification system and a dynamic intracellular iron reservoir. Recent advances in cryo-electron microscopy (cryo-EM) have transformed ferritin research by revealing its structural organization, molecular interactions, and functional states at high resolution. Additionally, beyond protein–protein interactions, cryo-EM now enables direct visualization of ferritin-mediated biomineralization, allowing in situ observation of iron nucleation, mineral growth, and core organization within intact nanocages. Together, these advances establish cryo-EM as a transformative tool for elucidating ferritin structure, dynamics, and function – reshaping our understanding of iron metabolism and guiding the rational design of ferritin-based nanomaterials for biomedical applications.},
	language = {en},
	number = {10},
	urldate = {2026-08-06},
	journal = {FEBS Letters},
	author = {Mastrangelo, Eloise and Di Pisa, Flavio},
	year = {2026},
	keywords = {bioinorganic interfaces, cryo-EM, ferritin, ferritinophagy, iron homeostasis, iron storage, transferrin receptor},
	pages = {1420--1443},
	file = {Full Text PDF:files/1753/Mastrangelo and Di Pisa - 2026 - Structural biology of ferritin nanocages.pdf:application/pdf;Snapshot:files/1754/1873-3468.html:text/html},
}

@article{furukawa_chemistry_2013,
	title = {The {Chemistry} and {Applications} of {Metal}-{Organic} {Frameworks}},
	volume = {341},
	url = {https://www.science.org/doi/10.1126/science.1230444},
	doi = {10.1126/science.1230444},
	abstract = {Crystalline metal-organic frameworks (MOFs) are formed by reticular synthesis, which creates strong bonds between inorganic and organic units. Careful selection of MOF constituents can yield crystals of ultrahigh porosity and high thermal and chemical stability. These characteristics allow the interior of MOFs to be chemically altered for use in gas separation, gas storage, and catalysis, among other applications. The precision commonly exercised in their chemical modification and the ability to expand their metrics without changing the underlying topology have not been achieved with other solids. MOFs whose chemical composition and shape of building units can be multiply varied within a particular structure already exist and may lead to materials that offer a synergistic combination of properties.},
	number = {6149},
	urldate = {2026-08-06},
	journal = {Science},
	publisher = {American Association for the Advancement of Science},
	author = {Furukawa, Hiroyasu and Cordova, Kyle E. and O’Keeffe, Michael and Yaghi, Omar M.},
	month = aug,
	year = {2013},
	pages = {1230444},
	file = {Full Text PDF:files/1756/Furukawa et al. - 2013 - The Chemistry and Applications of Metal-Organic Frameworks.pdf:application/pdf},
}

@article{mastrangelo_structural_2026-1,
	title = {Structural biology of ferritin nanocages},
	volume = {600},
	copyright = {© 2026 The Author(s). FEBS Letters published by John Wiley \& Sons Ltd on behalf of Federation of European Biochemical Societies.},
	issn = {1873-3468},
	url = {https://onlinelibrary.wiley.com/doi/abs/10.1002/1873-3468.70302},
	doi = {10.1002/1873-3468.70302},
	abstract = {Ferritin is a ubiquitous and evolutionarily conserved iron-storage protein that plays a fundamental role in cellular iron homeostasis. By catalyzing the oxidation of ferrous iron and sequestering it as a ferric mineral within a protein nanocage, ferritin prevents toxic accumulation of labile iron and reactive oxygen species that damage proteins, lipids, and DNA. In humans, ferritin assembles into a 24-subunit nearly spherical shell enclosing a central cavity that safely stores thousands of iron atoms. This organized architecture enables ferritin to act as both an efficient iron detoxification system and a dynamic intracellular iron reservoir. Recent advances in cryo-electron microscopy (cryo-EM) have transformed ferritin research by revealing its structural organization, molecular interactions, and functional states at high resolution. Additionally, beyond protein–protein interactions, cryo-EM now enables direct visualization of ferritin-mediated biomineralization, allowing in situ observation of iron nucleation, mineral growth, and core organization within intact nanocages. Together, these advances establish cryo-EM as a transformative tool for elucidating ferritin structure, dynamics, and function – reshaping our understanding of iron metabolism and guiding the rational design of ferritin-based nanomaterials for biomedical applications.},
	language = {en},
	number = {10},
	urldate = {2026-08-06},
	journal = {FEBS Letters},
	author = {Mastrangelo, Eloise and Di Pisa, Flavio},
	year = {2026},
	keywords = {bioinorganic interfaces, cryo-EM, ferritin, ferritinophagy, iron homeostasis, iron storage, transferrin receptor},
	pages = {1420--1443},
	file = {Full Text PDF:files/1758/Mastrangelo and Di Pisa - 2026 - Structural biology of ferritin nanocages.pdf:application/pdf;Snapshot:files/1759/1873-3468.html:text/html},
}

@article{cho_crystal_2009,
	title = {The {Crystal} {Structure} of {Ferritin} from \textit{{Helicobacter} pylori} {Reveals} {Unusual} {Conformational} {Changes} for {Iron} {Uptake}},
	volume = {390},
	issn = {0022-2836},
	url = {https://www.sciencedirect.com/science/article/pii/S0022283609005452},
	doi = {10.1016/j.jmb.2009.04.078},
	abstract = {The crystal structure of recombinant ferritin from Helicobacter pylori has been determined in its apo, low-iron-bound, intermediate, and high-iron-bound states. Similar to other members of the ferritin family, the bacterial ferritin assembles as a spherical protein shell of 24 subunits, each of which folds into a four-α-helix bundle. Significant conformational changes were observed at the BC loop and the entrance of the 4-fold symmetry channel in the intermediate and high-iron-bound states, whereas no change was found in the apo and low-iron-bound states. The imidazole rings of His149 at the channel entrance undergo conformational changes that bear resemblance to heme configuration and are directly coupled to axial translocation of Fe ions through the 4-fold channel. Our results provide the first structural evidence of the translocation of Fe ions through the 4-fold channel in prokaryotes and the transition from a protein-dominated process to a mineral-surface-dominated process during biomineralization.},
	number = {1},
	urldate = {2026-08-06},
	journal = {Journal of Molecular Biology},
	author = {Cho, Ki Joon and Shin, Hye Jeong and Lee, Ji-Hye and Kim, Kyung-Jin and Park, Sarah S. and Lee, Youngmi and Lee, Cheolju and Park, Sung Soo and Kim, Kyung Hyun},
	month = jul,
	year = {2009},
	keywords = {ferritin, 4-fold channel, biomineralization, iron uptake mechanism},
	pages = {83--98},
	file = {ScienceDirect Full Text PDF:files/1762/Cho et al. - 2009 - The Crystal Structure of Ferritin from Helicobacter pylori Reveals Unusual Conformational Cha.pdf:application/pdf;ScienceDirect Snapshot:files/1761/S0022283609005452.html:text/html},
}

@article{harrison_ferritins_1996,
	title = {The ferritins: molecular properties, iron storage function and cellular regulation},
	volume = {1275},
	issn = {0005-2728},
	shorttitle = {The ferritins},
	url = {https://www.sciencedirect.com/science/article/pii/0005272896000229},
	doi = {10.1016/0005-2728(96)00022-9},
	abstract = {The iron storage protein, ferritin, plays a key role in iron metabolism. Its ability to sequester the element gives ferritin the dual functions of iron detoxification and iron reserve. The importance of these functions is emphasised by ferritin's ubiquitous distribution among living species. Ferritin's three-dimensional structure is highly conserved. All ferritins have 24 protein subunits arranged in 432 symmetry to give a hollow shell with an 80 Å diameter cavity capable of storing up to 4500 Fe(III) atoms as an inorganic complex. Subunits are folded as 4-helix bundles each having a fifth short helix at roughly 60° to the bundle axis. Structural features of ferritins from humans, horse, bullfrog and bacteria are described: all have essentially the same architecture in spite of large variations in primary structure (amino acid sequence identities can be as low as 14\%) and the presence in some bacterial ferritins of haem groups. Ferritin molecules isolated from vertebrates are composed of two types of subunit (H and L), whereas those from plants and bacteria contain only H-type chains, where ‘H-type’ is associated with the presence of centres catalysing the oxidation of two Fe(II) atoms. The similarity between the dinuclear iron centres of ferritin H-chains and those of ribonucleotide reductase and other proteins suggests a possible wider evolutionary linkage. A great deal of research effort is now concentrated on two aspects of fenitin: its functional mechanisms and its regulation. These form the major part of the review. Steps in iron storage within ferritin molecules consist of Fe(II) oxidation, FE(III) migration and the nucleation and growth of the iron core mineral. H-chains are important for Fe(II) oxidation and L-chains assist in core formation. Iron mobilisation, relevant to ferritin's role as iron reserve, is also discussed. Translational regulation of mammalian ferritin synthesis in response to iron and the apparent links between iron and citrate metabolism through a single molecule with dual function are described. The molecule, when binding a [4Fe-4S] cluster, is a functioning (cytoplasmic) aconitase. When cellular iron is low, loss of the [4Fe-4S] cluster allows the molecule to bind to the 5′-untranslated region (5′-UTR) of the ferritin m-RNA and thus to repress translation. In this form it is known as the iron regulatory protein (IRP) and the stem-loop RNA structure to which it binds is the iron regulatory element (IRE). IREs are found in the 3′-UTR of the transferrin receptor and in the 5′-UTR of erythroid aminolaevulinic acid synthase, enabling tight co-ordination between cellular iron uptake and the synthesis of ferritin and haem. Degradation of ferritin could potentially lead to an increase in toxicity due to uncontrolled release of iron. Degradation within membrane-encapsulated ‘secondary lysosomes’ may avoid this problem and this seems to be the origin of another form of storage iron known as haemosiderin. However, in certain pathological states, massive deposits of ‘haemosiderin’ are found which do not arise directly from ferritin breakdown. Understanding the numerous inter-relationships between the various intracellular iron complexes presents a major challenge.},
	number = {3},
	urldate = {2026-08-06},
	journal = {Biochimica et Biophysica Acta (BBA) - Bioenergetics},
	author = {Harrison, Pauline M. and Arosio, Paolo},
	month = jul,
	year = {1996},
	keywords = {Dinuclear iron, Ferritin, Ferroxidase activity, Haemosiderin, Iron regulatory element (IRE), Iron regulatory protein (IRP), Iron storage},
	pages = {161--203},
	file = {ScienceDirect Full Text PDF:files/1765/Harrison and Arosio - 1996 - The ferritins molecular properties, iron storage function and cellular regulation.pdf:application/pdf;ScienceDirect Snapshot:files/1764/0005272896000229.html:text/html},
}

@article{lalandec_robert_guided_2026,
	title = {Guided progressive reconstructive imaging: {A} new quantization-based framework for low-dose, high-throughput and real-time analytical ptychography},
	volume = {285},
	issn = {0304-3991},
	shorttitle = {Guided progressive reconstructive imaging},
	url = {https://www.sciencedirect.com/science/article/pii/S0304399126001038},
	doi = {10.1016/j.ultramic.2026.114411},
	abstract = {By profiting from recent developments in detector technologies, making it possible to access a stream of detection events with few-ns time resolutions, a new ptychographic workflow is established. This methodological framework, referred to as guided progressive reconstructive imaging, relies on a quantization-based description of the acquired intensity, through an elementary derivation. Established direct phase retrieval solutions, such as the Wigner distribution deconvolution approach, can then be adapted to a continuous treatment of received counts, with no need for a dense data representation. Consequently, the result is obtained in the form of a progressively improving estimate, while providing immediate user feedback thanks to a processing speed high enough to surpass the acquisition bandwidth. This fast measurement is enabled by the cumulative usage of a pre-calculated library of kernel-limited functions, accumulating count-wise contributions as a function of the triggered detector pixel. Hence, the reconstruction offers the same advantages of direct phase retrieval methods, in particular a high dose-efficiency and the absence of complex convergence dynamics, with much less stringent restrictions on the field of view than is typical in current alternatives. Its implementation is also significantly more straightforward and flexible. Overall, this work constitutes a major evolution in the state-of-the-art, facilitating repeatable and low-dose experiments with high accessibility, and being applicable to electron-based imaging, X-ray diffraction and optical microscopy.},
	urldate = {2026-08-06},
	journal = {Ultramicroscopy},
	author = {Lalandec Robert, Hoelen L. and Annys, Arno and Chennit, Tamazouzt and Verbeeck, Jo},
	month = aug,
	year = {2026},
	keywords = {Ptychography, Event-driven detection, Low-dose imaging, Wigner distribution deconvolution},
	pages = {114411},
	file = {ScienceDirect Snapshot:files/1767/S0304399126001038.html:text/html},
}

@article{chennit_investigating_2025-1,
	title = {Investigating the convergence properties of iterative ptychography for atomic-resolution low-dose imaging},
	volume = {278},
	issn = {0304-3991},
	url = {https://www.sciencedirect.com/science/article/pii/S0304399125001433},
	doi = {10.1016/j.ultramic.2025.114245},
	abstract = {This study investigates the convergence properties of a collection of iterative electron ptychography methods, under low electron doses ({\textless}103 e−/Å2) and gives particular attention to the impact of the user-defined update strengths. We demonstrate that carefully chosen values for this parameter, ideally smaller than those conventionally met in the literature, are essential for achieving accurate reconstructions of the projected electrostatic potential. Using a 4D dataset of a thin hybrid organic–inorganic formamidinium lead bromide (FAPbBr3) sample, we show that convergence is in practice achievable only when the update strengths for both the object and probe are relatively small compared to what is found in literature. Additionally we demonstrate that under low electron doses, the reconstructions initial error increases when the update strength coefficients are reduced below a certain threshold emphasizing the existence of critical values beyond which the algorithms are trapped in local minima. These findings highlight the need for carefully optimized reconstruction parameters in iterative ptychography, especially when working with low electron doses, ensuring both effective convergence and correctness of the result.},
	urldate = {2026-08-06},
	journal = {Ultramicroscopy},
	author = {Chennit, Tamazouzt and Li, Songge and Robert, Hoelen L. Lalandec and Hofer, Christoph and Schrenker, Nadine J. and Manna, Liberato and Bals, Sara and Pennycook, Timothy J. and Verbeeck, Jo},
	month = dec,
	year = {2025},
	keywords = {Electron ptychography, 4D-STEM, Low-dose imaging, Beam-sensitive materials, Iterative ptychography methods},
	pages = {114245},
	file = {ScienceDirect Full Text PDF:files/1770/Chennit et al. - 2025 - Investigating the convergence properties of iterative ptychography for atomic-resolution low-dose im.pdf:application/pdf;ScienceDirect Snapshot:files/1769/S0304399125001433.html:text/html},
}

@article{egerton_damage-limited_2026,
	title = {Damage-limited resolution for {X}-ray and electron microscopy of organic specimens},
	volume = {82},
	issn = {2059-7983},
	url = {https://journals.iucr.org/paper?S2059798326002445},
	doi = {10.1107/S2059798326002445},
	abstract = {Analytical expressions for the damage-limited resolution (DLR) are developed and applied to X-ray and electron imaging of beam-sensitive specimens, allowing for variation of the characteristic radiation dose with spatial resolution. The dependence of DLR on specimen thickness is illustrated for the common modes of X-ray and transmission electron-microscope imaging. Similarities and differences between the radiolysis damage caused by electrons and X-rays are discussed. The meaning of a `Bragg boost' in diffracted intensity is discussed.},
	number = {5},
	urldate = {2026-09-08},
	journal = {Acta Crystallographica Section D Structural Biology},
	author = {Egerton, Ray F. and Nave, Colin},
	month = may,
	year = {2026},
	pages = {471--483},
	file = {Full Text PDF:files/1772/Egerton and Nave - 2026 - Damage-limited resolution for X-ray and electron microscopy of organic specimens.pdf:application/pdf},
}

@article{robert_dynamical_2022-1,
	title = {Dynamical diffraction of high-energy electrons investigated by focal series momentum-resolved scanning transmission electron microscopy at atomic resolution},
	volume = {233},
	issn = {03043991},
	url = {https://linkinghub.elsevier.com/retrieve/pii/S030439912100200X},
	doi = {10.1016/j.ultramic.2021.113425},
	language = {en},
	urldate = {2026-09-08},
	journal = {Ultramicroscopy},
	author = {Robert, H.L. and Lobato, I. and Lyu, F.J. and Chen, Q. and Van Aert, S. and Van Dyck, D. and Müller-Caspary, K.},
	month = mar,
	year = {2022},
	pages = {113425},
}

@article{moecking_cryo-em_2025,
	title = {Cryo-{EM} single-particle analysis expanding towards increasingly native samples},
	volume = {81},
	issn = {2059-7983},
	url = {https://journals.iucr.org/paper?S2059798325008332},
	doi = {10.1107/S2059798325008332},
	abstract = {The explosion of cryo-electron microscopy (cryo-EM) over the last decade has brought with it a range of new approaches for gaining high-resolution structural information on previously inaccessible biological systems. Cryo-EM single-particle analysis (SPA) approaches typically entail overexpression and purification of the target protein. Larger and more complex molecular assemblies often require extensive optimization of the expression, purification and reconstitution procedures. Additionally, prior knowledge of the composition of the structure of interest is required. Approaches employing cryo-focused ion beam (FIB) milling and cryo-electron tomography (cryo-ET) have proven incredibly useful for exploring protein structures within cells while maintaining near-native conditions. Such strategies avoid purification of the target protein or protein complex, yet are often still limited in throughput and achievable resolution. Here, we highlight recent studies demonstrating that the range of samples suitable for SPA is expanding towards increasingly more native samples. We specifically focus on studies investigating complex macromolecular assemblies where tailored sample-preparation strategies made them amenable for SPA, while still keeping them in close-to-native conditions. These examples show that SPA has become a discovery tool for
              de novo
              protein identification and complex stoichiometry in more complex and thicker samples.},
	number = {11},
	urldate = {2026-09-10},
	journal = {Acta Crystallographica Section D Structural Biology},
	author = {Moecking, Jonas and Zeev-Ben-Mordehai, Tzviya},
	month = nov,
	year = {2025},
	pages = {587--597},
	file = {Full Text PDF:files/1775/Moecking and Zeev-Ben-Mordehai - 2025 - Cryo-EM single-particle analysis expanding towards increasingly native samples.pdf:application/pdf},
}

@article{y_primer_2015,
	title = {A primer to single-particle cryo-electron microscopy},
	url = {https://pubmed.ncbi.nlm.nih.gov/25910204/},
	abstract = {Cryo-electron microscopy (cryo-EM) of single-particle specimens is used to determine the structure of proteins and macromolecular complexes without the need for crystals. Recent advances in detector technology and software algorithms now allow images of unprecedented quality to be recorded and struc …},
	language = {en},
	urldate = {2026-09-10},
	journal = {PubMed},
	author = {Y, Cheng and N, Grigorieff and Pa, Penczek and T, Walz},
	year = {2015},
	file = {Snapshot:files/1777/25910204.html:text/html},
}

@article{seinen_radiation_1994,
	title = {Radiation damage in {NaCl}. {II}. {The} early stage of {F}-center aggregation},
	volume = {50},
	url = {https://link.aps.org/doi/10.1103/PhysRevB.50.9787},
	doi = {10.1103/PhysRevB.50.9787},
	abstract = {We study the early stage of aggregation of F centers into colloids in pure NaCl under irradiation. The crystals have been electron irradiated with a dose rate of 2 Mrad/h up to doses of 1500 Mrad and measured by optical-absorption spectroscopy. The major bands, the F, M, and the colloid band, are analyzed qualitatively as well as quantitatively. We have observed a relationship between the concentrations of the F and the M centers, which changes from quadratic to linear. The colloid band appears to peak at two distinct wavelengths, indicating that two types of colloids are formed during the nucleation stage. The defect concentrations are determined as a function of the dose and the irradiation temperature and are discussed in terms of models which describe the kinetics of defect formation.},
	number = {14},
	urldate = {2026-09-10},
	journal = {Physical Review B},
	publisher = {American Physical Society},
	author = {Seinen, J. and Groote, J. C. and Weerkamp, J. R. W. and den Hartog, H. W.},
	month = oct,
	year = {1994},
	pages = {9787--9792},
	file = {APS Snapshot:files/1779/PhysRevB.50.html:text/html},
}

@article{izumi_effects_1969,
	title = {The {Effects} of {Irradiation} of 100 {keV} {Electrons} on {Alkali} {Halide} {Crystals}; {The} {Point} {Defect} {Coagulates} in {NaCl} and {KCl} {Irradiated} at {Room} {Temperature}},
	volume = {26},
	issn = {0031-9015},
	url = {https://journals.jps.jp/doi/10.1143/JPSJ.26.1451},
	doi = {10.1143/JPSJ.26.1451},
	abstract = {Thin films of KCl and NaCl single crystals prepared by chemical polishing were examined by an electron microscope. During the observation, the thin films were damaged by electron beam used for the image formation. As a result of electron irradiation at room temperature, plate shaped defects parallel to one of the \{100\} planes and elongated in one of the {\textless}100{\textgreater} directions were formed in both kinds of crystals. The maximum dimension of them was about 1 µ in long axis 1000∼3000 Å in short axis and ∼100 Å in thickness. It is supposed that these plates correspond to the small alkali metal colloids which were considered to be the origin of X-band absorption in the optical measurement, because the behaviour of the plates during heat treatment between room temperature and 300°C showed good agreement with that of the colloids.},
	number = {6},
	urldate = {2026-09-10},
	journal = {Journal of the Physical Society of Japan},
	publisher = {The Physical Society of Japan},
	author = {Izumi, Kunihidi},
	month = jun,
	year = {1969},
	pages = {1451--1461},
}

\end{document}